\documentclass[%
 reprint,
 floatfix,
superscriptaddress,
 amsmath,amssymb,
 aps,
]{revtex4-2}
\usepackage{graphicx}
\usepackage{bm}
\usepackage{hyperref}
\usepackage{subcaption}
\usepackage[margin=1.in]{geometry}
\usepackage{pifont}
\newcommand{\xmark}{\ding{53}}%
\usepackage[capitalise]{cleveref}
\usepackage{floatrow}
\usepackage{xcolor}
\usepackage[nohyperlinks,printonlyused,nolist]{acronym}
\usepackage{rotating}

\usepackage{isotope}
\usepackage{paralist}
\usepackage{soul}
\def\dashedrule#1#2#3{{%
\dimen1=#2 \divide\dimen1 by 2
\def\@ruledash{%
\rule{\dimen1}{0pt}%
\rule[0.5ex]{#1}{0.4pt}%
\rule{\dimen1}{0pt}}%
\count1=0
\loop%
\ifnum\count1<#3%
\advance\count1 by 1%
\@ruledash%
\repeat}}
\def\hdashline{\dashedrule{.5em}{.5em}{3}} 
\begin{document}

\preprint{APS/123-QED}

\title{Neutrino-nucleus CC0\ensuremath{\pi} cross-section tuning in GENIE v3}

\author{J\'{u}lia Tena-Vidal}
\email{publications@genie-mc.org}
\affiliation{University of Liverpool, Dept. of Physics, Liverpool L69 7ZE, UK}
\altaffiliation[Now at ]{Tel Aviv University}
\author{Costas Andreopoulos}
\affiliation{University of Liverpool, Dept. of Physics, Liverpool L69 7ZE, UK}
\affiliation{Science and Technology Facilities Council, Rutherford Appleton Laboratory, Particle Physics Dept., Oxfordshire OX11 0QX, UK}
\author{Adi Ashkenazi}
\affiliation{Tel Aviv University, Tel Aviv 69978, Israel}
\author{Joshua Barrow}
\affiliation{Massachusetts Institute of Technology, Dept. of Physics, Cambridge, MA 02139, USA}
\affiliation{Tel Aviv University, Tel Aviv 69978, Israel}
\author{Steven Dytman}
\affiliation{University of Pittsburgh, Dept. of Physics and Astronomy, Pittsburgh PA 15260, USA}
\author{Hugh Gallagher}
\affiliation{Tufts University, Dept. of Physics and Astronomy, Medford MA 02155, USA}
\author{Alfonso Andres Garcia Soto}
\affiliation{Harvard University, Dept. of Physics, Cambridge, MA 02138, USA}
\author{Steven Gardiner}
\affiliation{Fermi National Accelerator Laboratory, Batavia, Illinois 60510, USA}
\author{Matan Goldenberg}
\affiliation{Tel Aviv University, Tel Aviv 69978, Israel}
\author{Robert Hatcher}
\affiliation{Fermi National Accelerator Laboratory, Batavia, Illinois 60510, USA}
\author{Or Hen}
\affiliation{Massachusetts Institute of Technology, Dept. of Physics, Cambridge, MA 02139, USA}
\author{Timothy~J.~Hobbs}
\affiliation{Fermi National Accelerator Laboratory, Batavia, Illinois 60510, USA}
\affiliation{Department of Physics, Illinois Institute of Technology, Chicago, IL 60616, USA}
\author {Igor D. Kakorin}
\affiliation{Joint Institute for Nuclear Research (JINR), Dubna, Moscow region, 141980, Russia}
\author {Konstantin S. Kuzmin}
\affiliation{Joint Institute for Nuclear Research (JINR), Dubna, Moscow region, 141980, Russia}
\affiliation{Alikhanov Institute for Theoretical and Experimental Physics (ITEP) of NRC ``Kurchatov Institute'', Moscow, 117218, Russia}
\author{Anselmo Meregalia}
\affiliation{CENBG, Universit\'{e} de Bordeaux, CNRS/IN2P3, 33175 Gradignan, France}
\author {Vadim A. Naumov}
\affiliation{Joint Institute for Nuclear Research (JINR), Dubna, Moscow region, 141980, Russia}
\author{Afroditi Papadopoulou}
\affiliation{Massachusetts Institute of Technology, Dept. of Physics, Cambridge, MA 02139, USA}
\author{Gabriel Perdue}
\affiliation{Fermi National Accelerator Laboratory, Batavia, Illinois 60510, USA}
\author{Marco Roda}
\affiliation{University of Liverpool, Dept. of Physics, Liverpool L69 7ZE, UK}
\author{Alon Sportes}
\affiliation{Tel Aviv University, Tel Aviv 69978, Israel}
\author{Noah Steinberg}
\affiliation{Fermi National Accelerator Laboratory, Batavia, Illinois 60510, USA}
\author{Vladyslav Syrotenko}
\affiliation{Tufts University, Dept. of Physics and Astronomy, Medford MA 02155, USA}
\author{Jeremy Wolcott}
\affiliation{Tufts University, Dept. of Physics and Astronomy, Medford MA 02155, USA}


\collaboration{GENIE Collaboration}

\date{\today}

\begin{abstract}
This article summarizes the state of the art of $\nu_\mu$ and $\overline{\nu}_\mu$ CC0$\pi$ cross-section measurements on carbon and argon and discusses the relevant nuclear models, parametrizations and uncertainties in GENIE v3.
The CC0$\pi$ event topology is common in experiments at a few-GeV energy range. 
Although its main contribution comes from quasi-elastic interactions, this topology is still not well understood.
The GENIE global analysis framework is exploited to analyze CC0$\pi$ datasets from MiniBooNE, T2K and MINER$\nu$A.
A partial tune for each experiment is performed, providing a common base for the discussion of tensions between datasets.
The results offer an improved description of nuclear CC0$\pi$ datasets as well as data-driven uncertainties for each experiment.
This work is a step towards a GENIE global tune that improves our understanding of neutrino interactions on nuclei.
It follows from earlier GENIE work on the analysis of neutrino scattering datasets on hydrogen and deuterium.
\end{abstract}

\maketitle

\section{Introduction}
A major experimental program aims to measure neutrino-nucleus interactions over the few-GeV region.
MiniBooNE was the first neutrino experiment to provide a double-differential flux-integrated CC0$\pi$ cross-section measurement with high statistics on carbon~\cite{MiniBooNEDetector}.
Since then T2K~\cite{ABE2011106}, MicroBooNE~\cite{MicrobooneDetector} and MINER$\nu$A~\cite{MINERvA:2021wjs} have produced a large body of measurements on different nuclei, such as carbon or argon. However, a detailed quantitative understanding of neutrino-nucleus interactions is still missing.

\begin{figure}
    \centering
    \includegraphics[width=0.9\columnwidth]{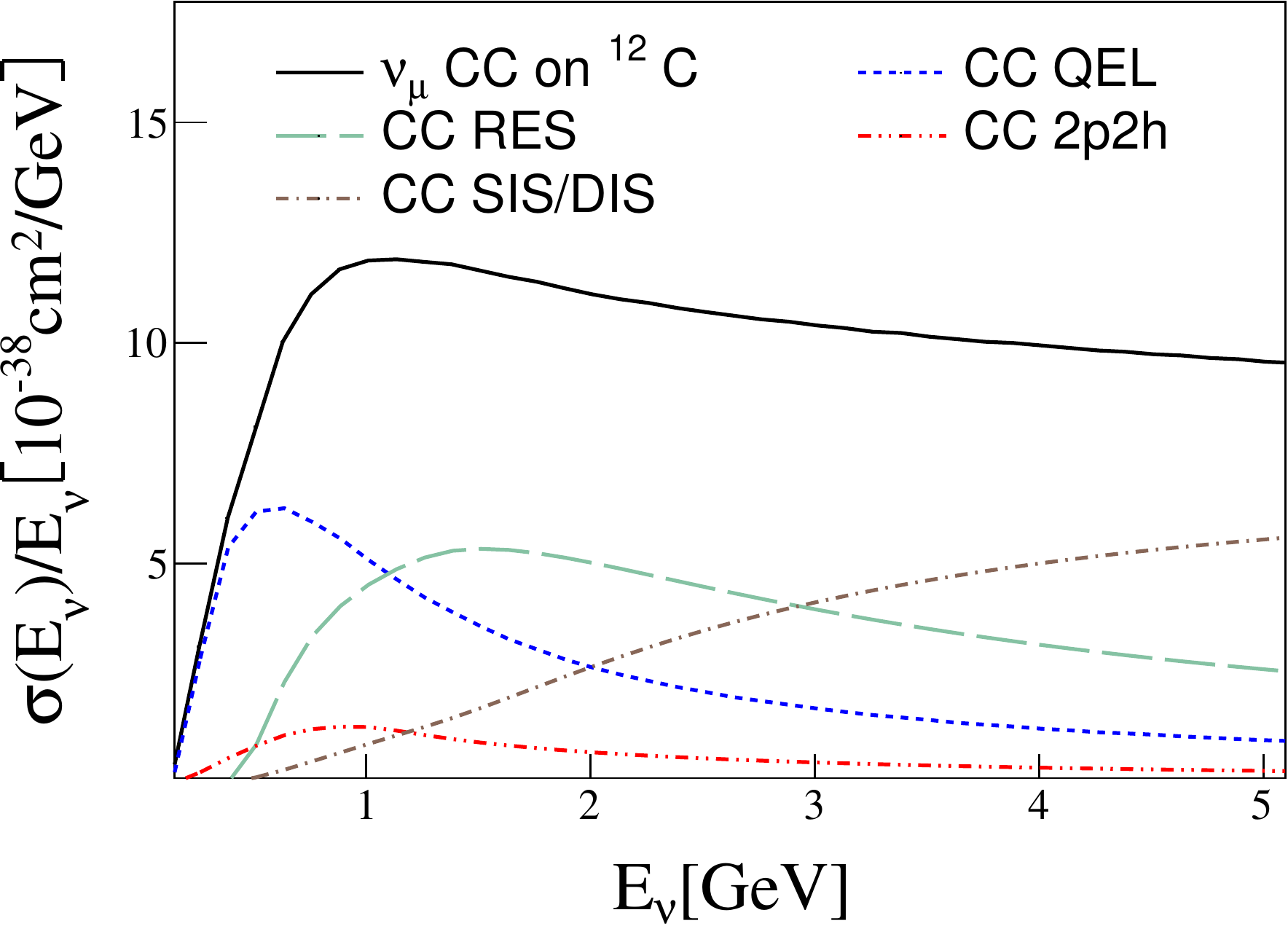}
    \includegraphics[width=0.9\columnwidth]{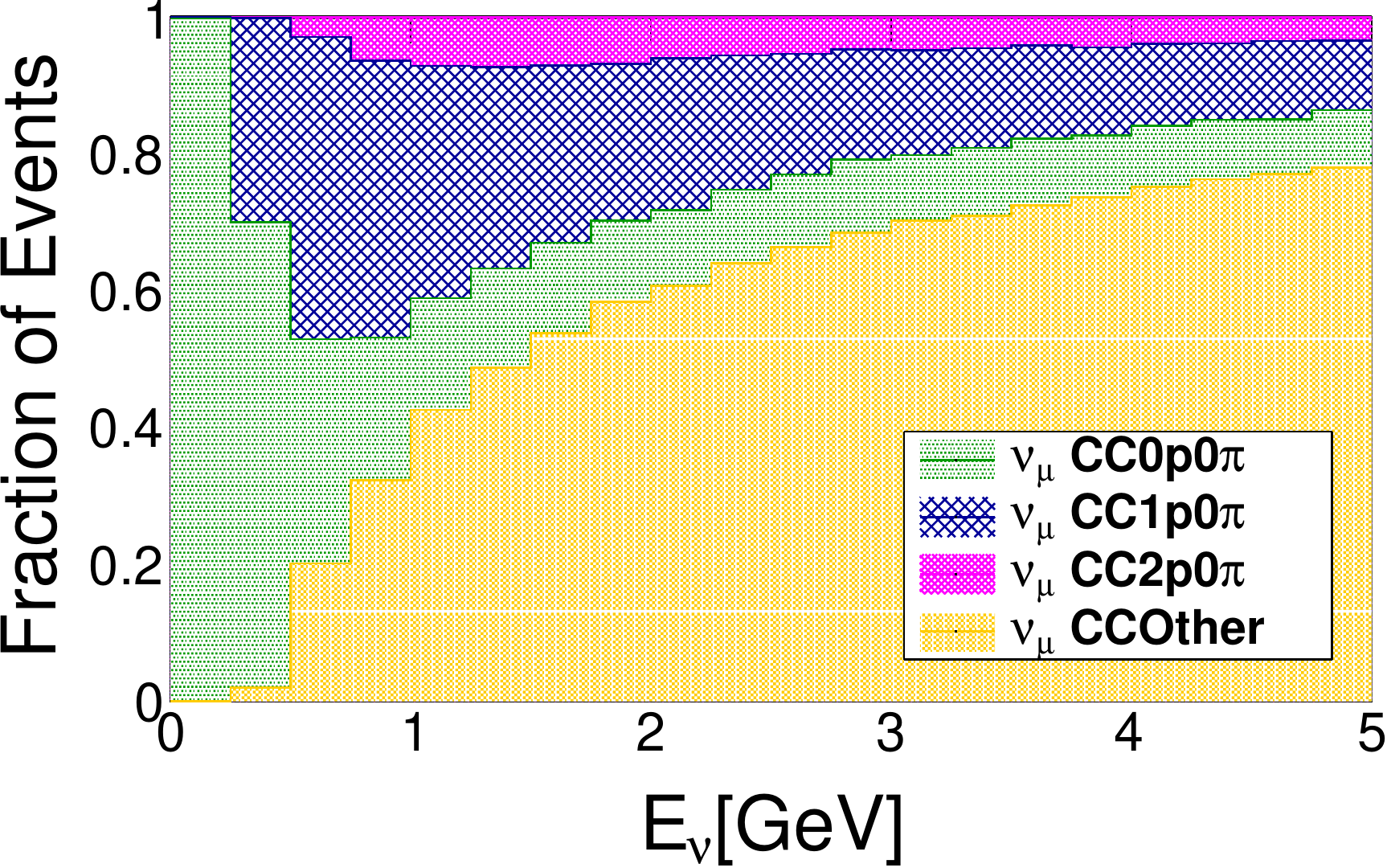}
    \includegraphics[width=0.9\columnwidth]{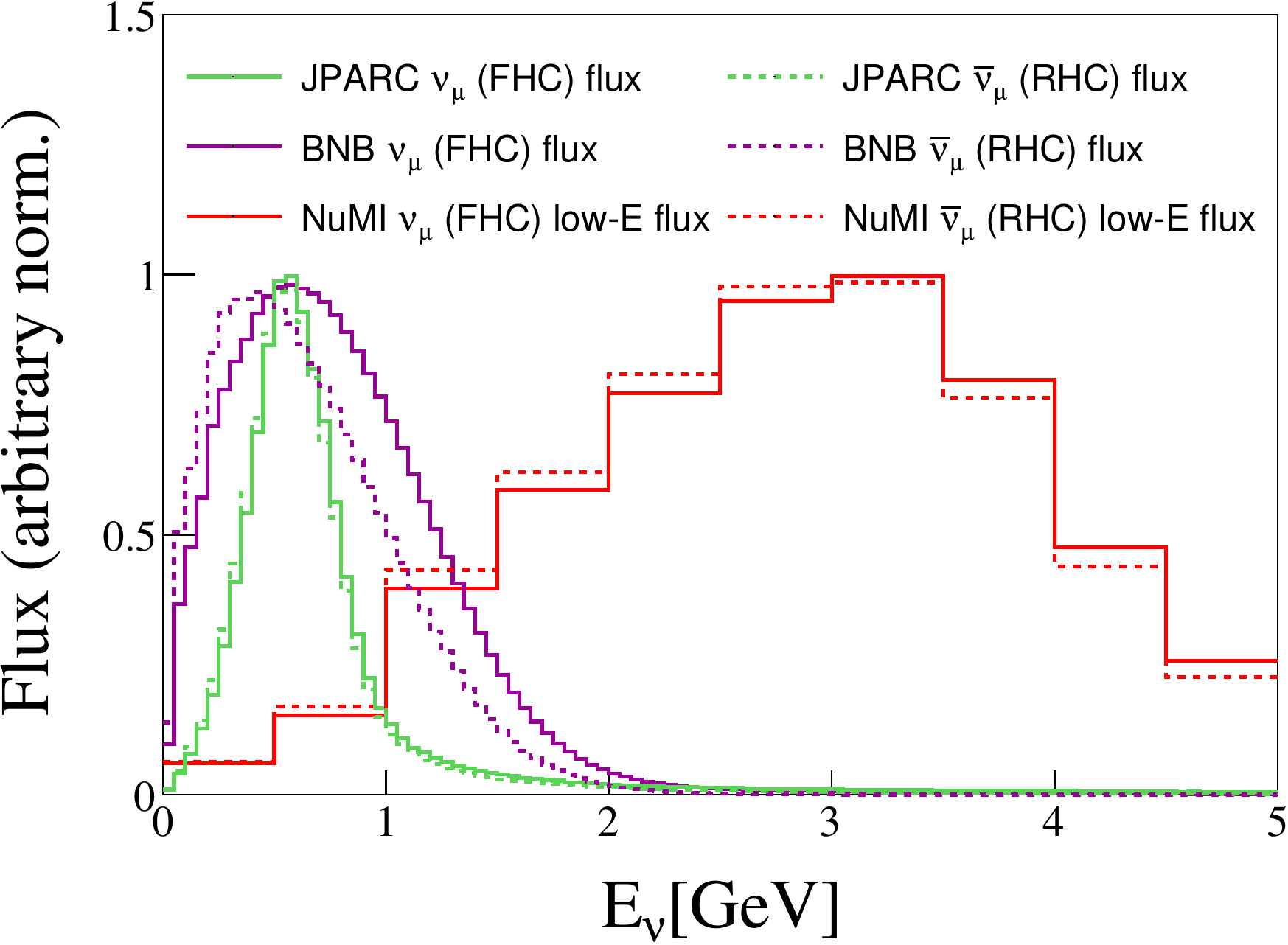}
   	\caption{(Top) Summary of contributions from each interaction process to the CC $\nu_\mu$ cross section on $^{12}$C as a function of neutrino energy, $E_\nu$.
   	(Middle) The corresponding fraction of the total $\nu_\mu$-$^{12}$C events arising from each of the $0\pi$ topologies. 
   	This plot assumes a momentum threshold for protons of $450$~MeV/c while the GENIE predictions are obtained with the \texttt{G18\_10a\_02\_11b} tune.
   	(Bottom) Summary of $\nu_\mu$ (continuous lines) and $\overline{\nu}_\mu$ (dashed lines) normalized flux distributions for T2K ND280 at JPARC~\cite{PhysRevD.87.012001}, MiniBooNE and MicroBooNE with the \ac{BNB}~\cite{BNBFlux}, and MINER$\nu$A with the \ac{NuMI}~\cite{PhysRevD.94.092005}.
   	The flux predictions for neutrino and antineutrino modes are refereed to as ``\ac{FHC}'' and ``\ac{RHC},'' respectively.}
    \label{fig:fluxes}
\end{figure}

In order to avoid biases in cross-section measurements due to theory assumptions, neutrino experiments focus on the study of specific topologies instead of interaction processes like \ac{QEL} scattering.
The most dominant event topology below the 1~GeV region is CC0$\pi$, which is usually defined as an event with one muon and no pions in the final state.
As a consequence of the nuclear medium, different interaction processes contribute to the CC0$\pi$ measurement.
Neutrino \ac{CC} \ac{QEL} interactions are the dominant contribution to this topology inside the few-GeV energy range.
\ac{2p2h} contributions have been shown to be crucial for the correct description of the data at these kinematics.
Adding to the complication, the \ac{SIS} process non-trivially intermixes with other underlying mechanisms;
this is due in part to the fact that pions produced after a \ac{CC} \ac{RES} interaction can be absorbed due to \ac{FSI}.
Moreover, \ac{DIS} can also contribute, with an interplay existing between the description of \ac{DIS} at slightly
higher energies and the treatment of the \ac{NRB} in the \ac{SIS} region.
In GENIE we refer to the \ac{NRB} as \ac{SIS}, see Ref.~\cite{mypaper_1} for details.
Figure~\ref{fig:fluxes} summarizes the $\nu_\mu$ $^{12}$C CC interaction processes and topologies of interest at the few-GeV region as a function of the neutrino energy.
In addition, the flux predictions used for the cross-section measurements of MiniBooNE, MicroBooNE, T2K ND280 and MINER$\nu$A are also provided.

The GENIE Collaboration is building a global analysis of the neutrino, charged-lepton and hadron-scattering data. 
This comprehensive analysis of the world's lepton-nuclear scattering data is being constructed in a staged manner, with recent efforts focused initially on the analysis of neutrino scattering on hydrogen and deuterium for the purpose of tuning aspects of the GENIE framework associated with the free-nucleon cross section: namely the \ac{SIS} region~\cite{mypaper_1} as well as tuning of hadronic multiplicities relevant for neutrino-induced hadronization models~\cite{hadronization_tune}.
The present work extends this analysis campaign to a second stage: an explicit tune of nuclear model parameters to recent nuclear data.

This work is further necessitated by outstanding discrepancies between GENIE predictions and more recent datasets, which use heavy nuclei as targets.
Several neutrino collaborations, such as MicroBooNE and MINER$\nu$A, tried to address these discrepancies by tuning GENIE against the $\nu_\mu$\ac{CC}$0\pi$ T2K and inclusive $\nu_\mu$ \ac{CC} MINER$\nu$A datasets, respectively~\cite{MicroBooNE:2021ccs,PhysRevD.101.112007,PhysRevD.103.112008}.
All these tunes simulate \ac{2p2h} interactions with the Valencia model~\cite{Nieves}.
In both cases, the results suggest an enhancement of the \ac{2p2h} cross section.
These tunes are not available for wider use within GENIE, and in some cases, these were performed with obsolete GENIE versions which differ substantially from the latest one. 

In this paper, we describe the GENIE analysis of the available $\nu_\mu$ and $\overline{\nu}_\mu$ CC0$\pi$ datasets from MiniBooNE, T2K, MINER$\nu$A and MicroBooNE.
The main goal is to provide improved simulations tuned to nuclear data and quantify the major sources of uncertainties in CC0$\pi$ measurements.
In order to do so, new degrees of freedom are developed within the GENIE \ac{MC} event generator in order to quantify the effect of variation away from the nominal models.
Most of the new degrees of freedom can be used to tune other available \ac{CMCs} in GENIE.
In this analysis we focus on the `retuning' of the \texttt{G18\_10a\_02\_11b} tune against $\nu_\mu$-$^{12}$C CC0$\pi$ data from MiniBooNE, T2K and MINER$\nu$A.
The \texttt{G18\_10a\_02\_11b} was previously tuned against free-nucleon data~\cite{mypaper_1}.
In this paper, we refer to \texttt{G18\_10a\_02\_11b} as the \emph{nominal} tune.

All predictions shown in this paper are calculated using the \texttt{G18\_10a\_02\_11b} tune.
\texttt{G18\_10a\_02\_11b} uses the Valencia model to simulate \ac{QEL} and \ac{2p2h} events in the nuclear medium, while
\ac{FSI}s are modeled using the \emph{hA} model and
the nuclear ground state is described with the \ac{LFG} model~\cite{GENIEHighlights}.
The other interaction processes are common with the free-nucleon recipe described in Ref.~\cite{mypaper_1}.
Tab.~\ref{tab:G18_10a_02_11a_details} details the full list of interaction processes associated with this \ac{CMC}.

\begin{table}
    \centering
    \begin{tabular}{@{\extracolsep\fill} c c}
    \hline\hline\noalign{\smallskip}
    Simulation domain         & Model \\
    \noalign{\smallskip}\hline\noalign{\smallskip}
    Nuclear model             & Local Fermi Gas~\cite{GENIEHighlights} \\
    QEL and \ac{2p2h}         & Valencia~\cite{Nieves,PhysRevC.70.055503}  \\
    QEL Charm                 & Kovalenko~\cite{Kovalenko:1990zi} \\
    QEL $\Delta S=1$          & Pais~\cite{PAIS1971361} \\
    RES                       & Berger-Sehgal~\cite{0709.4378}  \\
    SIS/DIS                   & Bodek-Yang~\cite{BODEK200270} \\
    DIS $\Delta S=1$          & Aivazis-Tung-Olness~\cite{aivazis1993nexttoleading} \\
    Coherent $\pi$ production & Berger-Sehgal~\cite{0709.4378} \\ 
    \noalign{\smallskip}\hline\noalign{\smallskip}
    Hadronization             & \text{AGKY}~\cite{Yang:2007zzt} \\
    \noalign{\smallskip}\hline\noalign{\smallskip}
    FSI                       & INTRANUKE hA~\cite{Andreopoulos:2015wxa} \\
    \noalign{\smallskip} \hline\hline
    \end{tabular}
    \caption{Complete list of models used for the \texttt{G18\_10a\_02\_11a} tune in GENIE v3~\cite{mypaper_1}.}
    \label{tab:G18_10a_02_11a_details}
\end{table}

We stress that \texttt{G18\_10a\_02\_11b} is only one of many \ac{CMC}s that can be tuned within GENIE.
Our main motivations behind this particular choice are: (1) we can use data-driven constraints from previous GENIE tunes on hydrogen and deuterium~\cite{mypaper_1}; (2) the \ac{QEL} and \ac{2p2h} processes are modeled with the Valencia model, a theory-based model which is used in most neutrino analyses; (3) \ac{FSI} interactions are modeled with the INTRANUKE \emph{hA} model, which is an easily tuned empirical model closely driven by hadron-nucleus scattering data.
Other \ac{CMC}s will be considered in future iterations of this work.

The GENIE global analysis software~\cite{mypaper_1} is used to perform a partial tune for each experiment using double-differential flux-integrated CC0$\pi$ cross-section measurements as a function of muon kinematics.
We further note that only carbon datasets are considered in this work. While a more expansive study of the nuclear
$A$ dependence will be a valuable aspect of future work, this choice carries the advantage of providing a consistent basis for the exploration of statistical tensions.
This work is a step closer to a global tune with neutrino-nucleus cross-section data, which can be performed using the same analysis strategy once all the tensions are well understood.
Future iterations of this work will also incorporate measurements on different topologies, such as CC1$\pi^\pm$.

This work is organized as follows: Sec.~\ref{sec:nucleardata} provides an overview of the available CC0$\pi$ data to date.
The newly developed GENIE parameters are discussed in Sec.~\ref{sec:UncertantiesParameterization}.
This is followed by a description of the tuning procedure in Sec.~\ref{sec:tuningprocedure} and a discussion of the tune results and tensions between \ac{CC}0$\pi$ and \ac{CC}Np0$\pi$ datasets in Sec.~\ref{sec:NuclearTuningResults}.
In addition, Sec.~\ref{sec:InvestigationTension} describes some modelling aspects relevant for the exploration of the \ac{CC}0$\pi$ and \ac{CC}Np0$\pi$ tension. 
The main conclusions of this paper are highlighted in Sec.~\ref{sec:NuclearConclusions}.

\section{Review of neutrino nucleus CC0\ensuremath{\pi} measurements}
\label{sec:nucleardata}

Cross-section measurements of the \ac{CC}0$\pi$ topology were carried out by SciBooNE~\cite{Alcaraz-Aunion:2009pzm}, NOMAD~\cite{NOMAD:2009qmu}, MiniBooNE~\cite{1002.2680,MinibooneAntineutrino}, T2K ND280~\cite{T2KCCQELikeMeasurement}, MINER$\nu$A~\cite{PhysRevD.99.012004,PhysRevLett.121.022504,MINERvA:2019ope} and MicroBooNE~\cite{uBooNEQE,PhysRevD.102.112013}.
This section provides with a review of neutrino-nucleus CC0$\pi$ data available to date.

\subsection{Scattering topologies and kinematics of experimental data}
\label{sec:topologies}

The CC0$\pi$ topology is usually defined as a CC event with no pions in the final state, regardless of the number of protons in the event.
However, the CC0$\pi$ topology definition is not universal as it varies between the different published measurements as a consequence of the different detection capabilities of each experiment.
In some analyses, its definition is optimized to study more exclusive final states with a specific proton multiplicity. 
The following nomenclature is adopted to avoid confusion for the reader:
analyses requiring one or more protons in the final state are referred to as \ac{CC}Np0$\pi$, where $\mathrm{N}\!\ge\! 1$.
If the analysis requires {\it exactly} zero or one proton in the final state, N is the replaced by the corresponding number, i.e.\ \ac{CC}0p0$\pi$ or \ac{CC}1p0$\pi$, respectively, for events with either no visible protons or precisely one.
In some cases, the topology definition requires at least two protons in the final state.
This is denoted as CC2p0$\pi$. We note that, in this case, CC2p0$\pi$ events include the very small probability to have
N$>\!2$ final-state protons --- a scenario which is challenging to isolate experimentally.
When there is no requirement on the proton multiplicity, the topology is refereed to as CC0$\pi$.
Fig.~\ref{fig:fluxes} presents the fraction of $\nu_\mu$ CC events as a function of the neutrino energy for different CC topologies. In this particular plot, the CC0$\pi$ topology contribution is broken down into more exclusive topologies depending on the proton multiplicity. It can be concluded that CC0$\pi$ events dominate the event rate for $E_\nu<1.5$~GeV. At higher energies, the contribution from events with pions in the final state (CC other) dominates.

Tab.~\ref{tab:datasummary} lists the available CC0$\pi$ and CCNp0$\pi$ cross section measurements to date. 
The table summarizes the information of interest for the evaluation of the GENIE predictions: the target type, neutrino flux mean energy and event topology definition.
The neutrino flux spectrum associated with each experiment is provided in Fig.~\ref{fig:fluxes} (bottom)~\cite{PhysRevD.87.012001,BNBFlux,PhysRevD.94.092005}.
We use the same neutrino flux prediction for MiniBooNE and MicroBooNE.

The \emph{kinematic quantity} column in Tab.~\ref{tab:datasummary} lists the kinematic quantities used to extract the cross-section measurements. 
The definition of each kinematic quantity is given in Appendix ~\ref{sec:Observables}.
Some of the available measurements are double-differential or triple-differential ones.
This is indicated by a comma-separated list for the kinematic quantities used in the corresponding analysis.
In addition, the year of the data release and the number of bins ($N_{\text{Bins}}$) for each dataset are specified.
The details on the analysis requirements for MiniBooNE, T2K ND280, MINER$\nu$A and MicroBooNE datasets as well as comparisons of the \texttt{G18\_10\_02\_11b} predictions to the data are presented in Appendix~\ref{sec:comparisons_data}.
The main observations from Appendix~\ref{sec:comparisons_data} are summarized in Sec.~\ref{sec:datasetconsiderations}.
For completeness, Tab.~\ref{tab:datasummary} includes measurements from SciBooNE and NOMAD which are not discussed further in this paper as their analysis strategy is limited with respect to the other measurements discussed in this work.
 
\begin{table*}
    \centering
    \resizebox{0.6\textwidth}{!}{%
    \begin{tabular}{@{\extracolsep\fill} c c c c c c c c c } 
\hline\hline\noalign{\smallskip}
Experiment & $\langle E_\nu \rangle$ & Target & Topology & Kinematic quantity & $N_{\text{Bins}}$ & Year & Ref. & \\
\noalign{\smallskip}\hline\noalign{\smallskip}
\multicolumn{9}{c}{$\nu_\mu$-A measurements} \\ 
\noalign{\smallskip}\hline\noalign{\smallskip}
SciBooNE   &   700~MeV & $^{12}$C  & CC0$\pi$   & $E_\nu^{\text{QEL}}$                              &   5 & 2006 & \cite{Alcaraz-Aunion:2009pzm} & \xmark     \\
\noalign{\smallskip}\hline\noalign{\smallskip}
NOMAD      &   23~GeV  & $^{12}$C  & CC0$\pi$   & $E_\nu^{\text{QEL}}$                              &  10 & 2009 & \cite{NOMAD:2009qmu}          & \xmark     \\ 
\noalign{\smallskip} \hline\noalign{\smallskip}
MiniBooNE  & $788$~MeV & $^{12}$C  & CC0$\pi$   & $T_\mu$, $\cos\theta_\mu$                         & 137 & 2010 & \cite{1002.2680}              & \checkmark \\
           &           &           &            & $Q^2_{\text{QEL}}$                                &  17 &      &                               &            \\
           &           &           &            & $E_{\nu}^{\text{QEL}}$                            &  14 &      &                               &            \\
\noalign{\smallskip}\hline\noalign{\smallskip}
T2K ND280  & $600$~MeV & $^{12}$C  & CC0p0$\pi$ & $p_\mu$, $\cos\theta_\mu$                         &  60 & 2018 & \cite{T2KCCQELikeMeasurement} & \checkmark \\ 
\hdashline & $600$~MeV & $^{12}$C  & CC1p0$\pi$ & $\cos\theta_\mu$, $\cos\theta_p$, $p_p$           &  40 & 2018 & \cite{T2KCCQELikeMeasurement} & \xmark     \\
\hdashline & $600$~MeV & $^{12}$C  & CC2p0$\pi$ & $-$                                               &   1 & 2018 & \cite{T2KCCQELikeMeasurement} & \xmark     \\
\hdashline & $600$~MeV & $^{12}$C  & CCNp0$\pi$ & $\delta p_T$                                      &   8 & 2018 & \cite{T2KCCQELikeMeasurement} & \xmark     \\ 
           &           &           &            & $\delta \phi_T$                                   &   8 &      &                                            \\ 
           &           &           &            & $\delta \alpha_T$                                 &   8 &      &                                            \\ 
           &           &           &            & $\Delta p_{p}$, $\cos\theta_\mu$, $p_\mu$         &  49 &      &                                            \\ 
           &           &           &            & $|\Delta\mathbf{p}_p|$, $\cos\theta_\mu$, $p_\mu$ &  49 &      &                                            \\ 
           &           &           &            & $\Delta\theta_p$, $\cos\theta_\mu$, $p_\mu$       &  35 &      &                                            \\ 
\noalign{\smallskip}\hline\noalign{\smallskip}
MINER$\nu$A& $3.5$~GeV & $^{12}$C  & CC0$\pi$   & $p^\mu_T$, $p^\mu_L$                              & 144 & 2019 & \cite{PhysRevD.99.012004} & \checkmark \\
           &           &           &            & $Q^{2}_{\text{QEL}}$                              &  16 &      &                               &            \\ 
           &           &           &            & $E_\nu^{\text{QEL}}$                              &  12 &      &                               &            \\
\hdashline & $3.5$~GeV & $^{12}$C  & CCNp0$\pi$ & $p_p$                                             &  25 & 2018 & \cite{PhysRevLett.121.022504} & \checkmark \\ 
           &           &           &            & $\theta_p$                                        &  26 &      &                               &            \\ 
           &           &           &            & $\delta p_T$                                      &  24 &      &                               &            \\ 
           &           &           &            & $\delta \alpha_T$                                 &  12 &      &                               &            \\ 
           &           &           &            & $\delta \phi_T$                                   &  18 &      &                               &            \\
\hdashline & $3.5$~GeV & $^{12}$C  & CCNp0$\pi$ & $\delta p_{Tx}$                                   &  32 & 2020 & \cite{MINERvA:2019ope}    & \xmark     \\
           &           &           &            & $\delta p_{Ty}$                                   &  33 &      &                               &            \\
\hdashline &   $6$~GeV & $^{12}$C  & CC0$\pi$   & $p^\mu_T$, $p^\mu_L$                              & 184 & 2020 & \cite{MINERvA:2019gsf}        & \xmark     \\
           &           &           &            & $Q^{2}_{\text{QEL}}$                              &  19 &      &                               &            \\
\hdashline &   $6$~GeV & $^{12}$C  & CCN0$\pi$  & $p^\mu_T$, $p^\mu_L$, $\sum T_p$                  & 660 & 2022 & \cite{MINERvA:2022mnw}        & \xmark     \\
           &           &           &            & $E_\mu$, $q_0^{\text{QEL}}$, $\sum T_p$           & 540 &      &                               &            \\  
\noalign{\smallskip}\hline\noalign{\smallskip} 
MicroBooNE & $800$~MeV & $^{40}$Ar & CC1p$0\pi$ & $p_\mu$                                           &   7 & 2020 & \cite{uBooNEQE}               & \xmark \\
           &           &           &            & $\cos\theta_\mu$                                  &   7 &      &                               &            \\
           &           &           &            & $p_p$                                             &   7 &      &                               &            \\
           &           &           &            & $Q^2_{\text{QEL}}$                                &   7 &      &                               &            \\
           &           &           &            & $E_{\nu}^{\text{cal}}$                            &   7 &      &                               &            \\
\hdashline & $800$~MeV & $^{40}$Ar & CCNp$0\pi$ & $p_\mu^{\text{reco}}$                             &  10 & 2020 & \cite{PhysRevD.102.112013}    & \xmark     \\
           &           &           &            & $\cos\theta_\mu^{\text{reco}}$                    &  12 &      &                               &            \\
           &           &           &            & $p_p^{\text{reco}}$                               &  10 &      &                               &            \\
           &           &           &            & $\cos\theta_p^{\text{reco}}$                      &   9 &      &                               &            \\
           &           &           &            & $\theta_{\mu p}^{\text{reco}}$                    &   6 &      &                               &            \\
\noalign{\smallskip}\hline\noalign{\smallskip}
\multicolumn{9}{c}{$\overline{\nu}_\mu$-A measurements} \\ 
\noalign{\smallskip}\hline\noalign{\smallskip}
NOMAD      &  $23$~GeV & $^{12}$C  & CC0$\pi$   & $E_\nu^{\text{QEL}}$                              &   6 & 2009 & \cite{NOMAD:2009qmu}          & \xmark     \\
\noalign{\smallskip}\hline\noalign{\smallskip}
MiniBooNE  & $665$~MeV & $^{12}$C  & CC0$\pi$   & $T_\mu$, $\cos\theta_\mu$                         &  78 & 2013 & \cite{MinibooneAntineutrino}  & \checkmark \\
           &           &           &            & $Q^{2}_{\text{QEL}}$                              &  16 &      &                               &            \\ 
           &           &           &            & $E_\nu^{\text{QEL}}$                              &  14 &      &                               &            \\  
\noalign{\smallskip}\hline\noalign{\smallskip}
T2K ND280 & 600~MeV & H$_2$O & CC0$\pi$ & $p_\mu$, $\cos\theta_\mu$& 19 & 2019 & \cite{PhysRevD.102.012007} & \xmark \\
T2K ND280 & 600~MeV & $^{12}$C & CC0$\pi$ & $p_\mu$, $\cos\theta_\mu$ & 57 & 2020 & \cite{PhysRevD.101.112001} & \xmark \\ 
T2K WAGASCI & 860~MeV & H$_2$0 & CC0p0$\pi$ & $\theta_\mu$ & 6 & 2021 & \cite{10.1093/ptep/ptab014} & \xmark \\
\hdashline  & 860~MeV & CH & CC0p0$\pi$ & $\theta_\mu$ & 6 & 2021 & \cite{10.1093/ptep/ptab014} & \xmark \\
\noalign{\smallskip}\hline\noalign{\smallskip}
MINER$\nu$A& $3.5$~GeV & $^{12}$C  & CC0p0$\pi$ & $p^\mu_T$, $p^\mu_L$                              &  60 & 2018 & \cite{PhysRevD.97.052002}     & \checkmark \\
           &           &           &            & $Q^{2}_{\text{QEL}}$                              &   8 &      &                               &            \\ 
           &           &           &            & $E^{\text{QEL}}_{\nu}$                            &  10 &      &                               &            \\ 
\noalign{\smallskip}\hline\hline
\end{tabular}}
\caption{Summary of CC0$\pi$ analyses of $\nu_\mu$ and $\overline{\nu}_\mu$ interactions on nuclei.
		For each analysis, information on the neutrino flux mean energy, target type and event topology is provided.
		The \emph{kinematic quantity} column specifies the list of kinematic quantities used in the cross-section measurement.
		Integrated cross-section measurements are denoted with a ``$-$''.
		All kinematic quantities are defined in Appendix.~\ref{sec:Observables}.
		The last column specifies whether the dataset is considered in the analysis.
\label{tab:datasummary}}
\end{table*}

There is a now large body of CC0$\pi$ data in the literature.
This work focuses on the tuning of double differential flux-integrated CC0$\pi$ and CC0p0$\pi$ cross-section measurements on carbon from MiniBooNE, T2K ND280 and MINER$\nu$A.
This is sufficient for an initial study. 
Additional single- and triple-differential CCNp0$\pi$ datasets are not considered in the first iteration of this work; these will be included in future iterations.
However, some comparisons are given in this paper.

It is important to note the differences between the measurements considered in this work.
These are highlighted in Appendix~\ref{sec:comparisons_data}.
A significant difference is the treatment of uncertainties.
Bin-to-bin correlation are not reported by MiniBooNE for many of their cross section measurements (including the data used in this work). In addition, flux uncertainty is given as a single normalization uncertainty of 10.7\% and 17.2\% for neutrino and antineutrino measurements on carbon respectively.
These treatments involve approximations from modern treatments and do not fully incorporate MiniBooNE uncertainties~\cite{Avanzini:2021qlx}.
Despite the statistical limitations of this measurement, MiniBooNE's datasets are included in the analysis for a complete study of CC0$\pi$ datasets on carbon.
In this work, an additional normalization systematic uncertainty is added to account for the missing flux correlation, as suggested by Ref.~\cite{1002.2680}. 

\subsection{Dataset overview and initial considerations}
\label{sec:datasetconsiderations}

The need for a tuning exercise for GENIE is clear.
A few comparisons of \texttt{G18\_10a\_02\_11b} against the available nuclear data are shown here. 
The remaining plots are in Appendix~\ref{sec:Observables}.

We observe in~\cref{fig:MinibooneDefY,fig:Minerva2DXprojBreakdown,fig:MinervaSTKYDef1} that CC0$\pi$ and CC0p0$\pi$ datasets are under-predicted, whilst the CCNp0$\pi$ datasets are in quite good agreement with the \texttt{G18\_10a\_02\_11b} predictions. 
As a consequence, a coherent global tune of CC0$\pi$ and CCNp0$\pi$ datasets is not possible. 
Hence, the analysis is mostly focused on CC0$\pi$ and CC0p0$\pi$ datasets.
Nonetheless, understanding the tension is essential for future tuning efforts. 
This tension is further explored in this paper.

MiniBooNE CC0$\pi$ (Fig.~\ref{fig:MinibooneDefY}) and T2K ND280 CC0p0$\pi$ data are both under-predicted at muon backward angles, where the contribution to the prediction is mostly from \ac{CC}\ac{QEL} events. At forward angles, where the contribution from non-\ac{CC}\ac{QEL} events is significant, the data are also under-predicted.
The disagreement with MINER$\nu$A CC0$\pi$ data are most significant in the region where \ac{2p2h} events dominate, $0.15<p_{T}<0.7$~GeV, see Fig.~\ref{fig:Minerva2DXprojBreakdown}.
\ac{STKI} variables \cite{STKIVariables} bring in new sensitivities and comparisons against MINER$\nu$A data are shown in Fig.~\ref{fig:MinervaSTKYDef1}.
Non-\ac{QEL} events and FSI contributions dominate the region of high $\delta p_T$ and $\delta \alpha_T$. These contributions are essential to describe the data.

\begin{figure*}
    \centering
        \includegraphics[width=\textwidth]{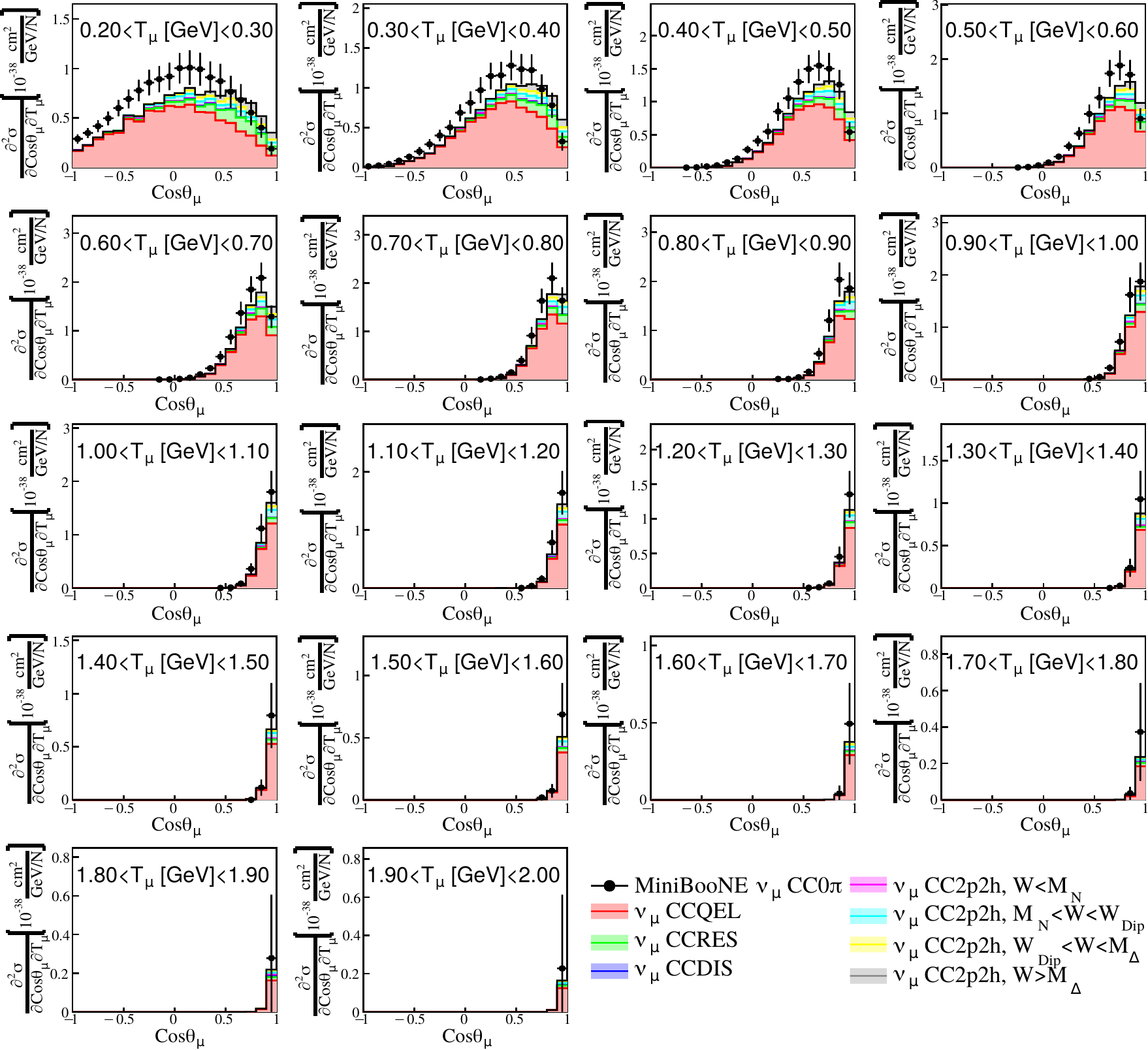}
        \caption{MiniBooNE $\nu_\mu$ \ac{CC}0$\pi$ double differential flux-averaged cross section as a function of the muon angle ($\theta_\mu$) and kinetic energy ($T_\mu$)~\cite{1002.2680}. The corresponding slices on $T_\mu$ are compared against the \texttt{G18\_10a\_02\_11b} tune. The GENIE prediction is divided into different interaction modes.}
        \label{fig:MinibooneDefY}
\end{figure*}

\begin{figure*}
\centering
    \includegraphics[width=\textwidth]{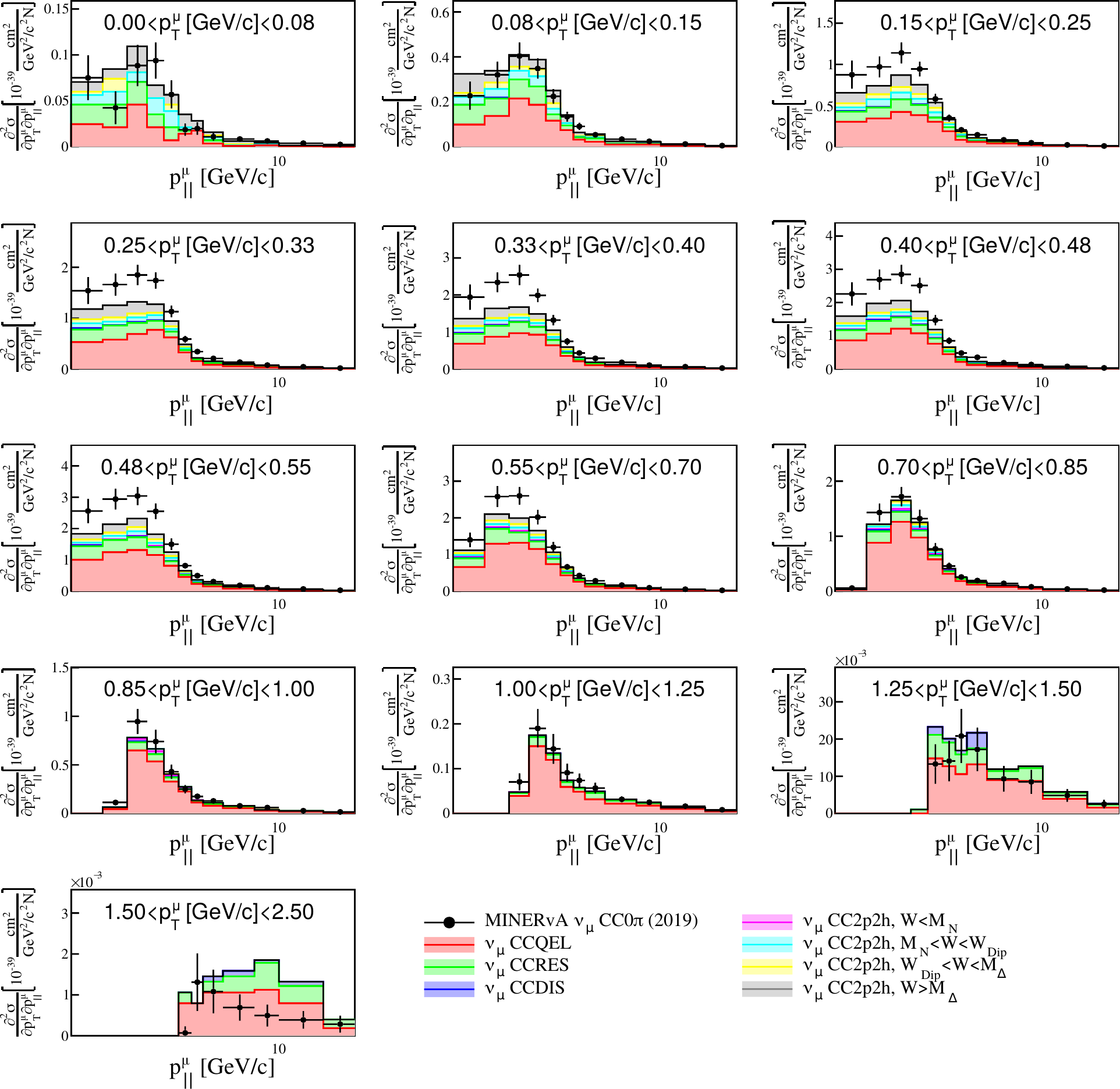} 
    \caption{MINER$\nu$A $\nu_\mu$ CC0$\pi$ double differential flux-averaged cross-section as a function of the muon longitudinal momentum, $p_\parallel$, and transverse momentum, $p_T$~\cite{PhysRevD.99.012004}. The corresponding slices on $p_T$ are compared against the \texttt{G18\_10a\_02\_11b} tune. The GENIE prediction is divided into different interaction modes.}
    \label{fig:Minerva2DXprojBreakdown}
\end{figure*}

\begin{figure}
    \centering
    \begin{subfigure}{0.9\textwidth}
    \includegraphics[width=\textwidth]{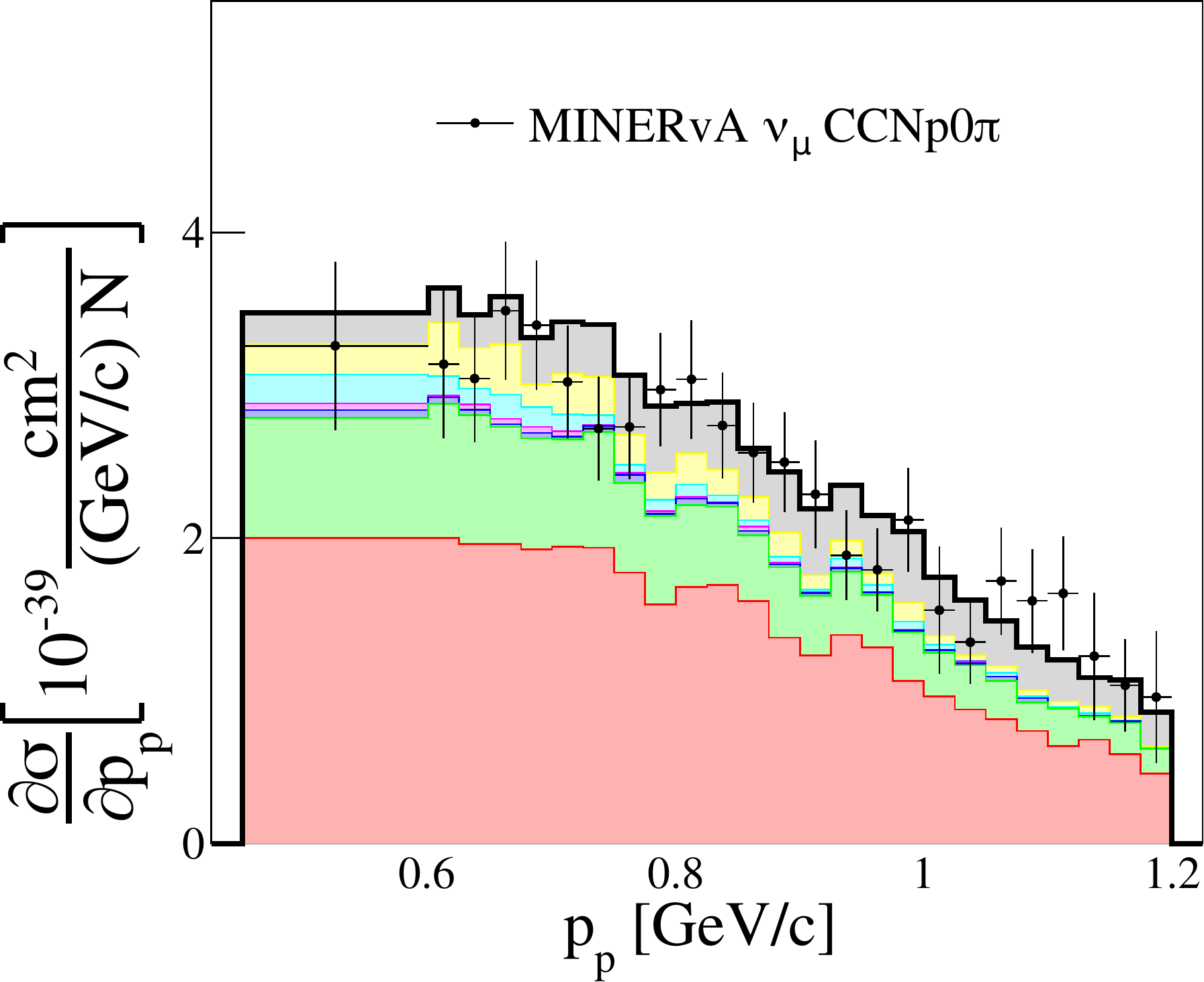}
    \caption{}
    \label{fig:MinervaBreakdownPmom}
    \end{subfigure}

    \begin{subfigure}{0.9\textwidth}
    \includegraphics[width=\textwidth]{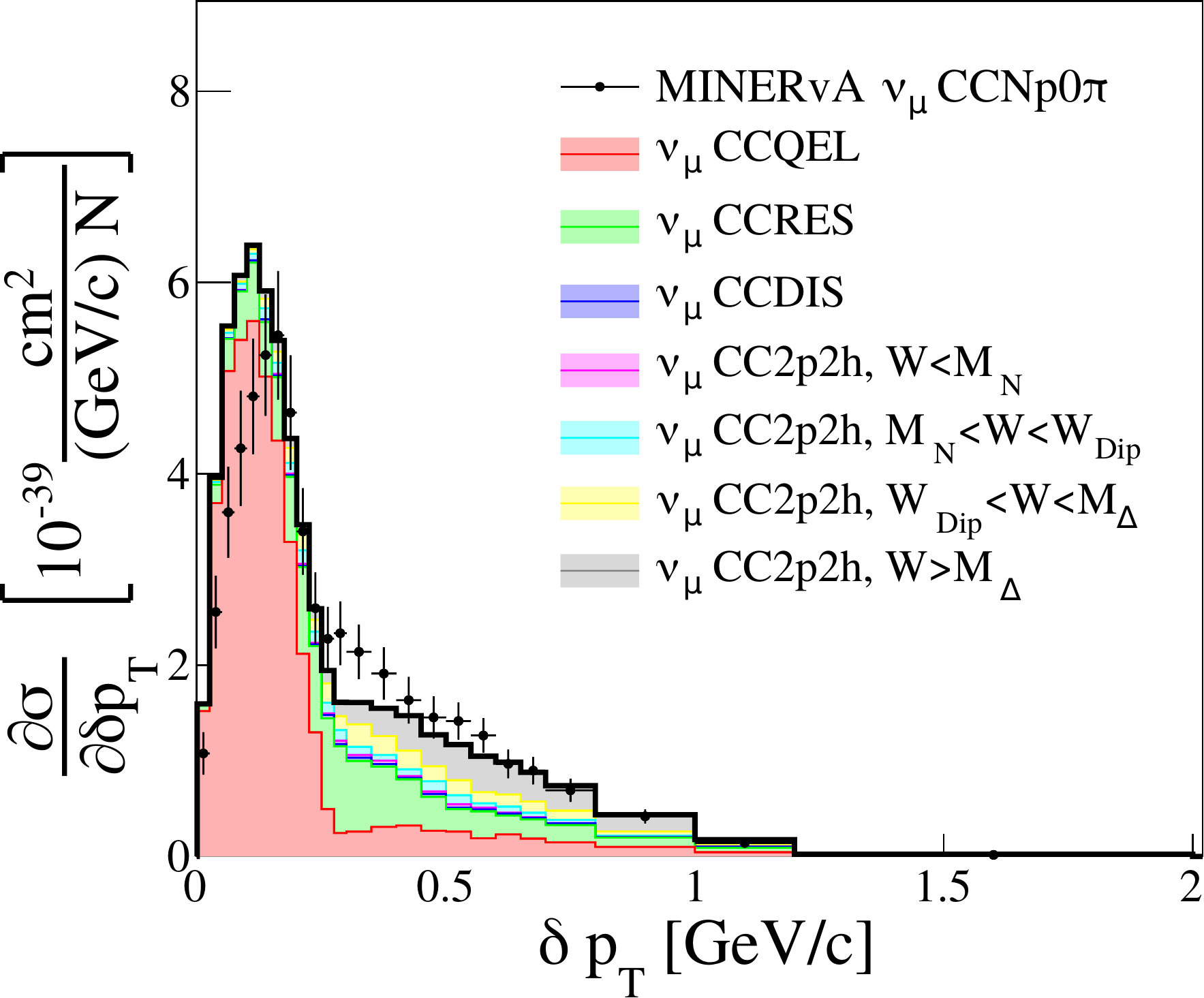}
    \caption{}
    \label{fig:MinervaBreakdowndpt}
    \end{subfigure}

    \caption{MINER$\nu$A $\nu_\mu$ CCNp0$\pi$ differential flux-averaged cross-section as a function of the leading-proton momentum, $\mathrm{p_p}$ (a), and the \ac{STKI} variable~\cite{ PhysRevLett.121.022504,MINERvA:2019ope} $\delta \mathrm{p_T}$ (b). The data are compared against the \texttt{G18\_10a\_02\_11b} tune. The GENIE prediction is divided into interaction modes.}
    \label{fig:MinervaSTKYDef1}
\end{figure}

\begin{figure}
	\centering
    \includegraphics[width=0.9\textwidth]{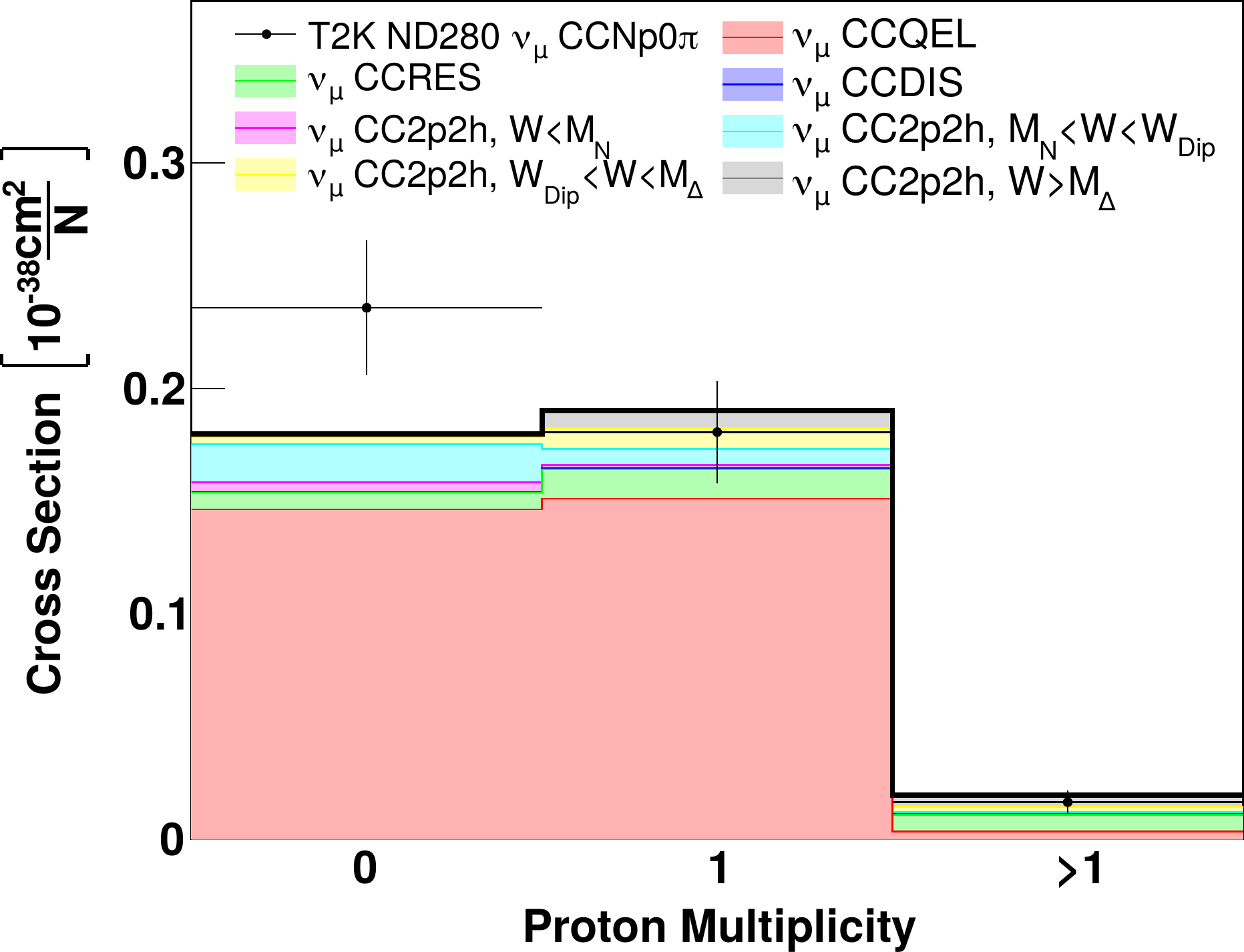}
    \caption{T2K ND280 flux-averaged $\nu_\mu$ \ac{CC}Np0$\pi$ differential cross section as a function of the proton multiplicity~\cite{PhysRevD.93.112012}. The data are compared against the \texttt{G18\_10a\_02\_11b} tune. The GENIE prediction is divided into different interaction modes.}
    \label{fig:T2KMultiplicity}
\end{figure}

The \texttt{G18\_10a\_02\_11b} predictions as a function of the leading proton momentum show a dependency of \ac{2p2h} with $W$: at high proton momentum, \ac{2p2h} events with $W>M_{\Delta}=1232~\text{MeV}/\text{c}^2$ dominate, whilst the opposite is true at low momentum. This is highlighted in Fig.~\ref{fig:MinervaBreakdownPmom}.
\ac{2p2h} events contributing to the T2K ND280 CC0p0$\pi$ sample (Fig.~\ref{fig:T2KMultiplicity}) have $W<W_{\text{Dip}}=1120~\text{MeV}/\text{c}^2$. Higher multiplicity samples have a significant contribution from \ac{2p2h} events with $W>W_{\text{Dip}}$.	
The contribution from \ac{2p2h} events with $W<M_N=938~\text{MeV}/\text{c}^2$ is negligible for all the analyses discussed in this paper.

For further comparisons to data, see Appendix~\ref{sec:comparisons_data}.

\section{Discussion of CC0\ensuremath{\pi} model implementation in GENIE}
\label{sec:UncertantiesParameterization}

This section describes the parameters available to most directly influence CC0$\pi$ predictions within \texttt{G18\_10a\_02\_11b}.
The parameters selected for this analysis are optimized for the \texttt{G18\_10a\_02\_11a} tune.
The complete list of parameters is shown in Tab.~\ref{tab:ParameterVariation}.
The parameter ranges of interest used for the Professor parametrization are also provided.
These can be grouped into five categories: \ac{CC}\ac{QEL}, \ac{CC}\ac{SIS}, \ac{CC}\ac{2p2h}, \ac{FSI} or nuclear model parameters. 

Not all the parameters from Tab.~\ref{tab:ParameterVariation} have been included in the analysis presented in this paper.
Only the parameters included in the final tune are described in this section.
Other parameters of interest to tune CC0$\pi$ data that have been excluded from this analysis are described in Appendix.~\ref{sec:AddlParams}. The reasons for excluding these parameters are summarized in Appendix~\ref{sec:FinalChoice}.

Most of these parameters can be applied to other \ac{CMC}s~\cite{mypaper_1}.
We strive to have as many common, model-independent parameters to allow for systematic comparison between \ac{CMC}s, but this is not always possible.
An extension of this work to other \ac{CMC} will be a subject of a future paper.

\begin{table}[]
\centering
\resizebox{\textwidth}{!}{
\begin{tabular}{cccc}
\hline\hline\noalign{\smallskip}
	Parameter                         &  Nominal Value & Range          & In Final Tune \\ 
\noalign{\smallskip}\hline\noalign{\smallskip}
	$M_A^{\text{QEL}}$~(GeV/c$^2$)    & $1.00\pm 0.01$ & $[0.97, 1.18]$ & \checkmark \\ 
	$S_{\text{QEL}}$                  & $1$            & $-$            & \xmark     \\ 
	$\omega_{\text{RPA}}$             & $1$            & $[-0.5,1.5]$   & \checkmark \\
	$\omega_{\text{No\,RPA}}$         & $0$            & $[-0.5,1.5]$   & \checkmark \\
	$M_{A}^{\text{RES}}$~(GeV/c$^2$)  & $1.09\pm0.014$ & $-$            & \xmark     \\
	$S_{\text{RES}}$                  & $0.84\pm0.03$  & $[0.5,1.5]$    & \checkmark \\
	$R_{\nu p}^{\text{CC}1\pi}$       & 0.008          & $-$            & \xmark     \\
	$R_{\nu n}^{\text{CC}1\pi}$       & $0.94\pm0.075$ & $-$            & \xmark     \\
	$R_{\nu p}^{\text{CC}2\pi}$       & $0.03\pm0.01$  & $-$            & \xmark     \\
	$R_{\nu n}^{\text{CC}2\pi}$       & $2.3\pm0.12$   & $-$            & \xmark     \\
	$S_{N}^{\text{2p2h}}$             & $1$            & $[0,2]$        & \checkmark \\	
	$S_{\Delta}^{\text{2p2h}}$        & $1$            & $[0,2]$        & \checkmark \\	
	$S_{PL}^{\text{2p2h}}$            & $1$            & $[0,2]$        & \checkmark \\
	$S_{Abs}^{\pi^\pm}$               & $1$            & $(1)$          & \xmark     \\
	$S_{\text{MFP}}^{\pi^\pm}$        & $1$            & $(1)$          & \xmark     \\
	$f^{\text{QEL}}$                  & $0$            & $(0)$          & \xmark     \\
	$f^{\text{2p2h}}$                 & $0$            & $(0)$          & \xmark     \\
\noalign{\smallskip}\hline\hline
\end{tabular}
\caption{Summary of parameters relevant for CC0$\pi$ analysis. The range of interest, nominal value in GENIE v3 is also shown. The range of interest corresponds to the parameter space used for the Professor parametrization~\cite{mypaper_1}. 
($-$) is used for parameters that are excluded in the analysis.
The range for the parameters considered in the Professor parametrization but not used in the final tune is not reported.
In such cases, the parameters are fixed to the corresponding nominal values (in parenthesis) in the final analysis, described in Sec.~\ref{sec:tuningprocedure}. 
The last column specifies whether the parameter is considered in the final analysis.}
\label{tab:ParameterVariation}}
\end{table}

\subsection{Charged-current quasi-elastic implementation}
\label{subsec:CCQELParam}
The \ac{QEL} cross section at the free-nucleon level is parametrized with the \ac{QEL} axial mass, $M_A^{\text{QEL}}$, and a \ac{QEL} scaling factor, $S_{\text{QEL}}$. 
Both parameters are common in the simulation of neutrino interactions on free nucleons and nuclei.
$M_A^{\text{QEL}}$ appears as the main degree of freedom in the widely-used dipole parametrization of the \ac{QEL} form factor.
We point out that more elaborate \ac{CMC}s based on the \mbox{z-expansion} model~\cite{Meyer:2016oeg} are now available in GENIE.
In this work, preference is given to tune $M_A^{\text{QEL}}$ as hydrogen and deuterium data provide informative priors
to help constrain this parameter~\cite{mypaper_1}.

\begin{figure}
    \centering
    \begin{subfigure}{\textwidth}
    \includegraphics[width=\textwidth]{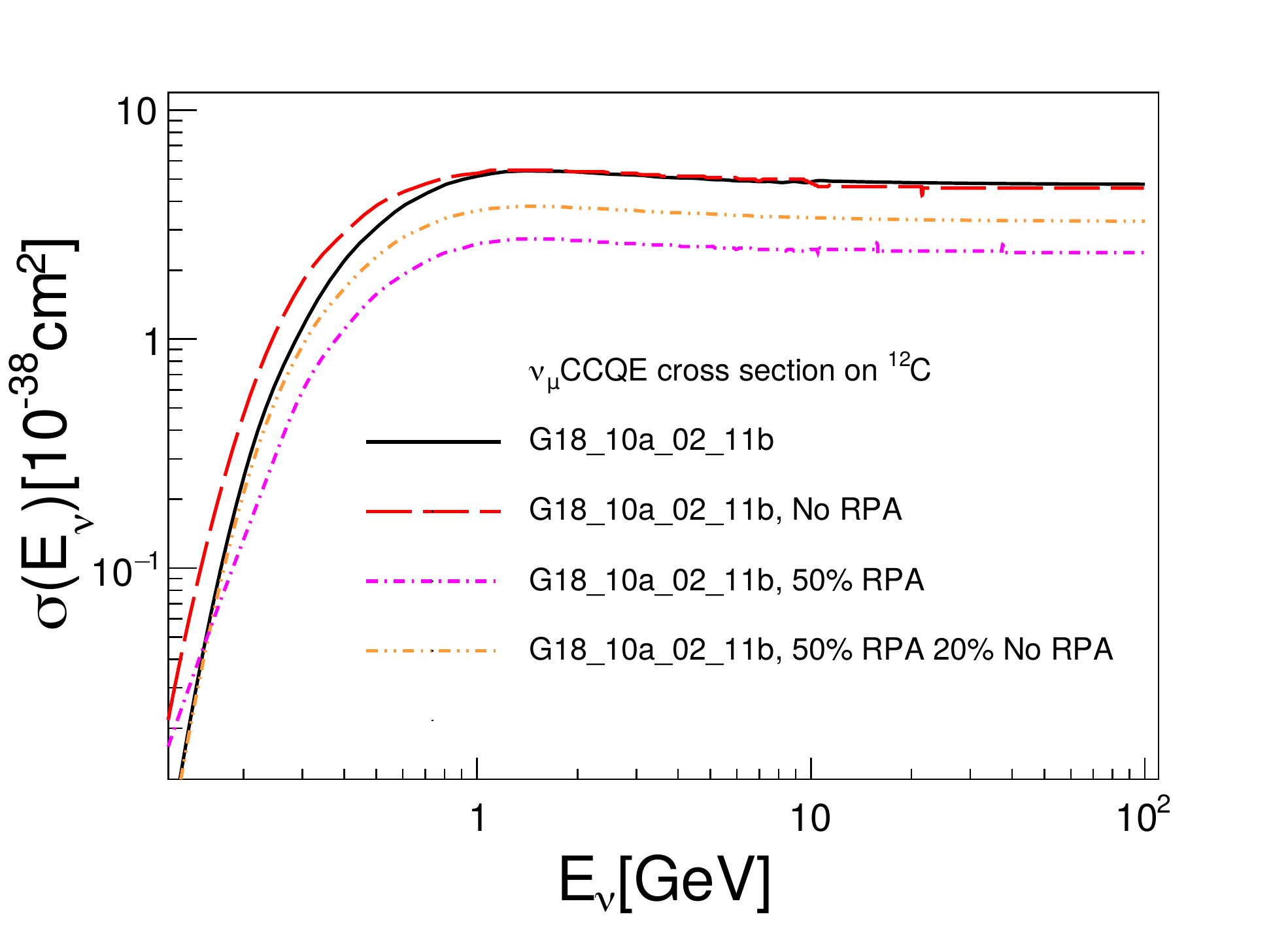}
    \caption{}
    \end{subfigure}

    \begin{subfigure}{0.95\textwidth}
    \includegraphics[width=\textwidth]{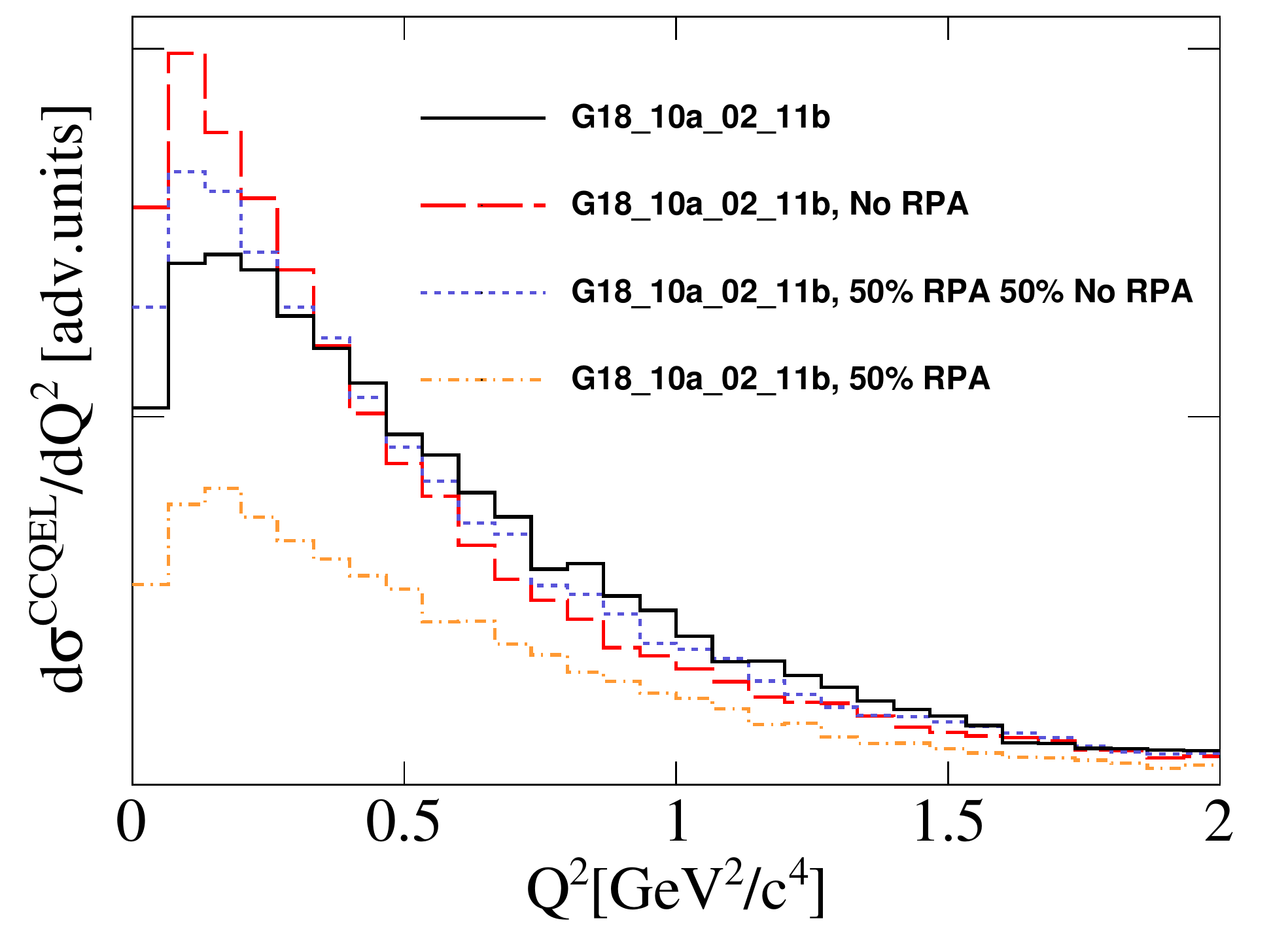}
    \caption{ } 
    \end{subfigure}
    \caption{Impact of the \ac{RPA} parametrization on the \ac{CC}\ac{QEL} cross section. The \texttt{G18\_10a\_02\_11b} prediction is shown in black. The other predictions are obtained with the same tune while changing the \ac{RPA} weight values. (a) Total \ac{CC}\ac{QEL} cross section for $^{12}$C. (b) Flux-integrated differential cross section as a function of $Q^2$. The prediction is obtained with the \ac{NuMI} flux in low-energy mode.}
    \label{fig:xsec_rpa_impactQ2}
\end{figure}

The \ac{QEL} cross section is affected by the dynamics of the nuclear medium.
We include long-range nucleon-nucleon correlations in our calculations with the \ac{RPA} correction~\cite{PhysRevC.70.055503}.
The main effect of the \ac{RPA} correction is a suppression of the \ac{QEL} cross section at low $Q^2$.
This correction is well supported by data and theory, but models differ in predicting its exact strength.
This uncertainty is incorporated in GENIE with two parameters:
one to scale the nominal \ac{QEL} cross-section prediction with \ac{RPA} corrections, $\omega_{\text{RPA}}$, and the other one to scale the \ac{QEL} cross section without \ac{RPA} corrections, $\omega_{\text{No\,RPA}}$.
The total \ac{QEL} cross section is calculated as a linear combination of the cross-section with and without \ac{RPA} corrections:
\begin{equation*}
\sigma^{\text{QEL}} = \omega_{\text{RPA}}\cdot \sigma^{\text{QEL}}_{\text{RPA}} + \omega_{\text{No\,RPA}}\cdot\sigma^{\text{QEL}}_{\text{No\,RPA}}.
\end{equation*}
This parametrization can be used to scale the \ac{QEL} cross section when $\omega_{\text{RPA}}+\omega_{\text{No\,RPA}}\neq 1$.
If $\omega_{\text{No\,RPA}}=0$, $\omega_{\text{RPA}}$ has the exact same effect as $S_{\text{QEL}}$.
Therefore, $S_{\text{QEL}}$ is not included in the tune.
One benefit of this approach is that possible scaling factors on the \ac{RPA} parametrization do not alter the agreement with free-nucleon data.
In addition, it reduces the analysis computing time.
In Fig.~\ref{fig:xsec_rpa_impactQ2}, the \ac{CC} \ac{QEL} cross section as a function of the neutrino energy is shown for different combinations of $\omega_{\text{RPA}}$ and $\omega_{\text{No\,RPA}}$.

\begin{figure}
    \centering
    \includegraphics[width=0.95\columnwidth]{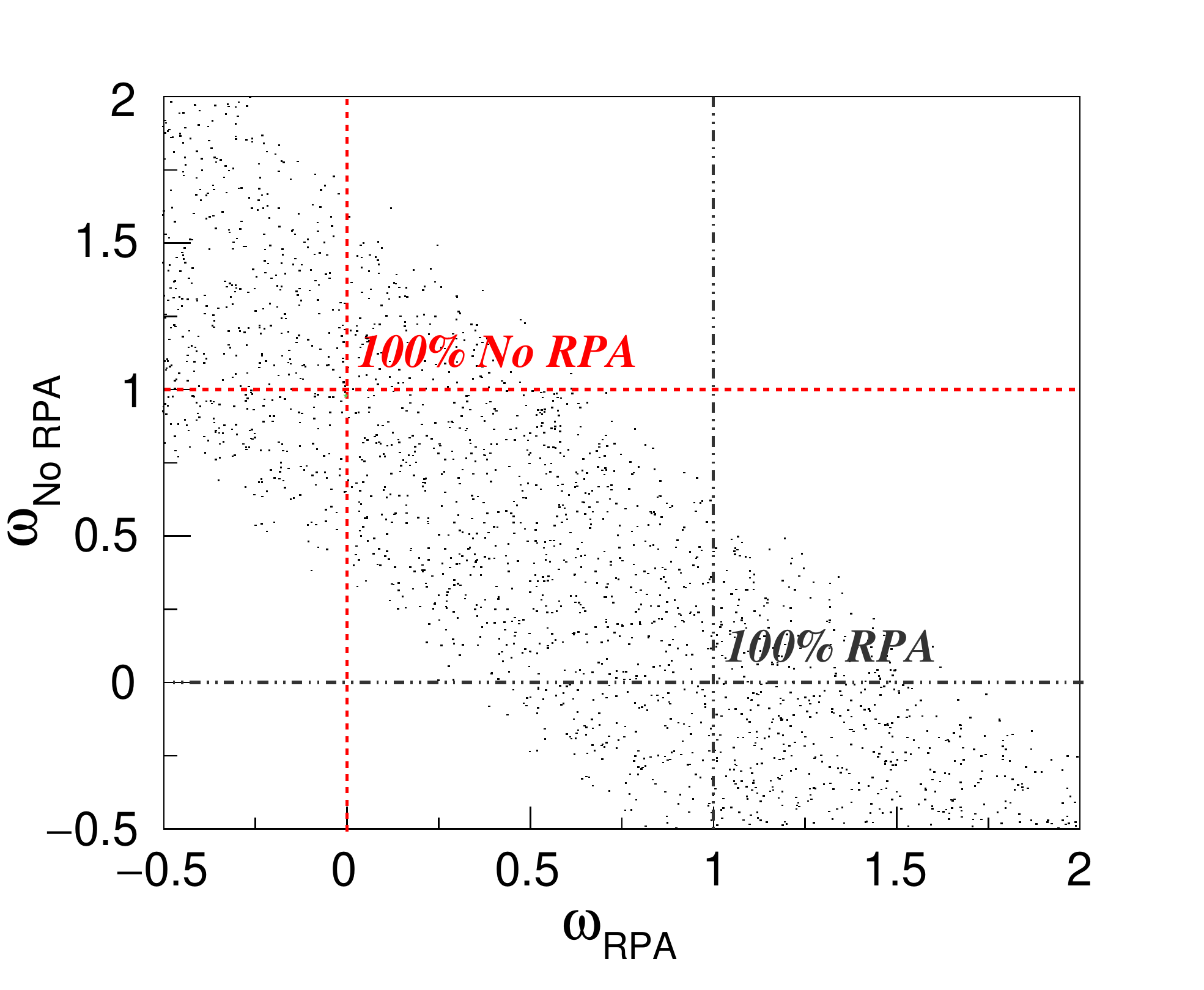}
    \caption{Distribution of scan points used for the GENIE tune in the $\omega_{\text{RPA}}$ vs $\omega_{No\,RPA}$ phase space. The grey (red) line intersection highlights the limit case in which we consider 100\% (0\%) of the \ac{RPA} strength.
    This plot shows a total of 2,050 scan points distributed uniformly. The parameter ranges considered are those from Tab.~\protect\ref{tab:ParameterVariation}.} 
    \label{fig:RPADistribution}
\end{figure}

Choosing each parameter range of interest is crucial for the correct evaluation of the post-fit uncertainties.
In some cases, such as for $\omega_{\text{No\,RPA}}$, we sample negative values to allow the best-fit result to be at its physical limit of 0.
In the case of the \ac{RPA} parametrization, we impose the additional condition that $0.4<\omega_{\text{RPA}}+\omega_{\text{No\,RPA}}<1.6$ in the sampling on the phase space so that $\sigma^{\text{QEL}}>0$.
Fig.~\ref{fig:RPADistribution} shows the distribution of sampled parameter values for $\omega_{\text{RPA}}$ and $\omega_{\text{No\,RPA}}$.
Notice that the two limit cases are at the centre of the phase space.

It is desirable to apply priors to $\omega_{\text{RPA}}$ and $\omega_{\text{No\,RPA}}$, as effectively, parameter combinations for which $\mathcal{S}_{\text{RPA}}\equiv \omega_{\text{RPA}}+\omega_{\text{No\,RPA}}\neq 1$ act as a scaling of the \ac{QEL} cross section.
Hydrogen and deuterium \ac{QEL} cross-section measurements are compatible with $\mathcal{S}_{\text{RPA}}=1$.
However, nuclear effects might introduce an uncertainty in the scaling.
A possible way to include this information is to consider uncorrelated priors on the sum, $\mathcal{S}_{\text{RPA}}$, and the difference, $\Delta_{\text{RPA}}\equiv\omega_{\text{RPA}}-\omega_{\text{No\,RPA}}$,
\begin{align*}
	\mathcal{S}_{\text{RPA}} = &\ 1\pm \sigma_{\mathcal{S}}, \\
	\Delta_{\text{RPA}}      = &\ 1 \pm\sigma_\Delta,
	\label{Eq:RPAPriors}
\end{align*}
with $\sigma_{\mathcal{S}}$ and $\sigma_{\Delta}$ being the variance associated with the priors on $\mathcal{S}_{\text{RPA}}$ and $\Delta_{\text{RPA}}$, respectively.
In terms of $\omega_{\text{RPA}}$ and $\omega_{\text{No\,RPA}}$, this approach includes a correlation between these parameters:
\begin{equation*}
\Sigma_{\text{RPA}} = \frac{1}{4}
\begin{pmatrix}
\sigma_{\mathcal{S}}^2+\sigma_{\Delta}^2 & \sigma_{\mathcal{S}}^2-\sigma_{\Delta}^2 \\
\sigma_{\mathcal{S}}^2-\sigma_{\Delta}^2 & \sigma_{\mathcal{S}}^2+\sigma_{\Delta}^2 \\
\end{pmatrix}
\end{equation*}
This correlation between $\omega_{\text{RPA}}$ and $\omega_{\text{No\,RPA}}$ is included in the tune.
The corresponding central values are $\mu_{\text{RPA}}=1$ and $\mu_{\text{No\,RPA}}=0$ respectively.
The $\sigma_{\mathcal{S}}$ and $\sigma_{\Delta}$ are determined from previous tune iterations, see Sec.~\ref{sec:FinalChoice}.
As concluded from Sec.~\ref{sec:nucleardata}, some flexibility in the \ac{QEL} scaling may be required to describe the data, hence, in this analysis $\sigma_{\mathcal{S}}=0.2$.
This method requires that we impose a prior on $\Delta_{\text{RPA}}$ as well.
Such prior affects the strength of the \ac{RPA} correction, which we aim to constrain from data.
In order to avoid strong constraints on $\Delta_{\text{RPA}}$, $\sigma_\Delta=5$. 

Alternative parameterizations of the \ac{RPA} correction uncertainty are available in the literature.
Theory driven uncertainties specific for the Nieves model are estimated in Ref.~\cite{VALVERDE2006325}.
Alternatively, T2K~\cite{PhysRevD.103.112008} and MicroBooNE~\cite{MicroBooNE:2021ccs} use empirical parameterizations to characterize the uncertainty on the \ac{RPA} correction.
For the first iteration of this work, we opted for a simple parameterization with two parameters to reduce the computational complexity of the tune.

Our method is similar to the \ac{RPA} parametrization used in the latest theory-driven MicroBooNE tune~\cite{MicroBooNE:2021ccs}.
The MicroBooNE Collaboration employed the GENIE ReWeight package to parametrize the \ac{RPA} effect as a linear combination from the \ac{QEL} cross section with the \ac{RPA} correction to the \ac{QEL} cross section without \ac{RPA} using a single parameter limited to [0,1]. 
We refer to this tune as $\mu$\texttt{BooNE} tune.
Both approaches are equivalent when $\mathcal{S}_{\text{RPA}}=1$.

\subsection{Charged-current multi-nucleon implementation}
\label{subsec:MECParameter}
The tuning of \ac{2p2h} models takes a central role in this work.
As discussed in Sec.~\ref{sec:nucleardata}, untuned GENIE CC0$\pi$ \texttt{G18\_10a\_02\_11b} predictions underestimate the data in regions where \ac{2p2h} events contribute.

Previous tuning attempts by other neutrino collaborations indicate a preference for a higher \ac{2p2h} cross section.
The simplest approach to enhance \ac{2p2h} is to use a global scaling factor.
We refer to this parameter as $S_{\text{2p2h}}$.
MINER$\nu$A opted for an empirical approach where they add an extra Gaussian contribution to enhance \ac{2p2h} interactions in $q_0$ and $q_3$. 
This is tuned to MINER$\nu$A \ac{CC} inclusive data.
This tune is known as \texttt{MnvGENIE v1} tune~\cite{RPAMINERVA,MINERvA:2015ydy}.
The $\mu$\texttt{BooNE} tune incorporates the \ac{2p2h} cross-section uncertainty with a linear extrapolation between the GENIE \ac{2p2h} Empirical and Valencia model to account for possible shape differences.
In addition, $S_{\text{2p2h}}$ is also considered.

Different GENIE \ac{2p2h} models predict a slightly different strength and shape for the \ac{2p2h} cross section~\cite{PhysRevD.101.033003}.
These differences motivated the development of a new parametrization that is able to modify the strength as well as the shape of the cross section in the $q_0$-$q_3$ space.
This is accomplished by scaling the \ac{2p2h} differential cross section a function of $W$:
\begin{equation*}
	\frac{d^2\sigma^{\text{2p2h}}}{dq_0dq_3}\rightarrow S(W)\cdot\frac{d^2\sigma^{\text{2p2h}}}{dq_0dq_3}
\end{equation*}
$S(W)$ is the scaling function and $d^2\sigma^{\text{2p2h}}/dq_0dq_3$ the \emph{nominal} double-differential cross section calculation.

\begin{figure}
    \centering
    \includegraphics[width=0.95\textwidth]{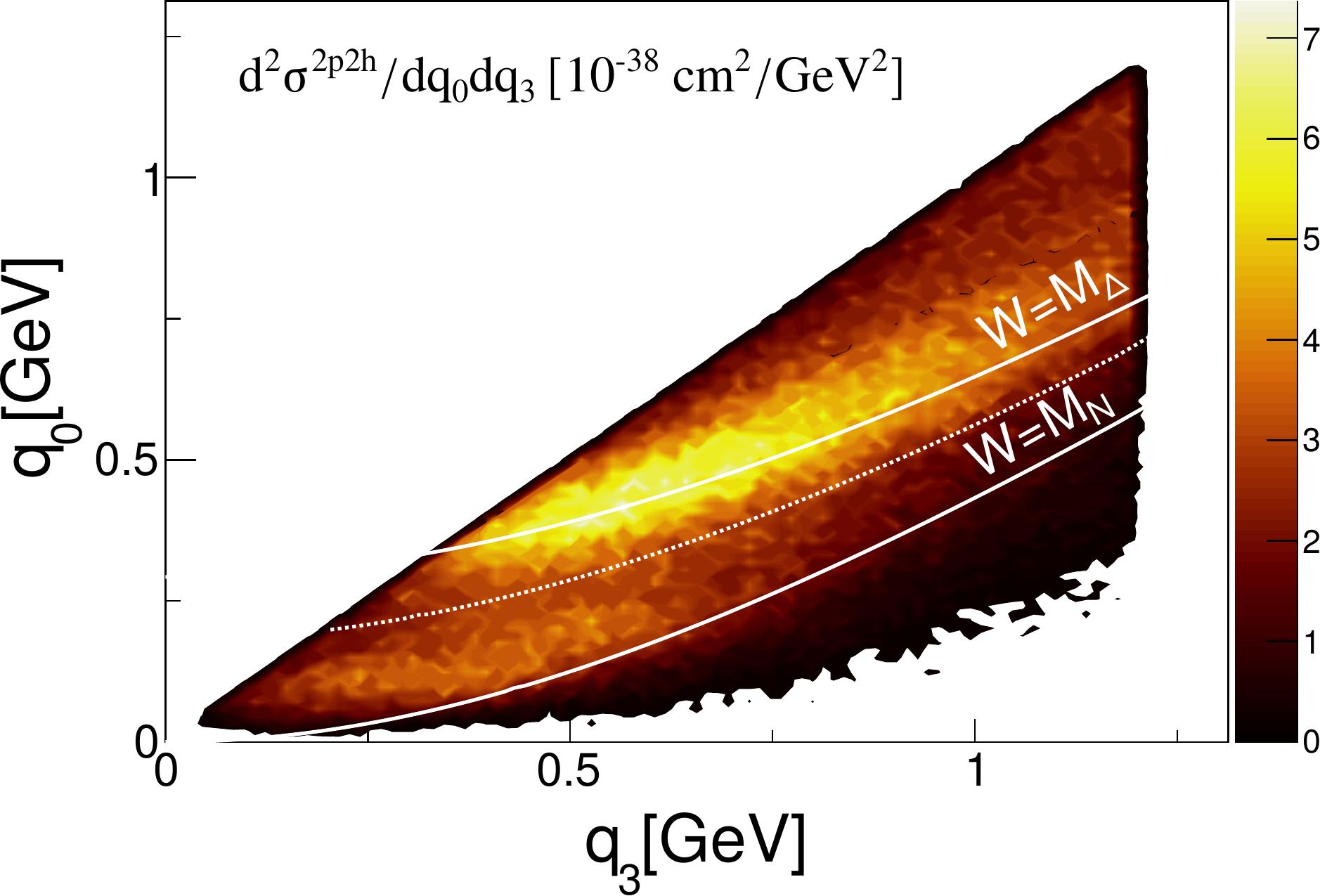}
    \caption{Double-differential $\nu_\mu$-$^{12}$C \ac{CC}\ac{2p2h} cross section from the Valencia model in GENIE. Lines of constant $W$ at $W=M_N=938$~MeV/c$^2$, $W=M_{\text{Dip}}=1120$ MeV/c$^2$ (dotted line) and $W=M_\Delta=1232$~MeV/c$^2$ are also shown.}
    \label{fig:ValenciaXSEC}
\end{figure}

The scaling function, $S(W)$, depends linearly on $W$.
In this work, the scaling function is optimized for the Valencia model which has two characteristic peaks in the $q_0$-$q_3$ space, as it can be seen in Fig.~\ref{fig:ValenciaXSEC}.
The peaks are situated at $W=M_N$ and $W=M_\Delta$.
The dip between the two peaks is at $W_{\text{Dip}}$.
This is implemented by imposing the following boundary conditions:
	\begin{itemize}
	  \setlength{\itemsep}{1pt}
	  \setlength{\parskip}{0pt}
	  \setlength{\parsep}{0pt}
        \item $S_{PL,\min}^{\text{2p2h}} \equiv S(W=W_{PL,\min})$
	    \item $S_{N}^{\text{2p2h}} \equiv S(W=M_{N})$
	    \item $S_{\text{Dip}}^{\text{2p2h}} \equiv S(W=W_{\text{Dip}})$
	    \item $S_{\Delta}^{\text{2p2h}} \equiv S(W=M_{\Delta})$
	    \item $S_{PL,\max}^{\text{2p2h}} \equiv S(W=W_{PL,\max})$
	\end{itemize}
The $S^{\text{2p2h}}$ parameters are referred to in this work as \ac{2p2h} scaling parameters.
The limits of the \ac{2p2h} phase space are defined by $W_{PL,\min}$ and $W_{PL,\max}$.
The upper limit is obtained by simply imposing $Q^2=0$.
The lower limit is parametrized as a function of $q_0$ and $q_3$.
This is an empirical approach that breaks the intrinsic microscopic model and it is only used to explore a possible dependency of the \ac{2p2h} cross section on $W$.
In all GENIE v3 \ac{CMC}s, the \ac{2p2h} scaling parameters are set to $1$.

Only three out of the five \ac{2p2h} scaling parameters are included in the tune: $S^{\text{2p2h}}_N$, $S^{\text{2p2h}}_\Delta$ and $S^{\text{2p2h}}_{PL,\max}$.
Events with $W<M_N$ are negligible for all CC0$\pi$ measurements of interest for this work, hence, $S^{\text{2p2h}}_{PL,\min}$ is not included in the tune.
In addition, $S^{\text{2p2h}}_{\text{Dip}}$ is also not included as the region between $N$ and $\Delta$ peaks is too narrow in $W$ and the data cannot be sensitive to such parameter.
In order to facilitate readability, the $S^{\text{2p2h}}_{PL,\max}$ parameter is redefined as $S^{\text{2p2h}}_{PL}$.
In the particular case of T2K ND280, variations of $S_{\Delta}^{\text{2p2h}}$ and $S_{PL}^{\text{2p2h}}$ do not affect the CC0p0$\pi$ predictions.
This is highlighted in Fig~\ref{fig:T2KMultiplicity}, where only events with $W<W_{\text{Dip}}$ contribute to the \ac{2p2h} cross section prediction with no protons above the detection threshold.
Therefore, these parameters are not included when tuning against T2K ND280 CC0p0$\pi$ data.

\begin{figure}
\includegraphics[width=\textwidth]{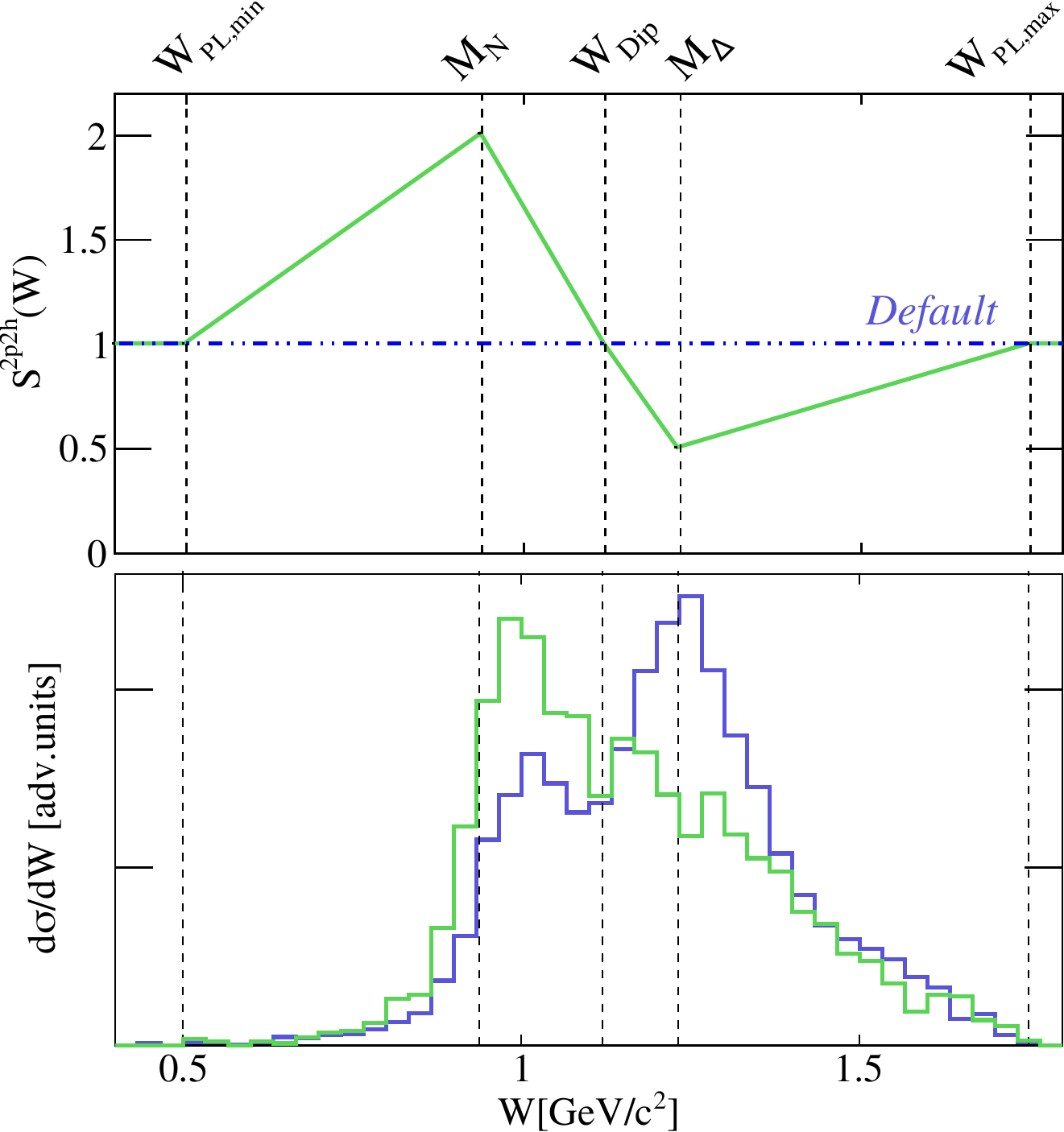}
\centering
    \caption{Graphic representation of the \ac{2p2h} scaling as a function of $W$. On the top, the \emph{default} parametrization (dashed blue) and an example scaling function (green) are shown. The highlighted dashed vertical lines correspond to the tunable scaling parameters for $W=W_{PL}^{\min}$, $M_{N},\,W_{\text{Dip}}$, $M_{\Delta}$, and $W_{PL}^{\max}$.
    The bottom figure shows the Valencia \ac{2p2h} flux-integrated cross section as a function of $W$ for the \texttt{G18\_10a\_02\_11b} tune in blue, and the same prediction scaled with the example scaling function in green.
    This plot is obtained simulating $\nu_\mu$ interactions on $^{12}$C with the \ac{NuMI} $\nu_\mu$ low energy configuration~\cite{PhysRevD.94.092005}.}
    \label{fig:MECScaleVsW}
\end{figure}

The dependency of the scaling function with $W$ for a particular set of parameters is shown in Fig.~\ref{fig:MECScaleVsW} (top).
This particular example enhances (suppresses) the \ac{2p2h} cross-section peak in the $W\!=\!M_N$ ($W\!=\!M_\Delta$) region.
The example scaling function considers
$S^{\text{2p2h}}(M_N)\!=\!2$, $S^{\text{2p2h}}(M_\Delta)\!=\!0.5$, and
$S^{\text{2p2h}}(W_{PL,\min})\! =\! S^{\text{2p2h}}(W_{\text{Dip}})\! =\! S^{\text{2p2h}}(W_{PL,\max})\!=\!1$. 
The effect on the predictions of interest for this paper depends on the neutrino energy, proton multiplicity and proton momenta, as discussed in Sec.~\ref{sec:nucleardata}.

\subsection{Charged-current shallow-inelastic implementation}

\ac{SIS} events also contribute to the CC0$\pi$ signal as pions can be absorbed by the nuclear medium.
Therefore, \ac{SIS} mismodeling impacts the interpretation of the measurements and must be considered in the tune.
The parameters available in GENIE to modify the \ac{RES} and \ac{NRB} background are:
\begin{enumerate}
	\item \ac{RES} axial mass, $M_{A}^{\text{RES}}$;
	\item \ac{RES} scaling factor, $S_{\text{RES}}$;
	\item \ac{SIS} scaling parameters that depend on the initial state, $R_{\nu p}^{\text{CC}1\pi}$, $R_{\nu n}^{\text{CC}1\pi}$, $R_{\nu p}^{\text{CC}2\pi}$, $R_{\nu n}^{\text{CC}2\pi}$.
\end{enumerate}
These parameters have been previously tuned against hydrogen and deuterium data~\cite{mypaper_1}.

This is a lesser issue for MiniBooNE and T2K ND280 CC0$\pi$ data, more significant for the higher energy MINER$\nu$A data.
Nuclear effects in SIS and \ac{DIS} remain imperfectly understood and are therefore an important open area, both for the current study as well as future neutrino-nuclear interaction research. Nuclear-medium effects were studied for pion and electron beams~\cite{Freedman:1982yp} and found to be moderately significant.

The $S_{\text{RES}}$ parameter is the only \ac{SIS} parameter included in the CC0$\pi$ tune.
\ac{NRB} parameters are not included: single pion \ac{NRB} parameters have a small impact on the CC0$\pi$ predictions. In addition, higher multiplicity \ac{SIS}/\ac{DIS} contributions are negligible.
In later instances, we refer to \ac{SIS}/\ac{DIS} contributions as \ac{DIS}.


\subsection{Discussion}
\label{sec:issues}
The choice of tuning parameters is always complicated as these must sample the core physics dependencies with minimal correlation. In the $\mu$\texttt{BooNE} tune~\cite{MicroBooNE:2021ccs}, only four parameters were used with an emphasis on \ac{RPA} and \ac{2p2h} modeling. Although the \ac{RPA} and \ac{2p2h} components are still important here, additional parameters are used to examine these aspects more fully. Since this exercise uses a broader range of neutrino energy, more parameters are needed to account for pion production. However, this contribution is small at neutrino energies $\sim$1 GeV and, although larger for MINER$\nu$A, we find that a single normalization parameter is sufficient to describe the CC0$\pi$ data included in this study. Additional potential parameters are introduced here and discussed more fully in Appendix~\ref{sec:AddlParams}.

Similarly, as we discuss in more detail in Sec.~\ref{sec:priors}, there can in principle be nonneglible correlations among the parameters associated with the nuclear models tuned in this current study and those associated with single-nucleon degrees of freedom as explored in Ref.~\cite{mypaper_1}. 
A possible approach is to fit both sets of parameters comprehensively. 
In the present work we concentrate on a more targeted partial tune of these nuclear parameters in order to map their relationship to the corresponding data taken on nuclear targets. 
This is further justified by the fact that the leading sensitivity to the nuclear parameters is provided by the nuclear data fitted here. 
Ultimately, however, performing nuclear tunes with frozen single-nucleon parameters can be expected to influence the resulting nuclear tune through the correlations mentioned above; systematically disentangling these correlations will require a more global comprehensive tune involving simultaneous fits of both types of data, an undertaking which will be informed by the present study with respect to model priors, methodology, and an understanding of compatibility of nuclear data sets explored in partial tunes as discussed below.

In terms of specific nuclear model choices, the nuclear binding energy is a complicated topic that we quantify through a single number in existing GENIE models which is independent of the momentum distribution. This is adequate for inclusive electron scattering~\cite{Moniz:1971mt}. In more sophisticated treatments of semi-exclusive data, the binding energy and the missing momentum are interrelated via spectral functions~\cite{BENHAR1994493}. Any binding energy parameters are found to be highly correlated with the other parameters chosen for tuning. We choose to leave this out of the tuning procedure and show the effect of these parameters in Appendix~\ref{sec:AddlParams}.

Similarly, FSI has been studied for many years and there are many disagreements about the proper treatment~\cite{dytman2021comparison}.
Although this is a natural aspect of a full tune, the CC0$\pi$ data are not particularly sensitive to this aspect;
\ac{FSI} parameters are most sensitive to CCNp0$\pi$ data.
A global analysis of CC0$\pi$ and CCNp0$\pi$ is out of the scope of this analysis and it is left for future iterations of this work.
We show some interesting CCNp0$\pi$ sensitivities in Appendix~\ref{sec:AddlParams}.

\section{Tuning procedure}
\label{sec:tuningprocedure}

This section summarizes the tuning procedure for the analysis.
The main goal is to tune GENIE against MiniBooNE, T2K ND280 and MINER$\nu$A CC0$\pi$ data.

\subsection{Construction of the GENIE prediction}
In order to build the prediction associated with each dataset specified in Sec.~\ref{sec:nucleardata}, we generate $\nu_\mu$ and $\overline{\nu}_\mu$ CC events for the experiment target using the neutrino fluxes from Fig.~\ref{fig:fluxes}.
In this work, the events are generated with the \texttt{G18\_10a\_02\_11b} tune~\cite{mypaper_1}.

To compute the prediction associated with the $i$th dataset, we generate $N^{\text{TOT}}_i$ events.
Events that do not satisfy the corresponding selection criteria specified in Sec.~\ref{sec:nucleardata} are rejected.
The number of accepted events in the $j$th bin is $N^i_{j}(\boldsymbol{\theta})$.
$\boldsymbol{\theta}$ is the vector of tunable parameters specified in Tab.~\ref{tab:ParameterVariation}.

We build the corresponding $n$-differential flux-integrated cross section prediction for a given set of observables, $O$, as
\begin{align*}
\left(\frac{\partial^n\sigma_{th}(\boldsymbol{\theta})}{\partial O^n}\right)^{i}_j
= &\ \frac{N^i_{j}(\boldsymbol{\theta})}{\Phi_i N^{\text{TOT}}_i \Delta O^i_{j}}  \\ 
  &\ \times\int dE_\nu \frac{d\phi_i}{dE_\nu} \sum_{T_i} R_{T_i} \sigma_{T_i}(E_\nu,\boldsymbol{\theta}),
\end{align*}
where $\Phi_i$ is the integrated flux for the $i$th dataset,
$\Delta O^i_{k}$ corresponds to the $j$th $n$-dimensional bin volume for the quantities used in the differential cross section calculation, and
$d\phi/dE_\nu$ is the expected flux at a given neutrino energy.
For a target mix, the averaged cross section is evaluated by summing over the nucleus type in the target mix, $T_i$.
The ratio of a specific nucleus type with respect to the total nuclei is $R_{T_i}$ and $\sigma_{T_i}(E_\nu)$ is the total cross section for a given nucleus type.

\subsection{Avoiding the Peele’s Pertinent Puzzle}
\label{sec:PPP}

The bin-to-bin covariance matrix provided by each experiment is considered in the evaluation of the $\chi^2$.
The T2K ND280, MicroBooNE and MINER$\nu$A datasets have highly correlated bin-to-bin covariance matrices.
Previous attempts to fit neutrino-nucleus data using the full covariance matrices result in a significant reduction of the cross section~\cite{Chakrani:2022tey,MicroBooNE:2021ccs,NuWroNuclearTune}.
These results are not surprising in highly correlated bins ($\rho>60\%$) in the Gaussian approximation~\cite{PPP2}.
This is known as \ac{PPP}~\cite{PPP1,PPP2}.

To avoid \ac{PPP}, we change our variables in order to reduce the correlation for the $i$th dataset using the following prescription:
\begin{equation*}
Z_{j}^i \equiv \begin{cases} \sum\limits_{k} D^i_{k} & j = 0, \\
\dfrac{D^i_{j}}{\sum\limits_{k} D^i_{k}} & 0 < j < N^i. \end{cases}
\label{eqn:new_variable}
\end{equation*}
$D^i_{j}$ corresponds to the $i$th dataset mean value at the $j$th bin.
The $j$th and $k$th indices run over the number of bins associated with the $i$th dataset.
This is known as \ac{NS} transformation.
After the \ac{NS} transformation, the integral is moved into the first bin of the $i$th dataset, whilst the rest describes the shape distribution.
This transformation is applied to both data and predictions.

The bin-to-bin covariance associated with the $i$th dataset, $\Sigma_{D}(\mathbf{D})^i_{jk}$, transforms as follows:
\begin{equation*}
   \Sigma_{NS}(\mathbf{Z})_{jk}^i \equiv \left[ \left( \frac{d \mathbf{Z}}{d \mathbf{D}}\right) \Sigma_{D}(\mathbf{D})
   \left(\frac{d\mathbf{Z}}{d\mathbf{D}}\right)^T\right]^i_{jk},
\end{equation*}
where
\begin{eqnarray*}
   \left(\frac{d\mathbf{Z}}{d\mathbf{D}}\right)_{ja}^i & = & \begin{cases} 1 & j = 0 \\
   \dfrac{\delta_{ja}\left(\sum\limits_{k} D_{k}^i\right)-D_{j}^i}{\left(\sum\limits_{k}D_{k}^i\right)^2} & 0 < j < N_i. \end{cases}
\end{eqnarray*}
After the \ac{NS} transformation the relative uncertainties are constant when the normalization changes. 

The same transformation is applied to the prediction mean values and covariance. 
Before the \ac{NS} transformation, the prediction covariance only has diagonal elements.
This is not true after the \ac{NS} transformation.
However, the off-diagonal elements on the prediction covariance are small and are neglected in this work.
The prediction central values and errors after the \ac{NS} transformation are denoted as $Y^i_{j}(\theta)$ and $\delta Y^i_{j}(\theta)$ respectively.


\subsection{Professor parametrization}

\begin{figure}
    \centering
    \includegraphics[width=0.92\textwidth]{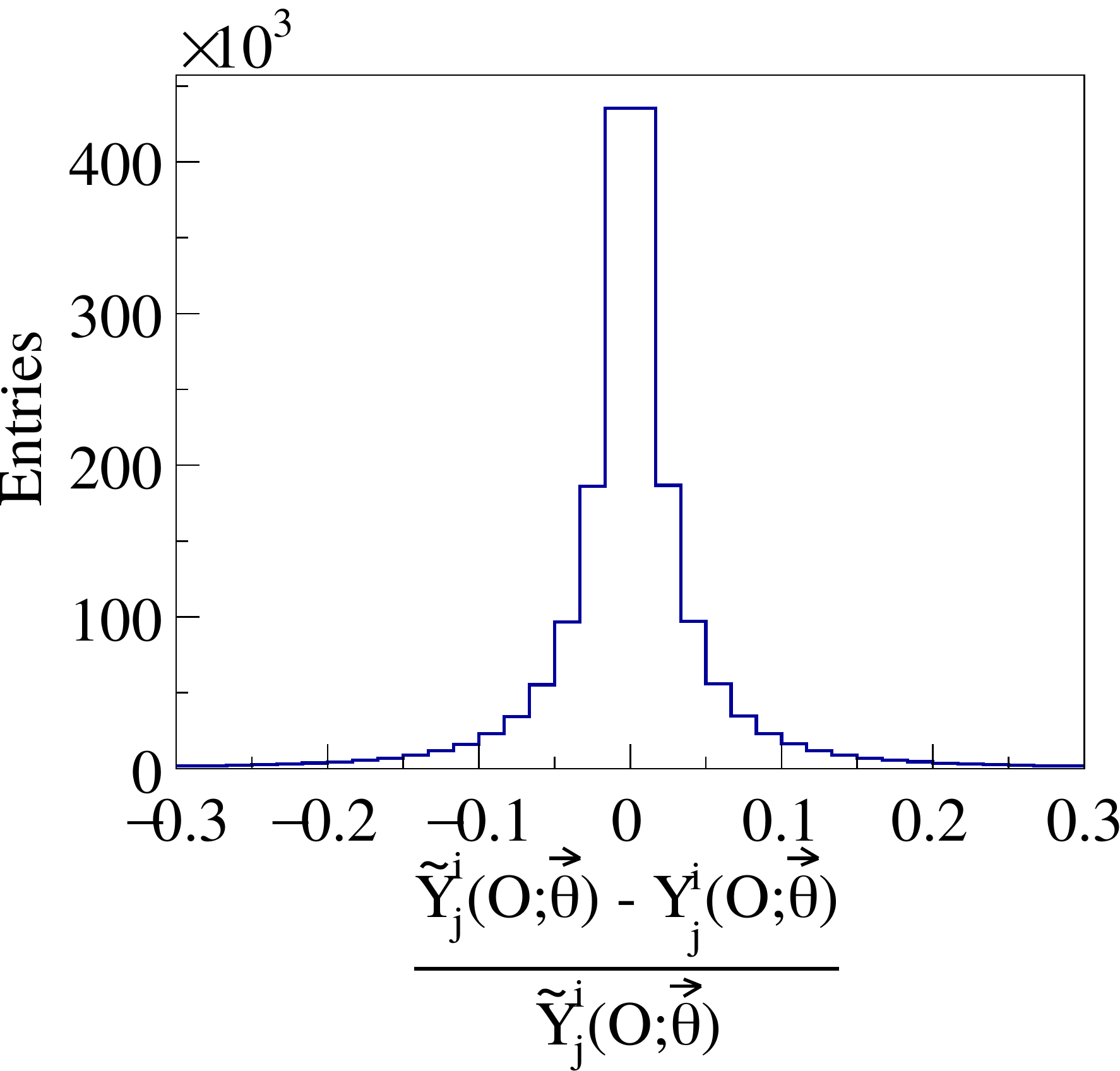}
    \caption{Fractional difference between true \ac{MC} predictions in the \ac{NS} space calculated with a given $\mathbf{\theta}$ parameter set.
    \label{fig:Residual}}
\end{figure}

Given that performing a multi-parameter brute-force scan is not feasible, we use Professor~\cite{Professor} to parametrize the behavior of our predicted cross section and error in each bin in the \ac{NS} space.
We refer to this quantities as $\tilde{Y}_{j}^i(\boldsymbol{\theta})$ and $\delta\tilde{Y}_{j}^i(\boldsymbol{\theta})$.
In this particular tune, we opted for a fourth-order parametrization.
This work, where originally eleven parameters were included in the analysis, requires a total of 2k event generations with $\mathbf{\theta}$ sampled across the ranges specified in Tab.~\ref{tab:ParameterVariation}.
The accuracy of the parametrization is shown in Fig.~\ref{fig:Residual}.
The distribution is centered at zero with a standard deviation of 0.05.
This distribution is similar to previous GENIE tunes~\cite{mypaper_1}.
This parametrization is used for the estimation of the best-fit values by minimizing the $\chi^2$.

The accuracy of the parameterization can be improved by increasing the order of the polynomial. 
However, an increase in the order of the polynomial is computationally expensive. 
For instance, a fifth-order polynomial requires 6k generations. 
A fourth-order polynomial is enough to describe the MC response in this work.

\subsection{Discussion of data-driven priors}
\label{sec:priors}

\begin{figure}
    \centering
    \includegraphics[width=0.95\textwidth]{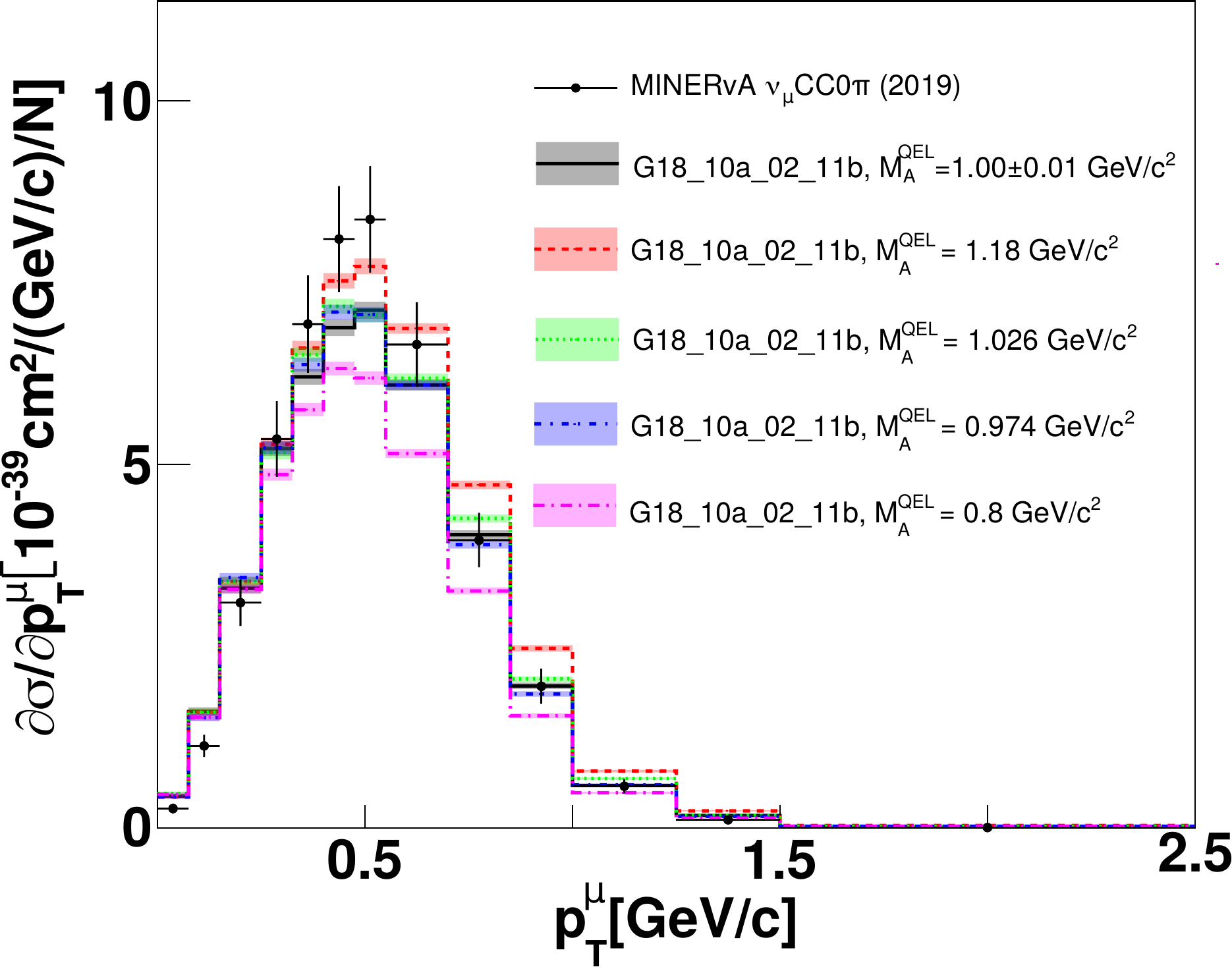}
    \caption{Impact of $M_A^{QEL}$ variations on MINER$\nu$A CC0$\pi$ flux-integrated differential cross section predictions as a function of $p_T$. The red line corresponds to the GENIE prediction computed with the $M_A^{QEL}$ best-fit value from the $\mu$\texttt{BooNE} tune~\cite{MicroBooNE:2021ccs}. No other parameters are modified from their nominal values.}
    \label{fig:effectMAQE}
\end{figure}

The basic structure of this tune is based on the model of separate nucleon and nucleus efforts.
Although the emphasis here is on neutrino-nucleus parameters, some of the parameters of interest were already tuned to neutrino-nucleon data~\cite{mypaper_1}.
Particularly, the \texttt{G18\_10a\_02\_11b} tune with hydrogen and deuterium data provided with data-driven constraints for $M_A^{\text{QEL}}$ and $S_{\text{RES}}$~\cite{mypaper_1}.
These parameters are crucial for the description of free-nucleon data and are strongly correlated with other aspects of the nuclear tune.
This correlation was observed in the $\mu$\texttt{BooNE} tune, leading to best-fit results with $M_A^{QE}=1.18\pm0.08$~GeV/c$^2$~\cite{MicroBooNE:2021ccs}.
The effect of varying $M_A^{QEL}$ on the MINER$\nu$A $\nu_\mu$CC0$\pi$ prediction is shown in Fig.~\ref{fig:effectMAQE}.
In this work, we chose to constrain $M_A^{QEL}$ and $S_{RES}$ using data driven priors from Ref.~\cite{mypaper_1}.
The information on the parameter priors central values as well as the correlation between the two parameters out of the free-nucleon tune is included in the $\chi^2$ minimization.
The complete information on the priors is provided in Tab.~\ref{tab:PriorsFreenucleonTune_NuclearTune}.
In this analysis we also include priors on $\omega_{\text{RPA}}$ and $\omega_{\text{No\,RPA}}$, as discussed in Sec.~\ref{subsec:CCQELParam}.

\begin{table}
	\centering
    \begin{subtable}{\textwidth}
    \centering
    \begin{tabular}{@{\extracolsep\fill} c c }
    \hline\hline\noalign{\smallskip}
    Parameter          & Prior                   \\
    \noalign{\smallskip}\hline\noalign{\smallskip}
    $M_A^{\text{QEL}}$ & $1.00\pm0.01$ GeV/c$^2$ \\
    $S_{\text{RES}}$   & $0.84\pm0.028$          \\
    \noalign{\smallskip}\hline\hline
    \end{tabular}
    \caption{}
\end{subtable}
\begin{subtable}{\textwidth}
    \centering
    \begin{tabular}{c c c} \hline\hline\noalign{\smallskip}
                                 & $M_A^{\text{QEL}}$   & $S_{\text{RES}}$     \\
    \noalign{\smallskip}\hline\noalign{\smallskip}
    $M_A^{\text{QEL}}$           & $1.8 \times 10^{-4}$ & $1.5 \times 10^{-4}$ \\
    $S_{\text{RES}}$             & $1.5 \times 10^{-4}$ & $6.0 \times 10^{-4}$ \\
    \noalign{\smallskip}\hline\hline
    \end{tabular}
    \caption{}
    \end{subtable}
    \caption{Priors~(a) and covariance matrix~(b) for $M_A^{\text{QEL}}$ and $S_{\text{RES}}$ obtained to the free-nucleon tune from Ref.~\protect\cite{mypaper_1}.}
    \label{tab:PriorsFreenucleonTune_NuclearTune}
\end{table}

\subsection{Evaluation of the \ensuremath{\chi^2}}

The complete form for our $\chi^2$ is:
\begin{align}
\chi^2 (\boldsymbol{\theta}) 
= &\ \sum_{i}^{\mathcal{N}}\sum_{j,k}^{N_i}\sqrt{w^i_{j}}\Delta\tilde{Y}^i_{j}(\boldsymbol{\theta})(\Sigma_{NS,\,jk}^{i})^{-1}\Delta \tilde{Y}^i_{k}(\boldsymbol{\theta})\sqrt{w^i_{k}} \nonumber \\ 
 &\ +\left(\boldsymbol{\theta}-\boldsymbol{\theta}_0\right)^T\Sigma^{-1}_{\theta}\left(\boldsymbol{\theta}-\boldsymbol{\theta}_0\right),
\label{eq:chi2NuclearTune}
\end{align}
being $i$ the index that runs over the $\mathcal{N}$ datasets considered in the fit.
$\Delta\tilde{Y}^i_{j}$ is the difference between the \ac{NS} parametrization prediction and the $i$th dataset at the $j$th bin, $\Delta\tilde{Y}^i_{j}(\boldsymbol{\theta}) \equiv \tilde{Y}^i_{j}(\boldsymbol{\theta})-Z^i_{j}$.
The $\omega_{ij}$ is the weight applied to $j$th bin from the $i$th dataset. 
In this work, weights are used to include or exclude data from the analysis. 
In other words, they are either 1 or 0.
The prediction errors, $\delta\tilde{Y}^i_{j}(\boldsymbol{\theta})$, are added in quadrature to $\Sigma_{NS}$.
The second term takes care of correlated priors in our fit. 
$\boldsymbol{\theta}_0$ and $\Sigma_\theta$ are the central values vector and the covariance matrix of the priors for the parameters of interest.
The details on the priors applied in this analysis are described in Sec.~\ref{sec:priors}.


\section{Tuning Results}
\label{sec:NuclearTuningResults}

We adopt the following naming scheme to characterise each of the partial GENIE tunes presented in this work:
\begin{center}
	\texttt{Gxx[a-d]}.
\end{center}
Here
\begin{itemize}
	\item[\texttt{G}] is a capital letter that stands for GENIE, highlighting the authorship of the	tunes.
	\item[\texttt{xx}] is a number assigned to each experiment, i.e., MiniBooNE (10), T2K ND280 (20) or MINER$\nu$A (30). When using antineutrino datasets, \texttt{xx} is increased by one unit. For CCNp0$\pi$ datasets, \texttt{xx} is increased by five units.
	\item[\texttt{[a-d]}] refers to the alternative intranuclear hadron model used in the analysis: (a) INTRANUKE/\emph{hA}, (b) INTRANUKE/\emph{hN}, (c) GEANT4/Bertini and (d) INCL++. 
\end{itemize}
Note that this is different from the standard naming scheme used for the tunes released through the GENIE platform. The standard naming convention from Ref.~\cite{mypaper_1} will be used if one or more of the tunes produced in this work or future iterations is prepared for release in GENIE.

In total, six partial tunes are performed: three tunes on neutrino CC0$\pi$ data, two tunes using antineutrino CC0$\pi$ data and one tune using $\nu_\mu$ CCNp0$\pi$ data.
The tunes on CC0$\pi$ data aim to explore avenues for improving the agreement between GENIE and data, consolidate the main elements of the GENIE CC0$\pi$ tuning methodology and provide a common ground for the discussion of tensions.
The tune on CCNp0$\pi$ data aims to highlight tensions between CC0$\pi$ and CCNp0$\pi$ datasets.
All of the tunes presented in this work consider carbon datasets only. 
Joint fits to all available data will be performed at a future iteration of this work, aiming to produce the tunes that will be publicly released through the GENIE platform.

In all CC0$\pi$ tunes, the analyses are carried out using double-differential CC0$\pi$ data as a function of muon kinematics.
Preference is given to datasets that do not require a minimum number of protons above detection threshold in the final state.
Whenever CC0$\pi$ datasets are not available for a particular experiment, the tune is performed using CC0p0$\pi$ datasets instead.

\texttt{G18\_10a\_02\_11b} is the starting point for all these tunes and provides the \emph{nominal} predictions.
The corresponding names assigned to each tune prepared for the purposes of this paper are the following:
\begin{itemize}
	\item[\texttt{G10a Tune}]: GENIE tune to MiniBooNE $\nu_\mu$CC0$\pi$ data~\cite{1002.2680}.
	\item[\texttt{G11a Tune}]: GENIE tune to MiniBooNE $\overline{\nu}_\mu$CC0$\pi$ data~\cite{MinibooneAntineutrino}.
	\item[\texttt{G20a Tune}]: GENIE tune to T2K ND280 $\nu_\mu$CC0p0$\pi$ data~\cite{T2KCCQELikeMeasurement}.
	\item[\texttt{G30a Tune}]: GENIE tune to MINER$\nu$A $\nu_\mu$CC0$\pi$ data~\cite{PhysRevD.99.012004}.
	\item[\texttt{G31a Tune}]: GENIE tune to MINER$\nu$A $\overline{\nu}_\mu$CC0p0$\pi$ data~\cite{PhysRevD.97.052002}.
	\item[\texttt{G35a Tune}]: GENIE tune to MINER$\nu$A $\nu_\mu$CCNp0$\pi$ data~\cite{PhysRevLett.121.022504}.
\end{itemize}
Other measurements, including MicroBooNE ones, are used for comparisons only.
Each partial tune is performed following the recipe described in Sec.~\ref{sec:tuningprocedure}.

\subsection{Discussion of partial CC0\ensuremath{\pi} tune results}

\begin{figure*}
    \centering
    \includegraphics[width=\textwidth]{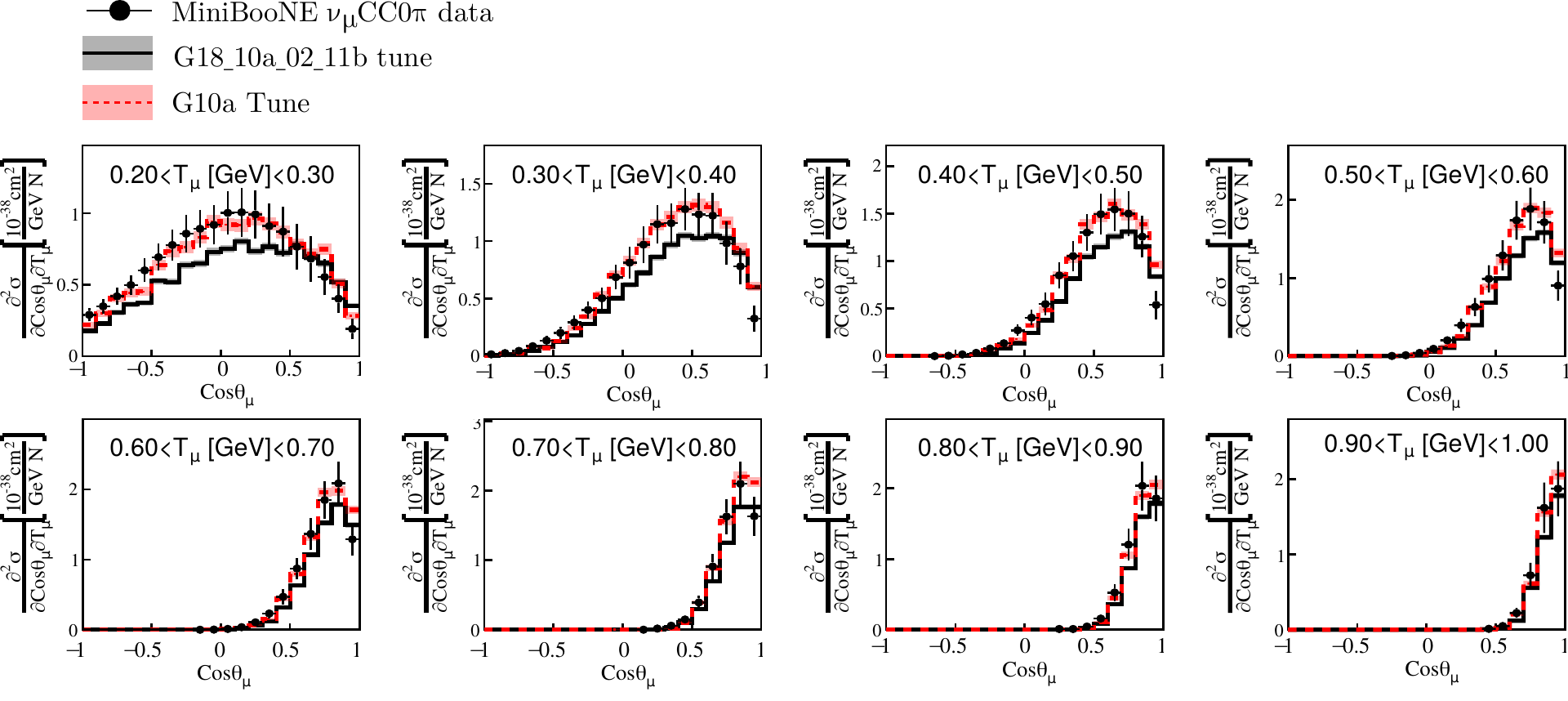}
    \caption{ Comparison of the \texttt{G18\_10a\_02\_11b} and \texttt{G10a} tunes against MiniBooNE $\nu_\mu$ CC0$\pi$ double differential data~\cite{1002.2680}. 
    The comparisons are restricted to the $0.2<T_\mu<1.0$~GeV phase space.
    The predictions are computed using the parameters specified in Tab.~\ref{tab:PartialTuneResults}. 
    The total $\chi^2$ associated with this dataset before and after the tune are reported in Tab.~\ref{tab:summarychi2_CC0piData}.
    \label{fig:MicroBooNEPartialTune}}
\end{figure*}

\begin{figure*}
    \centering
    \includegraphics[width=\textwidth]{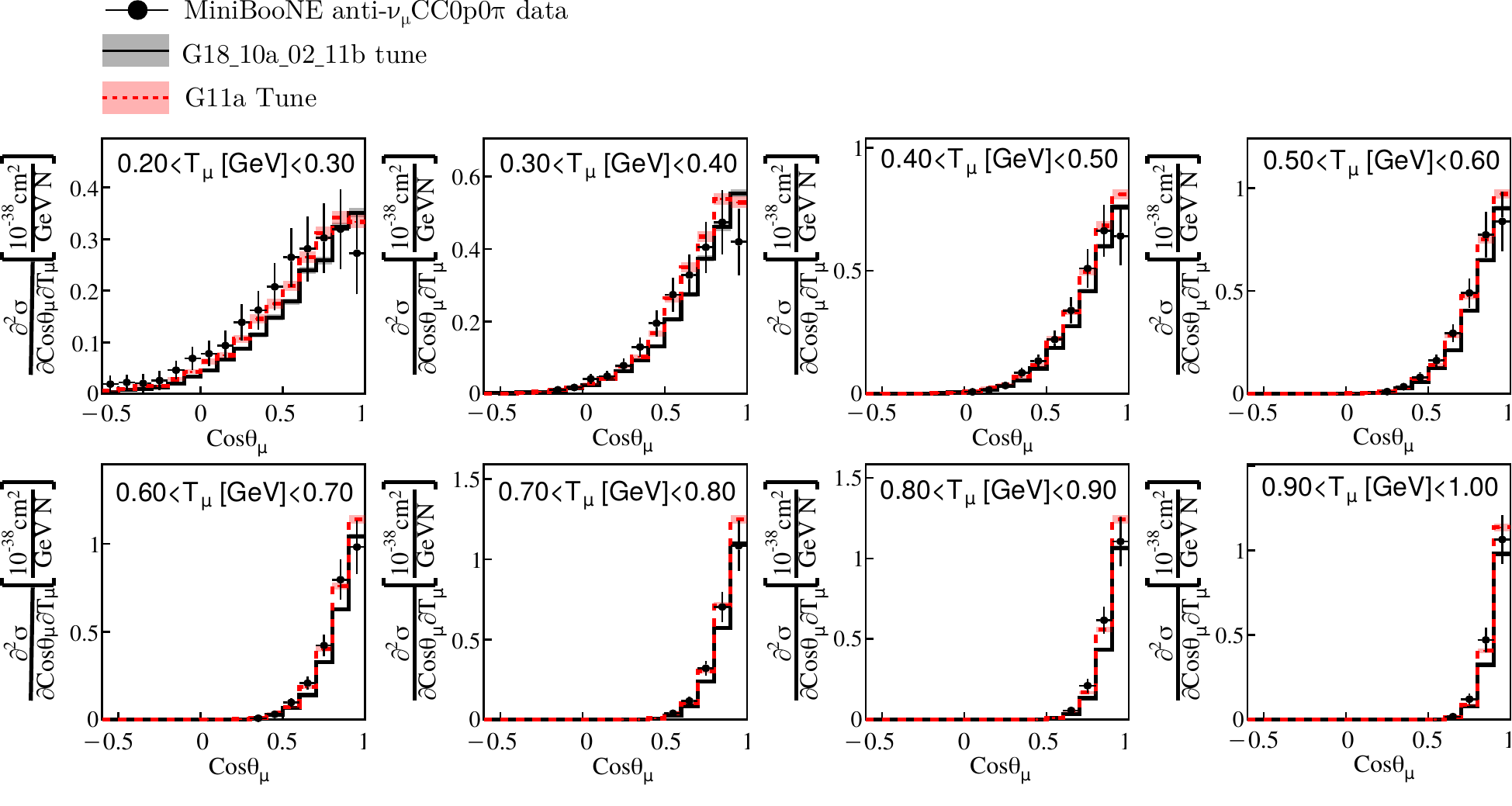}
    \caption{Comparison of the \texttt{G18\_10a\_02\_11b} and \texttt{G11a} tunes against MiniBooNE $\overline{\nu}_\mu$ CC0$\pi$ double differential data~\cite{MinibooneAntineutrino}. 
	The comparisons are restricted to the $0.2<T_\mu<1.0$~GeV phase space.
    The predictions are computed using the parameters specified in Tab.~\ref{tab:PartialTuneResults}. 
    The total $\chi^2$ associated with this dataset before and after the tune are reported in Tab.~\ref{tab:summarychi2_CC0piData}.
    \label{fig:MicroBooNEBarPartialTune}}
\end{figure*}

\begin{figure*}
    \centering
    \includegraphics[width=0.8\textwidth]{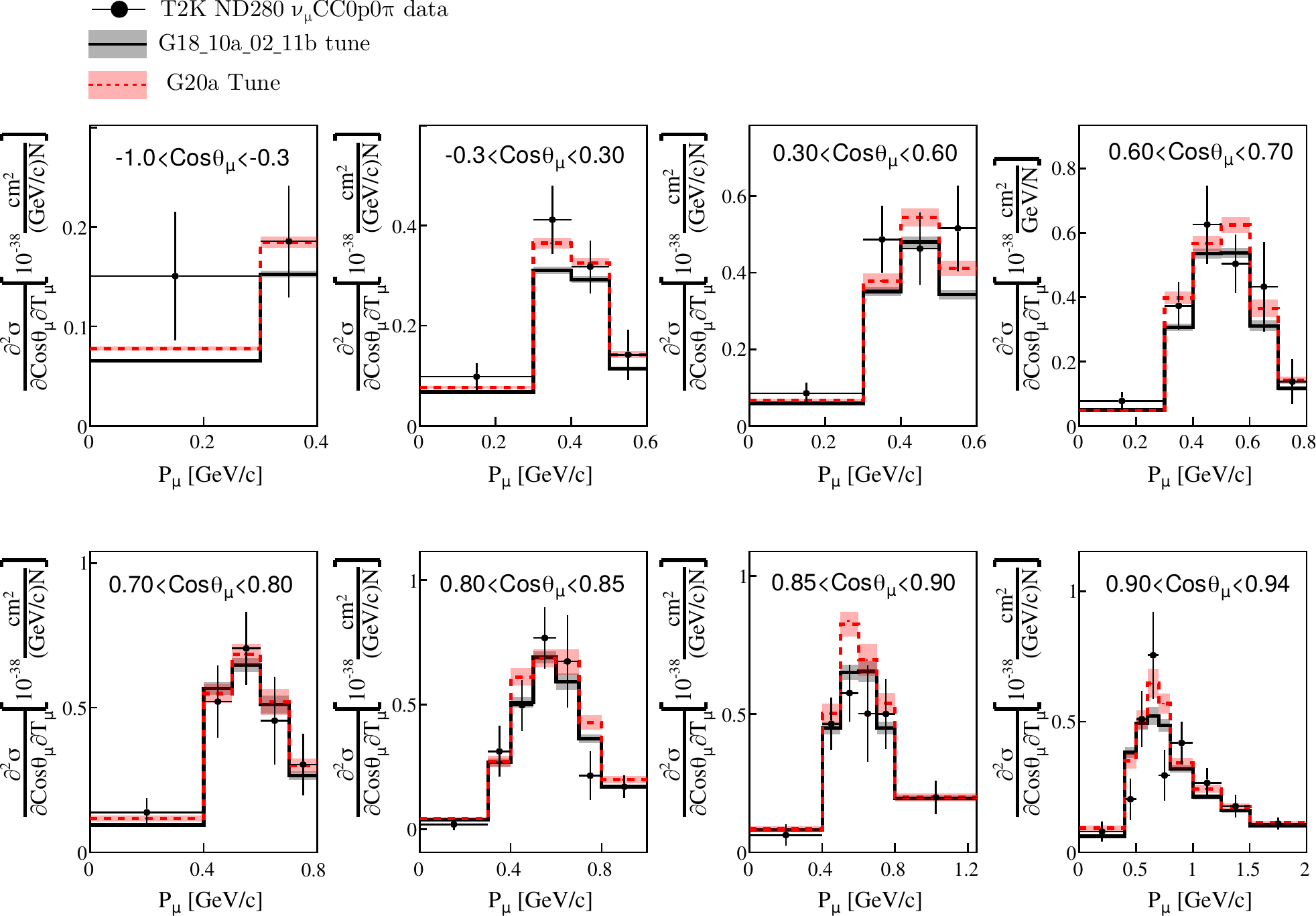}
    \caption{Comparison of the \texttt{G18\_10a\_02\_11b} and \texttt{G20a Tune} against T2K ND280 $\nu_\mu$ CC0p0$\pi$ double differential data~\cite{T2KCCQELikeMeasurement}. 
    The comparisons are restricted to the $-1.0<\cos\theta_\mu<0.94$ phase space.
    The predictions are computed using the parameters specified in Tab.~\ref{tab:PartialTuneResults}. 
    The total $\chi^2$ associated with this dataset before and after the tune are reported in Tab.~\ref{tab:summarychi2_CC0piData}.
    \label{fig:T2KPartialTune}}
\end{figure*}

\begin{figure*}
    \centering
    \includegraphics[width=\textwidth]{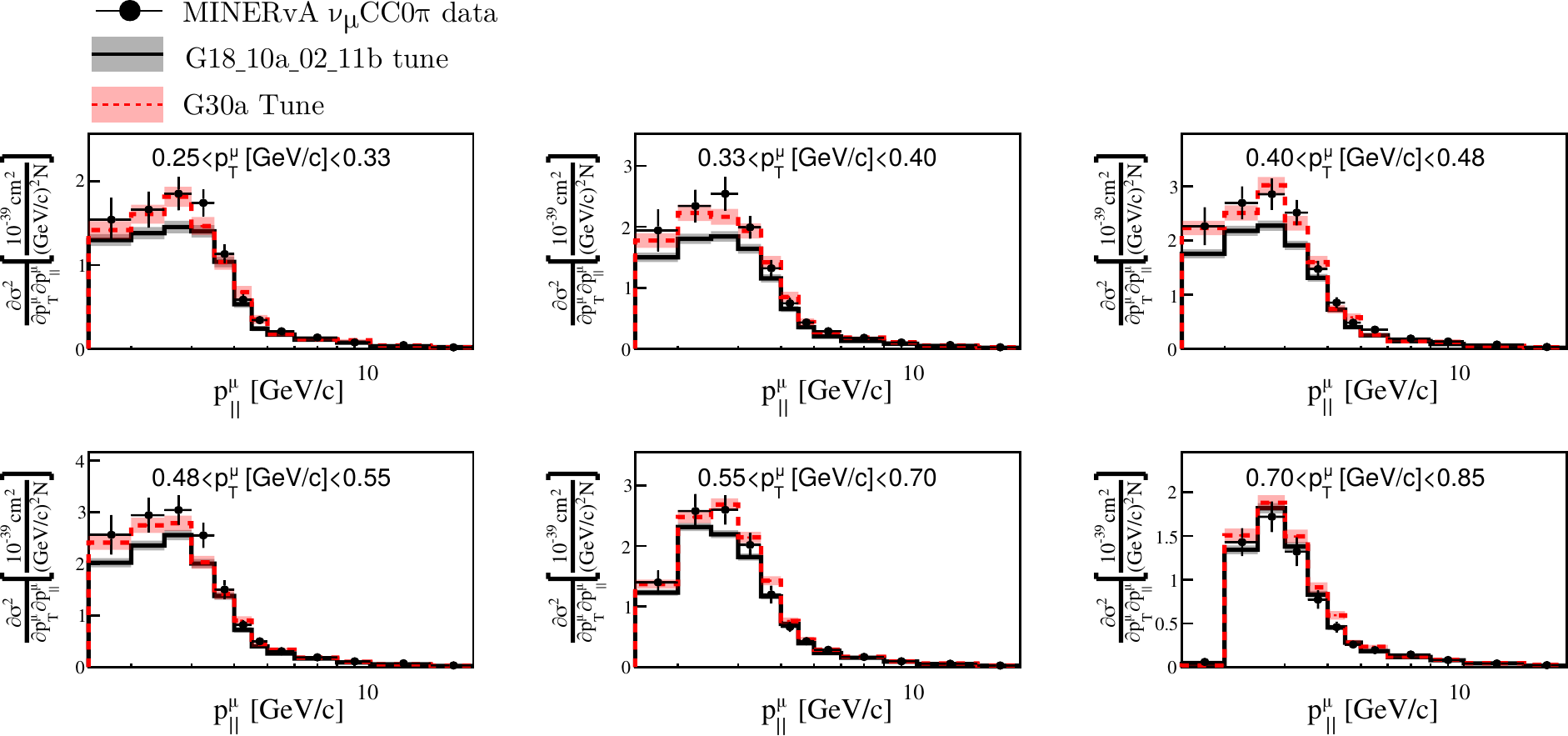}
    \caption{Comparison of the \texttt{G18\_10a\_02\_11b} and \texttt{G30a} tunes against MINER$\nu$A $\nu_\mu$ CC0$\pi$ double differential data~\cite{PhysRevD.99.012004}. 
    The comparisons are restricted to the $0.25<p_{T}<0.85$~GeV/c phase space.
    The predictions are computed using the parameters specified in Tab.~\ref{tab:PartialTuneResults}. 
    The total $\chi^2$ associated with this dataset before and after the tune are reported in Tab.~\ref{tab:summarychi2_CC0piData}.
    \label{fig:MINERvAPartialTune}}
\end{figure*}

\begin{figure*}
    \centering
    \includegraphics[width=0.8\textwidth]{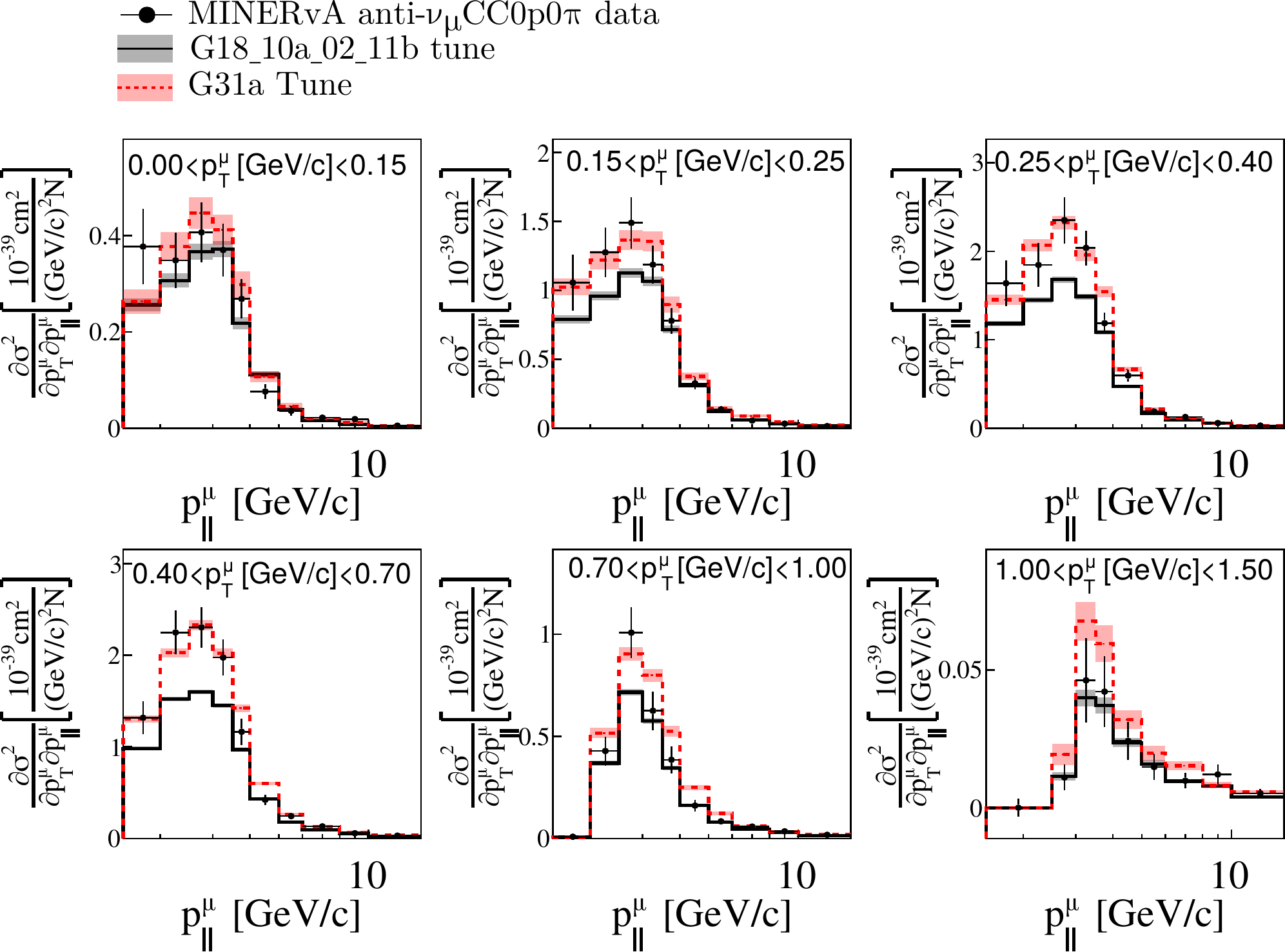}
    \caption{Comparison of the \texttt{G18\_10a\_02\_11b} and \texttt{G31a} tunes against MINER$\nu$A $\overline{\nu}_\mu$ CC0p0$\pi$ double differential data~\cite{MinibooneAntineutrino}. 
    The comparisons are restricted to the $0<p_{T}<1.5$~GeV/c phase space.
    The predictions are computed using the parameters specified in Tab.~\ref{tab:PartialTuneResults}. 
    The total $\chi^2$ associated with this dataset before (after) the tune reported in Tab.~\ref{tab:summarychi2_CC0piData}.
    \label{fig:MINERvABarPartialTune}}
\end{figure*}

\begin{figure}
    \centering
    \begin{subfigure}{0.9\textwidth}
    \centering
    \includegraphics[width=\textwidth]{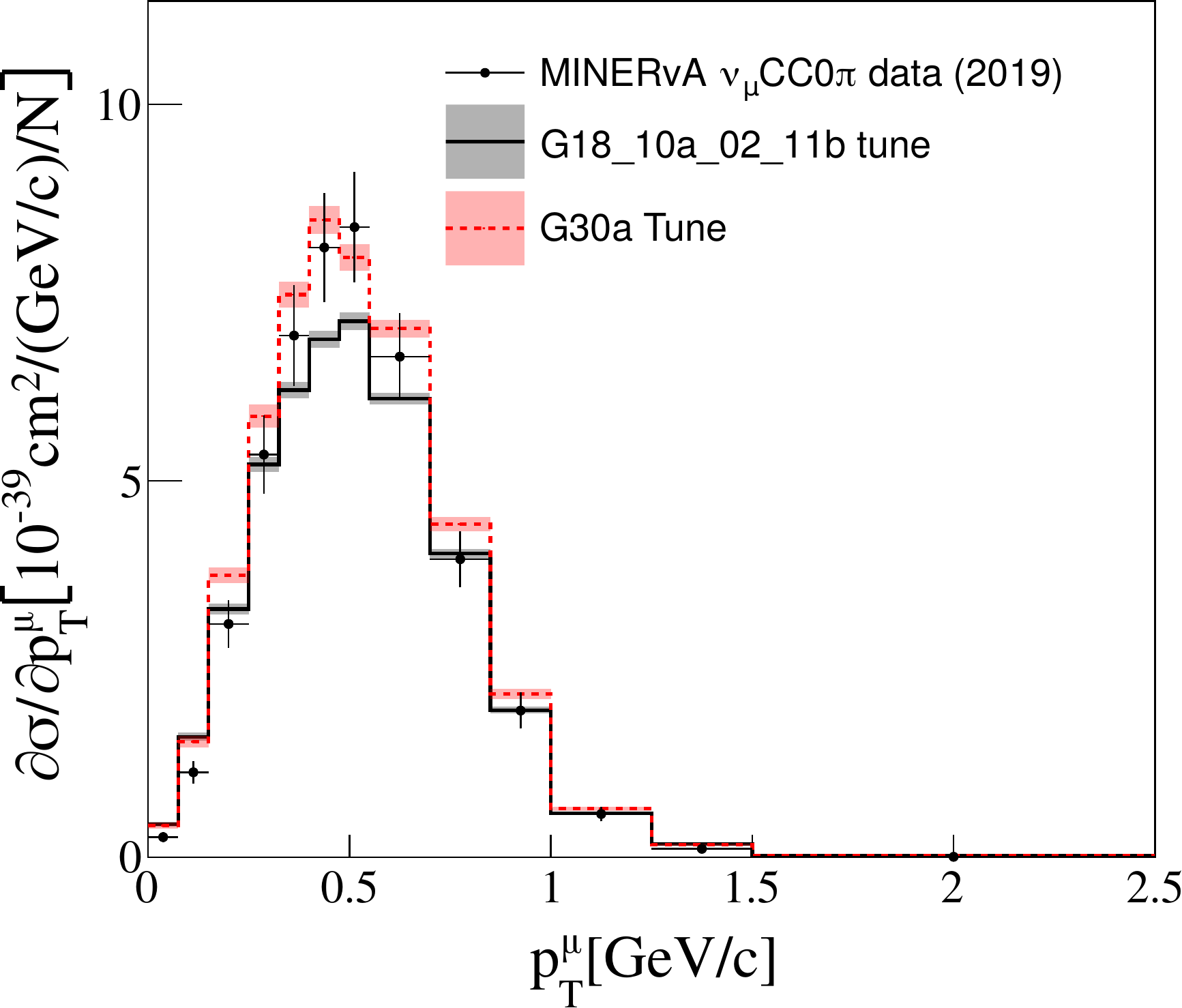}
    \caption{}
    \end{subfigure}
    \begin{subfigure}{0.8\textwidth}
    \end{subfigure}
    \caption{Comparison of the \texttt{G18\_10a\_02\_11b} and \texttt{G30a} tunes against MINER$\nu$A $\nu_\mu$ CC0p0$\pi$ single-differential data~\cite{PhysRevD.99.012004}. 
    The predictions are computed using the parameters specified in Tab.~\ref{tab:PartialTuneResults}.} 
    \label{fig:MINERvAPartialTuneReconstructed}
\end{figure}

\begin{figure}
    \centering
    \begin{subfigure}{0.8\textwidth}
    \centering
    \includegraphics[width=\textwidth]{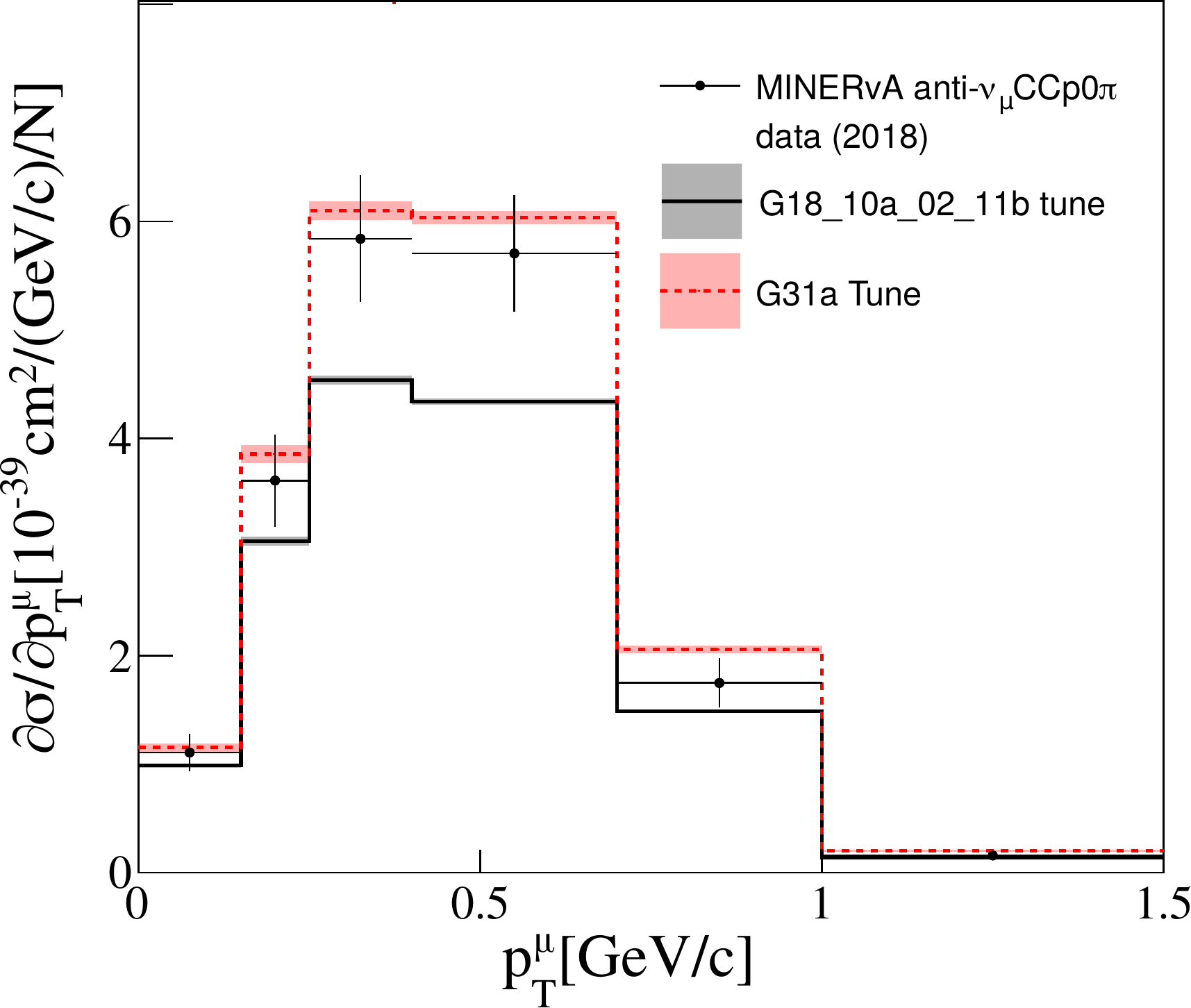}
    \caption{}
    \end{subfigure}
    \begin{subfigure}{0.8\textwidth}
    \centering
    \includegraphics[width=\textwidth]{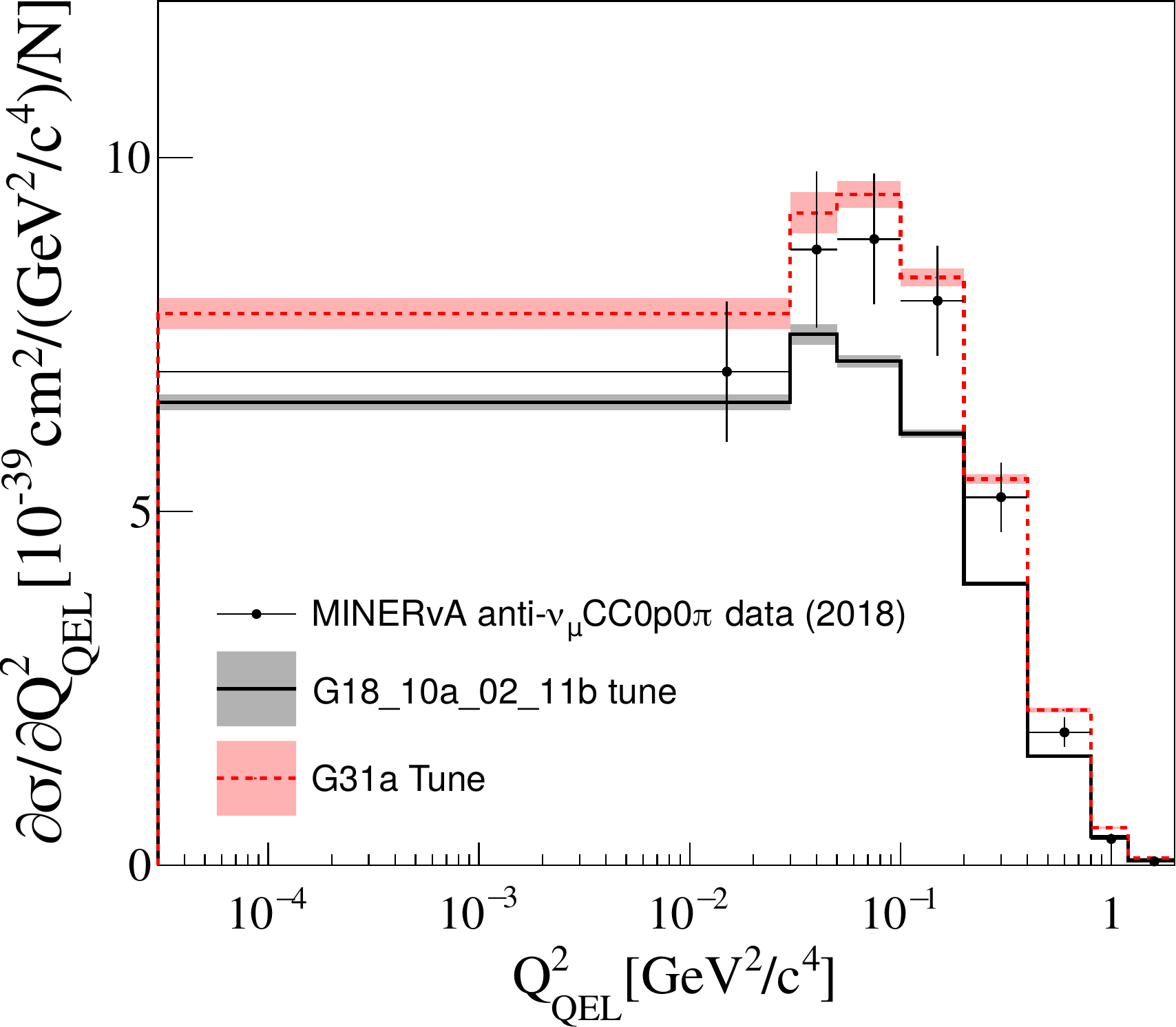}
    \caption{}
    \end{subfigure}
    \caption{Comparison of the \texttt{G18\_10a\_02\_11b} and \texttt{G31a} tunes against MINER$\nu$A $\overline{\nu}_\mu$ CC0p0$\pi$ single-differential data~\cite{MinibooneAntineutrino}. 
    The predictions are computed using the parameters specified in Tab.~\ref{tab:PartialTuneResults}.}
    \label{fig:MINERvABarPartialTuneReconstructed}
\end{figure}

Each tune's best-fit parameter values and the $\chi^2$ calculated with the Professor parametrization at the best-fit point are summarized in Tab.~\ref{tab:PartialTuneResults}.
The \emph{nominal} and best-fit predictions are shown in \Cref{fig:MicroBooNEPartialTune,fig:MicroBooNEBarPartialTune,fig:T2KPartialTune,fig:MINERvAPartialTune,fig:MINERvAPartialTuneReconstructed,fig:MINERvABarPartialTune,fig:MINERvABarPartialTuneReconstructed}.
Tab.~\ref{tab:summarychi2_CC0piData} provides the $\chi^2$ values computed with each tune's GENIE prediction and corresponding dataset.
In this case, the $\chi^2$ values are calculated with the \ac{NS} transformation with the GENIE predictions.
Notice that the $\chi^2$ values from Tab.~\ref{tab:summarychi2_CC0piData} are different to the ones provided in Tab.~\ref{tab:PartialTuneResults}.
This is a consequence of the Professor parametrization not being exact.

It is observed that the description of the data after the tune improved substantially.
For instance, the agreement with MINER$\nu$A $\nu_\mu$ CC0$\pi$ before the tune is $\chi^2_{\text{Nominal}}=626/144$~DoF.
After the tune, $\chi^2_{\text{Nominal}}=151/144$~DoF.
This is mainly a consequence of an improvement in the overall normalization for each partial tune.

\begin{table*}
    {\small
    \centering
    \begin{tabular}{@{\extracolsep\fill} c c c c c c c } \hline\hline\noalign{\smallskip}
    Parameters  & \texttt{G10a Tune} & \texttt{G11a Tune} & \texttt{G20a Tune} & \texttt{G30a Tune} & \texttt{G31a Tune} & \texttt{G35a Tune} \\
    \noalign{\smallskip}\hline\hline\noalign{\smallskip}    
    $M_A^{\text{QEL}}$(GeV/c$^2$) & $1.02\pm0.01$ & $1.01\pm0.01$ & $1.00\pm0.01$ & $1.00\pm0.02$ & $1.00\pm0.01$ & $0.99\pm0.01$ \\
    $\omega_{\text{RPA}}$         & $1.20\pm0.03$ & $1.14\pm0.06$ & $1.2 \pm0.2 $ & $0.9 \pm0.1 $ & $1.3 \pm0.2 $ & $0.75\pm0.3 $ \\
    $\omega_{\text{No\,RPA}}$     & $0.05\pm0.02$ & $0.09\pm0.05$ & $-0.1\pm0.1 $ & $0.2 \pm0.1 $ & $0.2 \pm0.2 $ & $0.09\pm0.3 $ \\
    $S_{\text{RES}}$              & $0.85\pm0.02$ & $0.86\pm0.05$ & $0.84\pm0.02$ & $0.84\pm0.03$ & $0.84\pm0.02$ & $0.84\pm0.02$ \\
    $S_{N}^{\text{2p2h}}$         & $1.5 \pm0.4 $ & $2.3 \pm0.01$ & $1.7\pm0.3$   & $1.2 \pm0.4 $ & $1.7 \pm0.5 $ & $0.33\pm0.2 $ \\
    $S_{\Delta}^{\text{2p2h}}$    & $0.7 \pm0.2 $ & $0.7 \pm0.3 $ & (1.00)        & $2.1 \pm0.2 $ & $2.3 \pm0.2 $ & $0.5 \pm0.4 $ \\
    $S_{PL}^{\text{2p2h}}$        & $0.4 \pm0.1 $ & $0.4 \pm0.1 $ & (1.00)        & $0.9 \pm0.2 $ & $0.4 \pm0.1 $ & $1.5 \pm0.4 $ \\
    \noalign{\smallskip} \hline\noalign{\smallskip} 
    $\chi^2$                      &  89/130       & 77/71         &   60/55       & 61/137        & 67/53         & 17/19         \\
    \noalign{\smallskip}\hline\hline
    \end{tabular}
    \caption{Best-fit parameter values for the different partial tunes. Parameter values within parenthesis are kept fixed during the fit. The $\chi^2$ values are calculated with the Professor parametrization, in accordance to Eq.~\ref{eq:chi2NuclearTune}.
    \label{tab:PartialTuneResults}}}
\end{table*}

\begin{table*}
    {\small
    \centering
    \begin{tabular}{@{\extracolsep\fill} c c c c c c c c c} \hline\hline\noalign{\smallskip}
     Dataset & $\chi_{\text{Nominal}}^2$ & $\chi_{\texttt{G10a}}^2$ & $\chi_{\texttt{G11a}}^2$ & $\chi_{\texttt{G20a}}^2$
             & $\chi_{\texttt{G30a}}^2$  & $\chi_{\texttt{G31a}}^2$ & $\chi_{\texttt{G35a}}^2$ & DoF                                  \\
    \noalign{\smallskip}\hline\hline\noalign{\smallskip}
     MiniBooNE $\nu_\mu$ CC0$\pi$       & 1817 & \textbf{121} & 160          & 314          & 379          & 1279        & 2727 & 137 \\
     MiniBooNE $\overline{\nu}_\mu$ CC0$\pi$ & 444  & 208          & \textbf{214} & 246          & 403          & 491         & 879  & 60  \\
     T2K ND280 $\nu_\mu$ CC0p0$\pi$     & 139  & 447          & 600          & \textbf{123} & 237          & 916         & 239  & 60  \\
     MINER$\nu$A   $\nu_\mu$ CC0$\pi$       & 626  & 252          & 202          & 270          & \textbf{151} & 360         & 953  & 144 \\
     MINER$\nu$A   $\overline{\nu}_\mu$ CC0$\pi$ & 2259 & 1837         & 1680         & 2232         & 1794         & \textbf{82} & 1810 & 78  \\
    \noalign{\smallskip}\hline\hline
    \end{tabular}
    \caption{Summary of $\chi^2$ values associated the CC0$\pi$ datasets specified in each row.
    The $\chi^2$ values are calculated using the \ac{NS} method for seven different GENIE predictions: \texttt{G18\_10a\_02\_11b}, \texttt{G10a}, \texttt{G11a}, \texttt{G20a}, \texttt{G30a}, \texttt{G31a} and \texttt{G35a}.
    The values highlighted in bold correspond to the best-fit $\chi^2$ for the partial tune predictions.
    \label{tab:summarychi2_CC0piData}}}
\end{table*} 

All carbon tunes show similar trends; whilst the tunes are in good agreement with the priors on $M_{A}^{\text{QEL}}$ and $S_{\text{RES}}$, the other parameters differ from the \emph{nominal} parameter values.
There is also a clear preference for \ac{QEL} with \ac{RPA} corrections.
In addition, the tunes prefer a higher \ac{QEL}, i.e.\ $\omega_{\text{RPA}}+\omega_{\text{RPA}}>1$, and \ac{2p2h} cross section.
Finally, the different tunes suggest an underlying energy dependence on the \ac{2p2h} cross section strength and shape: the \texttt{G10a}, \texttt{G11a} and \texttt{G20a} tunes enhance (suppress) the Valencia \ac{2p2h} cross section at the nucleon ($\Delta$) region.
Alternatively, the \texttt{G30a} and \texttt{G31a} tunes enhance the cross section at the nucleon and $\Delta$ region, with $S_{\Delta}^{2p2h}>S_{N}^{2p2h}$.
A hint of an underlying energy dependence was also observed in Ref.~\cite{Bourguille_2021}.

The enhancement of the \ac{QEL} cross section is crucial for the description of MiniBooNE CC0$\pi$ data at $\cos\theta_\mu<0$.
Particularly, the \texttt{G10a} and \texttt{G11a} tunes suggest an increase of the \ac{QEL} cross section of about 20\%.
Similar \ac{QEL} scalings have been observed by MicroBooNE~\cite{MicroBooNE:2021ccs} and recent Lattice \ac{QCD} calculations~\cite{meyer2022status}.
The increase (decrease) of the $S_{N}^{\text{2p2h}}$ ($S_{\Delta}^{\text{2p2h}}$ and $S_{PL}^{\text{2p2h}}$) is also crucial to correctly describe MiniBooNE $\nu_\mu$ and $\overline{\nu}_\mu$ CC0$\pi$ data.

The \texttt{G20a} tune also offers a better description of T2K ND280 CC0p0$\pi$ data. 
This tune suggests a scaling of $S_{N}^{\text{2p2h}}=1.7\pm0.3$, compatible with the results presented by MicroBooNE~\cite{MicroBooNE:2021ccs}.
In this particular case, the scaling of \ac{QEL} is around 10\%.
The post-fit value of $\omega_{\text{No\,RPA}}$, although negative, is compatible with zero.
This result is physical as $\omega_{\text{RPA}}+\omega_{\text{No\,RPA}}>0$, hence the total cross-section is positive. 
This scenario can be avoided by reducing the $\omega_{\text{No\,RPA}}$ range to [0,1.5].
However, the parameter range is not reduced further to allow a valid estimation of the error on $\omega_{\text{No\,RPA}}$.

\begin{figure}
    \centering
    \begin{subfigure}{0.8\textwidth}
    \centering
    \includegraphics[width=\textwidth]{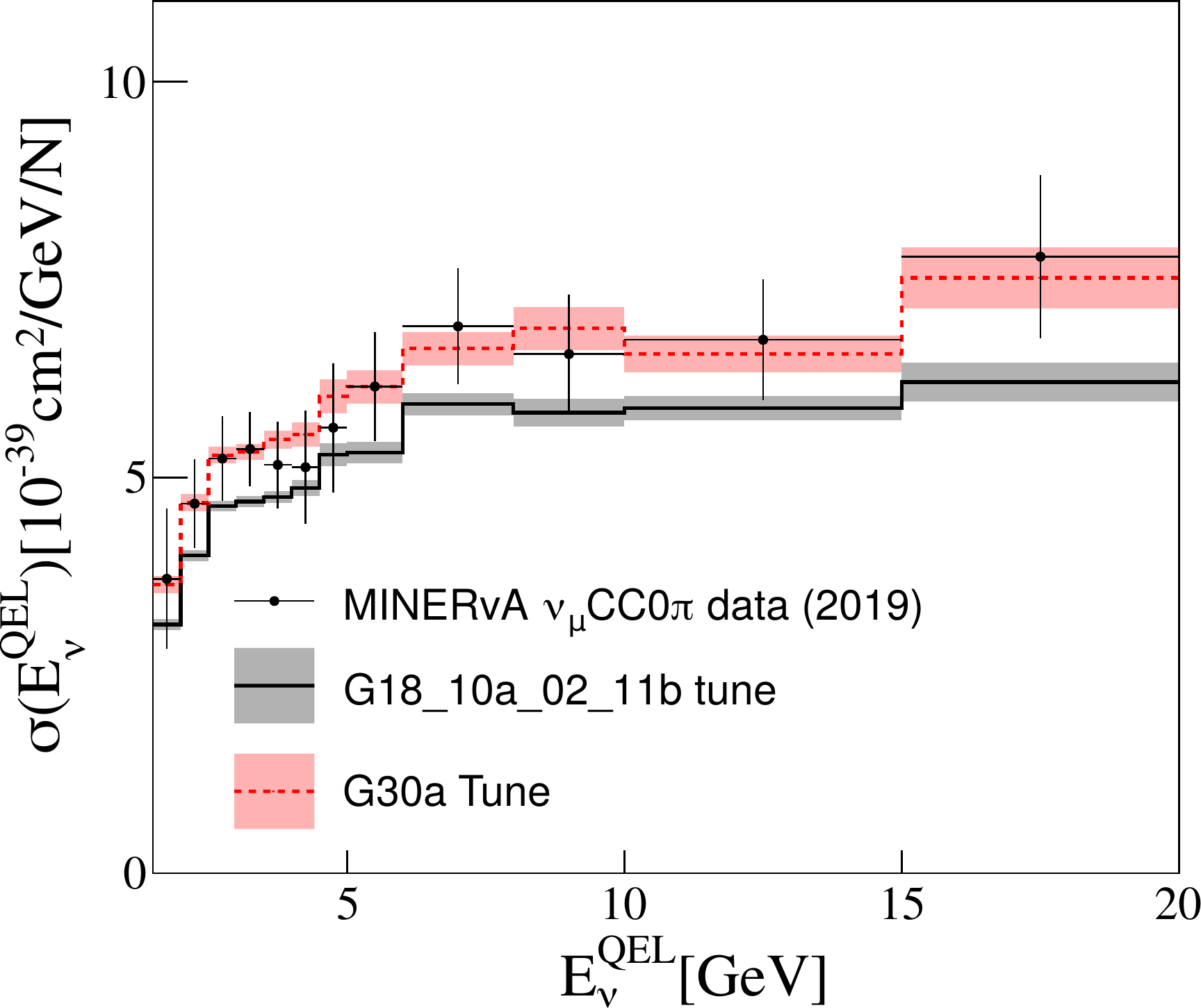}
    \caption{}
    \end{subfigure}
    \begin{subfigure}{0.8\textwidth}
    \centering
    \includegraphics[width=\textwidth]{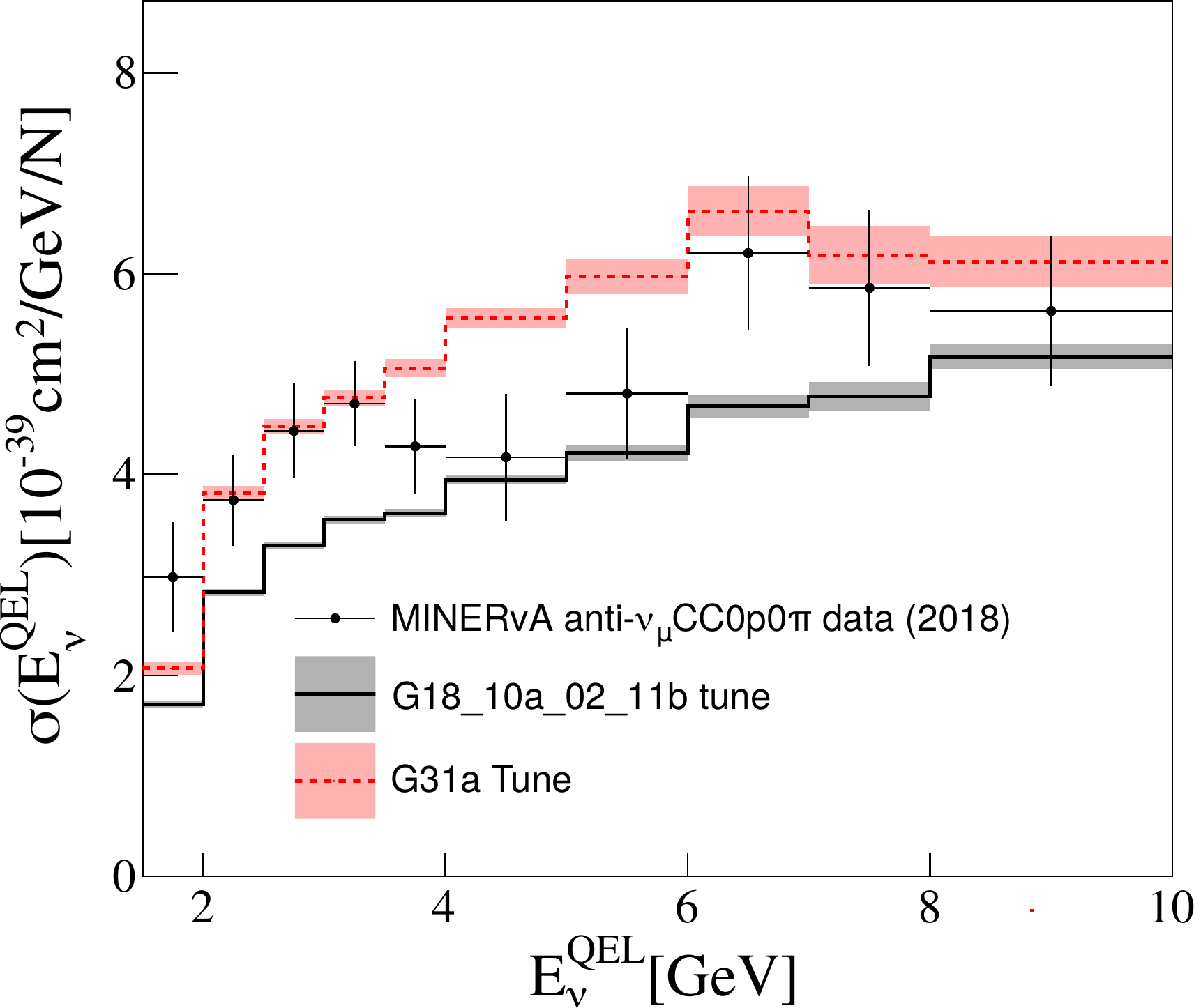}
    \caption{}
    \end{subfigure}
    \caption{Comparison of \texttt{G18\_10a\_02\_11b}, \texttt{G31a} (a) and \texttt{G31a} (b) against MINER$\nu$A $\nu_\mu$ and $\overline{\nu}_\mu$ CC0p0$\pi$ integrated cross section data~\cite{PhysRevD.99.012004,MinibooneAntineutrino}. 
    The predictions are computed using the parameters specified in Tab.~\ref{tab:PartialTuneResults}.}
    \label{fig:MINERvABarPartialTuneReconstructed2}
\end{figure}

Before the tune, the \texttt{G18\_10a\_02\_11b} prediction under-predicted MINER$\nu$A CC0$\pi$ data in the phase-space regions where \ac{2p2h} events dominate ($0.15<p_T<0.7$~GeV/c).
The results suggest that an enhancement of \ac{QEL}, as well as \ac{2p2h}, improves the agreement with data.
In fact, the \texttt{G30a} and \texttt{G31a} tunes provide with a better description of $\nu_\mu$ CC0$\pi$ and $\overline{\nu}_\mu$ CC0p0$\pi$ data respectively.
The improvement in the normalization of the cross section is reflected in the post-fit $\chi^2$ values from Tab.~\ref{tab:summarychi2_CC0piData}.
The same is true for the cross section as a function of the reconstructed neutrino energy, Fig.~\ref{fig:MINERvABarPartialTuneReconstructed}, and single-differential cross section data, Figs.~\ref{fig:MINERvAPartialTuneReconstructed} and \ref{fig:MINERvABarPartialTuneReconstructed2}. 
Both tunes over-predict the data at very low $Q^2_{\text{QEL}}$.

\subsection{Tension between CC0\ensuremath{\pi} partial tunes}

\begin{figure*}
    \centering
    \includegraphics[width=\textwidth]{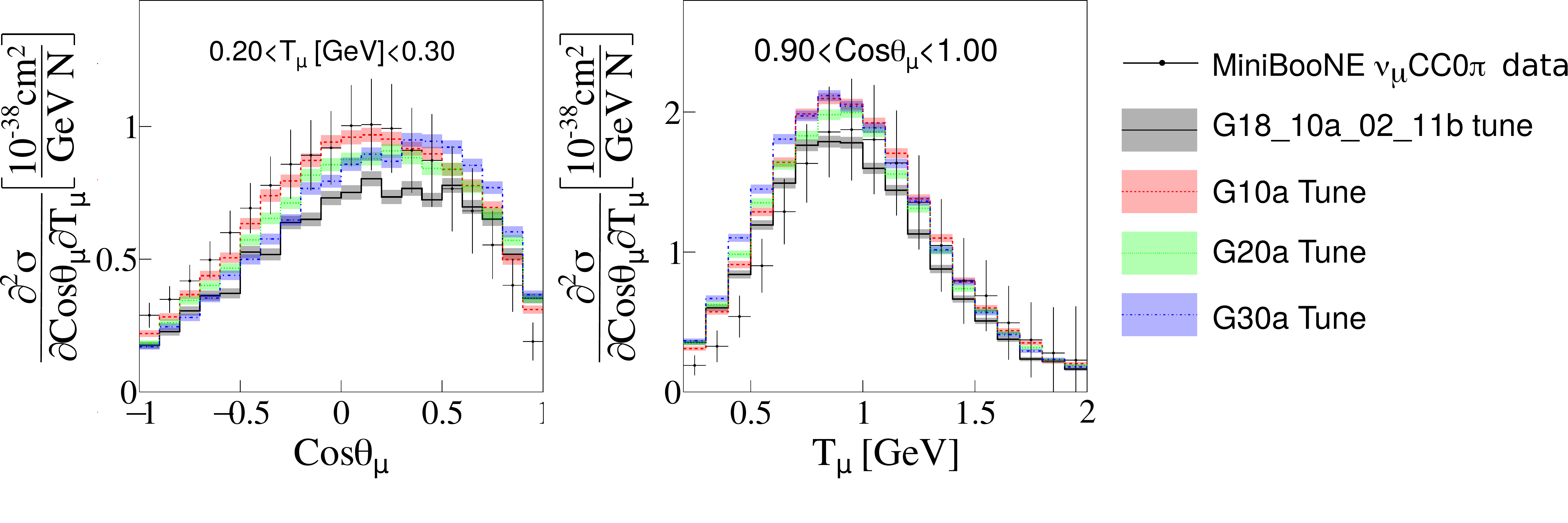}
    \caption{Comparison of the \texttt{G18\_10a\_02\_11b}, \texttt{G10a}, \texttt{G20a} and \texttt{G30a} tunes against MiniBooNE $\nu_\mu$ CC0p0$\pi$ integrated cross-section data~\cite{1002.2680}. 
    The predictions are computed using the parameters specified in Tab.~\ref{tab:PartialTuneResults}.}
    \label{fig:MINERvAPartialTuneComparison}
\end{figure*}

Tensions between datasets can be explored by comparing the different tunes.
Figure~\ref{fig:MINERvAPartialTuneComparison} compares the \texttt{G10a}, \texttt{G20a} and \texttt{G30a} predictions against MiniBooNE $\nu_\mu$ CC0$\pi$ data.
Even though the normalization of the three tunes is similar, differences in the predicted cross-section shape exist.
The \texttt{G10a} tune is the only one out of the three that successfully describes the shape of the data, as it can be seen in Fig.~\ref{fig:MINERvAPartialTuneComparison} (left).
The other tunes underestimate the cross-section at backward muon angles.
In addition, the \texttt{G30a Tune} over-predicts the cross section at forward angles as a consequence of the enhancement of the \ac{2p2h} cross section at the $\Delta$-region.
All tunes overestimate the cross section at forward muon angles and low muon kinetic energies, as demonstrated in Fig.~\ref{fig:MINERvAPartialTuneComparison} (right).

\begin{figure}
    \centering
    \begin{subfigure}{0.85\columnwidth} 
    \centering
    \includegraphics[width=\columnwidth]{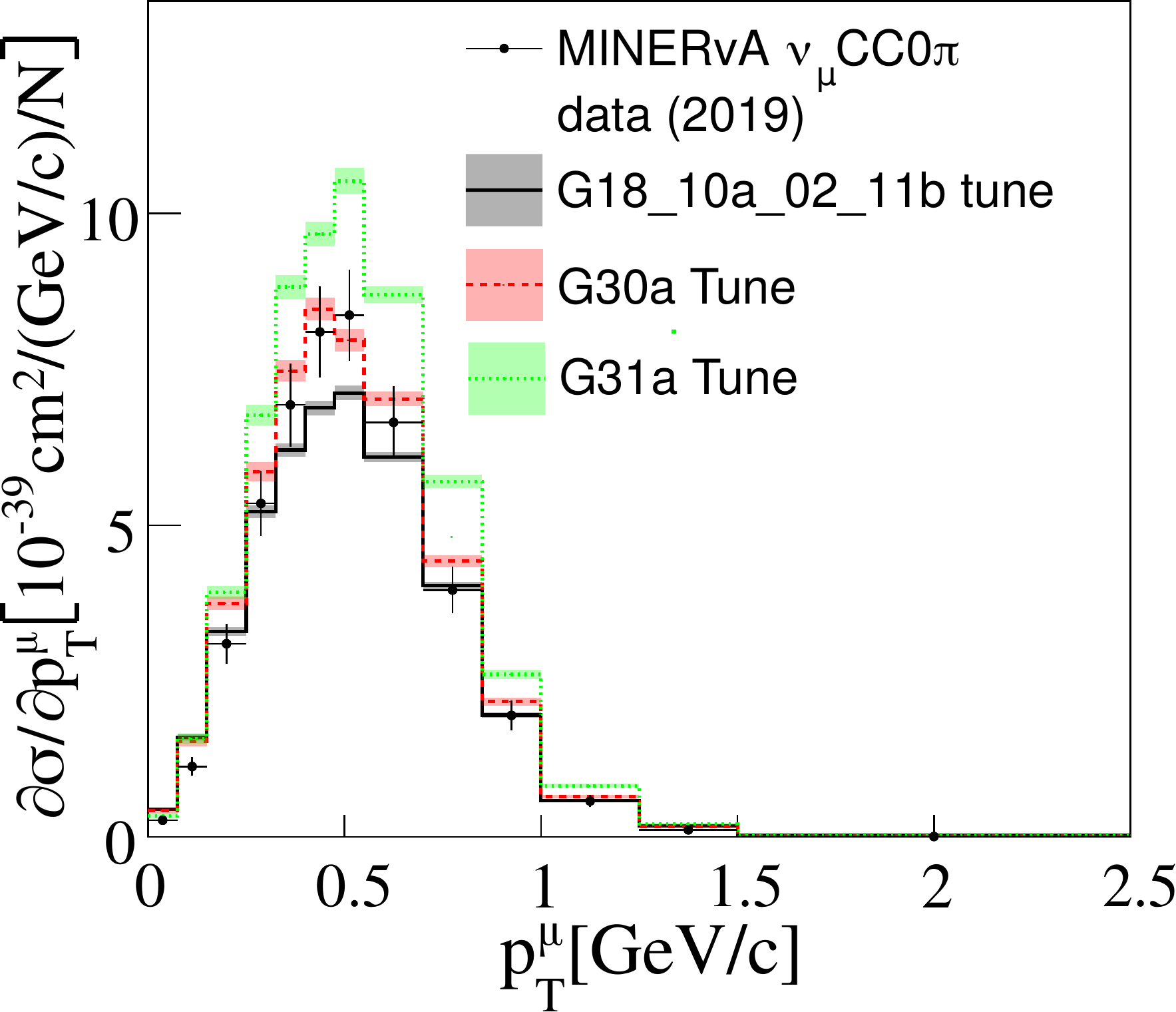}
    \caption{}
    \end{subfigure}
    \begin{subfigure}{0.9\columnwidth}
    \centering
    \includegraphics[width=\columnwidth]{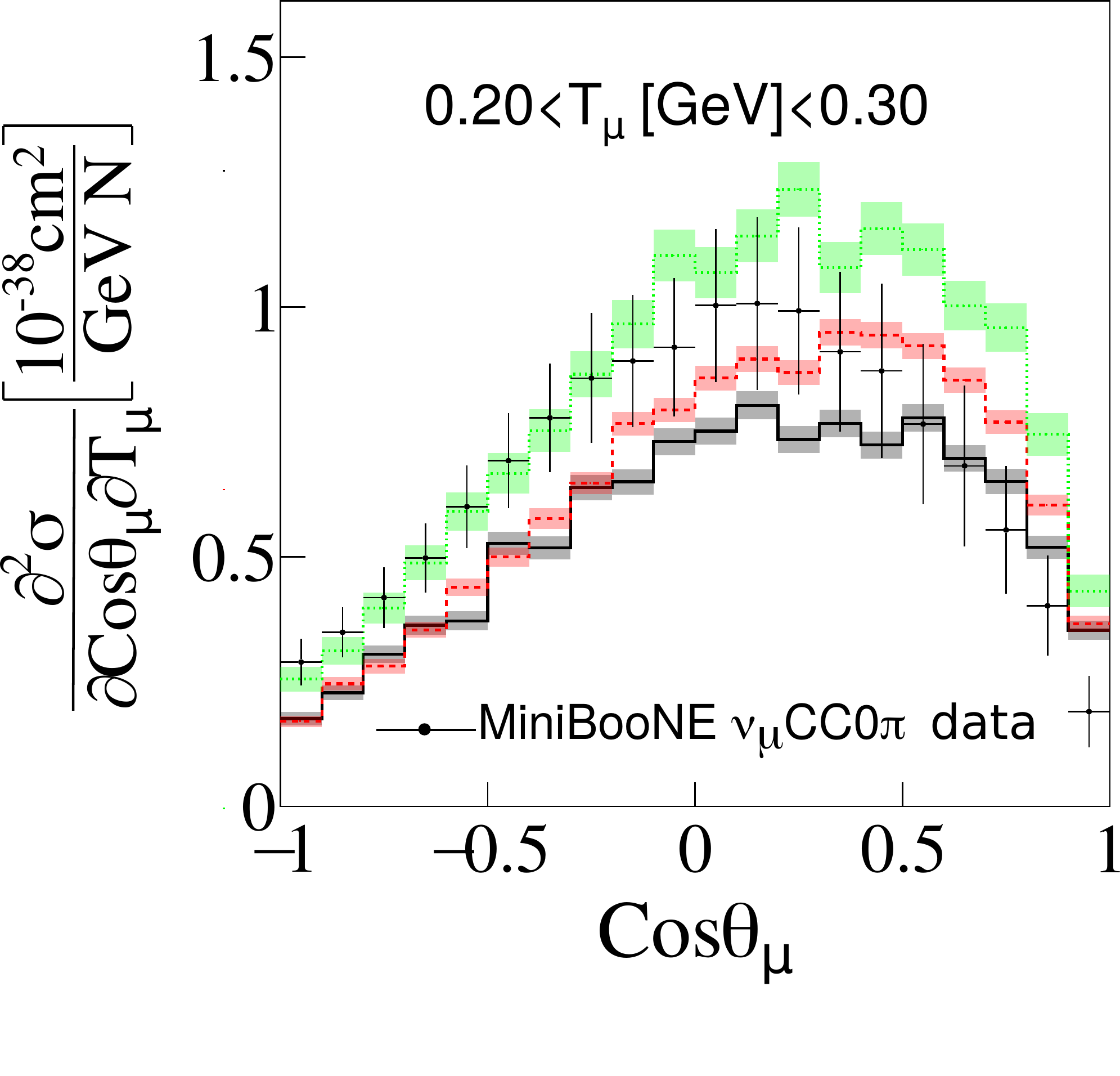}
    \caption{}
    \end{subfigure}
    \caption{Comparison of the \texttt{G18\_10a\_02\_11b}, \texttt{G30a} and \texttt{G31a} tunes against (a) MINER$\nu$A~\cite{PhysRevD.99.012004} and (b) MiniBooNE~\cite{1002.2680} $\nu_\mu$ CC0p0$\pi$ double-differential cross-section data. 
    The predictions are computed using the parameters specified in Tab.~\ref{tab:PartialTuneResults}.}
    \label{fig:MINERvAPartialTuneComparisonNumuAntinumu}
\end{figure}

The \texttt{G31a} tune is in clear tension with all the rest, including partial tunes performed with MINER$\nu$A neutrino data.
In comparison with the rest of the tunes, the \texttt{G31a} tune prefers higher \ac{QEL} and \ac{2p2h} cross sections. 
This leads to the over-prediction of all the other datasets.
The comparison of \texttt{G30a} and \texttt{G31a} against MINER$\nu$A and MiniBooNE $\nu_\mu$ CC0$\pi$ data are shown in Fig.~\ref{fig:MINERvAPartialTuneComparisonNumuAntinumu}.
The effect of this tension on the $\chi^2$ is reported in Tab.~\ref{tab:summarychi2_CC0piData}.

\begin{figure}
    \centering
    \includegraphics[width=0.8\columnwidth]{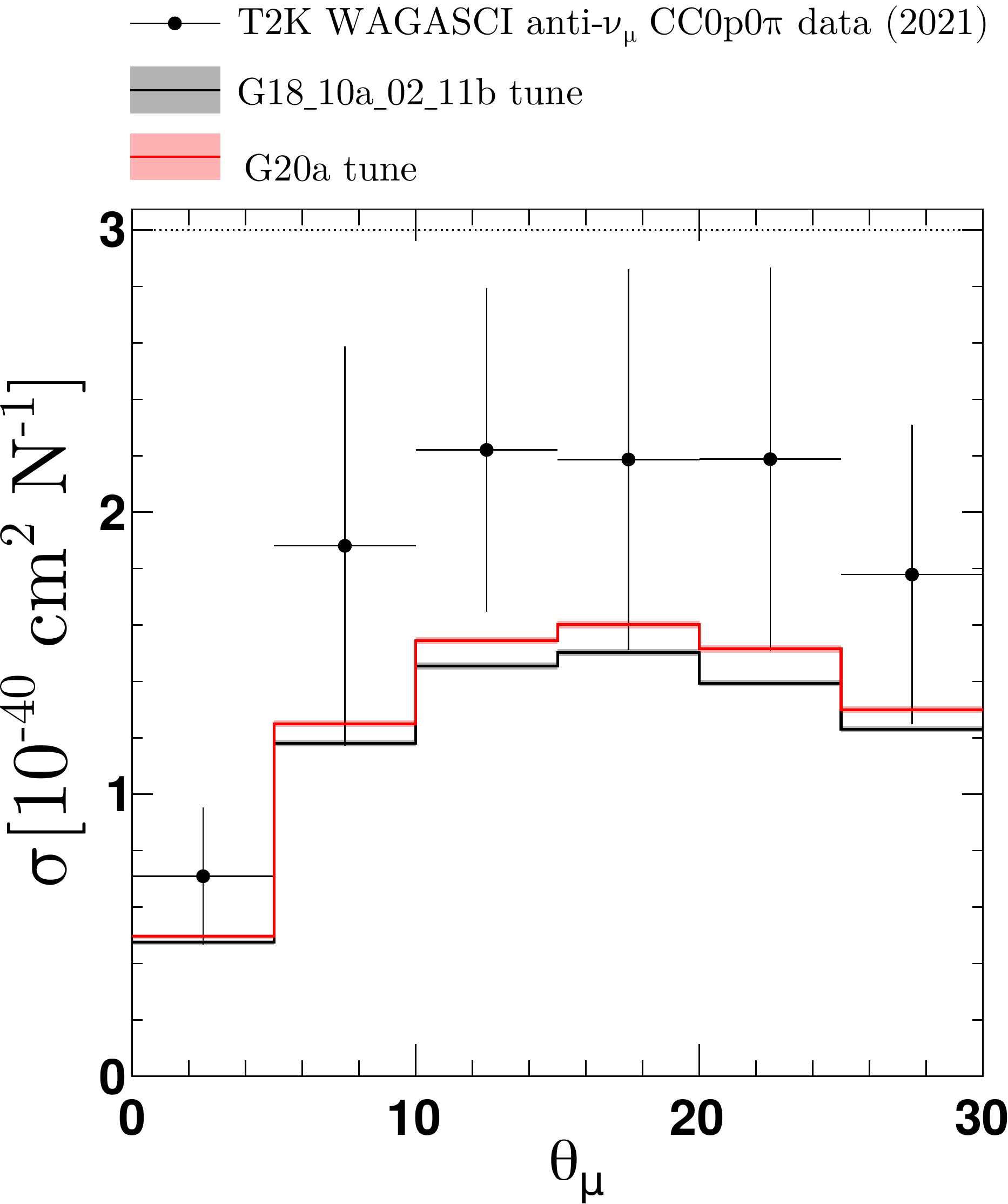}
    \caption{Comparison of the \texttt{G18\_10a\_02\_11b} and \texttt{G20a} tunes against T2K WAGASCI $\overline{\nu}_\mu$ CC0p0$\pi$ data.}
    \label{fig:WAGASCITune}
\end{figure}

The tension between the \texttt{G31a} tune and the rest can have different origins.
A possibility is that the model does not fully characterize the difference between neutrino and anti-neutrino fluxes. 
This is investigated comparing the \texttt{G20a} tune to the T2K WAGASCI antineutrino data.
Both the T2K WAGASCI and T2K ND280 analysis explore the $\overline{\nu}_\mu$ CC0p$0\pi$ topology, but these are exposed to different neutrino fluxes (see Appendix.~\ref{sec:T2KAnalysis}).
The impact of the \texttt{G20a} tune to these predictions is shown in Fig.~\ref{fig:WAGASCITune}.
It is observed that the \texttt{G20a} tune has little impact on the T2K WAGASCI predictions.
This indicates than an additional neutrino/antineutrino modeling uncertainty should be considered in a global tune of neutrino and antineutrino data. 
Another possible source of uncertainty is the different topology definition for MINER$\nu$A's $\overline{\nu}_\mu$ dataset, with requires no visible protons above $T_p=120$~MeV for the antineutrino sample.
The proton multiplicity uncertainty is explored further in Sec.~\ref{sec:TensionsNpDatsets}.

\subsection{Tensions between \ensuremath{\nu_\mu} CC0\ensuremath{\pi} and \ensuremath{\nu_\mu} CCNp0\ensuremath{\pi} datasets}
\label{sec:TensionsNpDatsets}

T2K ND280, MINER$\nu$A and MicroBooNE are the only experiments that released cross-section measurements for different proton multiplicities.
As discussed in Sec.~\ref{sec:datasetconsiderations}, CC0$\pi$ and CC0p0$\pi$ datasets are under-predicted, whilst the CCNp0$\pi$ datasets are slightly over-predicted by the nominal \texttt{G18\_10a\_02\_11b} prediction. 
This modeling limitation is also observed in Ref.~\cite{Bourguille_2021,PhysRevC.100.015505,T2KCCQELikeMeasurement}.

After the partial tunes using $\nu_\mu$ CC0$\pi$ data, the agreement with CCNp0$\pi$ data deteriorates. 
This is highlighted in Tab.~\ref{tab:summarychi2_CCNpData}, which summarizes the post-fit $\chi^2$ values associated with CCNp0$\pi$ datasets.
In all cases, the $\chi^2$ computed with each partial tune prediction increases with respect to the $\chi^2$ computed with the \emph{nominal} \texttt{G18\_10a\_02\_11b} tune.

\begin{table*}
    \centering
	\begin{tabular}{@{\extracolsep\fill} c c c c c c c c c}
	\hline\hline\noalign{\smallskip}
     Dataset                                & $\chi_{\text{Nominal}}^2$ & $\chi_{\texttt{G10a}}^2$ & $\chi_{\texttt{G11a}}^2$
     & $\chi_{\texttt{G20a}}^2$ & $\chi_{\texttt{G30a}}^2$ & $\chi_{\texttt{G31a}}^2$ & $\chi_{\texttt{G35a}}^2$ & \text{DoF} \\
    \noalign{\smallskip}\hline\noalign{\smallskip}
    \multicolumn{9}{c}{T2K ND280 CCNp0$\pi$ data} \\
    \noalign{\smallskip}\hline\noalign{\smallskip}
    $d\sigma/d\delta p_T$                   &  228 & 1741 & 1499 &  883 &  759 &   95 & 25          &  8 \\
    $d\sigma/d\delta \phi_T$                &  292 & 2489 & 2117 & 1190 & 1049 & 1950 & 16          &  8 \\
    $d\sigma/d\delta \alpha_T$              &   27 &   58 &   53 &   42 &   41 &   95 & 21          &  8 \\
    \noalign{\smallskip}\hline\noalign{\smallskip}
    \multicolumn{9}{c}{MINER$\nu$A CCNp0$\pi$ data}\\
    \noalign{\smallskip}\hline\noalign{\smallskip}
    $d\sigma/dp_p$                          &   21 &   22 &   25 &   32 &   36 &   58 & 27          & 25 \\
    $d\sigma/d\theta_p$                     &   58 &  153 &  150 &  113 &  129 &  226 & \textbf{20} & 26 \\
    $d\sigma/d\delta p_T$                   &  102 &  637 &  568 &  360 &  352 &  625 & 42          & 24 \\
	$d\sigma/d\delta\phi_T$                 &   87 &  505 &  467 &  314 &  354 &  566 & 18          & 23 \\
	$d\sigma/d\delta\alpha_T$               &   15 &   21 &   29 &   24 &   30 &   57 & 17          & 12 \\
    $d\sigma/d\delta p_{Tx}$                &  159 &  727 &  710 &  467 &  555 &  768 & 62          & 32 \\
    $d\sigma/d\delta p_{Ty}$                &  127 &  832 &  776 &  553 &  599 &  792 & 51          & 33 \\
    \noalign{\smallskip}\hline\noalign{\smallskip}
    \multicolumn{9}{c}{MicroBooNE CCNp0$\pi$ data}\\
    \noalign{\smallskip}\hline\noalign{\smallskip}
    $d\sigma/dp_\mu^{\text{reco}}$          &   71 &  402 &  413 &  245 &  251 & 1186 &  40         & 10 \\
    $d\sigma/d\cos\theta_\mu^{\text{reco}}$ &  413 &  238 &  236 &  210 &  245 &  471 & 149         & 12 \\
    $d\sigma/dp_p^{\text{reco}}$            &   33 &   96 &   97 &   73 &   76 &  267 &  20         & 10 \\
    $d\sigma/d\cos\theta_p^{\text{reco}}$   &  100 &  176 &  179 &  135 &  139 &  393 &  33         &  9 \\
	$d\sigma/d\theta_{\mu p}^{\text{reco}}$ &  549 &  186 &  196 &  199 &  218 &  304 & 136         &  6 \\
    \noalign{\smallskip}\hline\hline 
    \end{tabular}
    \caption{Summary of $\chi^2$ values associated the CCNp0$\pi$ datasets specified in each row.
    The $\chi^2$ values are calculated using the \ac{NS} method for seven different tunes: \texttt{G18\_10a\_02\_11b}, \texttt{G10a}, \texttt{G11a}, \texttt{G20a}, \texttt{G30a} and \texttt{G31a}. The values highlighted in bold correspond to the best-fit $\chi^2_{\texttt{G35a}}$ for the partial tune using the specified dataset.}
    \label{tab:summarychi2_CCNpData}
\end{table*} 

\begin{figure*}
    \centering
    \begin{subfigure}{0.4\columnwidth}
    \centering
    \includegraphics[width=\columnwidth]{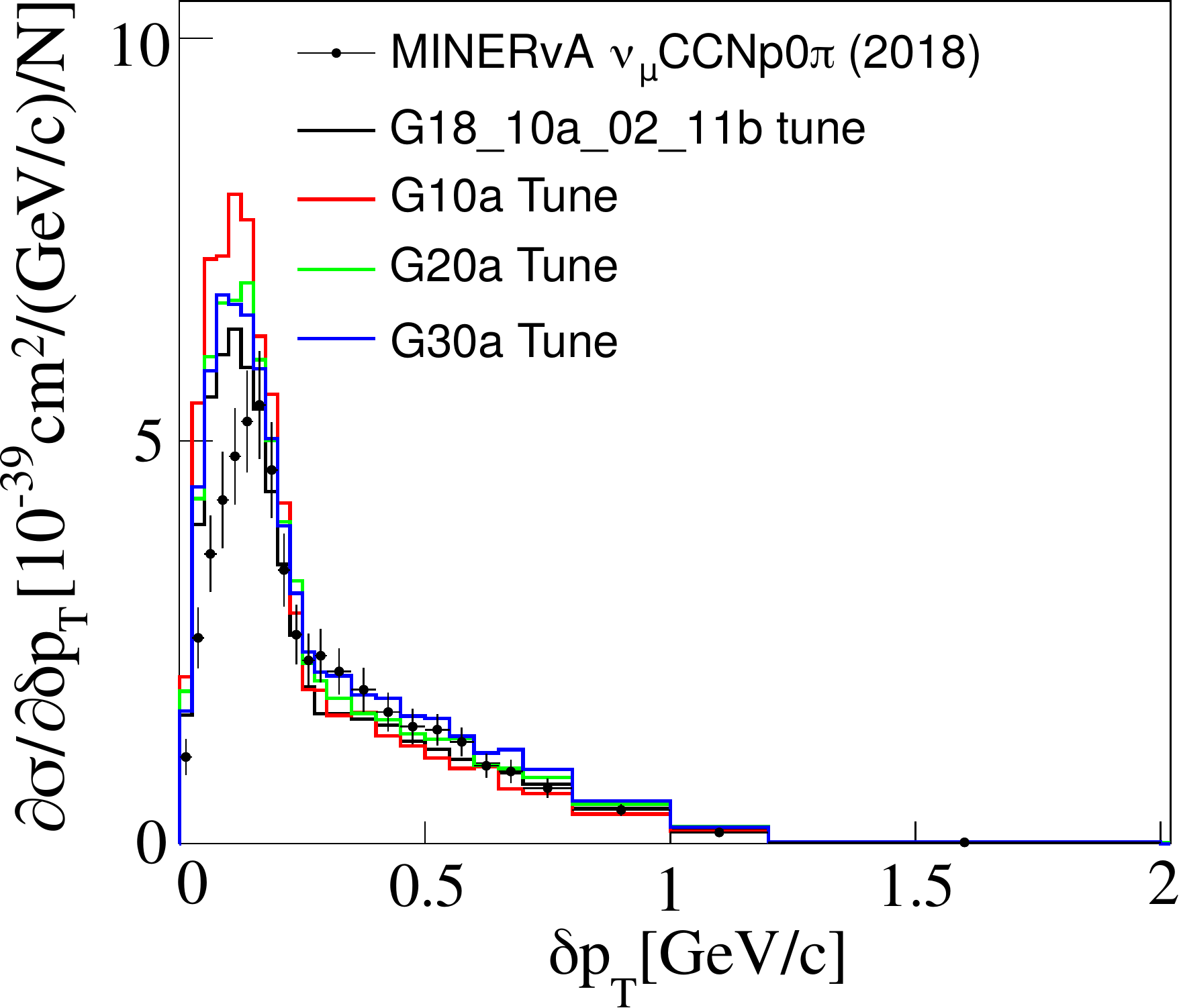}
    \caption{}
    \end{subfigure}
    \centering
    \begin{subfigure}{0.4\columnwidth}
    \centering
    \includegraphics[width=\columnwidth]{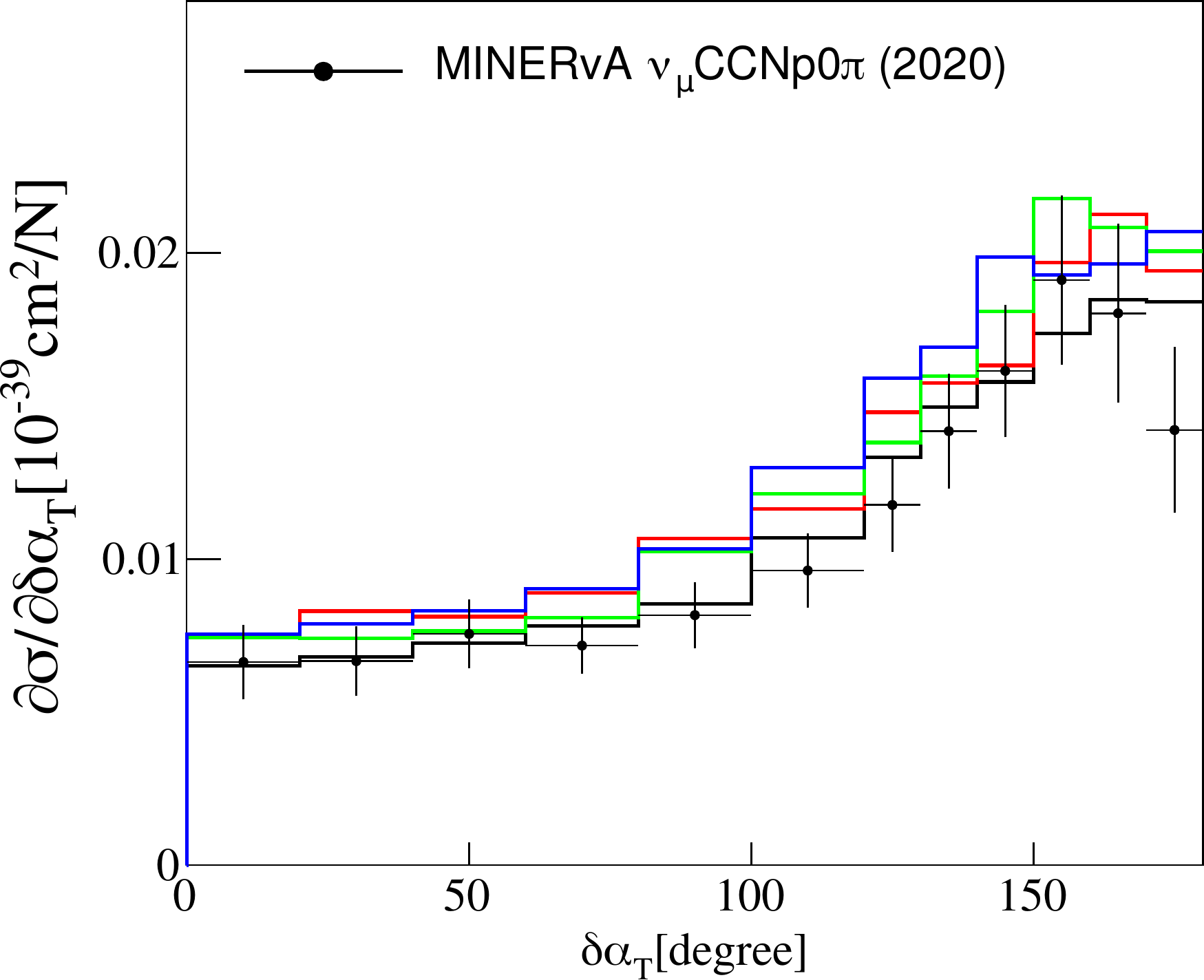}
    \caption{}
    \end{subfigure}

    \begin{subfigure}{0.4\columnwidth}
    \centering
    \includegraphics[width=\columnwidth]{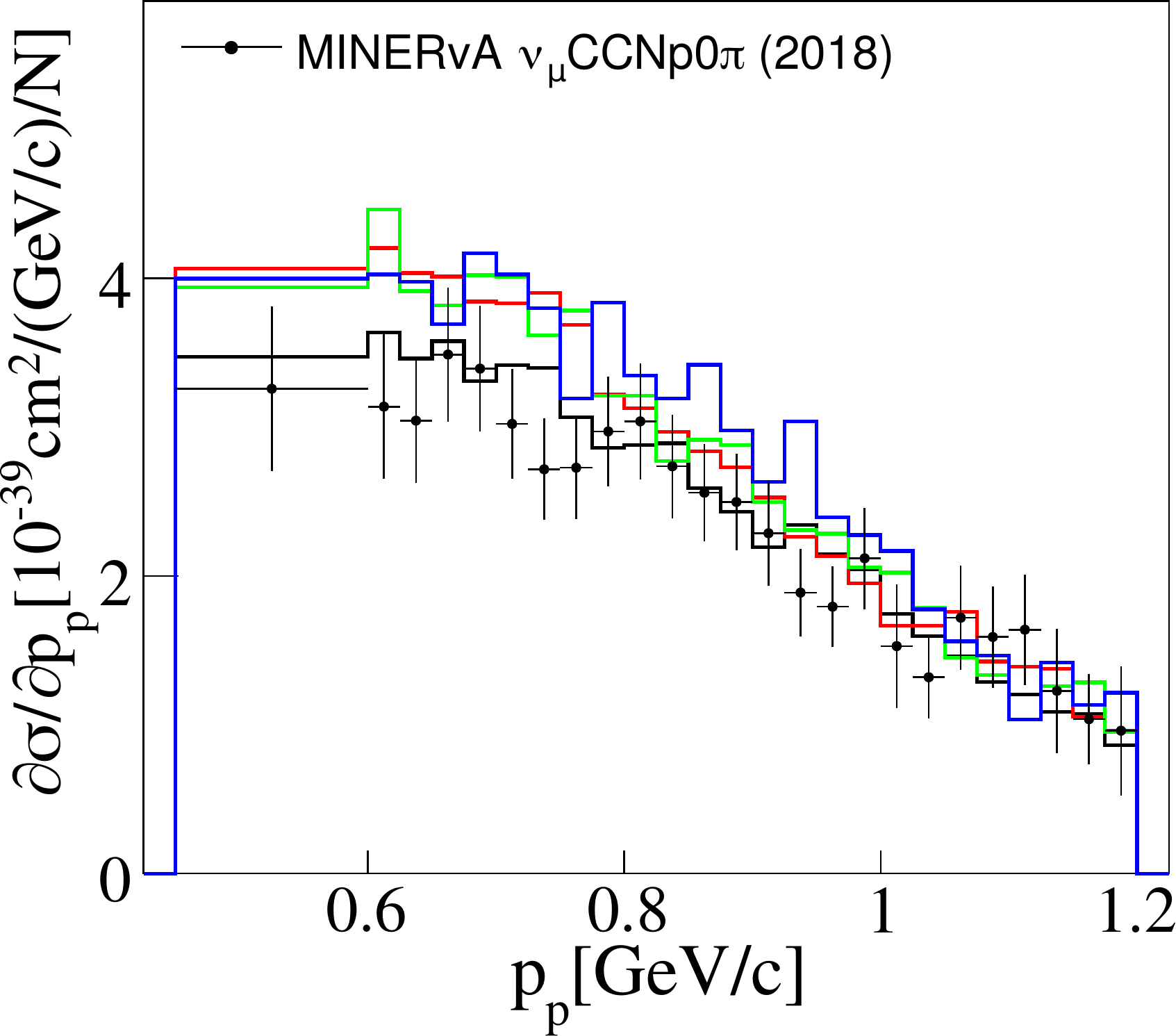}
    \caption{}
    \end{subfigure}
    \begin{subfigure}{0.4\columnwidth}
    \centering
    \includegraphics[width=\columnwidth]{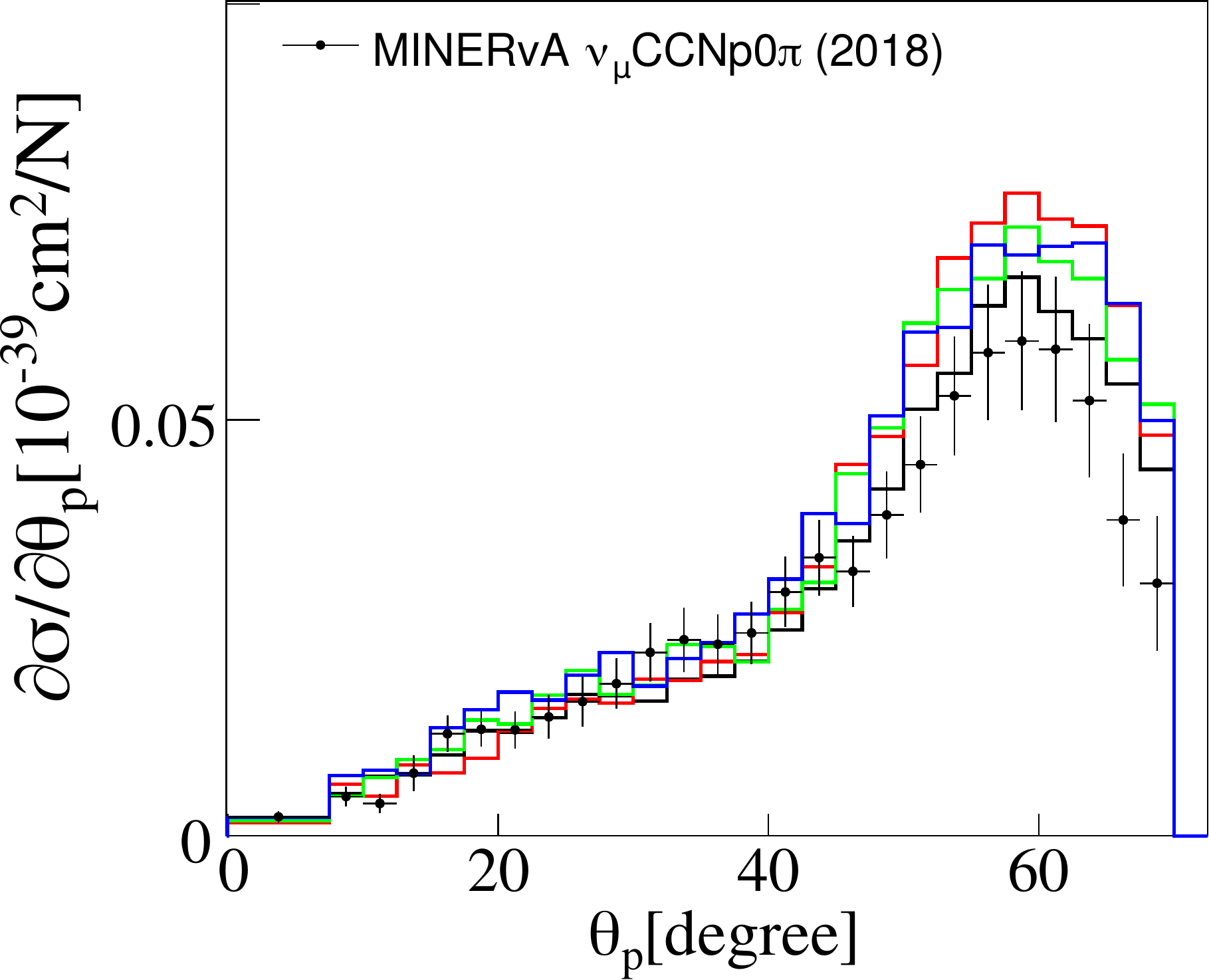}
    \caption{}
    \end{subfigure}
    \caption{Comparison of the \texttt{G18\_10a\_02\_11b}, \texttt{G10a}, \texttt{G20a} and \texttt{G30a} tunes against MINER$\nu$A $\nu_\mu$ CCNp0$\pi$ single-differential cross-section data as a function of (a) $\delta p_T$, (b) $\alpha_p$, (c) $p_p$ or (d) $\theta_p$. 
    In order to ease the readability of these plots, no statistical errors are shown.
    The predictions are computed using the parameters specified in Tab.~\ref{tab:PartialTuneResults}.}
    \label{fig:MINERvAPartialTuneComparisonNumuvsNpsamples}
\end{figure*}

\begin{figure}
    \centering
    \begin{subfigure}{0.8\columnwidth}
    \centering
    \includegraphics[width=\columnwidth]{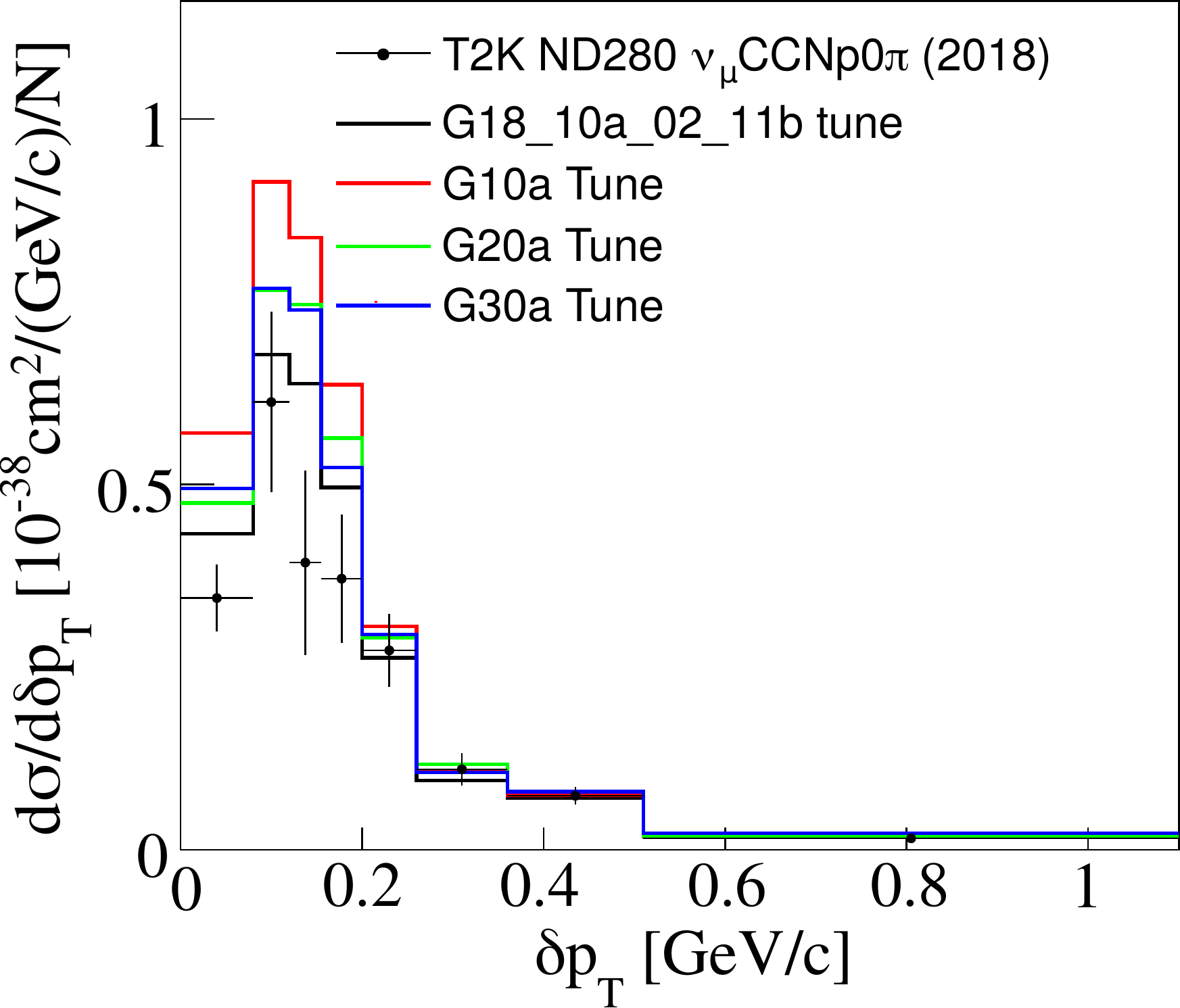}
    \end{subfigure}
    \begin{subfigure}{0.8\columnwidth}
    \centering
    \includegraphics[width=\textwidth]{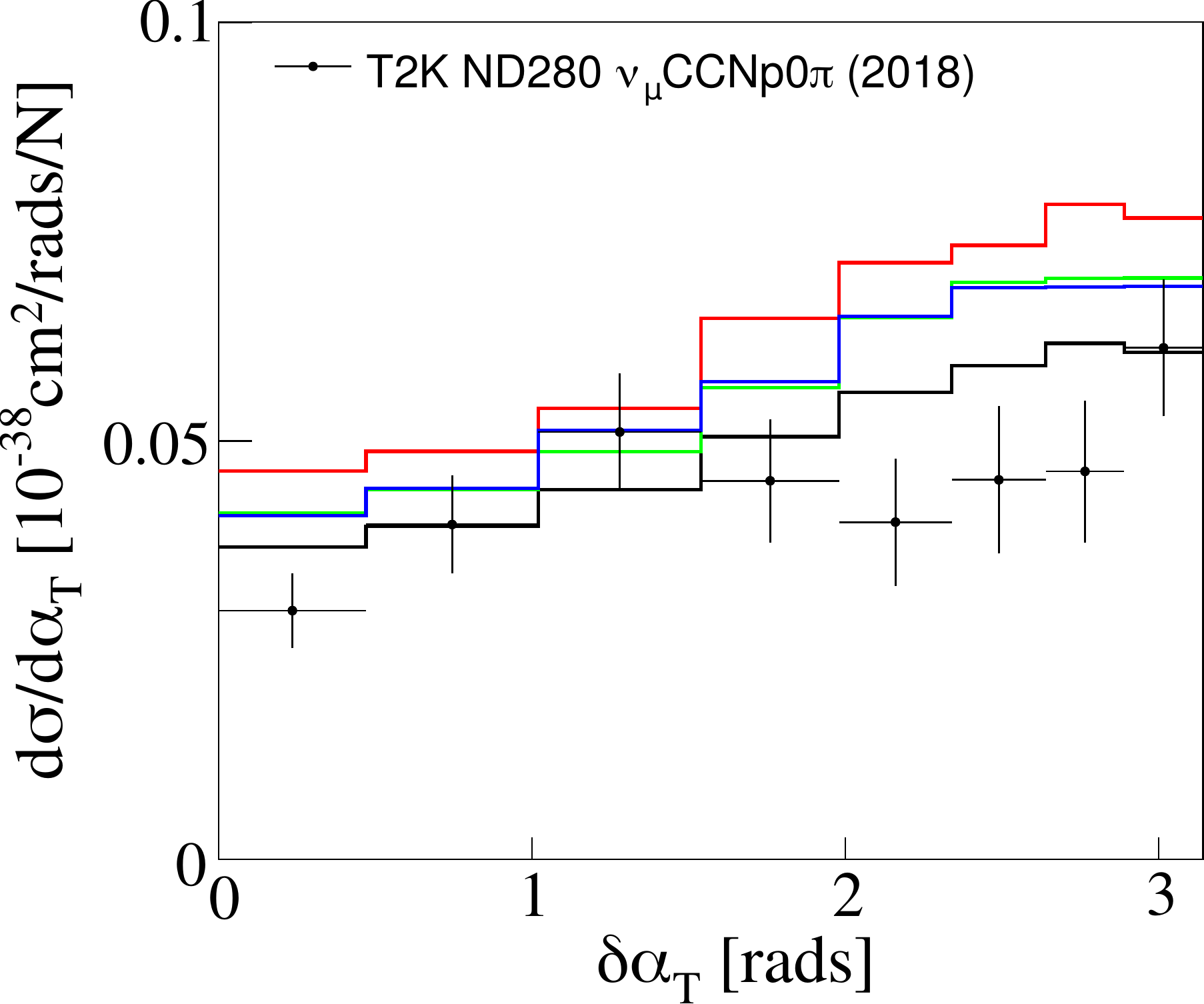}
    \end{subfigure}
    \caption{Comparison of the \texttt{G18\_10a\_02\_11b}, \texttt{G10a}, \texttt{G20a} and \texttt{G30a} tunes against T2K ND280 $\nu_\mu$ CCNp0$\pi$ single-differential cross-section data as a function of (a) $\delta p_T$ or (b) $\delta \alpha_T$~\cite{T2KCCQELikeMeasurement}.
    In order to ease the readability of these plots, no statistical errors are shown.
    The predictions are computed using the parameters specified in Tab.~\ref{tab:PartialTuneResults}.}
    \label{fig:MINERvAPartialTuneNumuvsNpsamplesSTKI}
\end{figure}

\begin{figure}
    \centering
    \begin{subfigure}{0.8\columnwidth}
    \centering
    \includegraphics[width=\columnwidth]{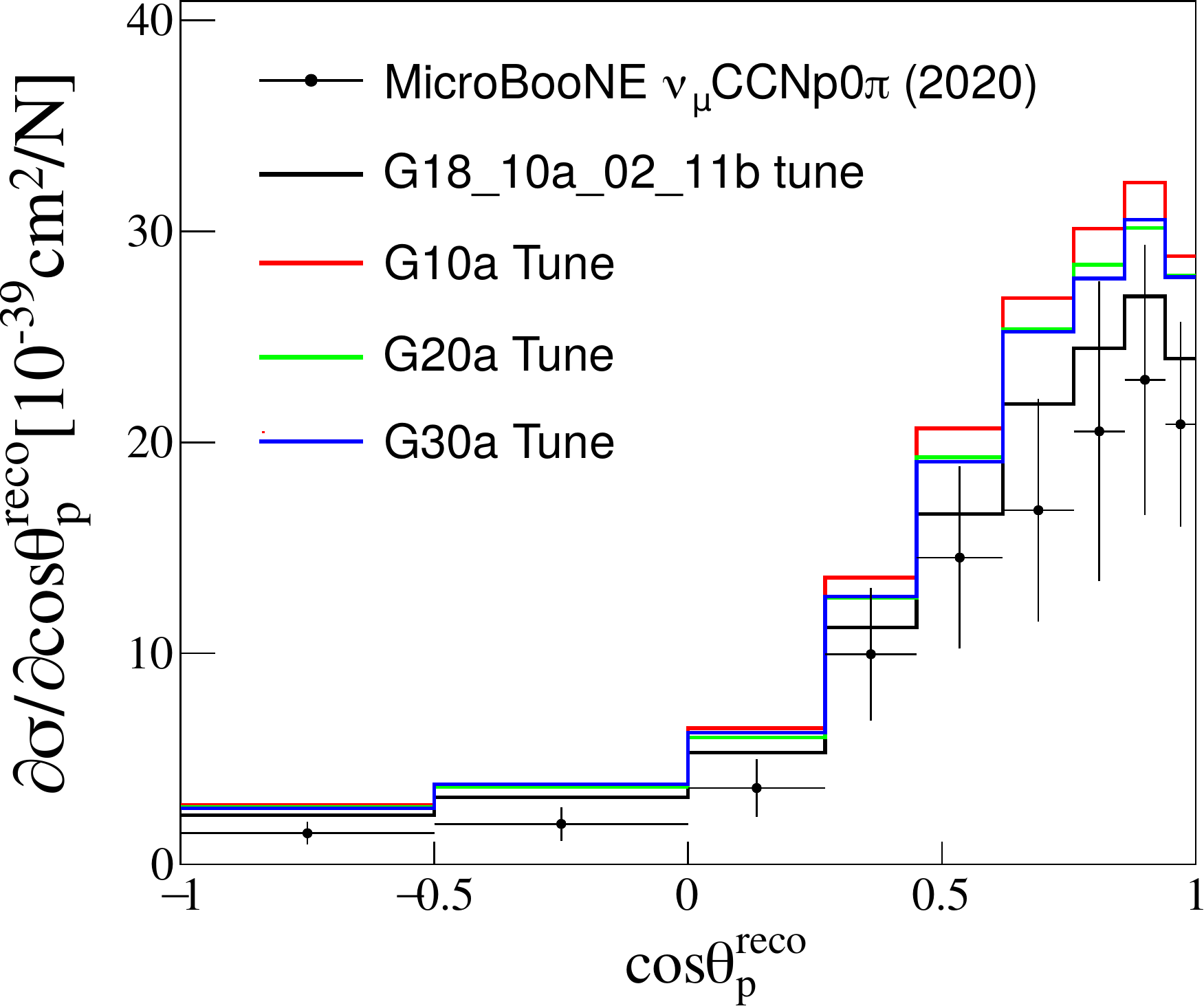}
    \caption{}
    \end{subfigure}
    \begin{subfigure}{0.8\columnwidth}
    \centering
    \includegraphics[width=\columnwidth]{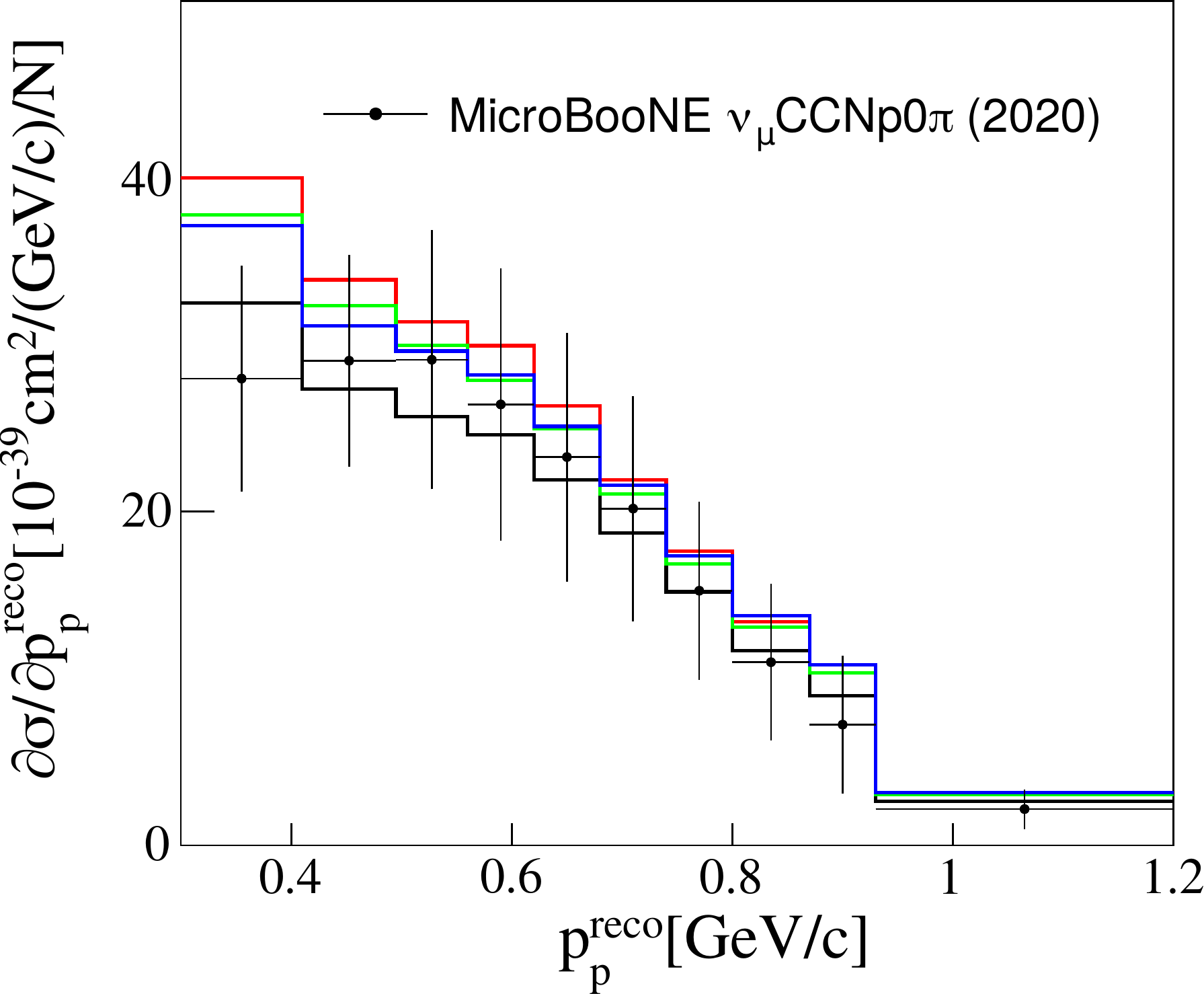}
    \caption{}
    \end{subfigure}
    \caption{Comparison of the \texttt{G18\_10a\_02\_11b}, \texttt{G10a}, \texttt{G20a} and \texttt{G30a} tunes against MicroBooNE $\nu_\mu$ CCNp0$\pi$ single-differential cross-section data as a function of (a) $\cos\theta_p^{\text{reco}}$ or (b) $p_p^{\text{reco}}$~\cite{PhysRevD.102.112013}. 
    In order to ease the readability of these plots, no statistical errors are shown.
    The predictions are computed using the parameters specified in Tab.~\ref{tab:PartialTuneResults}.}
    \label{fig:MINERvAPartialTunevsNpsamplesMicroBooNE}
\end{figure}

All \texttt{G10a}, \texttt{G20a} and \texttt{G30a} tunes overpredict $\nu_\mu$ CCNp0$\pi$ data.
Figure~\ref{fig:MINERvAPartialTuneComparisonNumuvsNpsamples} shows a comparison of the partial tune predictions against different single-differential CCNp0$\pi$ cross-section measurements from MINER$\nu$A. Figure~\ref{fig:MINERvAPartialTuneComparisonNumuvsNpsamples} shows that none of the available tunes can describe the peak at low $\delta p_T$ and that all partial tunes overestimate the cross section at low proton momentum and forward angles.
The same observations are made when comparing the tunes against T2K ND280 and MicroBooNE CCNp$0\pi$ data, see Figs.~\ref{fig:MINERvAPartialTuneNumuvsNpsamplesSTKI} and \ref{fig:MINERvAPartialTunevsNpsamplesMicroBooNE}, respectively.

To further explore this tension, an additional tune is performed using the MINER$\nu$A $\nu_\mu$ CCNp0$\pi$ dataset as a function of the proton angle.
Following the naming scheme described at the beginning of Sec.~\ref{sec:NuclearTuningResults}, this tune is referred to as \texttt{G35a}.
The best-fit results are listed in Tab.~\ref{tab:PartialTuneResults}.

The \texttt{G35a} tune suggests a significant reduction of the \ac{QEL} cross section.
In addition, the tune suppresses the Valencia cross-section peak prediction at $W=M_N$ and shifts the $\Delta$ peak to $W>M_\Delta$. 
This result contradicts the rest of the partial tunes presented in this article, reinforcing the fact that there is a strong tension between CC0$\pi$ and CCNp0$\pi$ datasets.
The summary of $\chi^2$ is reported in Tab.~\ref{tab:summarychi2_CC0piData} and Tab.~\ref{tab:summarychi2_CCNpData}.

An important observation is that the \texttt{G35a} tune also improves the agreement with MicroBooNE CCNp0$\pi$ data, suggesting that a possible A-dependency on the parameters does not play an important role.

The tension between CC0$\pi$ and CCNp0$\pi$ datasets needs to be resolved before attempting a global tune of CC0$\pi$ data that can describe all data available to date.
Some modelling aspects that may contribute to this tension are investigated in Sec.~\ref{sec:InvestigationTension}.

\section{Investigation of tensions between CC0\ensuremath{\pi} and CCNp0\ensuremath{\pi} datasets}
\label{sec:InvestigationTension}
This section offers an insight into possible modeling implementations that may contribute to the tension between CC0$\pi$ and CCNp0$\pi$ datasets and explores avenues of accommodating both within future joint tunes.
None of the uncertainties described in this section has a big impact on CC0$\pi$ datasets.

\subsubsection{Nuclear model variations}

The nuclear model determines the momentum and binding energy of the hit nucleon.
In GENIE, three nuclear models are available: \ac{RFG}, Local Fermi Gas (LFG) and \ac{CFG}~\cite{GENIEHighlights}.
By default, \texttt{G18\_10a\_02\_11b} uses the \ac{LFG}.

\begin{figure}
    \centering
	\begin{subfigure}{0.8\columnwidth}
	\centering
	\includegraphics[width=\columnwidth]{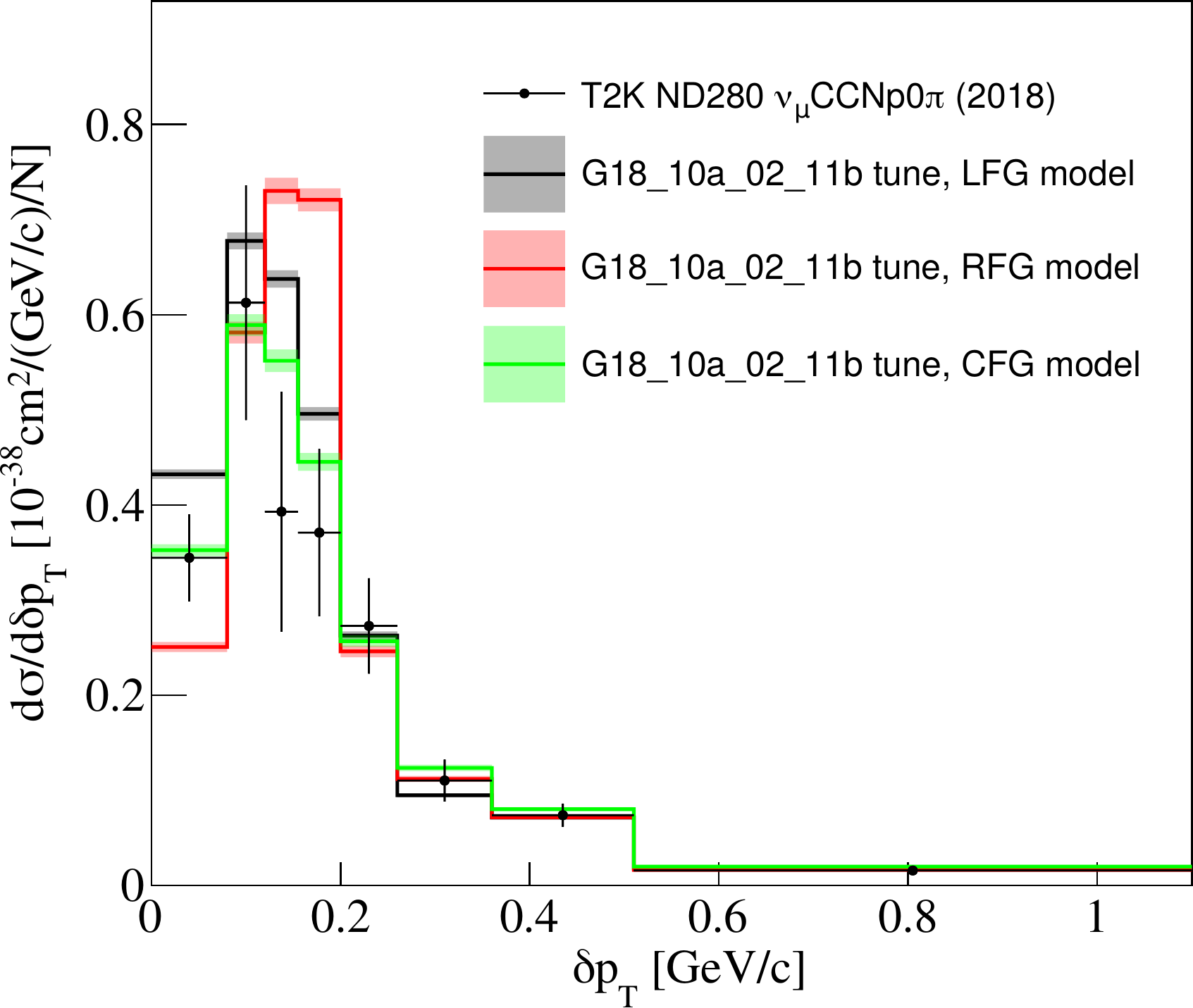}
    \caption{}
    \end{subfigure}
   	\begin{subfigure}{0.8\columnwidth}
	\centering
    \includegraphics[width=\columnwidth]{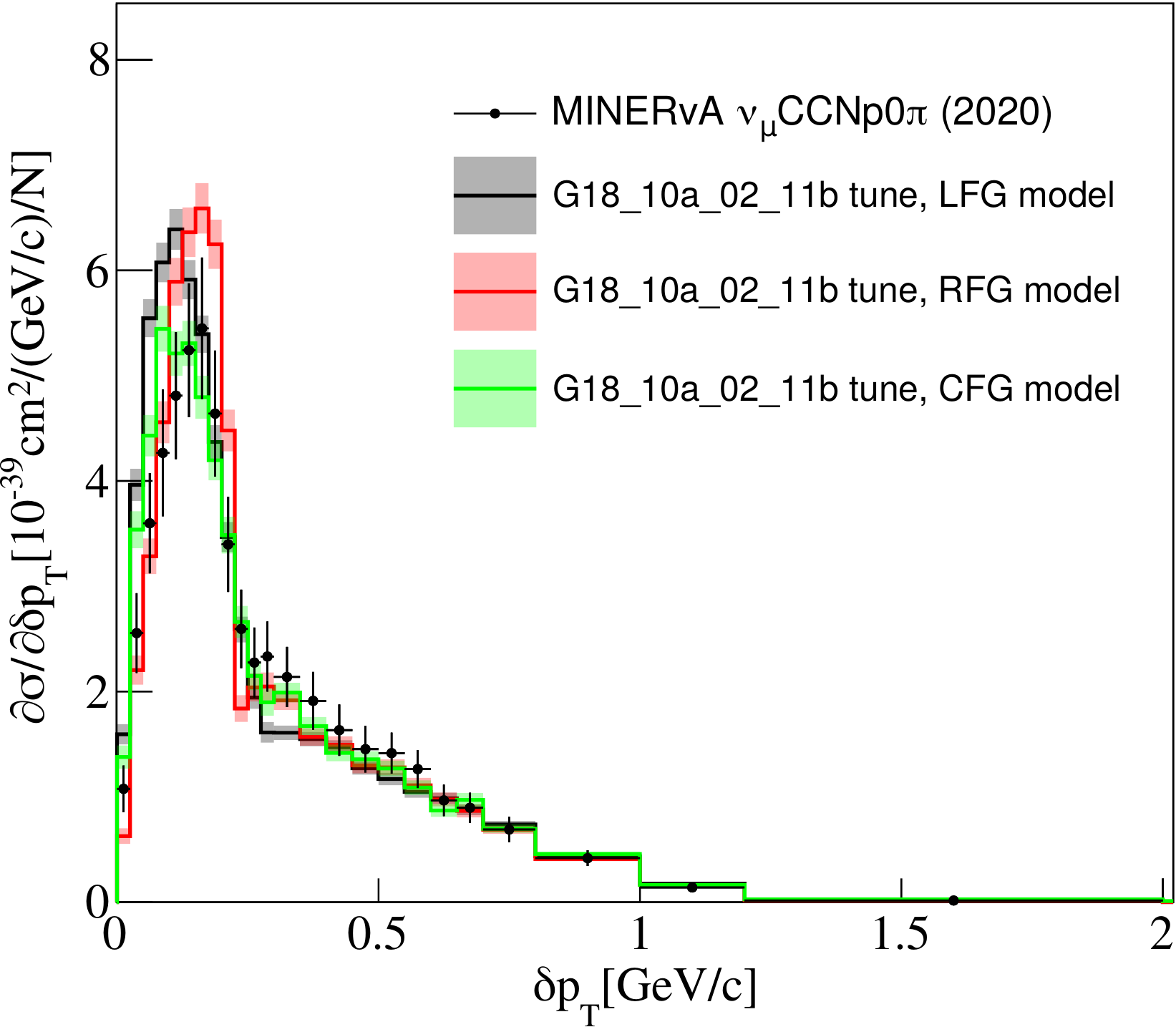}
    \caption{}
   	\end{subfigure}
   	\caption{Comparison of the \texttt{G18\_10a\_02\_11b} tune against T2K ND280~\cite{T2KCCQELikeMeasurement} (a) and MINER$\nu$A~\cite{MINERvA:2019ope} (b) $\nu_\mu$ CCNp0$\pi$ single-differential cross-section data as a function of $\delta \alpha_T$. 
    Three calculations are shown for different nuclear models: \ac{LFG} (black), \ac{RFG} (red) and \ac{CFG} (green). The errors correspond to statistical uncertainties only.}
    \label{fig:STKINuclearModel}
\end{figure}

The nuclear model choice affects the CCNp0$\pi$ predictions.
Figure~\ref{fig:STKINuclearModel} shows the impact of the underlying nuclear model against CCNp0$\pi$ single-differential cross-section measurements as a function of $\delta p_{T}$.
Differences between the models are significant for the cross-section peak prediction at low $p_{T}$.
The \ac{RFG} model is the only one out of the three that predicts the MINER$\nu$A data below the maximum.
However, it still over-predicts the cross section at the peak.
Alternatively, the \ac{CFG} model successfully predicts the peak normalization.
This is reflected in the $\chi^2_{\text{CFG}}$, reported in Tab.~\ref{tab:summarychi2_NModel_CCNpData}.

\begin{table}
    \centering
	\begin{tabular}{@{\extracolsep\fill} c c c c c}
	\hline\hline\noalign{\smallskip}
	\multicolumn{5}{c}{\texttt{G18\_10a\_02\_11b} } \\
    Dataset& $\chi_{\text{LFG}}^2$ & $\chi_{\text{RFG}}^2$ & $\chi_{\text{CFG}}^2$ & DoF \\
    \noalign{\smallskip}\hline\noalign{\smallskip}
    \multicolumn{5}{c}{T2K ND280 CCNp0$\pi$ data}\\
    \noalign{\smallskip}\hline\noalign{\smallskip}
    $d\sigma/d\delta p_T$                   & 228 & 149 &  27 &  8 \\
    $d\sigma/d\delta \phi_T$                & 292 &  29 &  20 &  8 \\
    $d\sigma/d\delta \alpha_T$              &  27 &  25 &  26 &  8 \\
    \noalign{\smallskip}\hline\noalign{\smallskip}
    \multicolumn{5}{c}{MINER$\nu$A CCNp0$\pi$ data}\\
    \noalign{\smallskip}\hline\noalign{\smallskip}
    $d\sigma/dp_p$                          & 21  &  23 &  15 & 25 \\
    $d\sigma/d\theta_p$                     &  58 &  35 &  34 & 26 \\
    $d\sigma/d\delta p_T$                   & 102 &  95 &  31 & 24 \\
	$d\sigma/d\delta\phi_T$                 &  87 &  32 &  18 & 23 \\
	$d\sigma/d\delta\alpha_T$               &  15 &  17 &  14 & 12 \\
    $d\sigma/d\delta p_{Tx}$                & 159 &  61 &  48 & 32 \\
    $d\sigma/d\delta p_{Ty}$                & 127 &  40 &  42 & 33 \\
    \noalign{\smallskip}\hline\noalign{\smallskip}
    \multicolumn{5}{c}{MicroBooNE CCNp0$\pi$ data}\\
    \noalign{\smallskip}\hline\noalign{\smallskip}
    $d\sigma/dp_\mu^{\text{reco}}$          &  71 & 35  &  32 & 10 \\
    $d\sigma/d\cos\theta_\mu^{\text{reco}}$ & 413 & 137 & 123 & 12 \\
    $d\sigma/dp_p^{\text{reco}}$            &  33 &  25 &  27 & 10 \\
    $d\sigma/d\cos\theta_p^{\text{reco}}$   & 100 &  49 &  42 &  9 \\
    $d\sigma/d\theta_{\mu p}^{\text{reco}}$ & 549 & 195 & 155 &  6 \\
    \noalign{\smallskip}\hline\hline
    \end{tabular}
    \caption{Summary of $\chi^2$ values associated with the CCNp0$\pi$ datasets specified in each row.
    The $\chi^2$ values are calculated using the \ac{NS} method for three GENIE predictions.
    The GENIE predictions are calculated with the \texttt{G18\_10a\_02\_11b} tune.
    Each prediction uses a different nuclear model: \ac{RFG}, \ac{LFG} or \ac{CFG}.
    \label{tab:summarychi2_NModel_CCNpData}}
\end{table} 

The main characteristic of the \ac{RFG} and the \ac{CFG} implementations in GENIE is that nucleons can have a momentum above the Fermi momentum in its ground state.
This tail in the momentum distribution is a consequence of nucleon correlations in the nuclear medium.
As a consequence of including those effects in the nuclear model, the description of the tail of the $\delta p_{T}$ distribution improves.
This study suggests that using a more elaborate nuclear model is key to describe CCNp0$\pi$ measurements.

\begin{figure}
    \centering
    \includegraphics[width=0.8\textwidth]{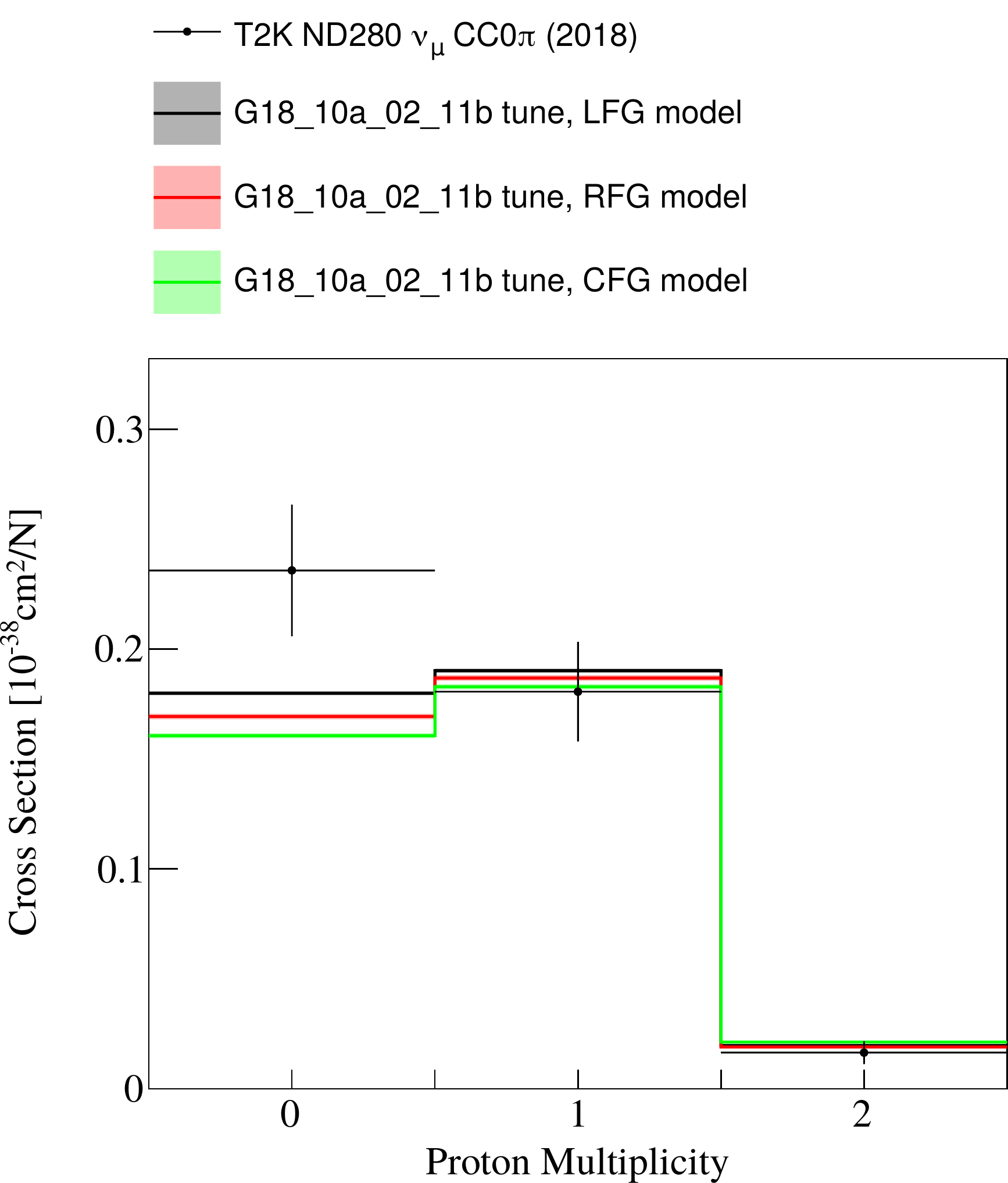}
    \caption{Comparison of \texttt{G18\_10a\_02\_11b} against T2K ND280 $\nu_\mu$ CCNp0$\pi$ total cross section data as a function of the proton multiplicity~\cite{T2KCCQELikeMeasurement}.
    Three calculations are shown for different nuclear models: \ac{LFG} (black), \ac{RFG} (red) and \ac{CFG} (green).
    \label{fig:PMultiplicityNuclearModel}}
\end{figure}

The differences between the three GENIE nuclear model predictions are not enough to explain the discrepancy between CC0$\pi$ and CCNp0$\pi$ data: all models predict a higher cross section for processes with protons in the final state concerning those with no protons in the final state. 
This is highlighted in Fig.~\ref{fig:PMultiplicityNuclearModel}.
This can be caused by FSI or initial state effects.
For example, more sophisticated nuclear models based on spectral functions were found to better describe CC0$\pi$ and CCNp0$\pi$ data, suggesting that a better nuclear model might be key to resolve the tension~\cite{Chakrani:2022tey}.

\subsubsection{Nucleon Final State Interaction model variations}
\label{sec:FSIImpactOnNsamples}

Mismodeling of nucleon \ac{FSI} can cause migration between CC0p0$\pi$ and CCNp0$\pi$ samples~\cite{PhysRevC.100.015505}.
Ref.~\cite{Dolan:2018zye} suggests increasing the nucleon mean-free path in cascade models might improve the agreement with CCNp0$\pi$ data from T2K ND280 and MINER$\nu$A.
This possibility is explored here. The effect of the mean-free path implementation is validated against proton transparency data for carbon.

A crucial test for \ac{FSI} models is to be able to reproduce nuclear transparency data from electron scattering experiments.
Transparency is defined as the probability for the knocked-out nucleon to not undergo \ac{FSI}s in the nuclear environment and it can be measured using electrons or neutrinos.
In transparency measurements, the final-state nucleon is produced inside the nucleus.
This feature is common with neutrino experiments, making transparency data extremely valuable to characterize and test \ac{FSI} modeling uncertainties.
Unfortunately, nuclear transparency measurements are scarce.
Few data points on proton transparency on carbon as a function of the proton momentum are available in Ref.~\cite{PhysRevC.45.780,ONEILL199587,PhysRevC.68.064603,PhysRevC.72.054602}.

Transparency can be easily calculated within \ac{MC} event generators as a ratio between the distribution of final-state protons which did or did not rescatter while leaving the nuclear environment.
Ref.~\cite{PhysRevC.100.015505} provided the first direct comparison of transparency calculations using a neutrino event generator.
This analysis took into account the experimental acceptances of the electron scattering experiments in the transparency definition.
Such an analysis could be replicated in GENIE; however, it is out of the scope of the present work.
To be able to compare GENIE's transparency calculations with data, we scale the GENIE predictions by the ratio between the transparency prediction from Ref.~\cite{PhysRevC.100.015505} with and without acceptance cuts.
This approach was used in Ref.~\cite{dytman2021comparison}.

\begin{figure}
    \centering
    \includegraphics[width=\textwidth]{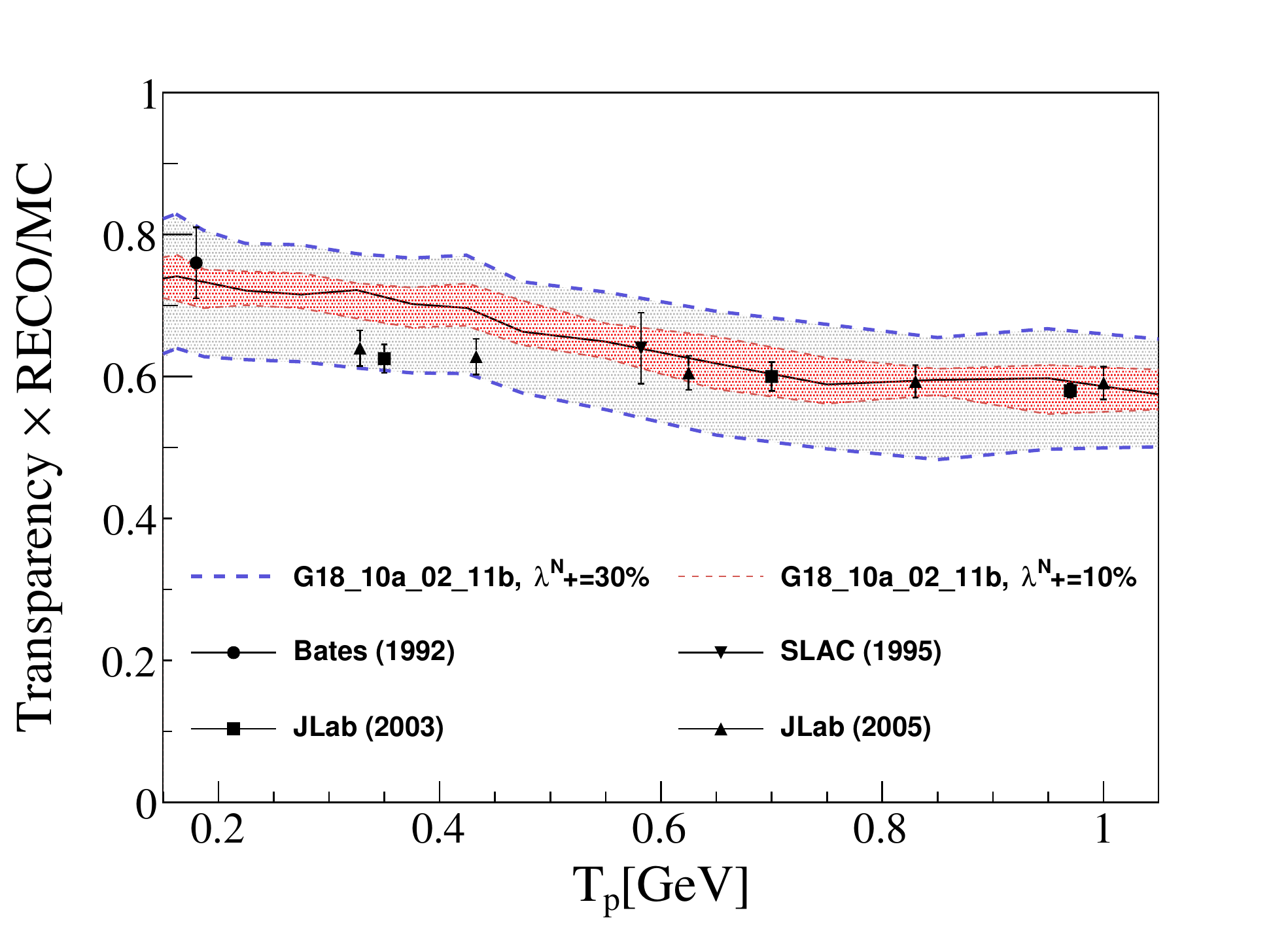}
    \caption{Comparison of proton transparency in carbon against the available electron scattering data~\cite{PhysRevC.45.780,ONEILL199587,PhysRevC.68.064603,PhysRevC.72.054602}.
	The GENIE prediction (black) is calculated using the \texttt{G18\_10a\_02\_11b}.
	The error bands correspond to the expected uncertainty when varying the nucleon mean free path by 10\% (red) and 30\% (grey).
	The predictions have been corrected according to experiments acceptance effects as determined in Ref.~\cite{PhysRevC.100.015505}.
    \label{fig:Transparency}}
\end{figure}

The effect on proton transparency calculations for carbon when varying the mean-free-path is shown in Fig.~\ref{fig:Transparency}.
The red and blue bands show the effect on the predictions when scaling up and down the nucleon mean-free path by 10\% and 30\% respectively.
The 10\% variation describes the data points with proton kinetic energies above 600~MeV within the 1$\sigma$ error bound.
The 30\% variation covers all the data available.
Fig.~\ref{fig:Transparency} suggests that at most a 30\% variation of the mean-free-path is feasible for low-momentum protons.
This is also supported by Ref.~\cite{dytman2021comparison}, where the authors observed a strong model dependency at low proton kinetic energies.
Another article~\cite{Bourguille_2021} finds that low momentum protons have a small re-scattering probability~\cite{Bourguille_2021}, reinforcing the need of a more realistic model for the nuclear ground state.

\begin{figure}
    \centering
    \includegraphics[width=0.9\textwidth]{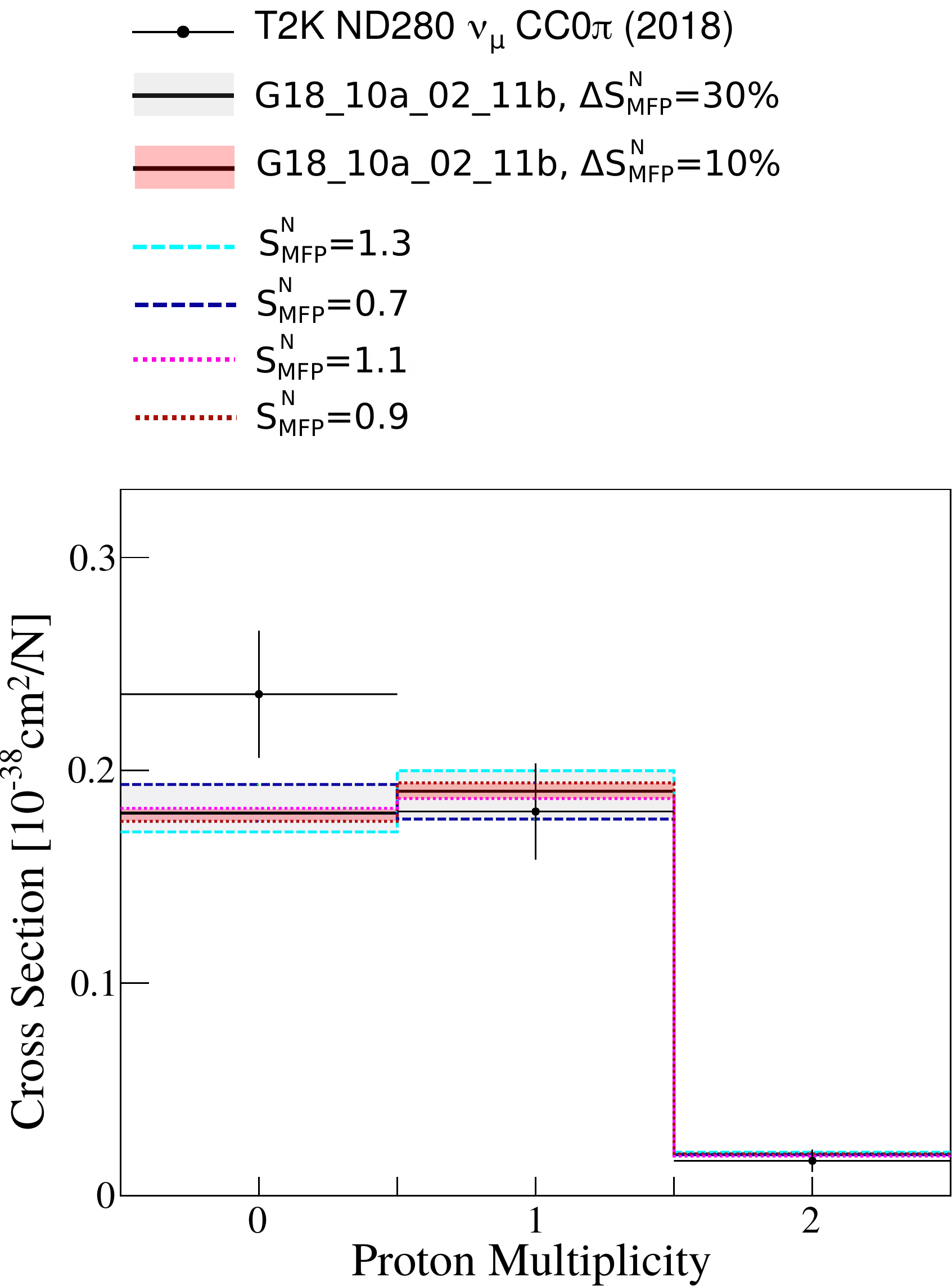}
    \caption{Comparison of the \texttt{G18\_10a\_02\_11b} tune against T2K ND280 $\nu_\mu$ CCNp0$\pi$ total cross-section data as a function of the proton multiplicity.
    The GENIE prediction (black) is calculated using the \texttt{G18\_10a\_02\_11b}.
	The error bands correspond to the expected uncertainty when varying the nucleon mean free path by 10\% (red) and 30\% (grey). 
    \label{fig:MFPEffectT2K}}
\end{figure}

The impact on the T2K ND280 cross section of variations in the nucleon mean-free path is shown in Fig.~\ref{fig:MFPEffectT2K} as a function of proton multiplicity.
It is observed that a higher nucleon mean-free path results in an increase of the proton multiplicity.
A higher cross-section is predicted for events with no protons above detection threshold when reducing the mean-free path. 
However, variations of the mean-free path are not enough to explain the observed tension.

Another possible line of study would be to determine whether more elaborate \ac{FSI} models can resolve the tension.
The \emph{hA} and \emph{hN} \ac{FSI} models are build on a simplistic view of the nuclear environment.
More complex approaches offer an improved description of CCNp0$\pi$ data~\cite{Ershova:2022jah,Nikolakopoulos:2022qkq,T2KCCQELikeMeasurement}.
Such a study is out of the scope of this paper.
		
\section{Conclusions}
\label{sec:NuclearConclusions}
This article describes the first neutrino-nucleus cross-section tuning effort within the GENIE Collaboration.
The goal of this work is to tune GENIE against CC0$\pi$ data and quantify the major sources of CC0$\pi$ modelling uncertainties.
In total, five partial tunes using double-differential flux-integrated $\nu_\mu$ or $\overline{\nu}_\mu$ CC0$\pi$ cross-section measurements on carbon as a function of the outgoing muon kinematics.
Each tune is performed with data from either MiniBooNE, T2K ND280 or MINER$\nu$A following the same analysis procedure.
Even though these experiments all use carbon as target, they are exposed to different neutrino beams, which peak at a different energy: the MiniBooNE and T2K fluxes peak below 1~GeV, whereas the MINER$\nu$A flux peaks at 3 GeV.
Hence, this work exploits tuning to study possible energy dependencies of components of the CC0$\pi$ cross section by comparing each of the partial tune results.
This analysis is based on the \texttt{G18\_10a\_02\_11b} CMC, which was previously tuned against free nucleon data~\cite{mypaper_1}. 

This tune confronted a number of new challenges with respect to previous GENIE free nucleon tuning efforts. This led to important changes in the GENIE tuning software.
In particular, modern nuclear data provide the full correlation between the data release bins due to systematic uncertainties. 
In order to incorporate this information in the analysis, the definition of the $\chi^2$ is modified to avoid the Peele’s Pertinent Puzzle, which leads to nonphysical normalization factors.

This analysis considers a total of seven parameters which attempt to capture the basic features of the component interactions - \ac{QEL}, \ac{2p2h}, and \ac{RES} as implemented in the \texttt{G18\_10a\_02\_11b} model set.
Some of the parameters used in this work affect the simulation of neutrino interactions on free nucleon.
We chose to constrain these parameters with correlated priors coming from previous GENIE tunes to bubble chamber data~\cite{mypaper_1}.
In addition, new parametrizations that encapsulate possible nuclear uncertainties were developed, using the Valencia model~\cite{Nieves} for the \ac{QEL} and \ac{2p2h} processes as a basis for choosing parameters.
These affect the strength of the RPA correction for \ac{CC}\ac{QEL} calculations in a nuclear environment as well as the strength and shape of the \ac{2p2h} cross section -- these are the main topic of this work.
Other relevant CC0$\pi$ parameterizations that affect the nuclear or \ac{FSI} models are discussed. These are found to be highly correlated with other aspects of the tune or not too sensitive to the CC0$\pi$ data considered in the tune.
For these reasons, these are not included in the tunes presented here.

All tunes present a common trend: the \ac{QEL} and \ac{2p2h} cross sections are enhanced and there is a preference for the \ac{QEL} cross section with \ac{RPA} corrections.
In addition, the tune results are in agreement with the priors imposed on free nucleon parameters.
Despite similarities, a clear energy dependence is observed for the \ac{2p2h} cross-section shape: the MiniBooNE and T2K tunes enhance the \ac{2p2h} cross section at the nucleon region, $W=M_N$, while suppressing it at the $\Delta$ region, $W=M_\Delta$.
Alternatively, both MINER$\nu$A tunes enhance the cross section in both regions, with an even higher scaling factor at the $\Delta$ region.
This suggests a dependence of the CC0$\pi$ cross section on the neutrino energy which is manifested in this work as a change in the shape and strength of the \ac{2p2h} cross section.

Tensions between the various CC0$\pi$ partial tunes and existing CCNp0$\pi$ data from T2K ND280, MINER$\nu$A and MicroBooNE exist. These results were shown in Table~\ref{tab:summarychi2_CC0piData} and are a key result of this work.
Parameters that give good agreement with one data set give poor $\chi^2$ values for the other data sets.
Since MiniBooNE and T2K are at very similar energies, this indicates tensions between the data sets of the two experiments.
In all cases, the tunes over-predict CCNp0$\pi$ data.
This tension was further investigated with a dedicated tune, which is performed using MINER$\nu$A $\nu_\mu$ CCNp0$\pi$ data.
The result suggests that a reduction of the \ac{QEL} and \ac{2p2h} cross sections would improve the agreement with all CCNp0$\pi$ data, contradicting the CC0$\pi$ partial tunes.
This tune also improves the agreement with MicroBooNE CCNp0$\pi$ data, suggesting that a possible $A$-dependence on the tune parameters is small, as was indicated in the MicroBooNE tune~\cite{MicroBooNE:2021ccs}.
The disagreement with CCNp0$\pi$ data is further explored in this paper, highlighting the importance of using a more realistic nuclear model and possible changes to FSI models to describe existing CCNp0$\pi$ data.
Tensions between neutrino and anti-neutrino tunes are also observed, suggesting the need of an additional modelling uncertainty. 
The observed tensions must be addressed before attempting to perform a global tune with all the available data.

\section{Acknowledgements}

We would like to thank Andy Buckley (University of Glasgow, UK) and Holger Schultz (Institute of Particle Physics Phenomenology, University of Durham, UK)
for their support interfacing the Professor tool with the software products that underpin the GENIE global analyses. 
We would like to thank the CC-IN2P3 Computing Center, as well as the Particle Physics Department at Rutherford Appleton Laboratory for providing computing resources and for their support. 
This work, as well as the ongoing development of several other GENIE physics tunes was enabled through a PhD studentship funded by STFC through LIV.DAT, 
the Liverpool Big Data Science Centre for Doctoral Training (project reference: 2021488).
The initial conceptual and prototyping work for the development of the GENIE / Professor interfaces, as well as for the development of the GENIE global analyses framework that, currently, underpins several analyses, was supported in part through an Associateship Award by the Institute of Particle Physics Phenomenology, University of Durham.
We thank: Stephen Dolan and Jaafar Chakrani for providing the method to correct for Peelle's Persistent Paradox; Daniel Ruterbories and Stephen Dolan for guidance to implement MINER$\nu$A's and T2K's analysis requirements in GENIE Comparisons; Xianguo Lu for discussion and comments on the paper.

\appendix

\section{Kinematic quantities of interest for CC0\ensuremath{\pi} measurements}
\label{sec:Observables}

Differential neutrino cross-section measurements are given as a function of different kinematic quantities. 
In this paper, these kinematic quantities are classified into \emph{direct}, \emph{inferred with an underlying process hypothesis} or \emph{inferred without an underlying process hypothesis}.

\subsection{Direct}
\label{sec:DirectObs}

Kinematic quantities that can be measured by the detector are classified as \emph{direct}.
For instance, an example of \emph{direct} quantity would be the muon momentum, $p_\mu$, or angle with respect to the beam-line axis, $\theta_\mu$.
In some cases, the muon kinetic energy, $T_\mu$, is used instead.
All cross-section measurements specified in Tab.~\ref{tab:datasummary} released data as a function of the muon kinematics.
Depending on the detector capabilities, \emph{direct} quantities can also be related to proton kinematics.
We refer to proton momentum and angle as $p_p$ and $\theta_p$ respectively.

Muon \emph{direct} quantities depend strongly on the neutrino energy and these are less sensitive to nuclear effects.
This motivated the recent efforts on the study of more exclusive topologies that allow measurement of the cross section as a function of the outgoing-proton kinematics~\cite{T2KCCQELikeMeasurement,PhysRevLett.121.022504,PhysRevD.102.112013,uBooNEQE}.
These depend weekly on the neutrino energy and are significantly altered by nuclear effects~\cite{STKIVariables}.

In some cases, the differential cross-section measurements are presented as a function of the reconstructed \emph{direct} quantities. 
Kinematic quantities in the reconstructed space are denoted with a '\text{reco}' superscript. For instance, $p_\mu^{\text{reco}}$ stands for reconstructed muon momentum.

\subsection{Inferred kinematic with an underlying process hypothesis}
\label{sec:RecoObs}

This category includes measurements that rely on the reconstruction of neutrino properties assuming a specific interaction type.
For instance, the kinematics of a CC0$\pi$ event can be reconstructed under the hypothesis that the initial nucleon was at rest and that there is no inelastic production of mesons in the final state (\ac{QEL} hypothesis).
Under this hypothesis, the reconstructed neutrino energy ($E_{\nu}^{\text{QEL}}$) and squared four-momentum transferred ($Q_{\text{QEL}}^2$) are:
\begin{gather}
\label{eq:EnuQE} 
E_{\nu}^{\text{QEL}} = \frac{M_{f}^{2}-(M_{i}-E_{b})^{2}-M^{2}_{\mu}+2(M_{i}-E_{b})E_{\mu}}{2(M_{i}-E_{b}-E_{\mu}+p_{\mu}\cos\theta_{\mu})} \\
\label{eq:Q2QE}
Q^2_{\text{QEL}} = 2E_{\nu,\text{QEL}}(E_{\mu}-p_{\mu}\cos\theta_{\mu})-M^2_{\mu}
\end{gather}
where $M_i$ ($M_f$) is the initial (final) nucleon mass, $M_\mu$ is the muon mass, and $E_\mu$ is the muon energy.
For the neutrino analysis, $M_i=M_n$ and $M_f=M_p$, whereas $M_i=M_p$ and $M_f=M_n$ for the antineutrino case.
The binding energy, $E_b$, depends on the target type.
Its specific value is provided in each analysis.

The main disadvantage of using these quantities is that the underlying hypothesis is uncertain. 
The presence of the nuclear environment complicates the characterization of event topologies: no single-event topology is produced only by a single underlying process.
This is highlighted by the T2K ND280 results on inferred kinematics~\cite{T2KCCQELikeMeasurement}.
In their analysis, they reconstruct the energy and momentum of the outgoing proton assuming a \ac{QEL} interaction.
Instead of presenting the cross-section measurement as a function of the \emph{inferred with kinematic with an underlying process} quantities, they used the difference between the \emph{direct} and the \emph{inferred} one.
In particular, T2K ND280 explored this quantity for the proton kinematics:
\begin{equation}
\label{eq:T2KInferred}
\begin{gathered}
	\Delta p_p            \equiv \left|\mathbf{p}_p^{\text{direct}}\right|-\left|\mathbf{p}_p^{\text{QEL}}\right|, \\
	\Delta \theta_p       \equiv \left|\theta_p^{\text{direct}}\right|-\left|\theta_p^{\text{QEL}}\right|, \\
	|\Delta \mathbf{p}_p| \equiv \left|\mathbf{p}_p^{\text{direct}}-\mathbf{p}_p^{\text{QEL}}\right|.
	\end{gathered}
\end{equation}
These are referred to as \emph{proton inferred kinematics} quantities.
Here, the superscript indicates whether the kinematic quantity is \emph{direct} (i.e $\mathbf{p}_p^{\text{direct}}$) or \emph{inferred} (i.e.\ $\mathbf{p}_p^{\text{QEL}}$) .
The reconstructed proton energy and momentum under the \ac{QEL} hypothesis are:
\begin{gather*}
E_p^{\text{QEL}} = E_\nu^{\text{QEL}} - E_\mu + M_p \\
\mathbf{p}_p^{\text{QEL}} = \left(-p_\mu^x, -p_\mu^y, -p_\mu^z+E_\nu^{\text{QEL}}\right)
\end{gather*}
These kinematic quantities can be used to highlight nuclear effects in CC0$\pi$ measurements, as the quantities defined in Eq.~\ref{eq:T2KInferred} deviate from zero when nuclear effects are present.


\subsection{Inferred without an underlying process hypothesis}

This category includes those kinematical quantities which are inferred from \emph{direct} ones but do not assume a specific underlying interaction process.
An example of interest for this work is the Single-Transverse Kinematic Imbalance (STKI) variables~\cite{STKIVariables}.
\ac{STKI} provide direct constraints on nuclear effects that, in some cases, have a weak dependence on the neutrino energy.
\ac{STKI} quantities are inferred from the muon and primary state hadron kinematics and only detectors capable of measuring low energy hadrons can provide such information.
So far, only T2K ND280 and MINER$\nu$A have released single-differential flux-integrated cross-section measurements as a function of these quantities~\cite{T2KCCQELikeMeasurement,PhysRevLett.121.022504,MINERvA:2019ope}.

The \emph{transverse momentum imbalance}, $\delta\mathbf{p}_T$, is defined as the sum of the transverse muon and proton momentum:
\begin{equation*}
	\delta \mathbf{p}_T \equiv \mathbf{p}^\mu_T + \mathbf{p}^p_T.
\end{equation*}
As the neutrino travels in the longitudinal direction, the transverse muon momentum is related to the transverse momentum transfer as $\mathbf{p}^\mu_T=-\mathbf{q}_T$.
The angle between $\delta\mathbf{p}_T$ and $-\mathbf{p}^\mu_T$ is known as \emph{boosting angle}, $\delta\alpha_T$:
\begin{equation*}
	\delta \alpha_T \equiv \arccos\left(\frac{-\mathbf{p}_T^\mu \cdot \delta \mathbf{p}_T}{p_T^{\mu}\delta p_T}\right).
\end{equation*}
The deflection of the nucleon with respect to $\mathbf{q}_T$ is measured with the $\delta\phi_T$ angle:
\begin{equation*}
	\delta \phi_T \equiv \arccos\left(\frac{-\mathbf{p}_T^\mu\cdot\mathbf{p}_T^p}{p_T^\mu p_T^p}\right).
\end{equation*}
A more recent study investigates the CC0$\pi$ cross-section dependency on the muon-proton momentum imbalances parallel ($\delta p_{Ty}$) and longitudinal ($\delta p_{Tx}$) to the momentum transfer in the transverse plane~\cite{MINERvA:2019ope}.
These quantities are mathematically defined as:
\begin{align*}
\delta p_{Tx} = &\ \left(\hat{\mathbf{p}}_\nu \times \hat{\mathbf{p}}_T^\mu\right) \cdot \delta \mathbf{p}_T \\
\delta p_{Ty} = &\ -\hat{\mathbf{p}}^\mu_T \cdot \delta \mathbf{p}_T
\end{align*}
given the Cartesian coordinate system defined with respect to the neutrino and muon kinematics.	
The neutrino direction is given by $\hat{\mathbf{p}}_\nu$.
All these quantities define what experiments refer to as \ac{STKI} variables.

\begin{figure}
	\centering
	\begin{subfigure}{0.45\textwidth}
		\centering
        \includegraphics[width=0.8\textwidth]{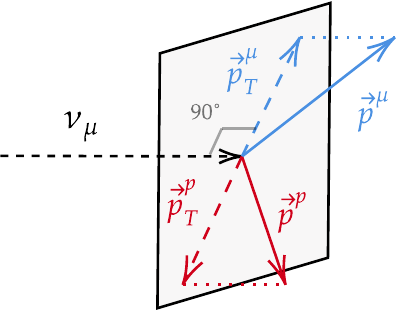}
        \caption{}
        \label{fig:STINoNuclearEffects}
	\end{subfigure}%
	\begin{subfigure}{0.55\textwidth}
		\centering
        \includegraphics[width=\textwidth]{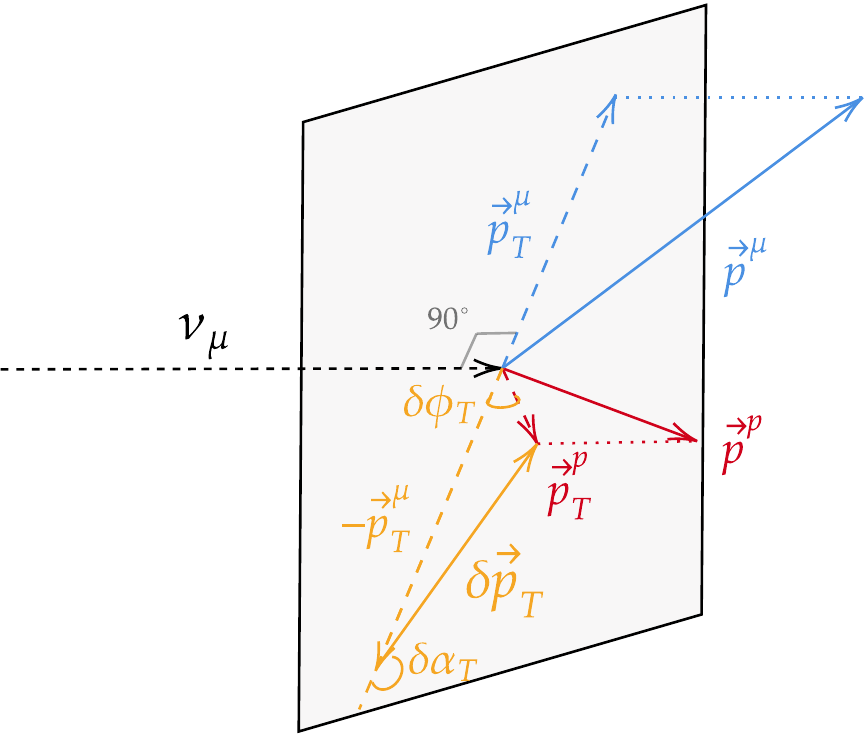}
        \caption{}
        \label{fig:NuclearEffectsSTI}
	\end{subfigure}
	\caption{Graphical definition of the \ac{STKI} variables in a $\nu_\mu$ \ac{CC}\ac{QEL} neutrino interaction on a nuclear target. The incoming neutrino, represented as a dashed arrow, interacts with a free nucleon at rest (a) or with a bound nucleon subject to Fermi motion (b). The outgoing muon (proton) is represented in blue (red). The transverse plane is represented in grey. The incoming neutrino is perpendicular to the transverse plane. Nuclear effects distortion the free-nucleon picture (a) creating an imbalance between the muon and nucleon transverse momentum (b). The \ac{STKI} variables that define this imbalance are highlighted in orange. 
	\label{fig:STI}}
\end{figure}

\begin{figure}
	\centering
	\begin{subfigure}{\textwidth}
		\centering
        \includegraphics[width=0.9\textwidth]{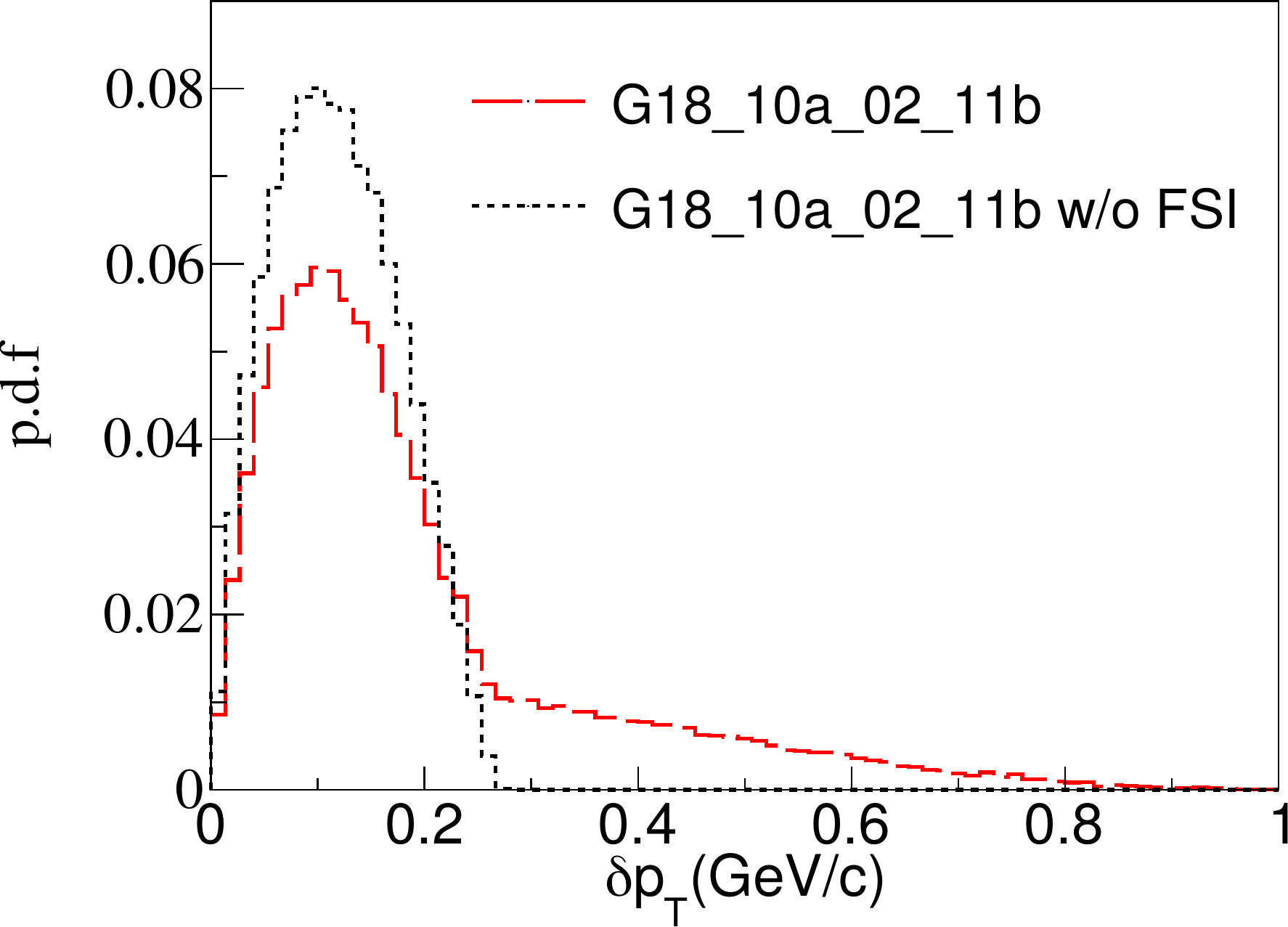}
        \caption{}
        \label{fig:STKIDptFSI}
	\end{subfigure}
	\begin{subfigure}{\textwidth}%
		\centering
        \includegraphics[width=0.9\textwidth]{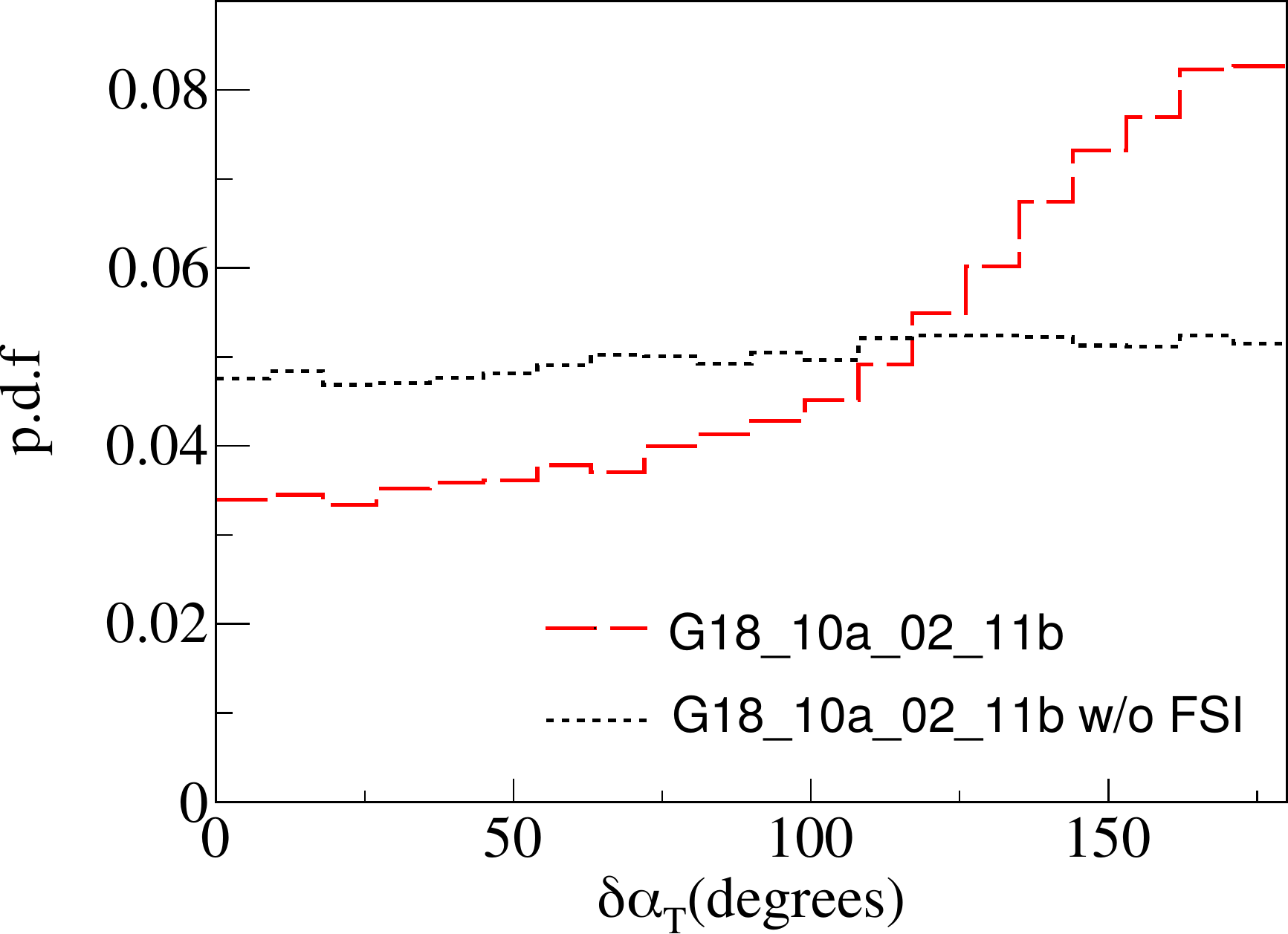}
        \caption{}
	\end{subfigure}
	\caption{Probability density function of $\delta p_T$ and $\delta\alpha_T$ for the \texttt{G18\_10a\_02\_11b} tune with (a) and without (b) \ac{FSI}. 
	Both predictions are obtained simulating $\nu_\mu$ \ac{CC}\ac{QEL} interactions only on $^{12}$C at 1~GeV with the \texttt{G18\_10a\_02\_11b} tune.}
\label{fig:STKIFSI}
\end{figure}

A graphical representation of the definition of the \ac{STKI} variables for a neutrino interaction with and without nuclear effects is shown in Fig.~\ref{fig:STKIFSI}.
When the interaction occurs with a static free nucleon, i.e.\ no nuclear effects, ${\mathbf{p}^\mu_T = - \mathbf{p}^p_T}$, $\delta \mathbf{p}_T = 0$ and $\delta\phi_T=0$, see Fig.~\ref{fig:STINoNuclearEffects}.
However, this picture is modified by Fermi motion, nucleon correlations, non-\ac{QEL} interactions and \ac{FSI}.
If \ac{FSI} effects and nucleon correlations are neglected, $\delta \mathbf{p}$ coincides with the initial nucleon momentum $\mathbf{p}_{N_i}$.
Moreover, $\delta\alpha_T$ is uniform due to the isotropic nature of the Fermi motion.
\ac{FSI} effects smear the $\delta \mathbf{p}_T$ distribution and modify the shape of the $\delta\alpha_T$ distribution.
In GENIE, the \emph{hA} \ac{FSI} model enhances the cross section at $\delta p_T>0.2$~GeV/c and $\delta\alpha_T\sim 180^{\circ}$; see Fig.~\ref{fig:STKIFSI}.
This region is refereed to as high-transverse kinematic imbalance region.

\begin{figure}
	\centering
	\begin{subfigure}{0.8\columnwidth}
		\centering
        \includegraphics[width=\columnwidth]{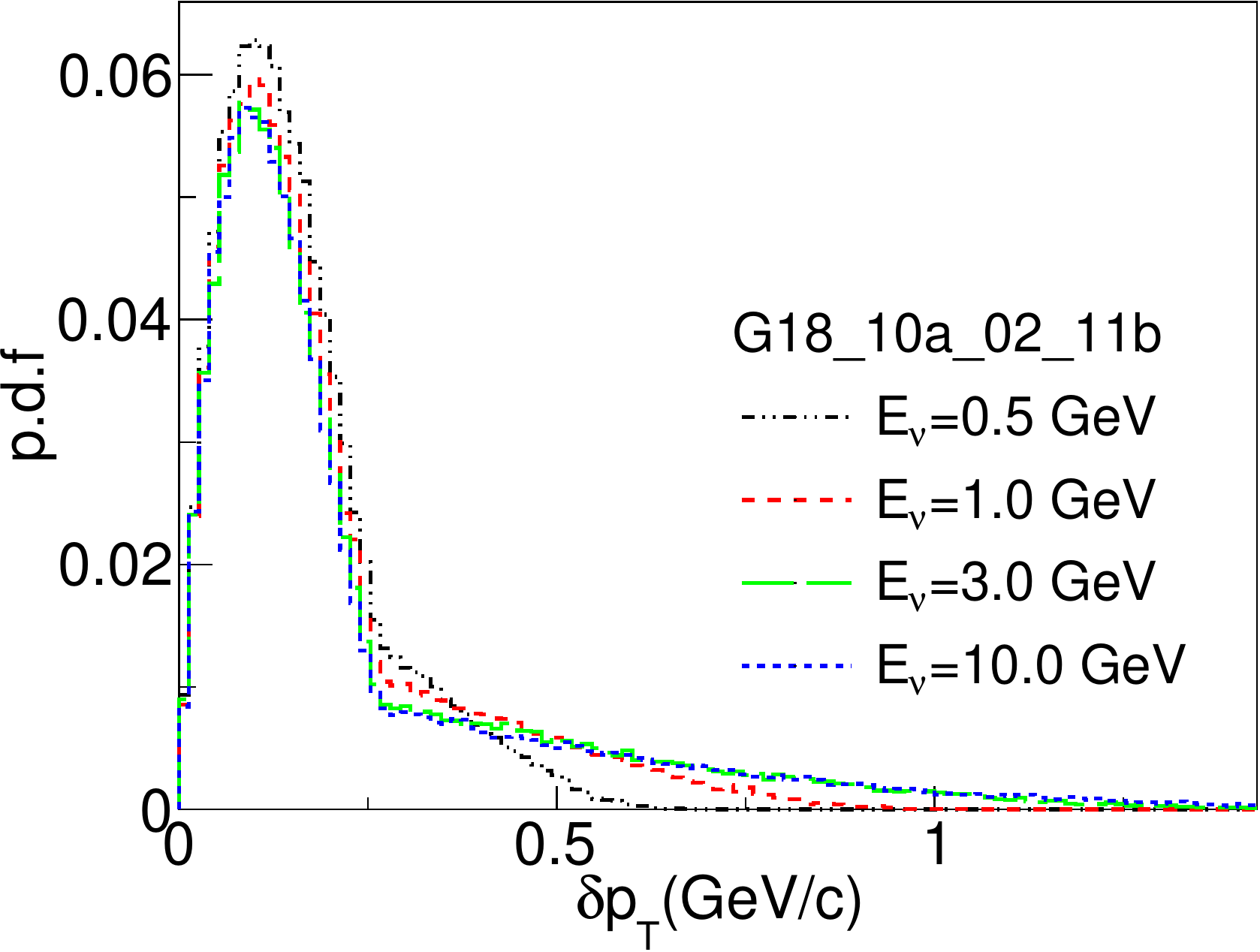}
        \caption{}
        \label{fig:STKIDptEnu}
	\end{subfigure}
	
	\begin{subfigure}{0.8\columnwidth}%
		\centering
        \includegraphics[width=\columnwidth]{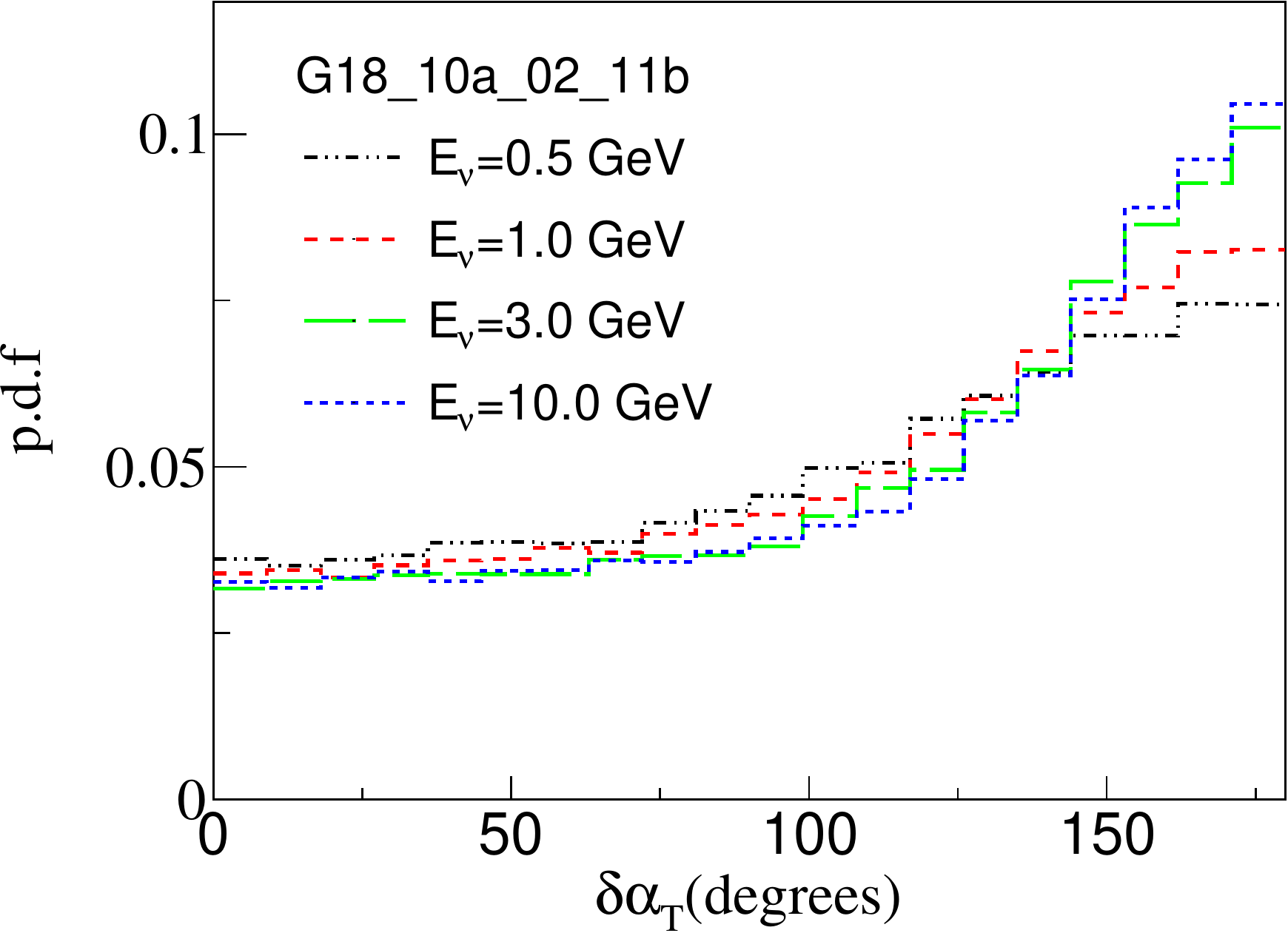}
        \caption{}
        \label{fig:STKIDatEnu}
	\end{subfigure}

		\begin{subfigure}{0.8\columnwidth}%
		\centering
        \includegraphics[width=\textwidth]{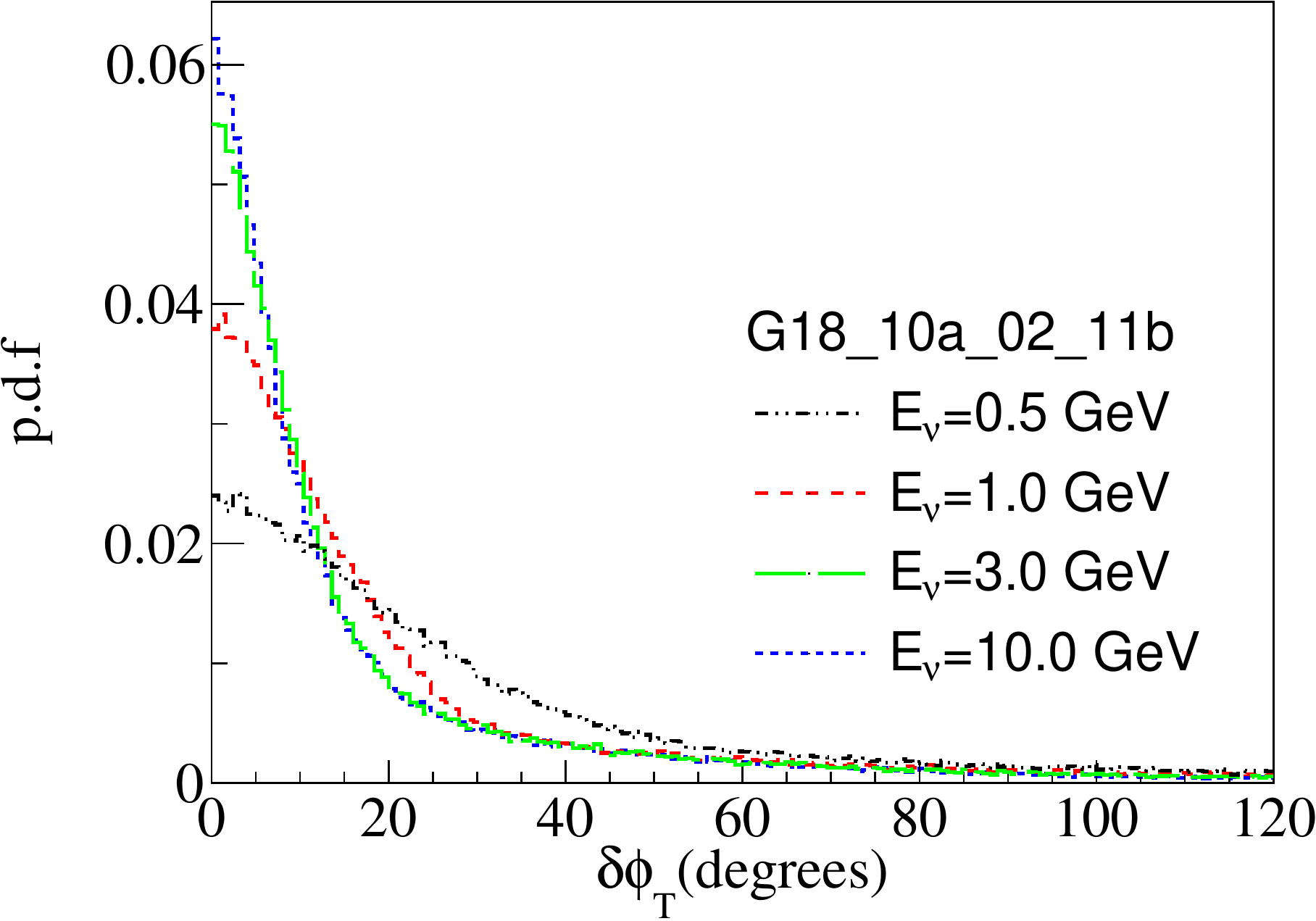}
        \caption{}
        \label{fig:STKIDphitEnu}
	\end{subfigure}
	\caption{Probability density function of \ac{STKI} variables.
			The predictions are obtained simulating $\nu_\mu$ \ac{CC}\ac{QEL} interactions only on $^{12}$C with the \texttt{G18\_10a\_02\_11b} tune at different neutrino energies.}
\label{fig:STKIEnu}
\end{figure}

Ref.~\cite{STKIVariables} demonstrated that the $\delta p_T$ and $\delta\alpha_T$ dependence on the neutrino energy is smaller than possible uncertainties due to \ac{FSI} modeling.
The $\delta\phi_T$ variable has a stronger dependence on the neutrino energy as it scales with $\delta p_T/p^\mu_T$: at higher neutrino energies, the distribution at small angles becomes narrower.
The dependency of the \ac{STKI} variables in GENIE with the neutrino energy is shown in Fig.~\ref{fig:STKIEnu}.
Changes in the neutrino energy affect mostly the tail of the $\delta p_T$ distribution and the $\delta\alpha_T$ distribution at backward angles.

\section{Comparisons \texttt{G18\_10\_02\_11b} against neutrino-nucleus CC0$\pi$ data}
\label{sec:comparisons_data}
This section offers with comparisons of GENIE against all CC0$\pi$ and CCNp0$\pi$ data available from MiniBooNE, T2K, MINER$\nu$A and MicroBooNE.
The corresponding GENIE predictions are obtained by replicating the analysis within GENIE: neutrino interaction events are simulated for each experiment given the neutrino flux, target material, and analysis cuts.
The normalized neutrino flux spectra is reported in Fig.~\ref{fig:fluxes}.
With this information, the GENIE prediction for the corresponding differential flux-integrated cross section is evaluated.

The format of all the comparisons with data reported in this appendix is common: the data and differential cross-section prediction are represented in black.
In addition, the contribution from different interaction models is shown for \ac{CC}\ac{RES}, \ac{CC}\ac{2p2h} and \ac{CC}\ac{DIS}/\ac{SIS}.
The contribution to the \texttt{G18\_10a\_02\_11b} predictions from \ac{CC}\ac{DIS}/\ac{SIS} events is really small at the neutrino energies considered in this work.
For this reason, the contribution is grouped into a single category (\ac{DIS}).
The \ac{2p2h} contribution is divided further into four categories that depend on the event invariant mass, $W$.
The $W$ regions are:
\begin{compactitem}
    \item $W<M_N=938$~MeV/c$^2$
    \item $M_N<W<W_{\text{Dip}}=1120$~MeV/c$^2$
    \item $W_{\text{Dip}}<W<M_{\Delta}=1232$~MeV/c$^2$
    \item $W>M_\Delta$
\end{compactitem}
The data error bars include statistical and systematic uncertainties.
The errors on the $x$-axis represent the bin width used in the original analysis.

\subsection{MiniBooNE CC0\ensuremath{\pi} cross-section measurement}
\label{sec:MiniBooNEAnalysis}

The MiniBooNE experiment studies neutrinos produced with the \ac{BNB}~\cite{BNBFlux}. 
MiniBooNE published the first-high statistics $\nu_\mu$ and $\overline{\nu}_\mu$ CC0$\pi$ flux-integrated double differential cross-section measurement on carbon, at $\langle E_\nu \rangle \sim 800$~MeV and $\langle E_{\overline{\nu}}\rangle \sim 500~\text{MeV}$ respectively~\cite{1002.2680,MinibooneAntineutrino}.
The flux-unfolded total cross section, $\sigma\left(E_\nu^{\text{QEL}}\right)$,
and the flux-integrated single differential cross section as a function of the squared four-momentum transferred, $d\sigma/dQ^2_{\text{QEL}}$, were also reported.

Both MiniBooNE analyses study \ac{CC}0$\pi$ events with a muon in the final state and no pions.
The signal topology of a muon in the detector is described in two sub-events: the first one associated with the primary Cherenkov light from the muon, and the second one, produced by the Cherenkov light from the Michel electron, which is produced in the muon decay.
This requirement provides a sample of mostly \ac{CC} events, as neutral-current events only have one sub-event.

Positively charged pions produced in the detector leave a distinct signature in the detector, as the $\pi^+$ decays immediately into a muon and a muon neutrino. 
The Cherenkov light from the $\pi^+$ contributes to the total light of the primary muon. 
This process can be distinguished from a \ac{CC}\ac{QEL} interaction as the muon produced from the pion decay will also decay into a Michel electron (three sub-events).
Negatively charged pions are absorbed by the nuclear environment and contribute to the CC0$\pi$ topology.
In the GENIE predictions, pion production events are removed by requiring no pions in the final state.

 Recoil protons also emit scintillation light. 
However, such scintillation light signal produced is either indistinguishable from the muon signal or its momentum below the Cherenkov threshold.
For this reason, no requirements based on the recoil proton are considered in the MiniBooNE analyses.

The analysis considers further model-dependent cuts to correct for backgrounds and extract the \ac{CC}\ac{QEL} cross-section from the CC0$\pi$ sample.
In the original publication, these are referred to as \emph{irreducible backgrounds}.
An example of irreducible background is CC1$\pi$ events that were not removed by the cut on the pion subevent topology
or pion production events in which the pion is absorbed.
This is corrected using a \ac{MC} simulation tuned to $\nu_\mu$ CC1$\pi$ MiniBooNE data.
Information on $\nu_\mu$ CC1$\pi^+$ sample is used to characterize this background and correct for single-pion events which were not removed by the CC0$\pi$ selection criteria in the neutrino and antineutrino analyses.
This procedure is one of the main limitations of this dataset as it incorporates strong biases in the reported measurement.
The contribution to the cross-section measurement from irreducible backgrounds is also reported, allowing the comparison against CC0$\pi$ data.

The quality of the MiniBooNE CC0$\pi$ data release is poor in comparison with the rest.
The MiniBooNE collaboration provided measurements in bins of $T_\mu$ and $\cos\theta_\mu$, but did not provide the bin-to-bin covariances for either of the two measurements.
Instead, they quoted a normalization systematic uncertainty of $\sim10.7$\% ($17.2$\%) for the neutrino (antineutrino) measurement.
As suggested by Ref.~\cite{1002.2680}, this error is added as a systematic in our database, effectively including a correlation between the bins.

\begin{figure*}
    \centering
        \includegraphics[width=\textwidth]{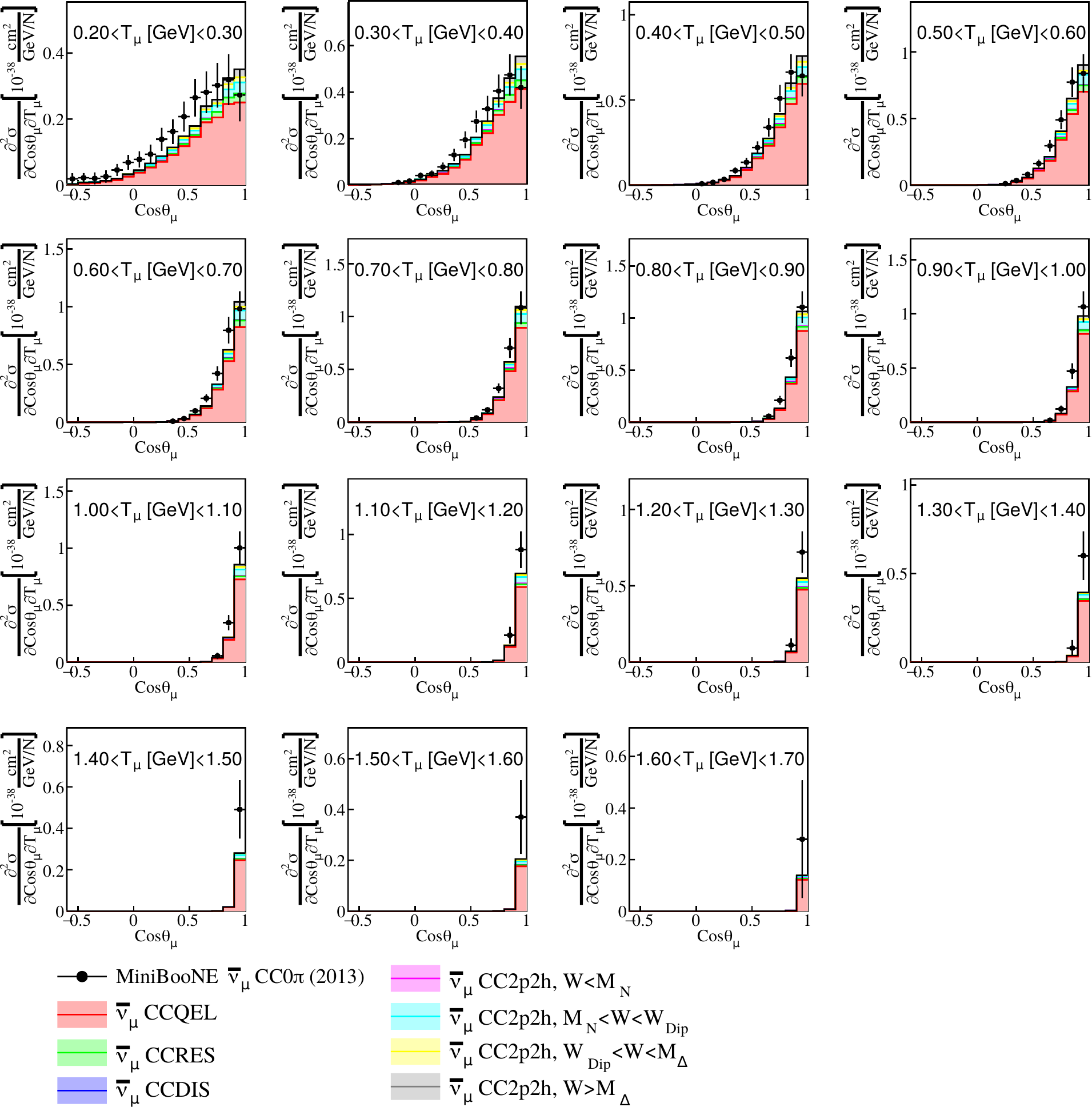}
        \caption{MiniBooNE $\overline{\nu}_\mu$ CC0$\pi$ double differential flux-averaged cross section as a function of the muon angle, $\theta_\mu$, and kinetic energy, $T_\mu$~\cite{MinibooneAntineutrino}. The corresponding slices on $T_\mu$ are compared against the \texttt{G18\_10a\_02\_11b} tune. The GENIE prediction is divided into different categories: \ac{CC}\ac{QEL}, \ac{CC}\ac{RES}, \ac{CC}\ac{2p2h} and \ac{CC}\ac{DIS}.}
        \label{fig:MinibooneBarDefY}
\end{figure*}

In Figs.~\ref{fig:MinibooneDefY} and \ref{fig:MinibooneBarDefY}, the flux-integrated double differential $\nu_\mu$ and $\overline{\nu}_\mu$ CC0$\pi$ cross section data as a function of $p_\mu$ and $T_\mu$ are compared against GENIE.
The main observation is that the GENIE tune under-predicts the data. 
In particular, the \texttt{G18\_10a\_02\_11b} disagreement with the data are more significant at backward angles, where the cross-section is determined by \ac{CC}\ac{QEL} events only.
The disagreement is also observed at forward angles, where there is a significant contribution from non-\ac{QEL} events.

\subsection{T2K CC0\ensuremath{\pi} cross-section measurements}
\label{sec:T2KAnalysis}

The Tokai-to-Kamioka (T2K) experiment is an accelerator-based long-baseline experiment that studies neutrino oscillations. 
Neutrinos are generated at the Japan Proton Accelerator Research Complex (J-PARC) facility~\cite{PhysRevD.87.012001}. The target for the neutrino beam is 280~m away from the T2K near detectors~\cite{ABE2011106}: INGRID, WAGASCI and ND280.
The T2K ND280 detector is used to measure neutrino interactions on carbon at $\langle E_\nu\rangle \sim 600$~MeV.
The WAGASCI module was recently added to the T2K ND facility and it measures neutrino interactions at 0.86 GeV.
Details on the detector setup can be found in Ref.~\cite{T2KCCQELikeMeasurement,10.1093/ptep/ptab014}.
Most measurements described here use the detector central tracker region, composed of three time projection chambers (TPC) and two fine-grained detectors (FGD1 and FDG2). 
The FGDs are the target mass and are also used to track charged particles.
Carbon measurements use the FGD1 as the target mass.
The central region is surrounded by an electromagnetic calorimeter (ECal), which is contained within a magnet.
This setup allows measuring the particle charge and momentum.
This information, together with energy deposition, is used to identify charged particles.

The first double-differential $\nu_\mu$ \ac{CC}0$\pi$ measurement provided by T2K ND280 was released back in 2015~\cite{PhysRevD.93.112012}.
This measurement is surpassed by Ref.~\cite{T2KCCQELikeMeasurement}, which considers improved constraints on systematic uncertainties.
Ref.~\cite{T2KCCQELikeMeasurement} provides additional measurements including double- and triple-differential measurements for different proton multiplicities as well as two CCNp0$\pi$ single-differential cross-section measurements as a function of \ac{STKI} and \emph{proton inferred kinematics} quantities. 

All measurements from Ref.~\cite{PhysRevD.93.112012} require one muon and no pions in the final state, regardless of the number of nucleons in the event.
Any event must contain at least one track in the TPC, which must be either a muon or a proton. 
If it is a proton, they look for a muon-like track in the FDG1 or ECal.
Other events with tracks that are not consistent with the muon-like or proton-like signature are rejected.
Events with low-momentum charged or neutral pions are removed by requiring no Michel electrons or photons.
At the \ac{MC} level, this is implemented by removing events with pions or photons in the final state, respectively.

The selected sample is divided further depending on the number of protons above the detection threshold of 500~MeV/c: no protons (CC0p0$\pi$), one proton (CC1p0$\pi$) or more than one visible proton (CC2p0$\pi$) in the final state.
The CC0p0$\pi$ and CC1p0$\pi$ are double- and triple-differential cross-section measurements as a function of the muon and muon and proton kinematics respectively.
The total CC2p0$\pi$ cross-section is also reported.
The \ac{STKI} and \emph{proton inferred kinematics} are obtained with the CCNp0$\pi$ sample: they require the presence of at least one visible proton ($p_p>500$~MeV/c).

\begin{table*}
    \centering
    \resizebox{\textwidth}{!}{
    \begin{tabular}{c c c c c} \hline\hline\noalign{\smallskip}
    T2K ND280 Analysis & $p_p$ & $\cos\theta_p$ &$p_\mu$ & $\cos\theta_\mu$ \\
    \noalign{\smallskip}\hline\hline\noalign{\smallskip}
	CC0p0$\pi$ & $<500$~MeV/c (or no proton) & & & \\
	CC1p0$\pi$ & $>500$~MeV/c & & & \\ 
	CCNp0$\pi$, \ac{STKI} & $450<p_p<1000$~MeV/c & $>0.4$ & $>250$~MeV/c & $>-0.6$ \\
	CCNp0$\pi$, proton inferred kinematics & $>450$~MeV/c & $>0.4$ & & \\

    \noalign{\smallskip}\hline\hline
    \end{tabular}}
    \caption{Phase-space restrictions for the T2K ND280 analyses from Ref.~\cite{T2KCCQELikeMeasurement}. 					The proton cuts are only applied to the highest energy proton.}
    \label{tab:T2KRequirements}
\end{table*}

Efficiency corrections for $\nu_\mu$ CCNp0$\pi$ events can be model dependent.
To avoid this, different kinematical restrictions are considered for each analysis, selecting regions in which the efficiency is flat or well understood.
These are specified in Tab.~\ref{tab:T2KRequirements}.
Events with more than one proton are reconstructed using the information from the highest energy one, which has to satisfy the kinematical limits of Tab.~\ref{tab:T2KRequirements}.
The samples are not corrected for events with protons below the detection threshold or any of the kinematical cuts considered in the analysis.
The same cuts are applied at the generator level when evaluating the GENIE predictions.

\begin{figure*}
\centering
    \includegraphics[width=0.92\textwidth]{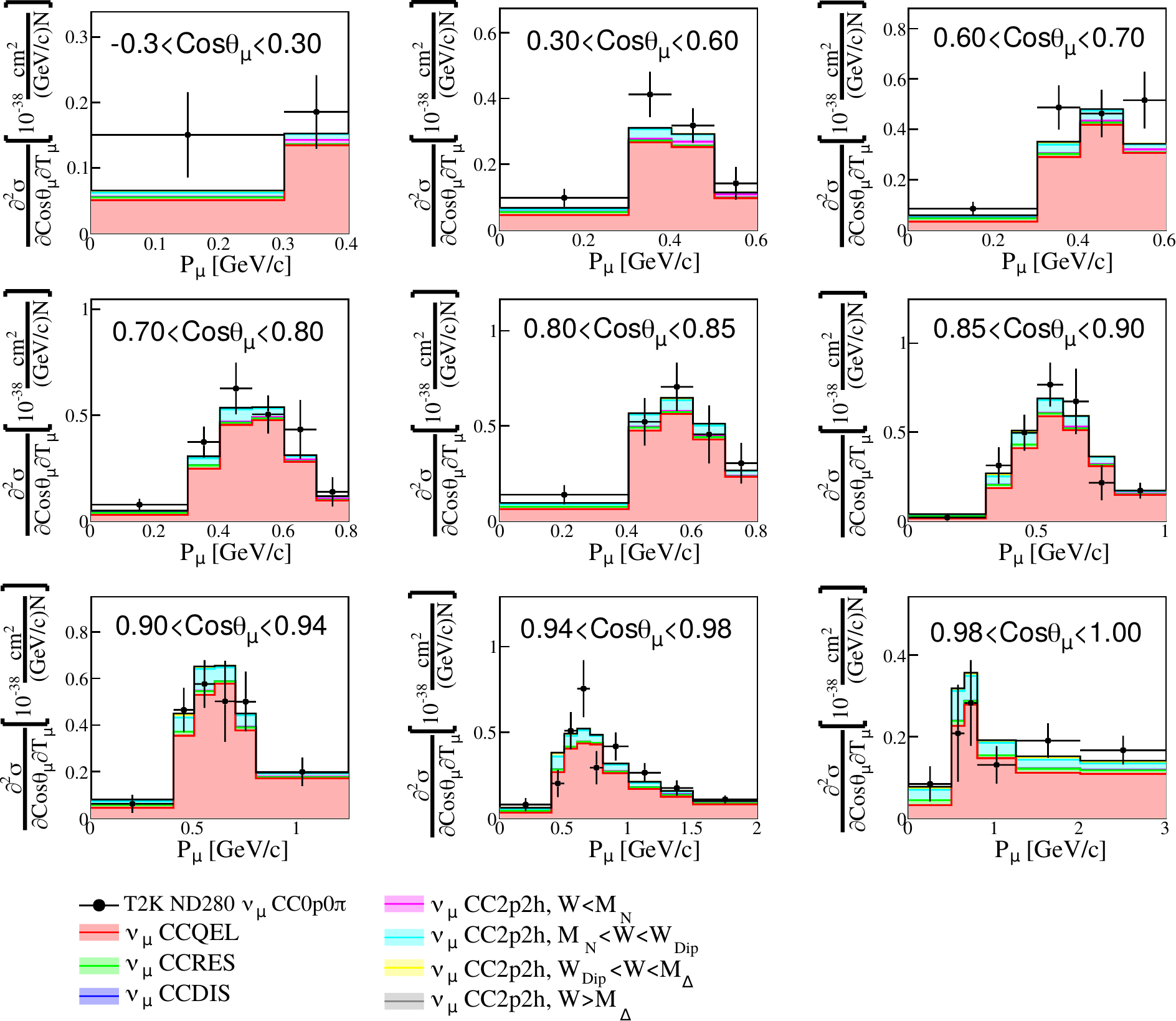}
    \caption{T2K ND280 flux-averaged $\nu_\mu$ \ac{CC}0p0$\pi$ differential cross section as a function of the proton multiplicity~\cite{PhysRevD.93.112012}. The data are compared against the \texttt{G18\_10a\_02\_11b} tune. The GENIE prediction is divided into different categories: \ac{CC}\ac{QEL}, \ac{CC}\ac{RES}, \ac{CC}\ac{2p2h} and \ac{CC}\ac{DIS}.  For readability, the \ac{CC}0p$0\pi$ high energy bins are not included in these plots.
    \label{fig:T2K0p}}
\end{figure*}

The GENIE comparison against the $\nu_\mu$ CC0p0$\pi$ double-differential cross section are presented in Fig.~\ref{fig:T2K0p}.
The main contribution to the CC0p0$\pi$ topology comes from \ac{CC}\ac{QEL} events.
The second contribution is from \ac{CC}\ac{2p2h} events with $M_N<W<W_{\text{Dip}}$.
The contribution from \ac{2p2h} events with $W<M_N$ or $W>W_{\text{Dip}}$ is negligible for the CC0p0$\pi$ measurement.
GENIE is under-predicting the data at backward angles.

This disagreement in the overall normalization is also observed in Fig.~\ref{fig:T2KMultiplicity}, which compares GENIE against the cross-section as a function of the proton multiplicity.
This observation conflicts with $\nu_\mu$ \ac{CC}1p0$\pi$ data, which is not under-predicted.
There are some outstanding differences between the GENIE predictions for CC0p0$\pi$ and CC1p0$\pi$ data.
Whilst the total contribution from \ac{2p2h} events is similar, the main \ac{2p2h} contribution comes from \ac{2p2h} events with $W>W_{\text{Dip}}$.
In addition, the fraction from \ac{RES} events is higher with respect to the CC0p0$\pi$ one.

\begin{figure}
    \centering
    \begin{subfigure}{0.95\textwidth}
        \includegraphics[width=\textwidth]{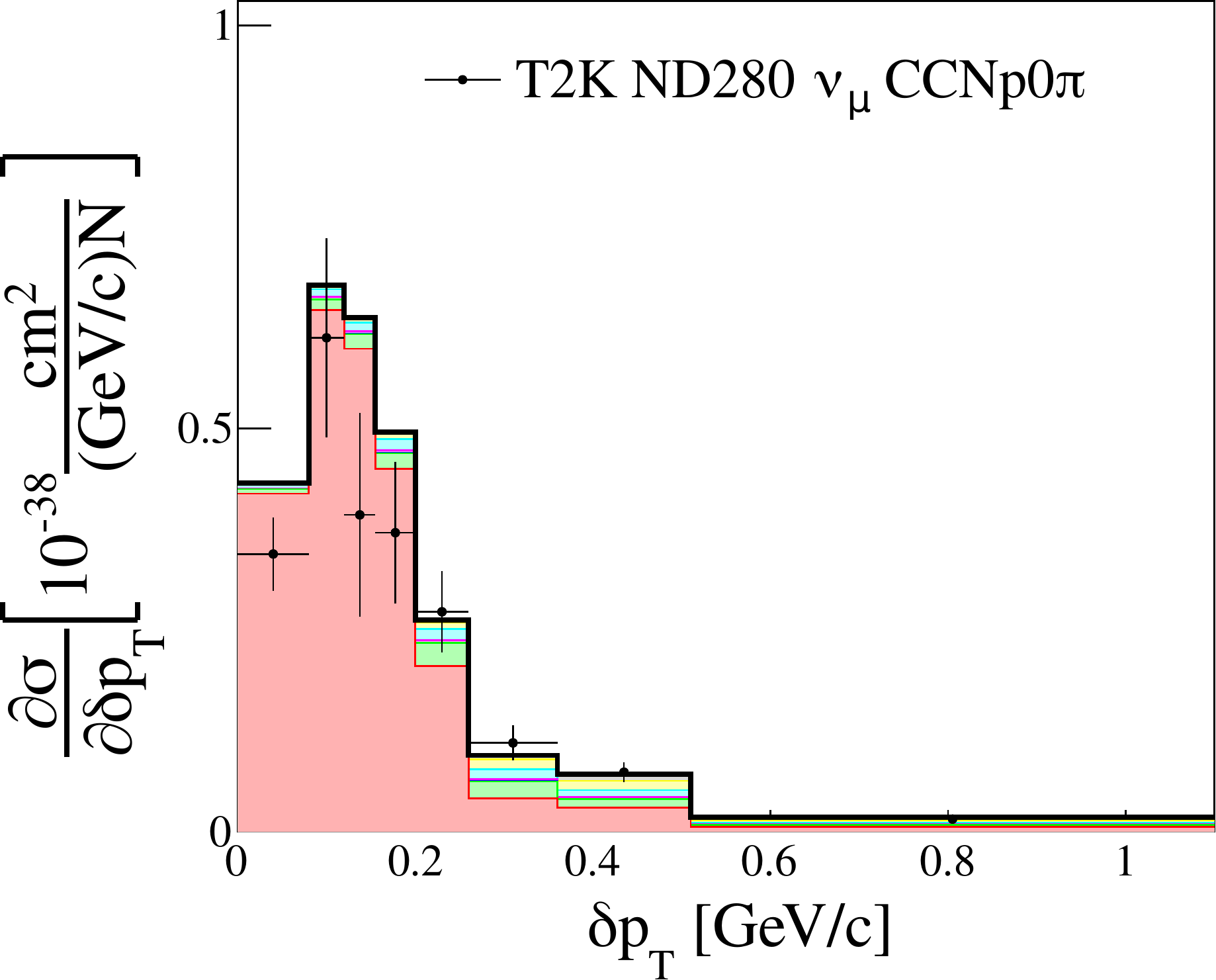}
        \caption{}
    \end{subfigure}
    \begin{subfigure}{0.95\textwidth}
        \includegraphics[width=\textwidth]{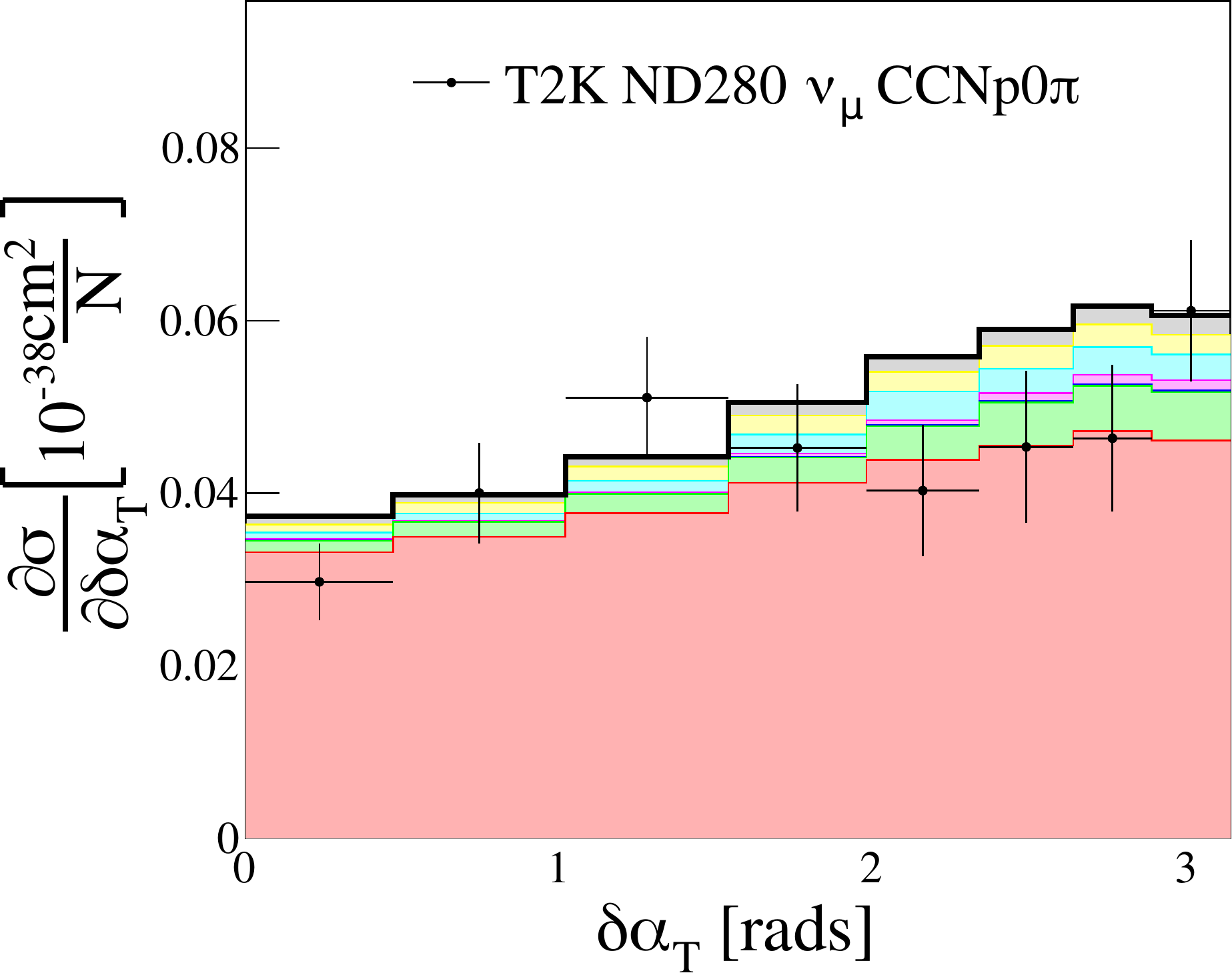}
        \caption{}
    \end{subfigure}
        \begin{subfigure}{0.95\textwidth}
        \includegraphics[width=\textwidth]{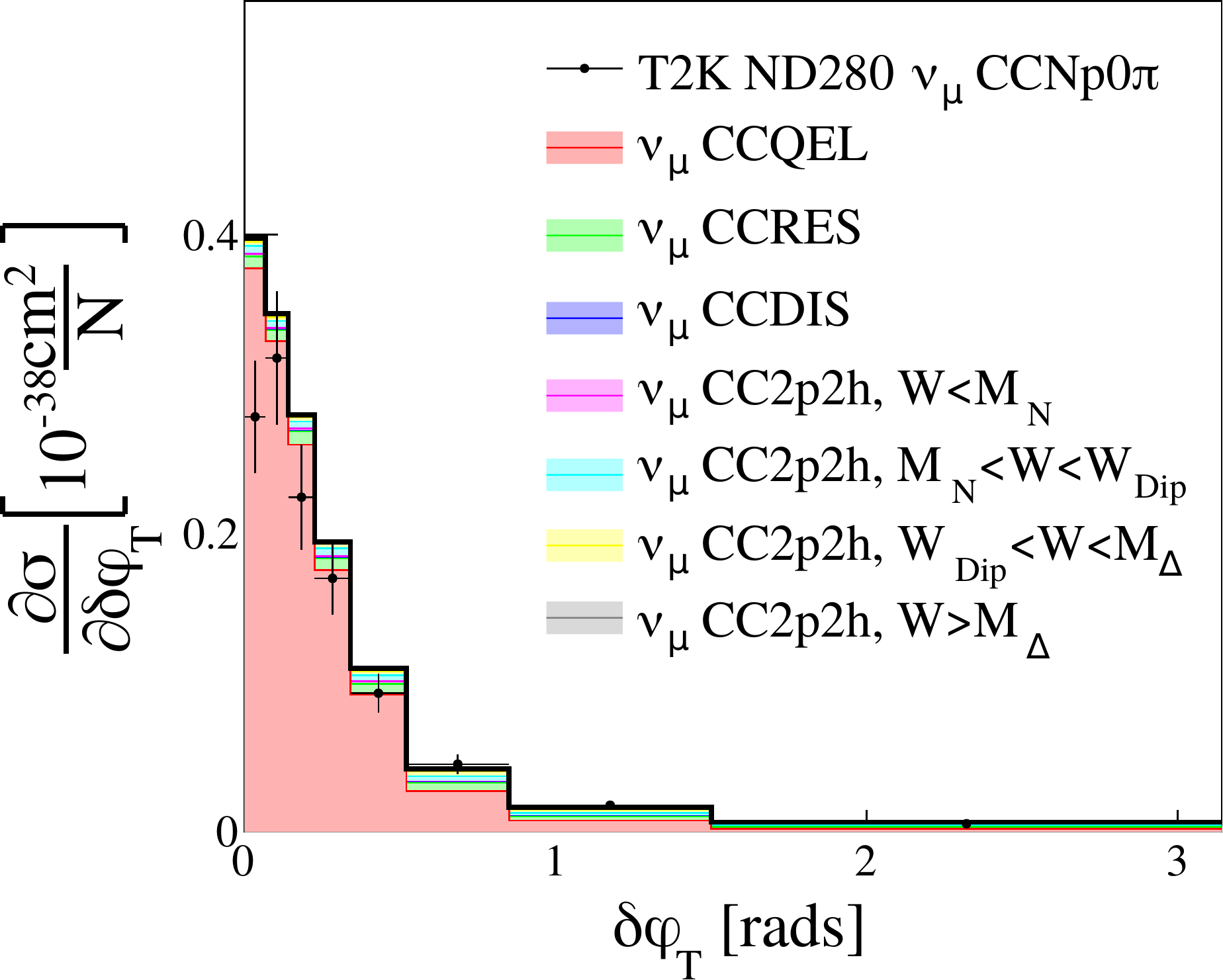}
        \caption{}
    \end{subfigure}
    \caption{T2K ND280 flux-averaged $\nu_\mu$ \ac{CC}Np0$\pi$ differential cross section as a function of \ac{STKI} variables~\cite{PhysRevD.93.112012}. The data are compared against the \texttt{G18\_10a\_02\_11b} tune. The GENIE prediction is divided into different interaction categories: \ac{CC}\ac{QEL}, \ac{CC}\ac{RES}, \ac{CC}\ac{2p2h} and \ac{CC}\ac{DIS}.}
    \label{fig:T2KNp}
\end{figure}

Fig.~\ref{fig:T2KNp} provides comparisons against $\nu_\mu$ CCNp0$\pi$ data as a function of the \ac{STKI} variables, concluding that non-\ac{QEL} interactions are essential to describe this data within regions of high transverse kinematic imbalance.

\begin{figure}
    \centering
    \includegraphics[width=\textwidth]{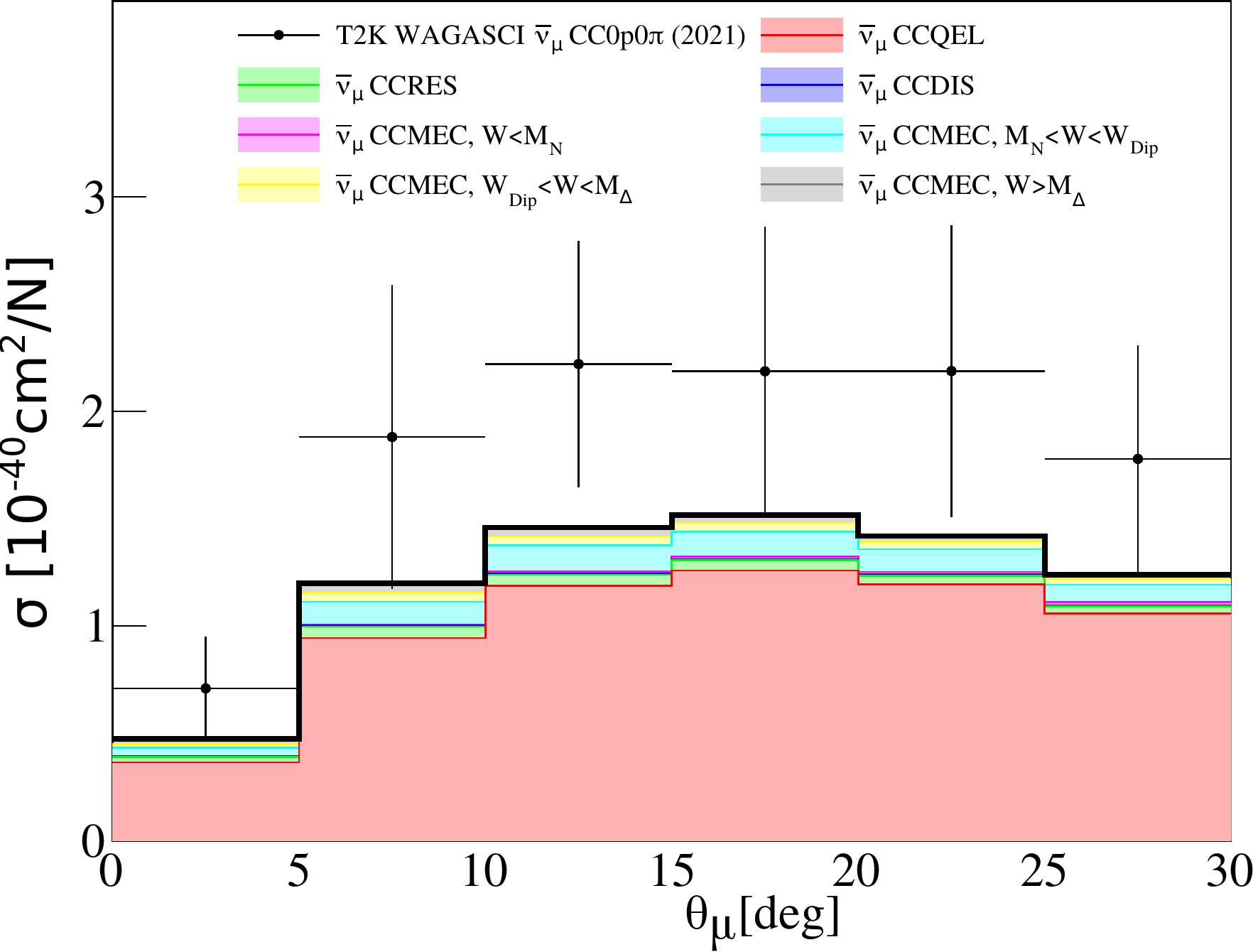}
    \caption{T2K WAGASCI flux-averaged $\overline{\nu}_\mu$ \ac{CC}0p0$\pi$ differential cross section as a function of the muon angle. The data are compared against the \texttt{G18\_10a\_02\_11b} tune. The GENIE prediction is divided into different interaction categories: \ac{CC}\ac{QEL}, \ac{CC}\ac{RES}, \ac{CC}\ac{2p2h} and \ac{CC}\ac{DIS}.}
    \label{fig:T2KWAGASCI}
\end{figure}

Antineutrino CC0$\pi$ cross section measurements are also available: $\overline{\nu}_\mu$ and $\overline{\nu}_\mu$+$\nu_\mu$CC0$\pi$ on water~\cite{PhysRevD.102.012007} and hydrocarbon at T2K ND280~\cite{PhysRevD.101.112001}, and $\overline{\nu}_\mu$ and $\overline{\nu}_\mu$+$\nu_\mu$CC0p0$\pi$ at the T2K WAGASCI, INGRID and Proton Module detectors~\cite{10.1093/ptep/ptab014}. 
Ref.~\cite{10.1093/ptep/ptab014} reports the $\overline{\nu}_\mu$ and the $\overline{\nu}_\mu+\nu_\mu$ CC0p0$\pi$ cross sections on water and hydrocarbon. For all measurements, the muon phase space is restricted to $p_\mu>400$~MeV/c and $\theta_\mu<30^\circ$. The analysis requires events with a muon and no visible pions ($p_\pi>200$~MeV/c, $\theta_\pi<70^\circ$) or protons ($p_p>600$~MeV/c, $\theta_p<70^\circ$). Fig.~\ref{fig:T2KWAGASCI} shows the agreement of the \texttt{G18\_10a\_02\_11b} tune for the T2K WAGASCI $\overline{\nu}_\mu$ CC0p0$\pi$ data. 
The GENIE prediction underpredicts the data as well when considering a higher neutrino flux ($\langle E_\nu \rangle=0.86$~GeV). At higher energies, the \ac{2p2h} contribution $W>W_{dip}$ is non-negligible. 

\subsection{MINER$\nu$A CC0\ensuremath{\pi} cross-section measurements}
\label{sec:MINERvAAnalysis}
MINER$\nu$A 
studies neutrino interactions on nuclear targets for neutrino and antineutrino interactions at $\sim 1-10$~GeV at Fermilab~\cite{ALIAGA2014130,MINERvA:2021csy}.
MINER$\nu$A's detector is composed of a segmented scintillator detector surrounded by electromagnetic and hadronic calorimeters.
The detector is situated 2.1~m upstream of the MINOS near detector~\cite{MICHAEL2008190}, which is a magnetized iron spectrometer.
MINOS is used to reconstruct the muon momentum and charge.

Neutrinos are generated at the \ac{NuMI} beamline~\cite{PhysRevD.94.092005}.
This beam has two configurations: \emph{low energy} flux ($\langle E_\nu \rangle \sim 3.5$~GeV) and \emph{medium energy} flux ($\langle E_\nu \rangle \sim 6$~GeV).
The \ac{NuMI} beam can operate in neutrino mode (\ac{FHC}) and antineutrino mode (\ac{RHC}).
The \ac{FHC} and \ac{RHC} \emph{low energy} flux predictions are shown in Fig.~\ref{fig:fluxes}.

MINER$\nu$A extracted several CC$0\pi$ and CCNp$0\pi$ measurements using the \ac{NuMI} \emph{low-energy} flux~\cite{PhysRevD.99.012004,PhysRevD.97.052002,PhysRevLett.121.022504,MINERvA:2019ope}.
A CC0$\pi$ measurement using the \ac{NuMI} \emph{medium energy} flux is also available~\cite{MINERvA:2019gsf}.
This review focuses on the CC0$\pi$ and CCNp0$\pi$ measurements obtained with the \emph{low-energy} flux.

The exact target mixture is composed of carbon (88.51\%), hydrogen (8.18\%), oxygen (2.5\%), titanium (0.47\%), chlorine (0.2\%), aluminium (0.07\%), and silicon (0.07\%).
In the calculation of the GENIE predictions, only the three most abundant targets are considered.
The relative mass abundances are renormalized to take this approximation into account. 
This simplifies the computing power and has a negligible effect on our predictions. 

\subsubsection{MINER$\nu$A $\nu_\mu$ and $\overline{\nu}_\mu$ CC0\ensuremath{\pi} cross-section measurement}
\label{subsec:MINERvACCQE}

MINER$\nu$A reported the CC0$\pi$ differential flux-integrated cross-section as a function of muon momentum in the transverse ($T$) and longitudinal ($\parallel$) direction relative to the neutrino beam~\cite{PhysRevD.99.012004,PhysRevD.97.052002}.
The differential cross-section as a function of $E_\nu^{\text{QEL}}$ and $Q^2_{\text{QEL}}$ are reported as well~\cite{PhysRevD.99.012004}.
The neutrino energy and the momentum transferred are reconstructed under the \ac{QEL} hypothesis, described in Sec.~\ref{sec:RecoObs}.
The binding energy used to reconstruct $E_\nu^{\text{QEL}}$ according to Eq.~\ref{eq:EnuQE} in their neutrino and antineutrino analysis is $E_b=34$~MeV and $E_b=30$~MeV respectively.

The $\nu_\mu$ CC0$\pi$ topology is defined as an event with one muon, $\mu^-$, any number of protons and neutrons, any photons below nuclear de-excitation energies, $E_\gamma\leq 10$~MeV, no mesons and no heavy or excited baryons in the final state.
The MINER$\nu$A detector is not able to measure the muon charge as it does not have a magnetic field.
For this reason, muons are identified by looking for tracks that have a match with the MINOS detector, which is used to determine the muon momentum and charge.
Because of geometric acceptance, both analyses require $\theta_\mu< 20^\circ$.
Events containing low-energy photons are accepted as they can arise from nuclear de-excitation.
Pions are removed by applying a cut on the recoil energy, $E_{\text{recoil}}\leq 500$~MeV, defined as the activity that is not coming from a muon or any tracked protons. $E_{\text{recoil}}$ is corrected for the calorimetric detector response~\cite{PhysRevD.97.052002}.
The recoil energy does not include energy deposited at less than 150~mm from the neutrino vertex as it could be due to proton absorption nearby the vertex.
Moreover, events with Michel electrons are removed, as they assume they come from a $\pi$ decay chain ($\pi\rightarrow\mu\rightarrow e$). 

The $\overline{\nu}_\mu$ CC0p0$\pi$ topology~\cite{PhysRevD.97.052002} is similar to the $\nu_\mu$CC0$\pi$ one, with some differences.
Due to the nature of this interaction, the muon must be positively charged.
Moreover, the analysis requires there are no visible protons in the final state, i.e.\ protons with kinetic energy above 120~MeV.
Finally, mesons are removed using the information on the recoil energy deposited outside the vertex region only.

The GENIE prediction is evaluated with \ac{MC} events that satisfy the criteria specified above with few exceptions: the removal of events with mesons in the final state is based on true information only. 
Baryons, are short living and decayed into mesons using the GENIE particle decayer.
The requirements on the removal energy are not implemented in our \ac{MC} analysis either as the data was already corrected for this effect.

\begin{figure}
    \centering
        \includegraphics[width=0.95\textwidth]{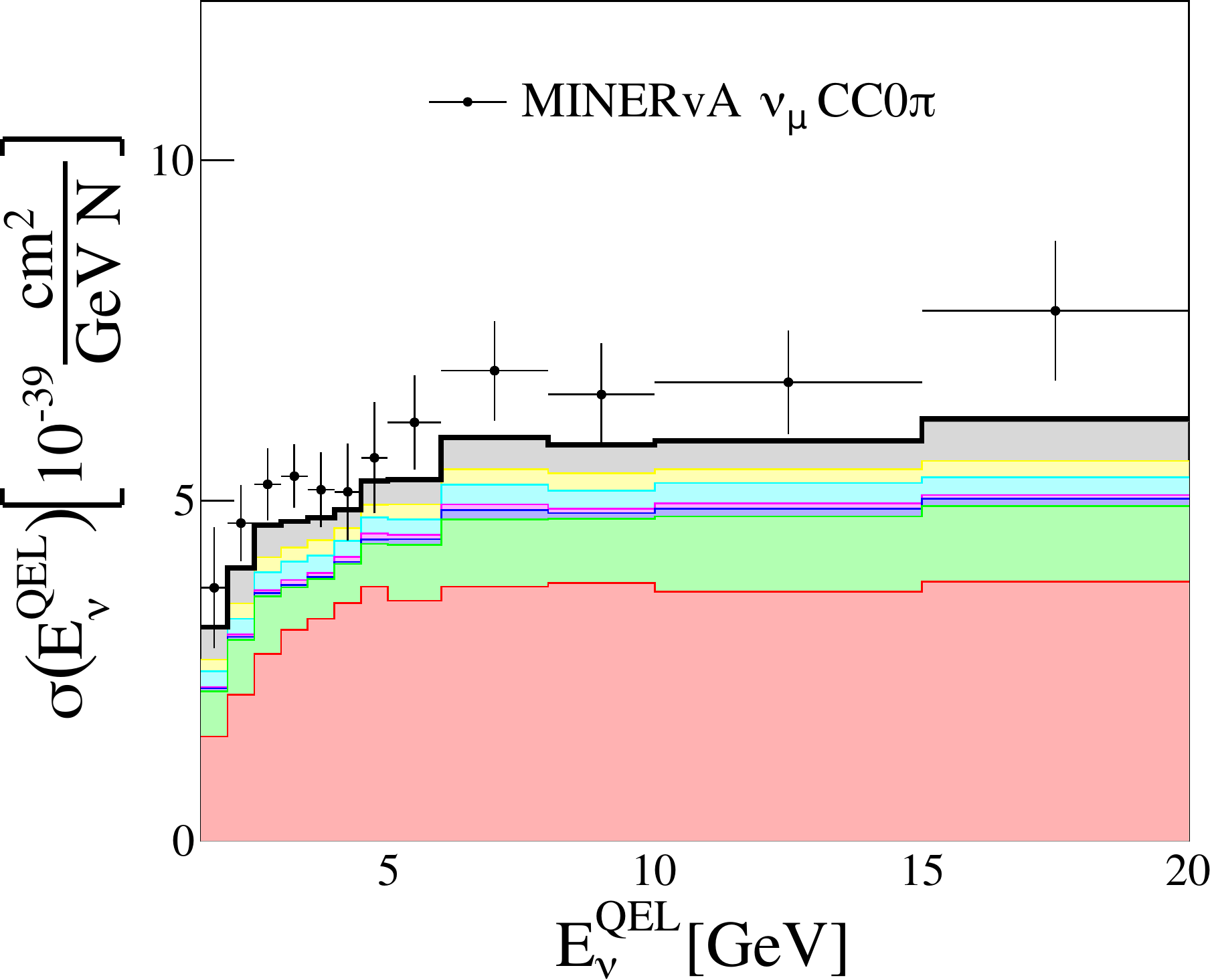}
        \caption{MINER$\nu$A $\nu_\mu$ CC0$\pi$ flux-averaged cross-section as a function of the reconstructed neutrino energy, $E_\nu^{\text{QEL}}$~\cite{PhysRevD.99.012004}. The data are compared against the \texttt{G18\_10a\_02\_11b} tune. The notation for the histogram is the same as in Fig.~\ref{fig:T2KNp}.}
    \label{fig:MinervaEnu}
\end{figure}

\begin{figure*}
\centering
    \includegraphics[width=\textwidth]{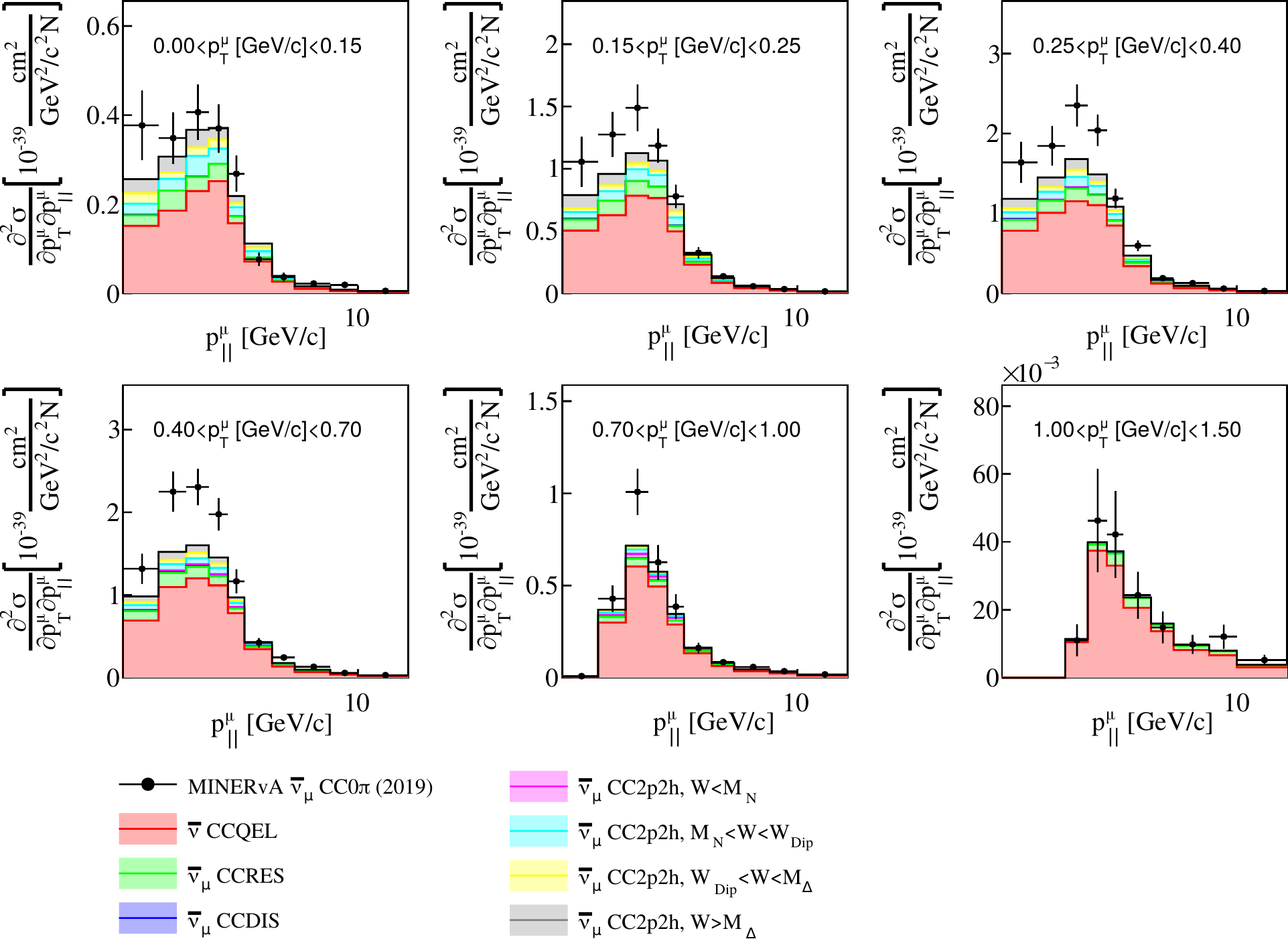} 
    \caption{MINER$\nu$A $\overline{\nu}_\mu$ CC0$\pi$ double differential flux-averaged cross-section as a function of the muon longitudinal momentum, $p_\parallel$, and transverse momentum, $p_T$~\cite{PhysRevD.99.012004}. The corresponding slices on $p_T$ are compared against the \texttt{G18\_10a\_02\_11b} tune. The GENIE prediction is divided into different interaction categories.}
    \label{fig:Minerva2DBarXProjBreakdown}
\end{figure*}

GENIE comparisons against the double-differential $\nu_\mu$ CC$0\pi$ measurement is shown in Figs.~\ref{fig:Minerva2DXprojBreakdown} and \ref{fig:Minerva2DBarXProjBreakdown} for $\overline{\nu}_\mu$ CC0p$0\pi$ data. 
For both $\nu_\mu$ and $\overline{\nu}_\mu$ data, the \texttt{G18\_10a\_02\_11b} underestimates the data.
This is true especially in the phase-space regions in which \ac{2p2h} events dominate. 
In high $p_T$ regions, where the contribution of \ac{2p2h} events is negligible, the agreement improves.
This can be seen for the $0.85<p_T<2.5$~GeV/c slices in Fig.~\ref{fig:Minerva2DXprojBreakdown}.
Consequently, the reconstructed neutrino energy is also under-predicted, as observed in Fig.~\ref{fig:MinervaEnu}.

\subsubsection{MINER$\nu$A \ensuremath{\nu_\mu} CCNp0\ensuremath{\pi} production cross-section measurement}
\label{subsec:MINERvACCQE_proton}

The MINER$\nu$A Collaboration released two CCNp0$\pi$ analyses: single-differential cross-section measurements as a function of $\delta p_T$, $\delta p$, $\delta\alpha_T$ and $\delta\phi_T$~\cite{PhysRevLett.121.022504}, and a single-differential measurement as a function of $\delta p_{Ty}$ or $\delta p_{Tx}$, respectively~\cite{MINERvA:2019ope}.

In both analyses, the topology is defined as events with a muon, no mesons and at least one proton in the final state that satisfy the following conditions:
\begin{gather*}
	1.5~\text{GeV}/c < p_\mu < 10~\text{GeV}/c \enskip\text{and}\enskip \theta_\mu<20^{\circ}, \\
	0.45~\text{GeV}/c < p_p < 1.2~\text{GeV}/c \enskip\text{and}\enskip \theta_p<70^{\circ},
\end{gather*}
where $p_\mu$ ($p_p$) and $\theta_\mu$ ($\theta_p$) are the muon (lead proton) momentum and opening angle with respect to the neutrino direction.
The lead proton is defined as the proton with the highest energy that satisfies the phase space cuts mentioned above.

\begin{figure*}
    \centering
    \begin{subfigure}{0.45\textwidth}
        \includegraphics[width=0.9\textwidth]{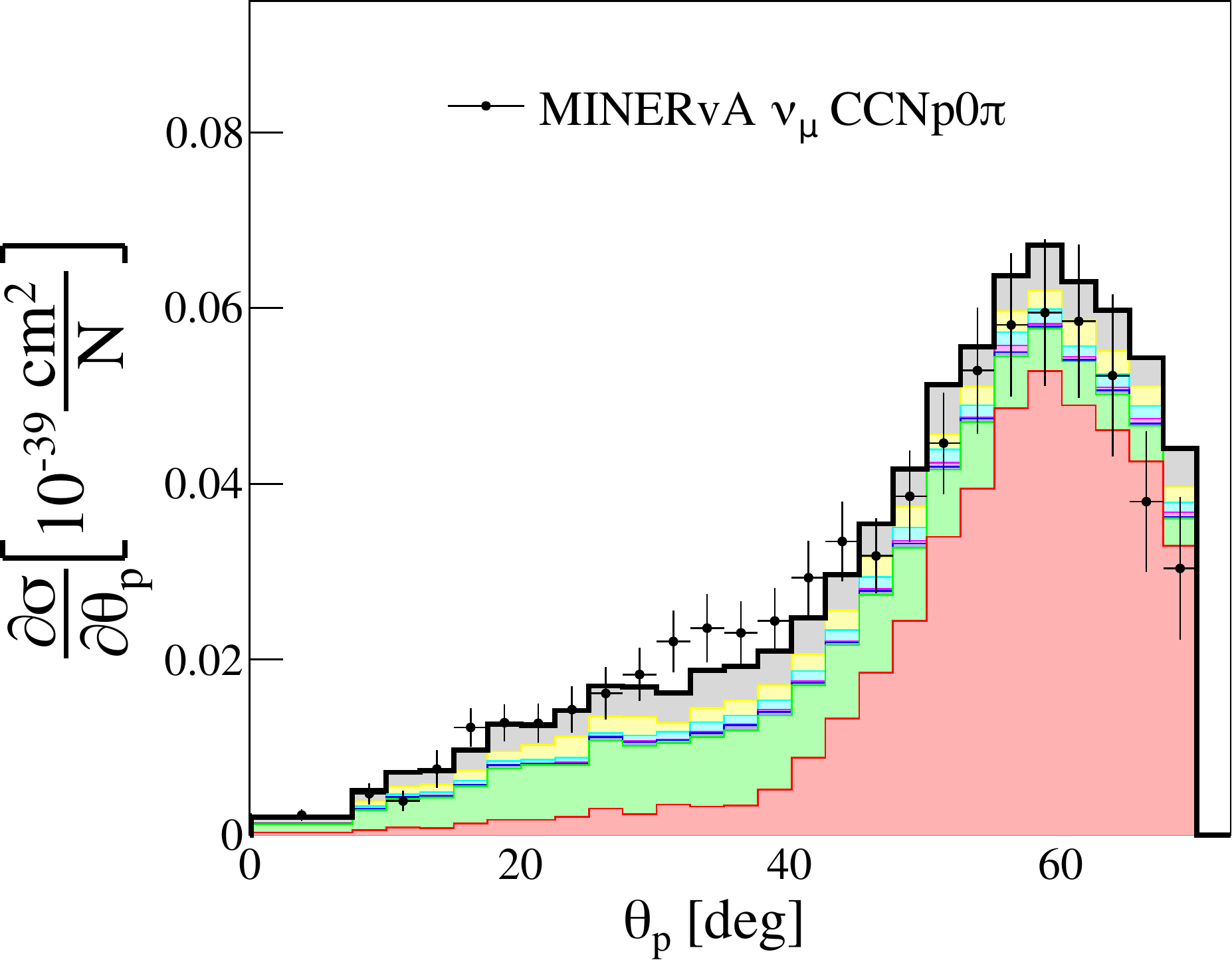}
        \caption{}
        \label{fig:MinervaBreakdownPtheta}
    \end{subfigure}
    \begin{subfigure}{0.45\textwidth}
        \includegraphics[width=0.9\textwidth]{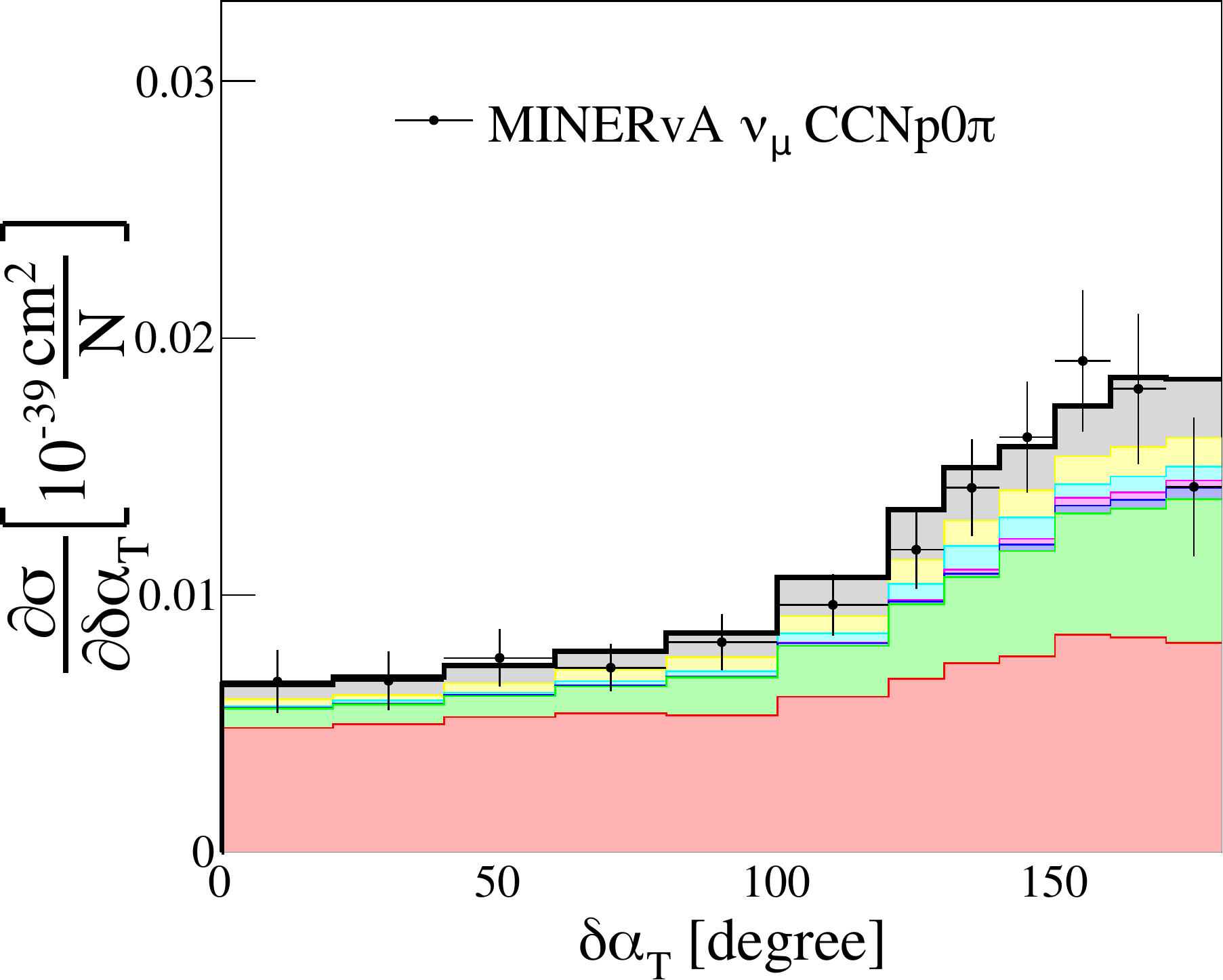}
        \caption{}
        \label{fig:MinervaBreakdownAlpha}
    \end{subfigure}
    
    \begin{subfigure}{0.45\textwidth}
        \includegraphics[width=0.9\textwidth]{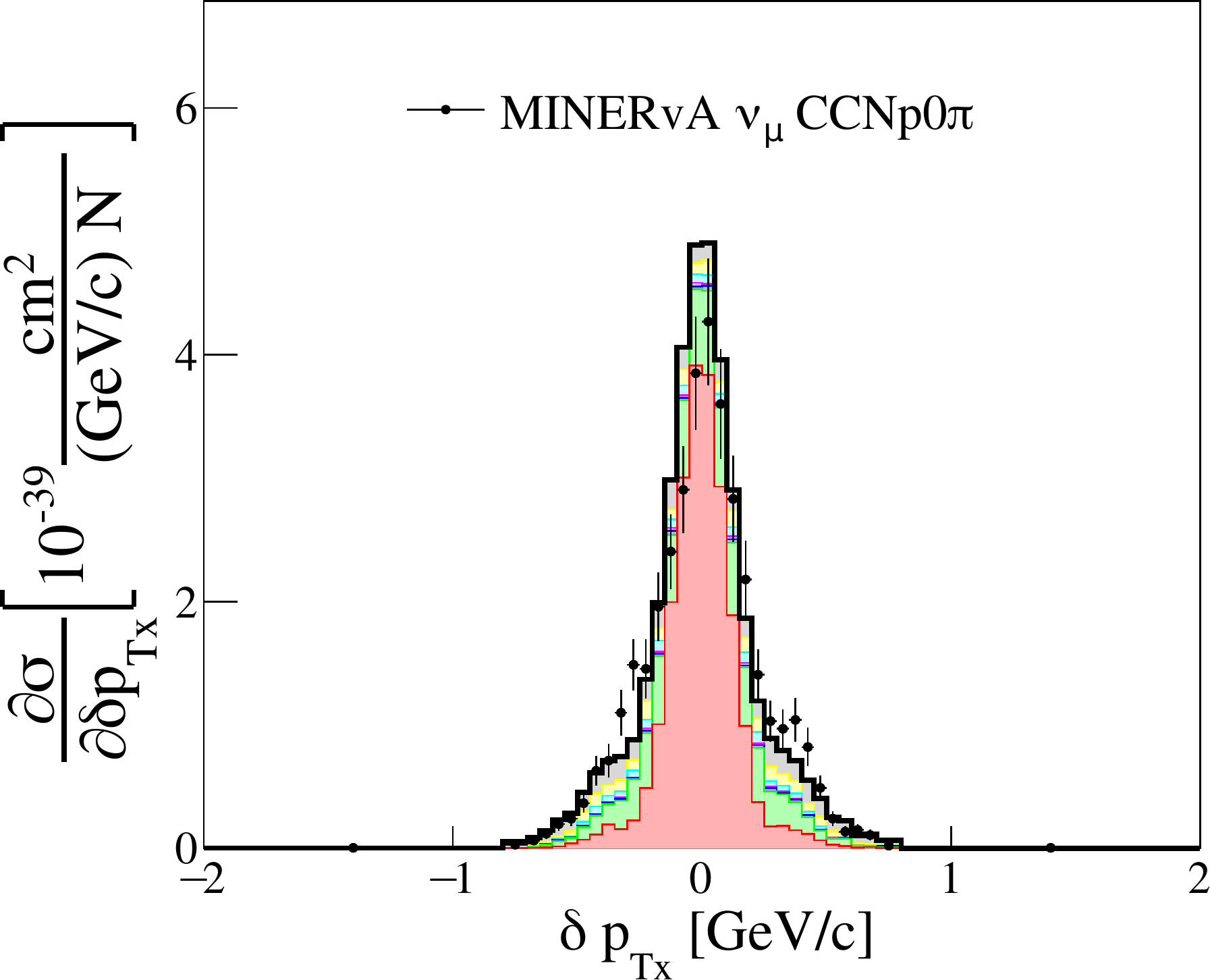}
        \caption{}
        \label{fig:MinervaBreakdowndptx}
    \end{subfigure}
    \begin{subfigure}{0.45\textwidth}
        \includegraphics[width=0.9\textwidth]{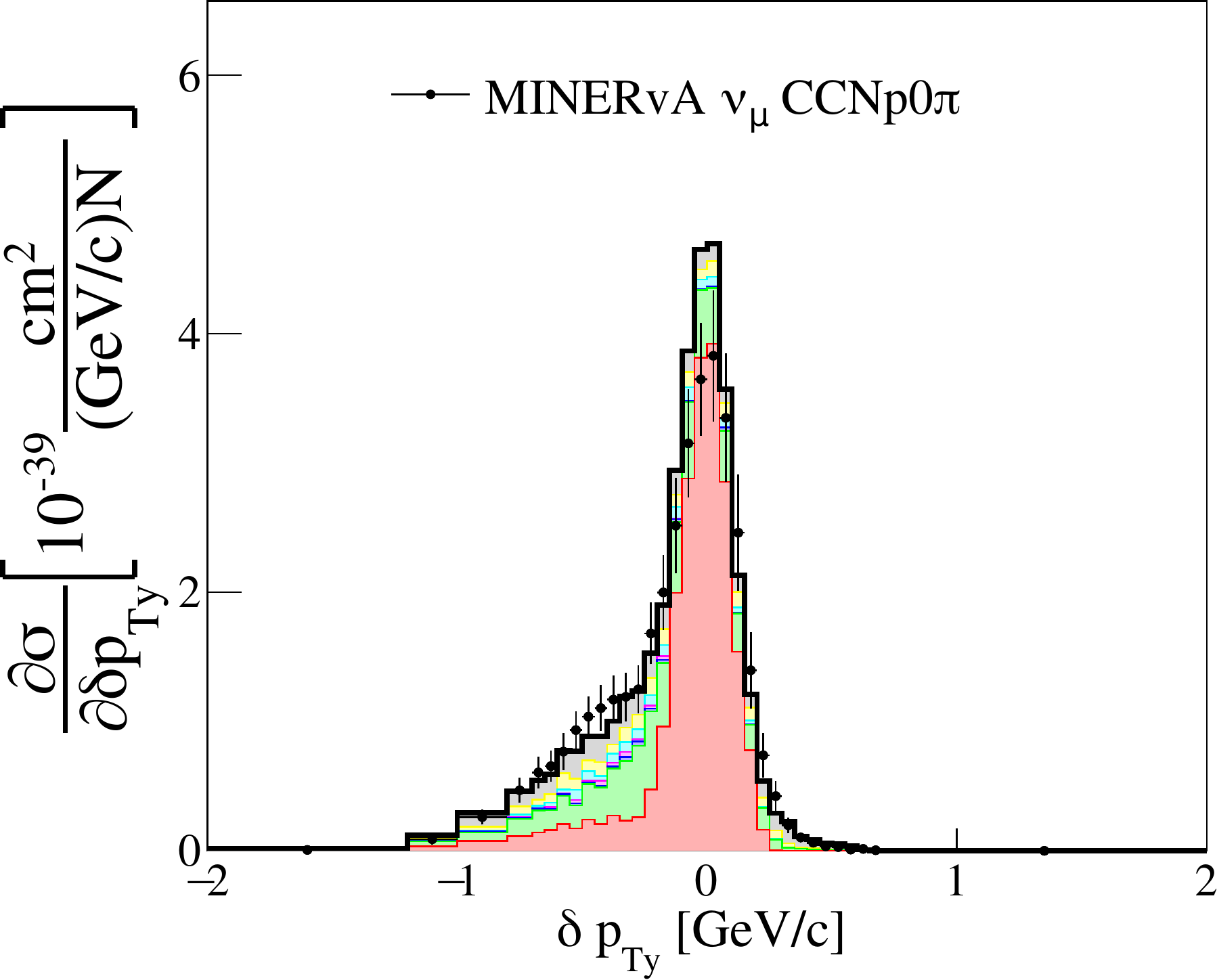}
        \caption{}
        \label{fig:MinervaBreakdowndpty}
    \end{subfigure}
    \caption{MINER$\nu$A $\nu_\mu$ CCNp0$\pi$ differential flux-averaged cross-section as a function of \ac{STKI} variables~\cite{ PhysRevLett.121.022504,MINERvA:2019ope}. The data are compared against the \texttt{G18\_10a\_02\_11b} tune. The GENIE prediction is divided into interaction modes. The notation for histograms is the same as in Fig.~\ref{fig:T2KNp}.}
    \label{fig:MinervaSTKYDef}
\end{figure*}

In this case, the \texttt{G18\_10a\_02\_11b} tune is not under-predicting the data, see Fig.~\ref{fig:MinervaSTKYDef1} and Fig.~\ref{fig:MinervaSTKYDef}.
This fact, already observed in T2K data, reflects a possible tension between CC$0\pi$ and CCNp$0\pi$ measurements.
The breakdown into different interaction modes highlights the \ac{2p2h} model dependence with $W$ at different proton momenta: \ac{2p2h} events with low (high) $W$ dominate at low (high) proton momentum.
The contribution from \ac{RES} events is most significant for low proton momenta, small proton angles and $\delta\alpha_T\sim 180^\circ$.
In fact, non-\ac{QEL} events dominate in regions of high transverse kinematic imbalance, such as $\delta p_T>0.2$~GeV/c.

\subsection{MicroBooNE CCNp0\ensuremath{\pi} cross-section measurement}
\label{sec:MicroBooNEAnalysis}

The MicroBooNE experiment is a \ac{LArTPC} detector situated 500~m away from the \ac{BNB} beam at Fermilab~\cite{MicrobooneDetector,BNBFlux}. 
\ac{LArTPC} detectors use complex software algorithms to reconstruct the neutrino event topology with excellent spatial resolution in the detector~\cite{Acciarri_2017,Adams_2018,MicroBooNE:2017xvs}.
For instance, MicroBooNE can reconstruct proton tracks of 2~cm with a $\sim26$\% efficiency~\cite{PhysRevD.102.112013}.
Different \ac{PID} algorithms, based on the characteristic signal of each particle in the detector, allow the identification of proton and $\mu/\pi$ candidates, but these methods fail to distinguish between muons and pions. 

MicroBooNE provides the first high-statistics cross-section measurements on argon: $\nu_\mu$ CC inclusive~\cite{PhysRevLett.123.131801}, $\nu_\mu$ CC1p0$\pi$~\cite{uBooNEQE}, $\nu_\mu$ CCNp0$\pi$~\cite{PhysRevD.102.112013}, and $\nu_\mu$ CC $\pi^0$ production~\cite{PhysRevD.99.091102}.
The detector is situated 500~m away from the \ac{BNB} beam at Fermilab~\cite{BNBFlux}.
In this section, we focus on the description of the CCNp0$\pi$ measurement~\cite{PhysRevD.102.112013}, given that the CC1p0$\pi$ measurement~\cite{uBooNEQE} is a subsample of the CCNp0$\pi$ one.

The CCNp0$\pi$ analysis presents a total of five single differential flux-integrated cross-section measurements.
The single differential cross sections are given in terms of the muon momentum ($p_\mu$), muon angle ($\theta_\mu$), leading proton momentum ($p_p$), leading proton angle ($\theta_p$), and the angle between the muon and the leading proton ($\theta_{\mu p}$).

The CCNp0$\pi$ topology is defined as an event with one muon, at least one visible proton, any number of neutrons and no pions in the final state.
In the analysis, the muon candidate is the longest track which is not identified as a proton.
Other tracks in the event must be compatible with the proton \ac{PID} hypothesis. 
In order to guarantee at least a 5\% efficiency in the momentum reconstruction, they require the muon (proton) to have a momentum of at least 100~MeV/c (300~MeV/c).
In addition, the leading proton candidate is must have a reconstructed momentum of less than 1.2~GeV/c. This cut avoids regions of the phase-space in which the proton candidate length is greater than the muon one.
These analysis criteria removes events with pions below 30~MeV/c, which are not reconstructed.
No corrections are applied to remove events with protons or pions below the detection threshold.
The same requirements are applied to the corresponding \ac{MC} predictions.

The differential cross-section measurements were not unfolded to true muon momentum and muon angle. 
Instead, the results are presented in terms of the reconstructed quantities. 
The smearing matrices that convert from the reconstructed to the truth quantities are provided in the data release and are used for the evaluation of the GENIE predictions in the reconstructed space~\cite{PhysRevD.102.112013}. 
This method is known as \emph{forward folding}.

\begin{figure*}
    \centering
    \begin{subfigure}{0.45\textwidth}
        \includegraphics[width=\textwidth]{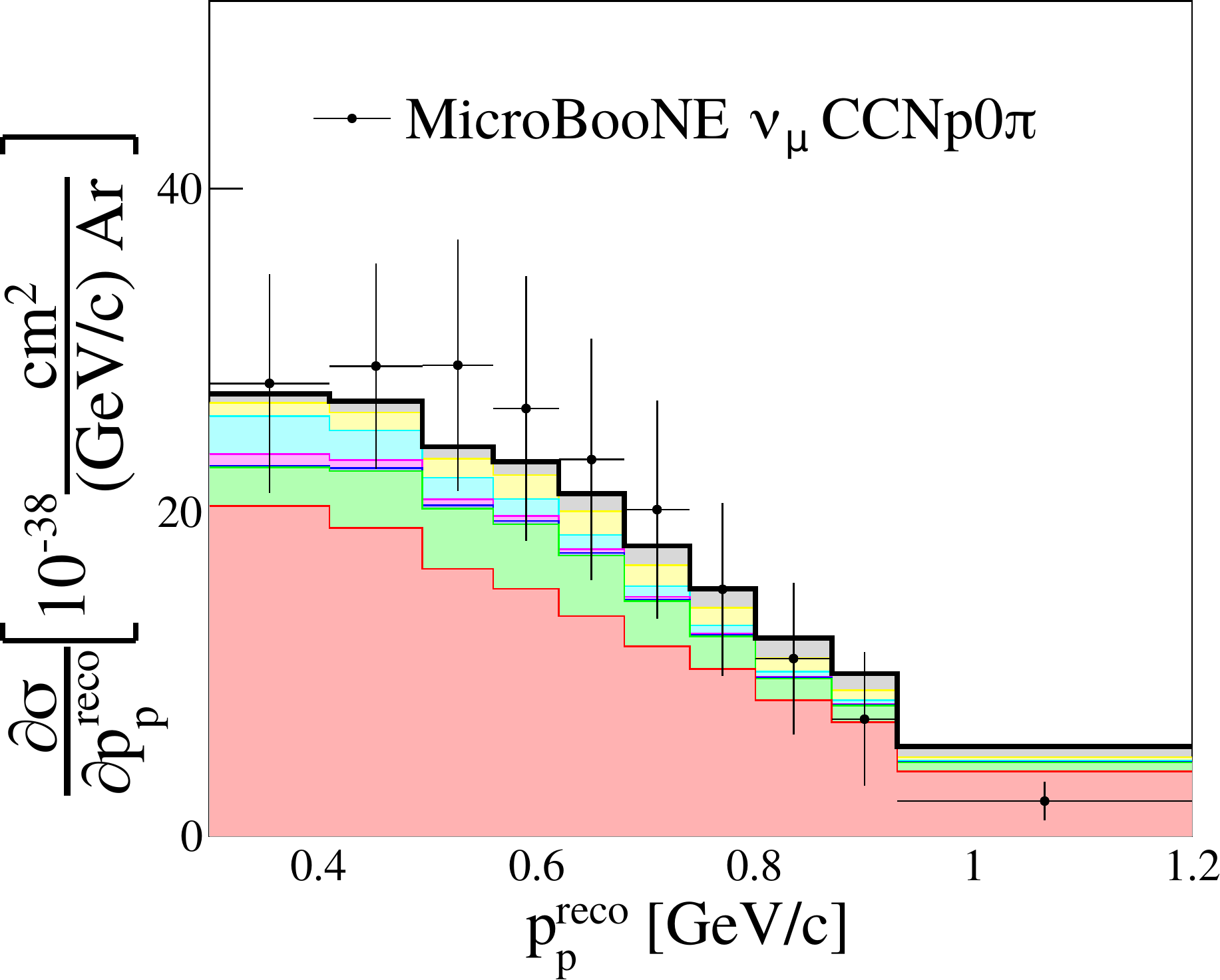}
        \caption{}
        \label{fig:MicroBooNEPmom}
    \end{subfigure}
    \begin{subfigure}{0.45\textwidth}
        \includegraphics[width=\textwidth]{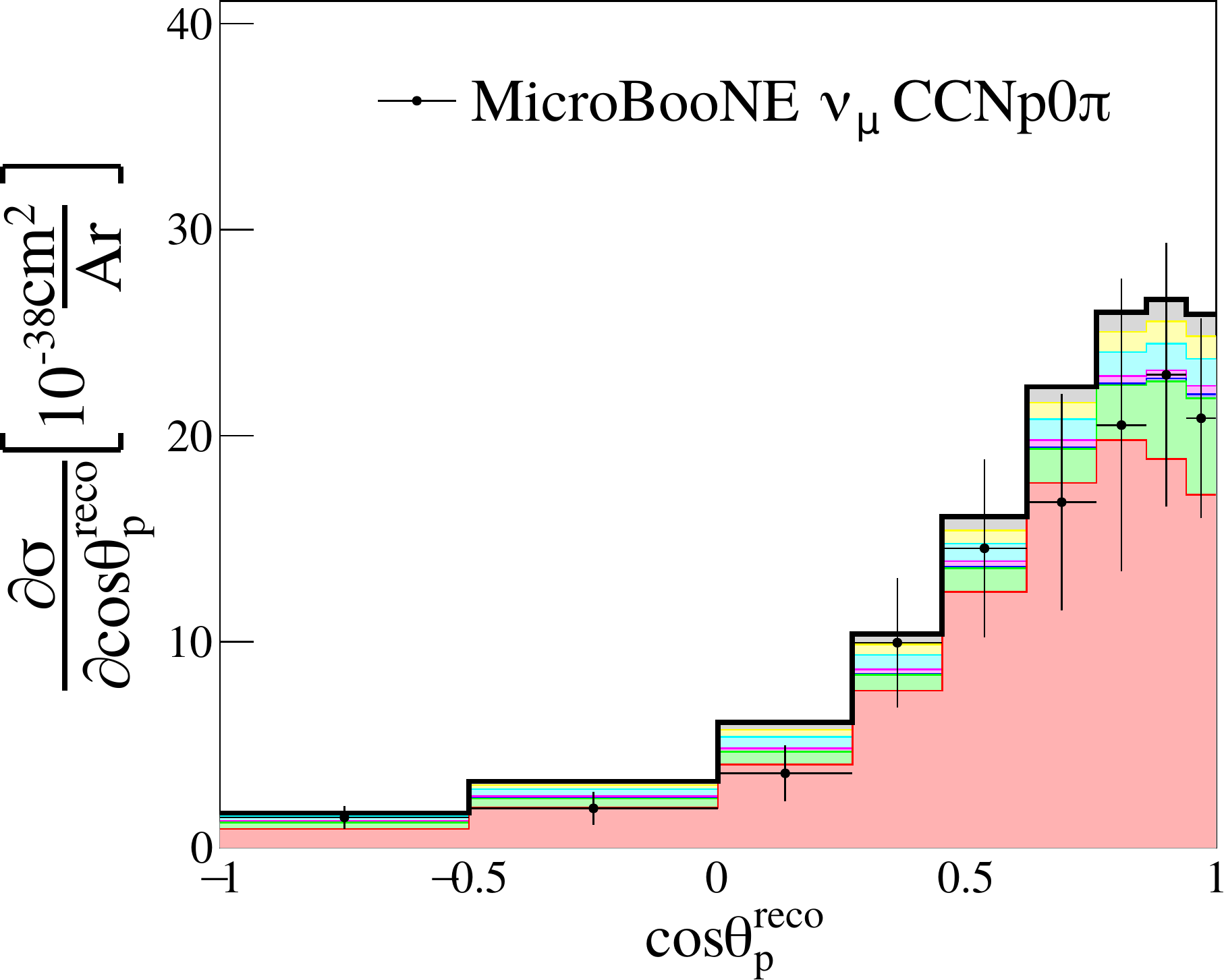}
        \caption{}
        \label{fig:MicroBooNEPAngle}
    \end{subfigure}
    \begin{subfigure}{0.45\textwidth}
        \includegraphics[width=\textwidth]{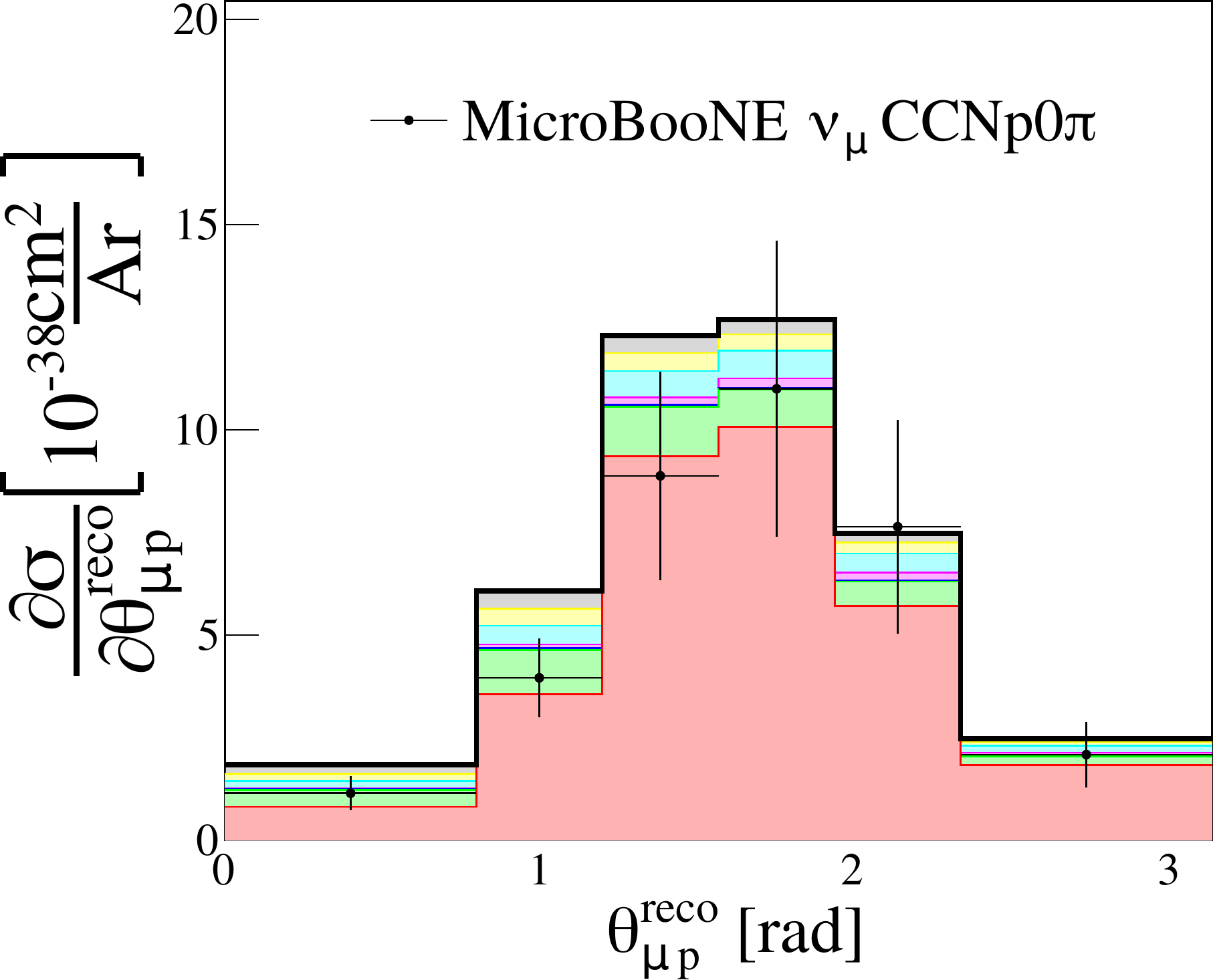}
        \caption{}
        \label{fig:MicroBooNEMuPAngle}
    \end{subfigure}
    \begin{subfigure}{0.45\textwidth}
        \includegraphics[width=\textwidth]{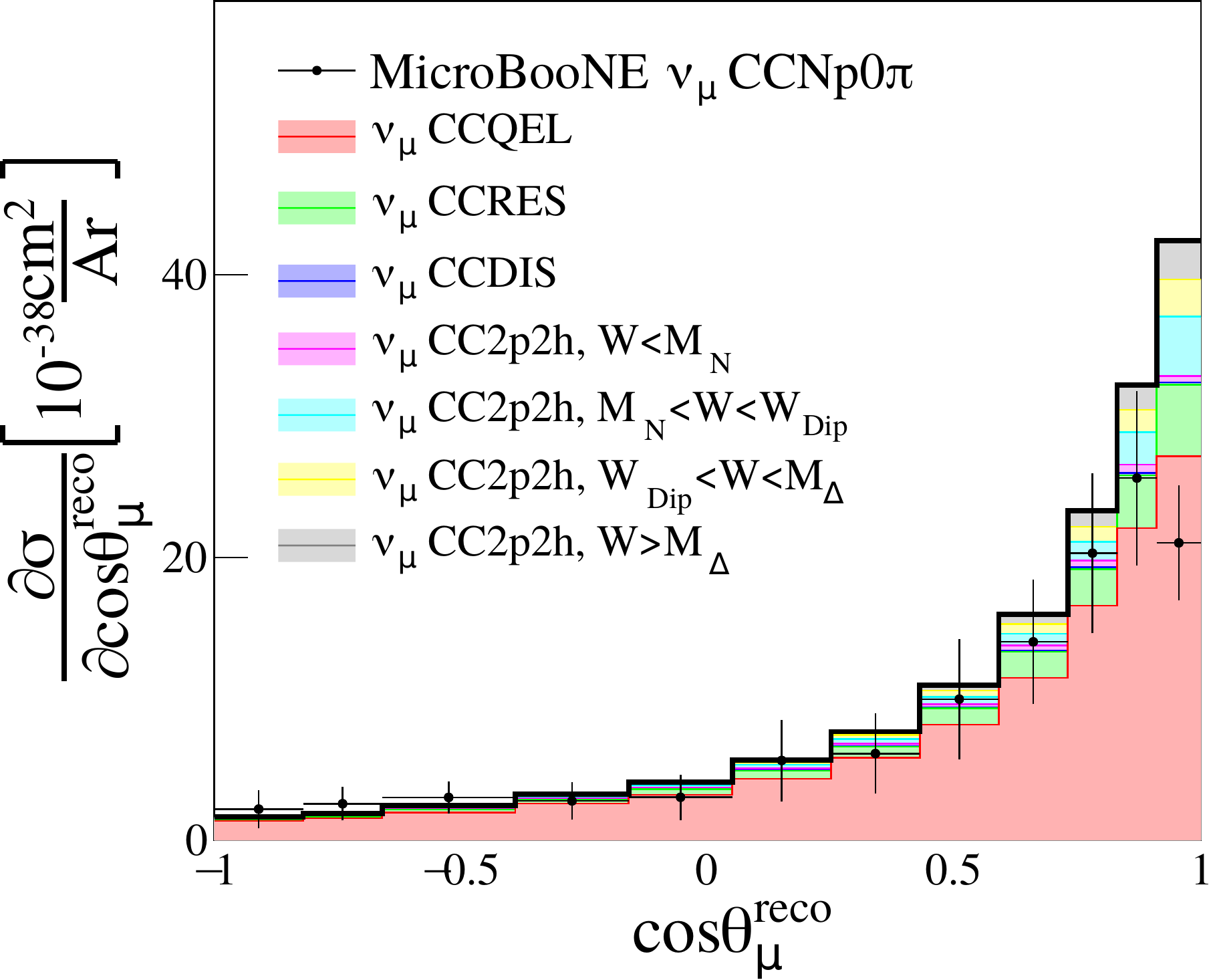}
        \caption{}
        \label{fig:MicroBooNEMuAngle}
    \end{subfigure}
    \caption{MicroBooNE $\nu_\mu$CCNp0$\pi$ flux-averaged differential cross section on $^{40}$Ar as a function of muon and proton kinematics. The GENIE prediction is obtained with the \texttt{G18\_10a\_02\_11b} tune. The nominal prediction is divided into interaction modes.}
    \label{fig:MicroBooNEDataDef}
\end{figure*}

Figure~\ref{fig:MicroBooNEDataDef} presents the comparison between the MicroBooNE data and the GENIE predictions.
The nominal agreement for the \texttt{G18\_10a\_02\_11b} tune is reasonably good, except for the bin at highest $\cos\theta_\mu^{\text{reco}}$, which is largely over-predicted. 
The contribution of non-\ac{QEL} interactions increases at forward muon and proton angles, see Figs.\ref{fig:MicroBooNEPAngle} and \ref{fig:MicroBooNEMuAngle}.
The \texttt{G18\_10a\_02\_11b} dependency on \ac{2p2h} events at different $W$ with the proton momenta is re-encountered.

\section{Additional Nuclear Uncertainties}
\label{sec:AddlParams}
Here, we explore modeling aspects that were not included in the tuning exercise.

\subsection{Nuclear model implementation}
\label{subsec:nuclearmodel}
Uncertainties in the nuclear model affect the dynamics of the outgoing muon and nucleon after a \ac{QEL} or a \ac{2p2h} interaction.

In the Valencia model implementation in GENIE, the differential cross section is evaluated at an effective energy
transfer $\tilde{q}_0$, which takes into account the nucleon removal energy. 
The implementation in the \ac{QEL} and \ac{2p2h} processes is slightly different. 
The effective energy transfer $\tilde{q}_0$ used in the Valencia \ac{QEL} model implementation is:
\begin{equation}
	\tilde{q}_0 = q_0+E_{N_i}-E_p = E_{N_f}-E_{p}
	\label{eq:q0Tilde1}
\end{equation}
$E_{N_i}$ is the energy of the off-shell initial nucleon, which is bound with a binding energy $E_b$.
$E_p$ is the energy of the initial nucleon on-shell with a momentum $\mathbf{p}$, $E_p = \sqrt{M_N^2+\mathbf{p}^2}$.
$E_{N_f}$ is the energy of the nucleon produced after the \ac{QEL} interaction, which is on-shell.
In other words, the effective energy transfer is reduced relative to the ordinary one by the amount of energy needed to put the initial nucleon on the mass shell.
The binding energy and initial nucleon momentum are determined by the corresponding nuclear model. 
In this work, for \ac{QEL} interactions we refer to $\tilde{q}_0$ as $\tilde{q}^{\text{QEL}}_0$.
Notice that $\tilde{q}^{\text{QEL}}_0$ depends on the event kinematics.

In the Valencia \ac{2p2h} model implementation, the effective energy transfer is calculated as:
\begin{equation*}
	\tilde{q}_0 = q_0-q^{\text{2p2h}}_{\text{shift}}, \enskip \text{where} \enskip q^{\text{2p2h}}_{\text{shift}} \equiv M(A_{Z+1})-M(A_Z).
\end{equation*}
In this case, $q^{\text{2p2h}}_{\text{shift}}$ is independent of the event kinematics. For a carbon target, $q^{\text{2p2h}}_{\text{shift}}(^{12}\text{C})=16.8$~MeV, whilst $q^{\text{2p2h}}_{\text{shift}}(^{40}\text{Ar})=0.99$~MeV for argon.

Shifts on $\tilde{q}_0$ are effective modifications of the binding energy in the nuclear model. 
It is possible to apply relative shift to $\tilde{q}_0$ for both \ac{QEL} and \ac{2p2h} calculations by modifying $q^{\text{QEL}}_0$ and $ q^{\text{2p2h}}_{\text{shift}}$.
This modification translates as: 
\begin{eqnarray*}
   \tilde{q}^{\text{QEL}}_0\rightarrow\tilde{q}^{\text{QEL}}_0( 1 + f^{\text{QEL}})\\
   q^{\text{2p2h}}_{\text{shift}}\rightarrow q^{\text{2p2h}}_{\text{shift}}(1 + f^{\text{2p2h}})
\end{eqnarray*}
$f^{\text{QEL}}$ and $f^{\text{2p2h}}$ are two dimensionless parameters.
In the GENIE v3 version, both parameters default to 0. 
Both $f^{\text{QEL}}$ and $f^{\text{2p2h}}$ parameters are included in the initial iteration of this analysis.

\begin{figure}
    \centering
    \includegraphics[width=\textwidth]{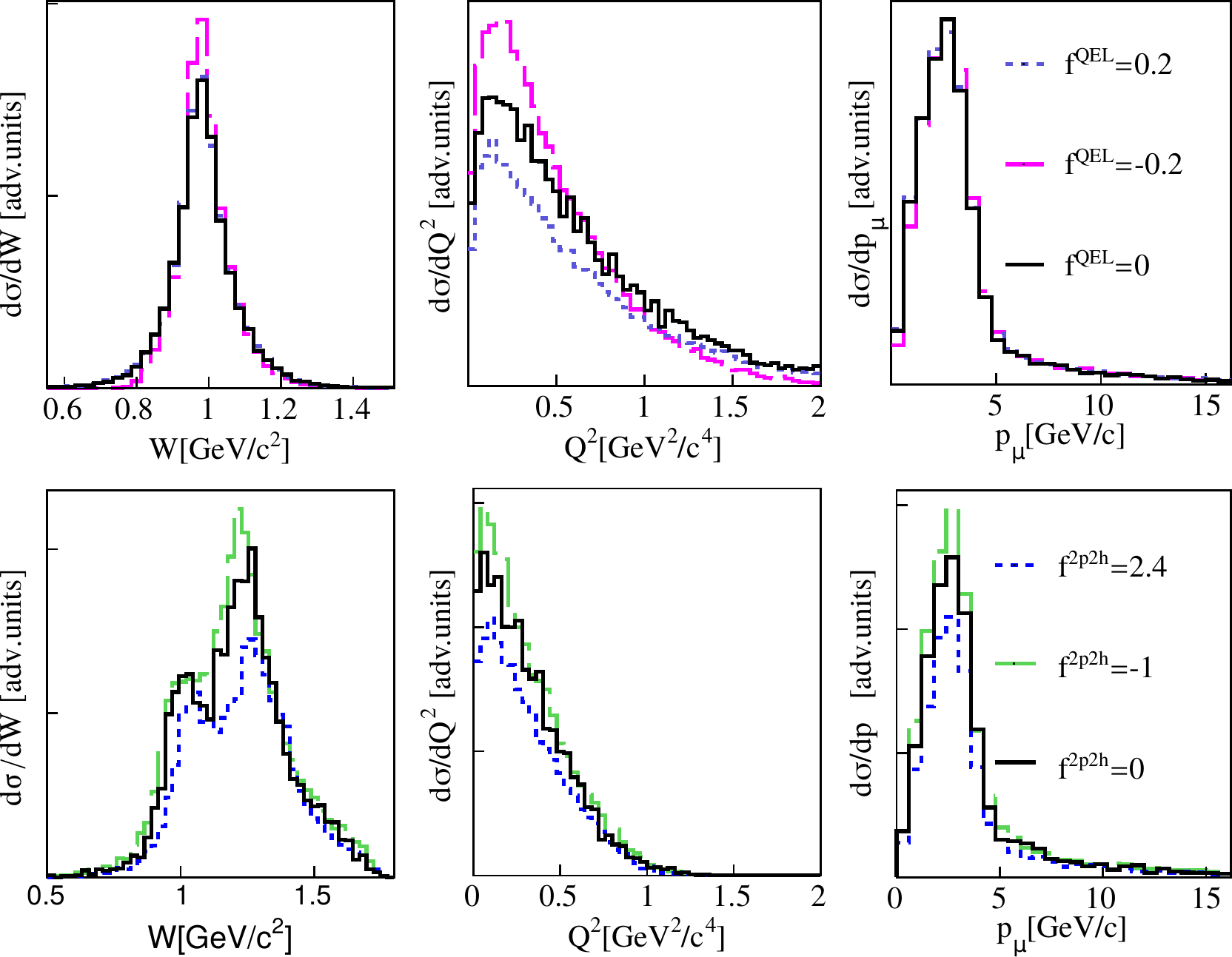} 
    \caption{Flux-integrated differential $\nu_\mu ^{12}$C \ac{CC}\ac{2p2h} cross section dependence with $W$, $Q^2$ or $p_\mu$. 
    Events are generated with the G18\_10a\_02\_11b tune and the \ac{NuMI} $\nu_\mu$ low energy flux~\cite{PhysRevD.94.092005}.
	The top (bottom) three plots show the \ac{CC}\ac{QEL} (\ac{CC}\ac{2p2h}) differential cross section as a function of $W$, $Q^2$ or $p_\mu$.
    The black prediction corresponds to the GENIE v3 case, where no shifts on $\tilde{q}^{\text{QEL}}_0$ and $q^{\text{2p2h}}_{\text{shift}}$ are considered.
    The variations considered for the $f^{\text{QEL}}$ and $f^{\text{2p2h}}$ parameters correspond to an absolute shift to $\tilde{q}^{\text{QEL}}_0$ and $q^{\text{2p2h}}_{\text{shift}}$ of 20~MeV for \ac{QEL} interactions and of 40~MeV for \ac{2p2h} interactions.}
    \label{fig:Kine_Qvalue}
\end{figure}

Ref.~\cite{BodeckRemovalQValue} suggests that shifts on $\tilde{q}^{\text{QEL}}_0$ ($q^{\text{2p2h}}_{\text{shift}}$) of $0-20$~MeV ($0-40$~MeV) for \ac{QEL} (\ac{2p2h}) are in reasonable agreement with electro-scattering data.
The effect of such variations on the \ac{2p2h} cross-section prediction is shown in Fig.~\ref{fig:Kine_Qvalue}. 
The biggest variation is observed on $d\sigma/dQ^2$ for both \ac{QEL} and \ac{2p2h}. 
For the \ac{2p2h} cross section, this systematic shifts peaks position in $W$.

\subsection{Final state interaction implementation}

Final-state interactions (FSI) are crucial for modeling nuclear cross sections as they affect the event topology and kinematics of an event.
There are different models available in GENIE to simulate \ac{FSI}~\cite{GENIEHighlights,dytman2021comparison}.
In particular, \texttt{G18\_10a\_02\_11b} models \ac{FSI} with the INTRANUKE \emph{hA} model~\cite{Andreopoulos:2015wxa}.

INTRANUKE \emph{hA} is an empirical model that considers a single interaction which is based on hadron-nucleus data~\cite{dytman2021comparison}.
In particular, pion-nucleus data are used to determine the inelastic (Inel), absorption (Abs), charge-exchange (CEx) and pion production ($\pi$Prod) fractions ($f_i$).
The fractions depend on the pion kinetic energy and the nuclear atomic number. 
These fractions satisfy that $\sum_i f_i^{\pi^\pm} = 1$ (unitarity condition), where $i$ is an index that runs over the available processes aforementioned.

Two parameters are introduced to be able to modify the $f_{\text{Abs}}^{\pi^{\pm}}$ and $f_{\text{Abs}}^{\pi^{0}}$ while preserving unitarity:
\begin{equation*}
    f_{\text{Abs}}^{' \,\,\pi^\pm} = \frac{S_{\text{Abs}}^{\pi^\pm}\cdot f_{\text{Abs}}^{\pi^{\pm}}}{f_{\text{Inel}}^{\pi^{\pm}}+S_{\text{Abs}}^{\pi^\pm}\cdot f_{\text{Abs}}^{\pi^{\pm}}+f_{\text{CEx}}^{\pi^{\pm}}+f_{\pi \text{Prod}}^{\pi^{\pm}}}
\end{equation*}
The other fractions are also modified as a consequence of this scaling.
Notice that variations of $S_{\text{Abs}}^{\pi^\pm}$ do not scale $f_{\text{Abs}}^{' \,\,\pi^\pm}$ linearly.

Similarly, a scaling parameter is introduced to scale the charged pion mean-free path.
This is referred to as $S_{\text{MFP}}^{\pi^\pm}$.
The same approach can be applied to other processes and to nucleon fractions.

\begin{figure}
    \centering
    \begin{subfigure}{\textwidth}
    \centering\includegraphics[width=\columnwidth]{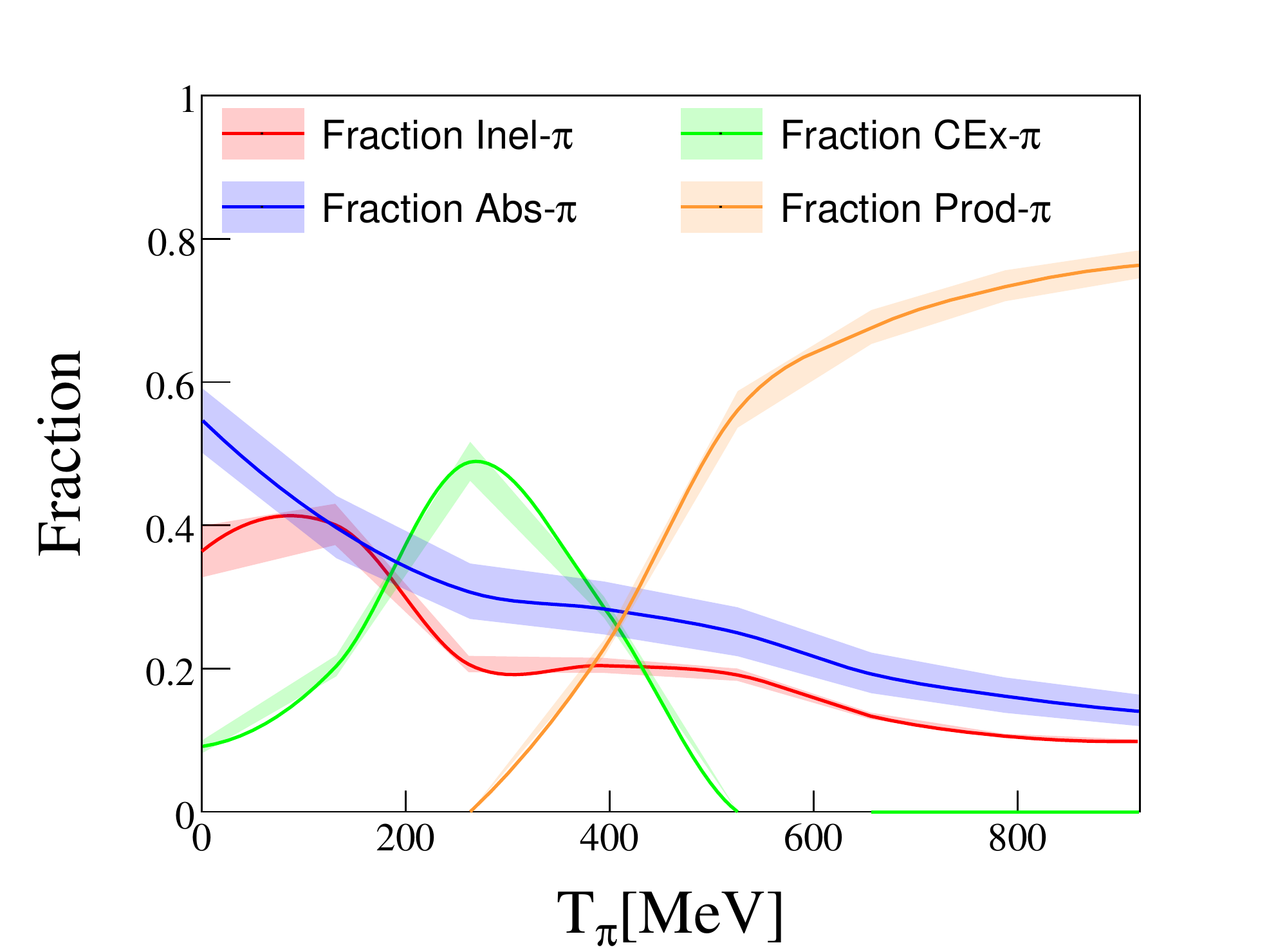}
    \caption{}
    \end{subfigure}
    \begin{subfigure}{\textwidth}
    \centering\includegraphics[width=\columnwidth]{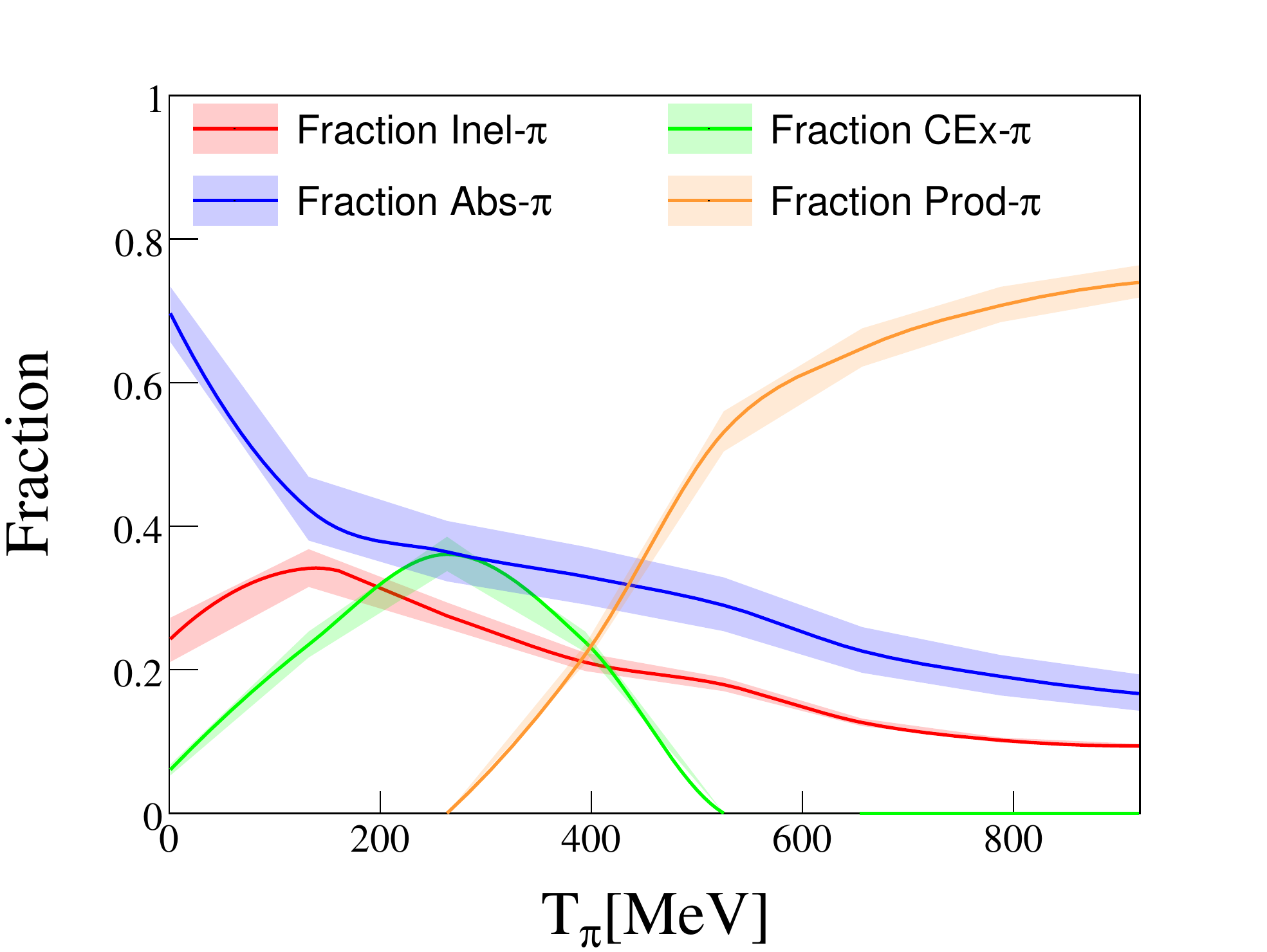}
    \caption{}
    \end{subfigure} 
    \caption{\emph{hA} FSI pion fractions for (a) $^{12}$C and (b) $^{40}$Ar as a function of the pion kinetic energy. The error bands represent the fraction variation when applying a $S_{\text{Abs}}^{\pi^\pm} = 1.2$ on the pion absorption fraction, which corresponds to a variation of $\sim15$\% for the pion absorption fraction on carbon at $T_\pi=200$~MeV.
    \label{fig:fracPi}}
\end{figure}

Figure~\ref{fig:fracPi} shows the dependence of each \emph{hA} fraction as a function of the pion kinetic energy ($T_\pi$) for carbon and argon targets.
The \ac{FSI} fractions and their uncertainty are extracted from fits to hadron-nucleus scattering data~\cite{PhysRevD.99.052007,dytman2021comparison}.
The uncertainty associated with $f_{\text{Abs}}^{\pi^\pm}$ is 15\%. 

\begin{figure}
    \centering
    \includegraphics[width=\textwidth]{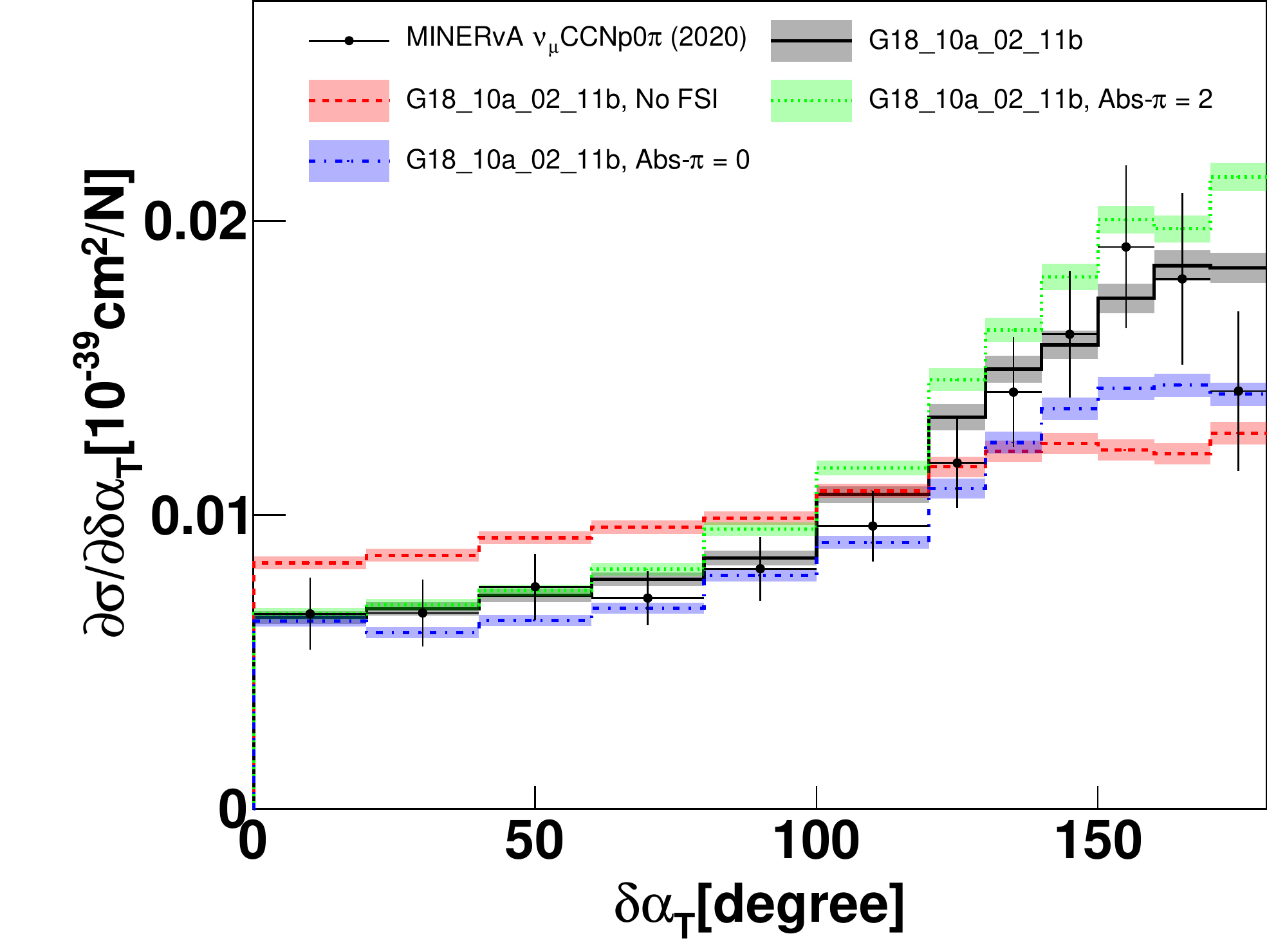}
    \caption{Impact of $S_{\text{Abs}}^{\pi^\pm}$ on MINER$\nu$A CCNp0$\pi$ flux-integrated differential cross section predictions as a function of $\delta\alpha_T$.
    \label{fig:FSIdatMinerva}}
\end{figure}

Variations of the \ac{FSI} parameters considered in this work result in the migration of CC1$\pi$ events into the CC0$\pi$ sample.
The effect on the prediction depends on the topology definition.
For CC0$\pi$ samples, it mostly affects the overall normalization of the cross-section.
The measurement most sensitive to this variation is the $\nu_\mu$ CCNp0$\pi$ MINER$\nu$A differential cross section as a function of $\delta\alpha_T$, see Fig.~\ref{fig:FSIdatMinerva}.
A decrease in $S_{\text{Abs}}^{\pi^\pm}$ reduces the cross section at $\delta\alpha_T\sim180^{\circ}$.
In addition, this model variation also affects the slope of the distribution.

In this tune, only parameters related to charged pion absorption are included: $S_{\text{Abs}}^{\pi^\pm}$ and $S_{\text{MFP}}^{\pi^\pm}$.
Pion inelastic fractions are not relevant at the energies of interest for this work. 
Nucleon \ac{FSI} parameters are relevant for the study of exclusive cross-section measurements with protons in the final state. 
Ideally, to perform a global tune with CC0$\pi$ and CCNp0$\pi$ data, nucleon \ac{FSI} parameters must be considered in the analysis.
Including these parameters in the analysis substantially increases the computing time.
In addition, it is desirable to first understand the tensions between CC0$\pi$ and CCNp0$\pi$ measurements.
Therefore, it is therefore convenient to reduce the complexity of the analysis and focus on CC0$\pi$ datasets only.
Nucleon \ac{FSI} parameters will be included in future iterations of this work.

\subsection{Final choice of parameters for the CC0\ensuremath{\pi} tune}
\label{sec:FinalChoice}

A series of preliminary tunes were performed using different priors or parameter sets.
The goal of this study is to determine which parameters to include in the final tune.

Nuclear effects in the \ac{QEL} cross section are tweaked with the \ac{RPA} parametrization.
Free-nucleon cross-section data suggests that the \ac{QEL} cross section should not be scaled.
This condition can be incorporated in our analysis by imposing a more restrictive prior on $S_{\text{RPA}}$ of $\sigma_S=0.01$.
Tunes performed using this prior result in worse goodness of fit, suggesting that a less restrictive prior on the sum is desired to improve the agreement with the data.
This motivated our choice for a prior on the sum of $\sigma_S=0.2$, as described in Sec.~\ref{subsec:CCQELParam}.

\ac{FSI} interactions are important to describe CC0$\pi$ measurements.
Additional parameters must be consistent with previous data~\cite{dytman2021comparison}, making this tricky.
The results of test cases with FSI suggest that variations of these parameter that respect pion-nucleus scattering data do not have a big impact on the tune results.
Consequently, these parameters are not included in the final analysis.

Various choices were made to get a more representative result.
Although full coverage in parameters can be sought, that is not always possible or desirable.
For this study, Professor allows a large parameter set which don't have to be ReWeight variables.
A study of the $f^{\text{QEL}}$ and $f^{\text{2p2h}}$ parameters in the tune was made.
Strong correlations are observed between the $f^{\text{QEL}}$, $f^{\text{2p2h}}$ and $\omega_{\text{RPA}}$ and $S^{\text{2p2h}}$ parameters.
In some cases, these correlations lead to unphysical values for $f^{\text{QEL}}$ and $f^{\text{2p2h}}$.
For this reason, these parameters are excluded from the analysis.

The final-parameter set used in this work is summarized in Tab.~\ref{tab:ParameterVariation}.

\begin{acronym}[nolist]
    \acro  {MC}    [MC]    {Monte Carlo}
    \acro  {SM}    [SM]    {Standard Model}                                 
    \acro  {CKM}   [CKM]   {Cabibbo-Kobayashi-Maskawa}                      
    \acro  {CC}    [CC]    {Charged-Current}
    \acro  {NC}    [NC]    {Neutral-Current}                                
    \acro  {QEL}   [QEL]   {Quasi-ELastic}
    \acro  {RES}   [RES]   {REsonance Scattering}                           
    \acro  {DoF}   [DoF]   {Degree of Freedom}                              
    \acro  {NRB}   [NRB]   {Non-Resonant Background}
    \acro  {2p2h}  [2p2h]  {Two-particles--two-holes}
    \acro  {SIS}   [SIS]   {Shallow-Inelastic Scattering}
    \acro  {DIS}   [DIS]   {Deep-Inelastic Scattering}
    \acro  {FSI}   [FSI]   {Final-State Interactions}
    \acro  {QCD}   [QCD]   {Quantum-ChromoDynamics}
    \acro  {RFG}   [RFG]   {Relativistic Fermi Gas}
    \acro  {LFG}   [LFG]   {Local Fermi Gas}
    \acro  {CFG}   [CFG]   {Correlated Fermi Gas}
    \acro  {RPA}   [RPA]   {Random-Phase Approximation}
    \acro  {AGKY}  [AGKY]  {Andreopoulos-Gallagher-Kehayias-Yang}
    \acro  {CMC}   [CMC]   {Comprehensive Model Configuration}
    \acro  {CMCs}   [CMCs] {Comprehensive Model Configurations}
    \acro  {STKI}  [STKI]  {Single-Transverse Kinematic Imbalance}
    \acro  {BNB}   [BNB]   {Booster Neutrino Beam}
    \acro  {NuMI}  [NuMI]  {Neutrino at the Main Injector}
    \acro  {FHC}   [FHC]   {Forward Horn Current}
    \acro  {RHC}   [RHC]   {Reverse Horn Current}
    \acro  {LArTPC}[LArTPC]{Liquid Argon Time Projection Chamber}
    \acro  {PPP}   [PPP]   {Peele’s Pertinent Puzzle}
    \acro  {NS}    [NS]    {Norm-Shape}
    \acro  {PID}   [PID]   {Particle IDentification}
\end{acronym}

\clearpage

\bibliography{bibliography.bib}
\end{document}